\documentclass[12pt,a4paper,twoside]{book}
\usepackage{siunitx}
\usepackage[T1]{fontenc}
\usepackage[table]{xcolor}
\usepackage[utf8]{inputenc}
\usepackage{mathptmx} 
\usepackage{fancyhdr} 
\usepackage{longtable}
\usepackage{multicol}
\usepackage{sectsty}
\usepackage[ngerman,english]{babel}
\usepackage{makeidx} 
\usepackage[nottoc,notindex]{tocbibind} 
\usepackage{tocloft}
\usepackage{ifthen} 
\usepackage{substr} 
\newcommand{\rpm}{\raisebox{.2ex}{$\scriptstyle\pm$}}
\usepackage[final]{pdfpages}
\usepackage{graphicx}
\usepackage[printonlyused]{acronym}
\usepackage{changepage}  
\usepackage{tabularx}  
\usepackage[caption=false]{subfig}
\usepackage{array}
\usepackage{booktabs}
\usepackage{colortbl}

\usepackage[hang,splitrule]{footmisc} 
\usepackage{titlesec}
\usepackage{geometry}
\usepackage{makecell}
\usepackage{amssymb}
\usepackage{pifont}
\usepackage{tabularx}
\newcolumntype{L}[1]{>{\raggedright\arraybackslash}p{#1}} 
\newcolumntype{C}[1]{>{\centering\arraybackslash}p{#1}} 
\newcolumntype{R}[1]{>{\raggedleft\arraybackslash}p{#1}} 

\usepackage{subscript}
\usepackage{rotating}
\usepackage{multirow}

\usepackage{marvosym}
\usepackage{amsmath}
\usepackage{amsfonts}   
\usepackage{wrapfig}
\usepackage{afterpage}
\usepackage{csquotes}
\usepackage{tabularx}

\definecolor{Gray}{gray}{0.9}

\usepackage{blindtext}

\usepackage{pgfplots}
\pgfplotsset{compat=1.12}
\usepackage{tikz}
\usetikzlibrary{arrows, decorations.pathmorphing, decorations.markings, backgrounds, positioning, fit, petri, shapes, calc}

\usepackage[gen]{eurosym}

\usepackage{wrapfig}

\label{Color Definition}
\usepackage{xcolor}
\definecolor{platinum}{HTML}{E5E4E2}
\colorlet{diagrambg}{platinum}

\usepackage{mdframed}

\usepackage{csvsimple}

\usepackage{pifont}
\makeindex

\usepackage[square,numbers]{natbib}

\newcommand{\mytitle}{Workload-Aware Systems and Interfaces for Cognitive Augmentation}
\newcommand{\mysubtitle}{}
\newcommand{\myname}{Thomas-Andreas Kosch}

\newcommand{\path}{../Figures}

\newcommand{\submissiondate}{München, den 28.01.2020}
\newcommand{\urldate}{2020-01-28}

\usepackage{natbib}
\usepackage{tikz}
\usepackage{lipsum}

\label{Quotes}

\label{Footnotes}


\makeatletter
\renewenvironment{theindex}
  {\if@twocolumn
      \@restonecolfalse
   \else
      \@restonecoltrue
   \fi
   \setlength{\columnseprule}{0pt}
   \setlength{\columnsep}{20pt}
   \begin{multicols}{2}[\chapter*{\indexname}]
   \markboth{\MakeUppercase\indexname}%
            {\MakeUppercase\indexname}%
   \thispagestyle{plain}
   \setlength{\parindent}{0pt}
   \setlength{\parskip}{0pt plus 0.3pt}
   \relax
   \let\item\@idxitem}%
  {\end{multicols}\if@restonecol\onecolumn\else\clearpage\fi}
\makeatother

\AtBeginDocument{%
	\definecolor{gray}{rgb}{0.8,0.8,0.8}
	\definecolor{darkgray}{RGB}{102,102,102}
	\definecolor{lightyellow}{RGB}{250,250,210}
}

\newcommand{\clearemptydoublepage}{%
  \ifthenelse{\boolean{@twoside}}{\newpage{\pagestyle{empty}\cleardoublepage}}%
  {\clearpage}}

\usepackage{ifpdf}

\ifpdf
  \usepackage{color}
  \definecolor{darkred}{rgb}{.25,0,0}
  \definecolor{darkgreen}{rgb}{0,.2,0}
  \definecolor{darkmagenta}{rgb}{.2,0,.2}
  \definecolor{darkcyan}{rgb}{0,.15,.15}
  \definecolor{headings}{rgb}{0,0,.3}
  \definecolor{primaryColor}{HTML}{A2C2DE}
  \definecolor{secondaryColor}{HTML}{E1EAF4}
  \definecolor{tertiaryColor}{HTML}{cbd6e5}
  \definecolor{MyGray}{rgb}{0.86,0.87,0.88}
  
  \definecolor{ReferenceGray}{rgb}{0.91,0.91,0.91}
  \definecolor{darkgray}{rgb}{.4,.4,.4}
  \definecolor{lightgray}{rgb}{.85,.85,.85}
  
  \usepackage[bookmarks=true,bookmarksopen=true,colorlinks=true,
    linkcolor=black,citecolor=black,filecolor=darkmagenta,
    menucolor=darkred,
    urlcolor=darkcyan,plainpages=false]{hyperref}
  \usepackage{graphicx}
	
  \renewcommand{\\}{}
    \hypersetup {
    pdfpagemode = {UseOutlines}, 
    pdftitle = {\mytitle},
    pdfsubject = {Dissertation of \myname},
    pdfauthor = {\myname},
    pdfdisplaydoctitle = true
  }

\else
  \usepackage[draft]{hyperref}
  \usepackage{graphicx}
\fi

\graphicspath{{../../Figures/General}}

\usepackage[scaled=0.85]{berasans}
\makeatletter

\label{Chapter Head}
\def\@makechapterhead#1{%
\thispagestyle{empty}
  \vspace*{-70\p@}%
  {\parindent \z@ \raggedright \normalfont
    \ifnum \c@secnumdepth >\m@ne
      \if@mainmatter
      		\par\vspace*{\fill}
      		\IfSubStringInString{: }{#1}{%
			\begin{flushright}
			\fontsize{50}{40}\fontfamily{qhv}\textbf{\thechapter }\par\nobreak
			\vspace{10pt}
			\fontsize{20}{80}\fontfamily{qhv}\textbf{\BeforeSubString{: }{#1}}\par\nobreak
			\fontsize{20}{80}\fontfamily{qhv}\textbf{\BehindSubString{: }{#1}}\par\nobreak
			  \end{flushright}
		}{
				
				\begin{flushright}
				\fontsize{50}{40}\fontfamily{qhv}\textbf{\thechapter }\par\nobreak
				\vspace{10pt}
				\fontsize{20}{80}\fontfamily{qhv}\textbf{#1}\par\nobreak
			     \end{flushright}
				\par\nobreak
        }
      \fi
  
        \vskip 30\p@
    
  }}
  
\label{Chapter Synopsis}

\label{Subsections}
\renewcommand\section{\@startsection {section}{1}{\z@}%
                                   {-3.5ex \@plus -1ex \@minus -.2ex}%
                                   {2.3ex \@plus.2ex}%
                                   {\fontfamily{qhv}\Large\bfseries}}
\renewcommand\subsection{\@startsection{subsection}{2}{\z@}%
                                     {-3.25ex\@plus -1ex \@minus -.2ex}%
                                     {1.5ex \@plus .2ex}%
                                     {\fontfamily{qhv}\large\bfseries}}
\renewcommand\subsubsection{\@startsection{subsubsection}{3}{\z@}%
                                     {-1.75ex\@plus -1ex \@minus -.2ex}%
                                     {0.1ex \@plus .1ex}%
                                     {\fontfamily{qhv}\normalsize\bfseries}}
\renewcommand\paragraph{\@startsection{paragraph}{4}{\z@}%
                                    {1.75ex \@plus1ex \@minus.2ex}%
                                    {0.1ex \@plus .1ex}%
                                    {\normalfont\normalsize\itshape\rmfamily}}
\renewcommand\subparagraph{\@startsection{subparagraph}{5}{\parindent}%
                                       {3.25ex \@plus1ex \@minus .2ex}%
                                       {-1em}%
                                      {\normalfont\normalsize\sffamily}}

\let\oldtabular\tabular
\renewcommand{\tabular}{\small\oldtabular}

\def\@part[#1]#2{%
    \ifnum \c@secnumdepth >-2\relax
      \refstepcounter{part}%
      \addcontentsline{toc}{part}{\thepart\hspace{1em}#1}%
    \else
      \addcontentsline{toc}{part}{#1}%
    \fi
    \markboth{}{}%
    {\centering
     \interlinepenalty \@M
     \normalfont
     \ifnum \c@secnumdepth >-2\relax
       \textcolor{gray}{\fontsize{240}{288}\mdseries\textrm{\thepart}}%
       \par
     \fi
     \centering \normalfont
     \fontsize{30}{36}\mdseries\scshape\textrm{#2}\par}%
    \@endpart}
\def\@spart#1{%
    {\centering
     \interlinepenalty \@M
     \normalfont \Huge \bfseries \SS@parttitlefont {#1}\par}%
    \@endpart}

\label{Header}
\makeatletter
\def\@fancyhead#1#2#3#4#5{#1\hbox to\headwidth{\fancy@reset
  \@fancyvbox\headheight{\hbox
   {\rlap{\parbox[b]{\headwidth}{\fontsize{10}{14}\rmfamily\mdseries\raggedright#2}}\hfill
      \parbox[b]{\headwidth}{\fontsize{12}{14}\rmfamily\mdseries\centering#3}\hfill
      \llap{\parbox[b]{\headwidth}{\fontsize{10}{14}\rmfamily\mdseries\raggedleft#4}}}\headrule}}#5}
\makeatother







\newlength{\RoundedBoxWidth}
\newsavebox{\GrayRoundedBox}
\newenvironment{GrayBox}[1][\dimexpr\textwidth-4.5ex]%
    {\setlength{\RoundedBoxWidth}{\dimexpr#1}
     \begin{lrbox}{\GrayRoundedBox}
        \begin{minipage}{\RoundedBoxWidth}}%
    {   \end{minipage}
     \end{lrbox}
     \begin{center}
     \begin{tikzpicture}%
        \draw node[draw=black,fill=black!10,rounded corners,%
              inner sep=2ex,text width=\RoundedBoxWidth]%
              {\usebox{\GrayRoundedBox}};
     \end{tikzpicture}
     \end{center}}



\newcommand{\publicationsbegin}{ 
	\begin{list}{\raise .5ex\hbox{\scriptsize$\bullet$}}
		{ \setlength{\itemsep}{5pt}      \setlength{\parsep}{3pt} 
			\setlength{\topsep}{6pt}       \setlength{\partopsep}{0pt}
			\setlength{\leftmargin}{2.5em} \setlength{\labelwidth}{1em}
			\setlength{\labelsep}{0.7em} } }
	
	\newcommand{\publicationsend}{
\end{list}  }

\makeatletter
\long\def\@makecaption#1#2{%
  \vskip\abovecaptionskip
  \vskip2mm
  \bfseries
  \centering
  \sbox\@tempboxa{#1: #2}%
	\ifdim \wd\@tempboxa >0.95\textwidth
	  \begin{minipage}{0.95\textwidth}
    	\bfseries\small{#1:} \normalfont\small #2\par
    \end{minipage}
  \else
    \global \@minipagefalse
    \bfseries\small{#1:} \normalfont\small #2\par
  \fi
  \vskip\belowcaptionskip}
\makeatother

\usepackage{bibentry}
\usepackage{paralist} 
\usepackage{epigraph}
\usepackage{multirow, tabularx}
\usepackage{rotating}
\usepackage{txfonts}
\usepackage{mathptmx}
\usepackage{booktabs}
\usepackage{ccicons}
\usepackage{wrapfig} 
\usepackage{array}
\usepackage{eqnarray}
\usepackage{booktabs}
\usepackage{color}
\usepackage{colortbl}
\usepackage{verbatim} 
\usepackage[hang,splitrule]{footmisc} 
\usepackage{titletoc} 
\usepackage{graphicx}
\usepackage{enumitem} 
\usepackage{overpic}
\usepackage{units}
\usepackage{longtable,array,ragged2e}

\makeatletter
\newcommand*{\rom}[1]{\expandafter\@slowromancap\romannumeral #1@}

\def\@makechapterhead#1{%
	\vspace*{-35\p@}%
	{\parindent \z@ \raggedright \normalfont
		\ifnum \c@secnumdepth >\m@ne
		\if@mainmatter
		\fontsize{20}{24}\mdseries\textsf{\@chapapp}\space\fontsize{80}{96}\mdseries\textsf{\thechapter}
		\hrulefill
		\par\nobreak
		\vskip 30\p@
		\fi
		\else
		
		\fi
		\interlinepenalty\@M
		
		\IfSubStringInString{? }{#1}{%
			\fontsize{36}{50}\mdseries\textsf{\BeforeSubString{? }{#1}}?\par\nobreak
			\fontsize{22}{26}\mdseries\textsf{\BehindSubString{? }{#1}}\par\nobreak
		}{%
			\fontsize{26}{40}\bfseries\textsf{#1}\par\nobreak
		}
		
		\vskip 40\p@
}}

\label{Page Heading}
\renewcommand{\chaptermark}[1]{\markboth{ #1}{}}
\renewcommand{\sectionmark}[1]{\markright{ #1}}
\newcommand{\phstyle}{\fontfamily{qhv}\fontsize{10}{14}\bfseries}
\author{\myname\\\small mariam.hassib@ifi.lmu.de}
\title{\mytitle\\
\mysubtitle\\}
\date{\svndate\ \small(rev. \svnrevision)}

\newmdenv[
  topline=false,
  rightline=false,
  bottomline=false,
  skipabove=\topsep,
  skipbelow=\topsep,
  linewidth=2pt,
  linecolor=darkgray
]{definitionbar}

\nobibliography*
\makeatletter

	{\end{multicols}\if@restonecol\onecolumn\else\clearpage\fi}
\makeatother

\newcommand*{\compress}{\@minipagetrue}
\makeatother

\label{Anecdote}

\label{Shortcuts}
\usepackage{xspace}

\begin{document}
\frontmatter

{%
\pagestyle{empty}%
\addtolength{\oddsidemargin}{11mm}
\addtolength{\evensidemargin}{11mm}
\addtolength{\topmargin}{11mm}


\newgeometry{inner=30pt,outer=30pt,top=30pt}
\noindent\begin{minipage}{188mm} 

\begin{picture}(0,0)%
  \put(258.7,-858.8){\includegraphics[width=140mm]{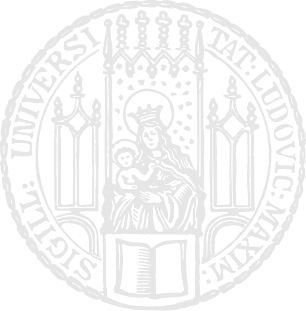}} 
\end{picture}

\setlength{\fboxsep}{0mm}%
\pagestyle{empty}%
\noindent

\vspace*{23mm}

\begin{center}
\vspace*{20mm}
\rule{\textwidth}{0.5pt}
\vskip 5mm
\fontsize{28}{34}\rmfamily\scshape\mytitle
\par
\vskip 15pt
\rule{\textwidth}{0.5pt}
\vskip 20mm
\huge \textbf{Dissertation}\\ 
\vspace*{5mm}

\normalfont\Large an der Fakultät für Mathematik, Informatik und Statistik\\
der Ludwig-Maximilians-Universität München\\
\vspace*{10mm}
vorgelegt von\\
\huge{\scshape{\textbf{\myname}}}\\
\normalfont\Large M.Sc.~Software Engineering\\
\vspace*{20mm}
\normalfont\Large München, den 28. Januar 2020 
\end{center}
\end{minipage}

\cleardoublepage


\clearpage

}
\pagestyle{fancy}%

\renewcommand{\chaptermark}[1]{%
	\markboth{\thechapter~~#1}{}}
	

\begin{picture}(0,0)%
  \put(181.7,-773.8){\includegraphics[width=140mm]{./img/general/lmu-siegel-grey.pdf}} 
\end{picture}

\vspace*{80mm}
{\large\noindent
\centering{
\begin{tabular}{ll}
\large{Erstgutachter:}	& \large{Prof.\ Dr.\ Albrecht Schmidt}\\[1ex]
\large{Zweitgutachter:} & \large{Prof.\ Dr.\ Fabio Paternò}\\[1ex]
\large{Drittgutachter:}	& \large{Prof.\ Dr.\ Jonathan Lazar}\\
\end{tabular} 
}\\
}
\vspace*{50mm}
\begin{center}
{\large\noindent
Tag der mündlichen Prüfung: 11.05.2020}
\end{center}

\cleardoublepage

\setlength{\parindent}{0pt} 
\setlength{\parskip}{7pt plus 2pt minus 1pt}

\selectlanguage{english}
\markboth{Abstract}{Abstract}
\section*{\LARGE\rmfamily\bfseries\scshape{Abstract}}

In today's society, our cognition is constantly influenced by information intake, attention switching, and task interruptions. This increases the difficulty of a given task, adding to the existing workload and leading to compromised cognitive performances. The human body expresses the use of cognitive resources through physiological responses when confronted with a plethora of cognitive workload. This temporarily mobilizes additional resources to deal with the workload at the cost of accelerated mental exhaustion. 

We predict that recent developments in physiological sensing will increasingly create user interfaces that are aware of the user’s cognitive capacities, hence able to intervene when high or low states of cognitive workload are detected. In this thesis, we initially focus on determining opportune moments for cognitive assistance. Subsequently, we investigate suitable feedback modalities in a user-centric design process which are desirable for cognitive assistance. We present design requirements for \textit{how} cognitive augmentation can be achieved using interfaces that sense cognitive workload.

We then investigate different physiological sensing modalities to enable suitable real-time assessments of cognitive workload. We provide empirical evidence that the human brain is sensitive to fluctuations in cognitive resting states, hence making cognitive effort measurable. Firstly, we show that electroencephalography is a reliable modality to assess the mental workload generated during the user interface operation. Secondly, we use eye tracking to evaluate changes in eye movements and pupil dilation to quantify different workload states. The combination of machine learning and physiological sensing resulted in suitable real-time assessments of cognitive workload. The use of physiological sensing enables us to derive \textit{when} cognitive augmentation is suitable.

Based on our inquiries, we present applications that regulate cognitive workload in home and work settings. We deployed an assistive system in a field study to investigate the validity of our derived design requirements. Finding that workload is mitigated, we investigated how cognitive workload can be visualized to the user. We present an implementation of a biofeedback visualization that helps to improve the understanding of brain activity. A final study shows how cognitive workload measurements can be used to predict the efficiency of information intake through reading interfaces. Here, we conclude with use cases and applications \textit{which} benefit from cognitive augmentation.

This thesis investigates how assistive systems can be designed to implicitly sense and utilize cognitive workload for input and output. To do so, we measure cognitive workload in real-time by collecting behavioral and physiological data from users and analyze this data to support users through assistive systems that adapt their interface according to the currently measured workload. Our overall goal is to extend new and existing context-aware applications by the factor \textit{cognitive workload}. We envision \textit{Workload-Aware Systems} and \textit{Workload-Aware Interfaces} as an extension in the context-aware paradigm. To this end, we conducted eight research inquiries during this thesis to investigate how to design and create workload-aware systems. 

Finally, we present our vision of future workload-aware systems and workload-aware interfaces. Due to the scarce availability of open physiological data sets, reference implementations, and methods, previous context-aware systems were limited in their ability to utilize cognitive workload for user interaction. Together with the collected data sets, we expect this thesis to pave the way for methodical and technical tools that integrate workload-awareness as a factor for context-aware systems.

\clearemptydoublepage

\markboth{Zusammenfassung}{Zusammenfassung}
\section*{\LARGE\rmfamily\bfseries\scshape{Zusammenfassung}}
\selectlanguage{ngerman}

Tagtäglich werden unsere kognitiven Fähigkeiten durch die Verarbeitung von unzähligen Informationen in Anspruch genommen. Dies kann die Schwierigkeit einer Aufgabe durch mehr oder weniger Arbeitslast beeinflussen. Der menschliche Körper drückt die Nutzung kognitiver Ressourcen durch physiologische Reaktionen aus, wenn dieser mit kognitiver Arbeitsbelastung konfrontiert oder überfordert wird. Dadurch werden weitere Ressourcen mobilisiert, um die Arbeitsbelastung vorübergehend zu bewältigen.

Wir prognostizieren, dass die derzeitige Entwicklung physiologischer Messverfahren kognitive Leistungsmessungen stets möglich machen wird, um die kognitive Arbeitslast des Nutzers jederzeit zu messen. Diese sind in der Lage, einzugreifen wenn eine zu hohe oder zu niedrige kognitive Belastung erkannt wird. Wir konzentrieren uns zunächst auf die Erkennung passender Momente für kognitive Unterstützung welche sich der gegenwärtigen kognitiven Arbeitslast bewusst sind. Anschließend untersuchen wir in einem nutzerzentrierten Designprozess geeignete Feedbackmechanismen, die zur kognitiven Assistenz beitragen. Wir präsentieren Designanforderungen, welche zeigen \textit{wie} Schnittstellen eine kognitive Augmentierung durch die Messung kognitiver Arbeitslast erreichen können.

Anschließend untersuchen wir verschiedene physiologische Messmodalitäten, welche Bewertungen der kognitiven Arbeitsbelastung in Realzeit ermöglichen. Zunächst validieren wir empirisch, dass das menschliche Gehirn auf kognitive Arbeitslast reagiert. Es zeigt sich, dass die Ableitung der kognitiven Arbeitsbelastung über Elektroenzephalographie eine geeignete Methode ist, um den kognitiven Anspruch neuartiger Assistenzsysteme zu evaluieren. Anschließend verwenden wir Eye-Tracking, um Veränderungen in den Augenbewegungen und dem Durchmesser der Pupille unter verschiedenen Intensitäten kognitiver Arbeitslast zu bewerten. Das Anwenden von maschinellem Lernen führt zu zuverlässigen Echtzeit-Bewertungen kognitiver Arbeitsbelastung. Auf der Grundlage der bisherigen Forschungsarbeiten stellen wir Anwendungen vor, welche die Kognition im häuslichen und beruflichen Umfeld unterstützen. Die physiologischen Messungen stellen fest, \textit{wann} eine kognitive Augmentierung sich als günstig erweist.

In einer Feldstudie setzen wir ein Assistenzsystem ein, um die erhobenen Designanforderungen zur Reduktion kognitiver Arbeitslast zu validieren. Unsere Ergebnisse zeigen, dass die Arbeitsbelastung durch den Einsatz von Assistenzsystemen reduziert wird. Im Anschluss untersuchen wir, wie kognitive Arbeitsbelastung visualisiert werden kann. Wir stellen eine Implementierung einer Biofeedback-Visualisierung vor, die das Nutzerverständnis zum Verlauf und zur Entstehung von kognitiver Arbeitslast unterstützt. Eine abschließende Studie zeigt, wie Messungen kognitiver Arbeitslast zur Vorhersage der aktuellen Leseeffizienz benutzt werden können. Wir schließen hierbei mit einer Reihe von Applikationen ab, \textit{welche} sich kognitive Arbeitslast als Eingabe zunutze machen.

Die vorliegende wissenschaftliche Arbeit befasst sich mit dem Design von Assistenzsystemen, welche die kognitive Arbeitslast der Nutzer implizit erfasst und diese bei der Durchführung alltäglicher Aufgaben unterstützt. Dabei werden physiologische Daten erfasst, um Rückschlüsse in Realzeit auf die derzeitige kognitive Arbeitsbelastung zu erlauben. Anschließend werden diese Daten analysiert, um dem Nutzer strategisch zu assistieren. Das Ziel dieser Arbeit ist die Erweiterung neuartiger und bestehender kontextbewusster Benutzerschnittstellen um den Faktor \textit{kognitive Arbeitslast}. Daher werden in dieser Arbeit \textit{arbeitslastbewusste Systeme} und \textit{arbeitslastbewusste Benutzerschnittstellen} als eine zusätzliche Dimension innerhalb des Paradigmas kontextbewusster Systeme präsentiert. Wir stellen acht Forschungsstudien vor, um die Designanforderungen und die Implementierung von kognitiv arbeitslastbewussten Systemen zu untersuchen. 

Schließlich stellen wir unsere Vision von zukünftigen kognitiven arbeitslastbewussten Systemen und Benutzerschnittstellen vor. Durch die knappe Verfügbarkeit öffentlich zugänglicher Datensätze, Referenzimplementierungen, und Methoden, waren Kontextbewusste Systeme in der Auswertung kognitiver Arbeitslast bezüglich der Nutzerinteraktion limitiert. Ergänzt durch die in dieser Arbeit gesammelten Datensätze erwarten wir, dass diese Arbeit den Weg für methodische und technische Werkzeuge ebnet, welche kognitive Arbeitslast als Faktor in das Kontextbewusstsein von Computersystemen integriert.

\clearemptydoublepage
\selectlanguage{english}  
\cleardoublepage

\chapter*{ACKNOWLEDGMENTS}
The last years were a challenging journey which I was happy to take. I had the opportunity to meet many interesting and joyful people throughout my studies that shaped me as a person I am today. First and foremost, I thank \textbf{Albrecht Schmidt} for warmly welcoming me to his group and showing me the privileged world of academia. I thank you for the freedom, your constant support that went beyond work, and the inspiration I received from you throughout my research projects. Meeting you and your group was a fortunate event I will never forget. I also thank \textbf{Fabio Paternò} and \textbf{Jonathan Lazar} who took their time to serve as external supervisors for this thesis. Further thanks go to \textbf{Anja Mebus}, our executive Christmas card overseer, for always having open ears whenever I stumbled through her door. Not to forget, our ``King'' \textbf{Rainer Fink} for his inexhaustible technical engagement and his efforts for providing us with desk and an internet connection when we arrived from planet Stuttgart. \textbf{Franziska Schwamb}, \textbf{Christa Feulner}, and \textbf{Doris Steiner} who were always there for me whenever bureaucracy was around the corner. \textbf{Murielle Naud-Barthelmeß} for hunting my students for timesheets. Another thanks go to \textbf{Benjamin Jillich}, \textbf{Manuel Jerger}, \textbf{Patrick Hilsbos}, \textbf{Michael Hilsbos}, and \textbf{Deborah Mash} for providing me a research internship at Neuromore. I learned a lot from you while I was working in Miami. \textbf{Mariam Hassib} for supervising my master thesis while doing ``rolling'' eye tracking and brain research with me.

I also owe my gratitude to \textbf{Tilman Dingler} for benignly stripping me away from the industry world and introducing me to the free-spirited world of academia. \textbf{Markus Funk} and \textbf{Stefan Schneegaß} for sharing their offices with me. Together with their drones and Nerf guns. \textbf{Bastian Pfleging} for sharing his train commuting experience. \textbf{Pascal Knierim}, BMBF friend and professional van driver, for showing our amazing projects throughout Germany while withstanding my taste in music. I will never forget your snowboard lessons and our mutual, but joyful, last-minute paper writing sessions. \textbf{Jakob Karolus}, non-baked milk protein refuser, and cocktail adorer, for being a friend who always had my back. \textbf{Matthias Hoppe}, Swabian carpet, banana lover, and ramen soulmate, for the drone projects we have done together. \textbf{Paweł Woźniak} for being a friend who took me by the hand whenever I was confronted with qualitative research. \textbf{Sebastian Feger}, our CERN acquaintance, for sheltering us whenever we shook France and Switzerland. \textbf{Florian Lang} for being a very German person, known for his legendary Schnitzel preparation and neighbor lifting skills. \textbf{Niels Henze} for being the devil's advocate. \textbf{Fiona Draxler} for teaching me web programming. \textbf{Tonja Machulla} for not being too tired to run three-hour interview sessions with our participants. \textbf{Lewis Chuang} for having a lot of patience with me and his encouragement to use my system 2. And his mediocre singing. \textbf{Ville Mäkelä} for sharing his gaming experience with me. \textbf{Tobias Benz}, \textbf{Adam Nowak}, and \textbf{Robin Welsch} for sharing their enthusiasm for metal while sharing an office with me. \textbf{Sven Mayer} for his barbecue preparations and introduction to the HCI community. \textbf{Lars Lischke} for various Xorg display debug sessions and his coffee list management. \textbf{Huy Le} and \textbf{Dominik Weber} for our adventurous, yet unforgotten, road trips. \textbf{Rufat Rzayev} for sharing his RSVP experience. \textbf{Alexandra Voit} for being a friend and an enjoyable (Dagstuhl) companion. \textbf{Passant El.Agroudy} for providing her social media experience and Windows PC when important events were around the corner. \textbf{Francisco Kiss} for showing me the taste of Mezcal. \textbf{Mauro Avila} for traveling with me to Greece. \textbf{Romina Poguntke} and \textbf{Patrick Bader} (the P. stands for president) for our haptic stress projects. \textbf{Yomna Abdelrahman} for traveling to Maui to present our work. \textbf{Florian Alt} who was never tired to work on projects with me and providing feedback during nighttime hours. \textbf{Daniel Buschek} for introducing me to machine learning. \textbf{Katrin Wolf} for being our official Berlin ambassador. \textbf{Valentin Schwind} for his uncanny cat research. \textbf{Miriam Greis} who resolved my uncertainty. \textbf{Gesa Wiegand} who was not afraid of organizing a group colloquium in Vienna with me. \textbf{Christina Schneegaß} for being patient with my last-minute analysis and writing. \textbf{Henrike Weingärtner} for supervising 24 Spaghetti Carbonara cooking sessions.

Further thanks go to the new generation of the Munich HCI troop: \textbf{Jesse Grootjen} and his Nerf arsenal, \textbf{Lauren Thevin} for being a metal concert companion and introducing a French flair, \textbf{Steeven Villa} for his ideas to hack high-performance treadmills, \textbf{Luke Haliburton} for destressing us, and \textbf{Francesco Chiossi} for resembling a stereotypical italian person and his effort to organize the inspirational NiCE meetings. And not to forget all the people who brightened my Munich days: \textbf{Saragon Bartsch}, \textbf{Florian Bemmann}, \textbf{Michael Braun}, \textbf{Heiko Drewes}, \textbf{Martin Eberl}, \textbf{Malin Eiband}, \textbf{Heinrich Hußmann}, \textbf{Yanhong Li}, \textbf{Sylvia Rothe}, \textbf{Matthias Schmidmaier}, \textbf{Nađa Terzimehić}, \textbf{Sarah Völkel}, \textbf{Thomas Weber}, \textbf{Andreas Butz}, \textbf{Michael Chromik}, \textbf{Dennis Dietz}, \textbf{David Englmeier}, \textbf{Linda Hirsch}, \textbf{Kai Holländer}, \textbf{Jingyi Li}, \textbf{Yong Ma}, \textbf{Carl Oechsner}, \textbf{Changkun Ou}, \textbf{Beat Rossmy}, \textbf{Daniel Ullrich}, \textbf{Alexander Wiethoff}, \textbf{Mohamed Khamis}, \textbf{Tobias Seitz}, \textbf{Christian Mai}, \textbf{Renate Häuslschmid}, \textbf{Hanna Schneider}, \textbf{Sarah Prange}, \textbf{Yasmeen Abdrabou}, and \textbf{Lukas Mecke}, the gentle giant. The presented research would not be possible without the fabuluous people who worked with me on the BMBF project \textit{KoBeLU}: \textbf{Thomas Grote}, \textbf{Oliver Korn}, \textbf{Benjamin Treptow}, \textbf{Oliver Flade}, \textbf{Klaus Klein}, \textbf{Jürgen Waser}, and \textbf{Martin Gmür}.

I would not be able to be here without my teachers and friends that thoroughly supported me during my journey: \textbf{Franz Leitermann} for being a great teacher and mentor who guided me to the righteous path. \textbf{Paul Schwezov}, \textbf{Lukas Jedrychowski}, \textbf{Jens Strobel}, \textbf{Martin Sauter}, \textbf{Evlyn Bayer}, \textbf{Klaus Mittrich}, \textbf{Martin Langlouis}, \textbf{Sascha Wizemann}, and \textbf{Kevin Borrmann} who stood firm by my side during good and bad times.

Finally, I want to thank \textbf{Jaqueline Walter} for her inexhaustible patience with me when deadlines where around the corner, countless business travels, night shifts, and project meetings during odd times. Thank you for your trust and unconditional love.

\cleardoublepage

\makeatletter
	\setlength{\cftbeforepartskip}{7ex}
	\setlength{\cftbeforechapskip}{5.5ex}
	\setlength{\cftbeforesecskip}{0.75ex}
	\setlength{\cftbeforesubsecskip}{0.1ex}
	\setlength{\cftbeforetoctitleskip}{-5mm}
	\setlength{\cftbeforeloftitleskip}{-5mm}
	\setlength{\cftbeforelottitleskip}{-5mm}
	\renewcommand{\cfttoctitlefont}{
	  \LARGE\usefont{OT1}{ptm}{b}{sc}\selectfont
	}
	\renewcommand{\cftloftitlefont}{
	  \LARGE\usefont{OT1}{ptm}{b}{sc}\selectfont
	}
	\renewcommand{\cftlottitlefont}{
	  \LARGE\usefont{OT1}{ptm}{b}{sc}\selectfont
	}
	\renewcommand{\cftpartfont}{
	  \fontsize{14}{18}\usefont{OT1}{ptm}{b}{sc}\selectfont
	}
	\renewcommand{\cftchapfont}{
	  \fontsize{13}{17}\usefont{OT1}{ptm}{b}{n}\selectfont
	}	

  \setcounter{tocdepth}{2}
  \setcounter{secnumdepth}{2}
  \pdfbookmark[1]{Table of Contents}{table}
  \markboth{Table of Contents}{Table of Contents}
  \settocname{Table of Contents}
  \tableofcontents
	\cleardoublepage	
	%
%

\chapter*{List of Acronyms}\markboth{LIST OF ACRONYMS}{}
\addcontentsline{toc}{chapter}{List of Acronyms}

\begin{acronym}[MyAbbMyAbb] 
	\setlength{\itemsep}{-0.06cm}
	\acro{AI}{Artificial Intelligence}
	\acro{ANOVA}{Analysis of Variance}
	\acro{AR}{Augmented Reality}
	\acro{BCI}{Brain-Computer Interface}
	\acro{BMBF}{German Federal Ministry of Education and Research}
	\acro{CNN}{Convolutional Neural Network}
	\acro{DAEM}{Designing Assistive Environments for Manufacturing}
	\acro{DALI}{Driver Activity Load Index}
	\acro{DL}{Deep Learning}
	\acro{ECoG}{Electrocorticography}
	\acro{EDA}{Electrodermal Activity}
	\acro{EEG}{Electroencephalography}
	\acro{ELSI}{Ethical, Legal and Social Implications}
	\acro{EMG}{Electromyography}
	\acro{EOG}{Electrooculography}
	\acro{ERP}{Event-Related Potential}
	\acro{FBN}{Functional Brain Network}
	\acro{FFT}{Fast Fourier Transform}
	\acro{fMRI}{Functional Magnetic Resonance Imaging}
	\acro{fNIRS}{Functional Near-Infrared Spectroscopy}
	\acro{GDPR}{General Data Protection Regulation}
	\acro{GWW}{Gemeinnützige Werkstätten und Wohnheime GmbH}
	\acro{HbO}{Oxygenated Hemoglobin}
	\acro{HbR}{Deoxygenated Hemoglobin}
	\acro{HCI}{Human-Computer Interaction}
	\acro{HMD}{Head-Mounted Display}
	\acro{HR}{Heart Rate}
	\acro{HRV}{Heart Rate Variability}
	\acro{Hz}{Hertz}
	\acro{IAF}{Individual Alpha Frequency}
	\acro{KoBeLU}{Context-Aware Learning Environment}
	\acro{ML}{Machine Learning}
	\acro{NASA-TLX}{NASA Task Load Index}
	\acro{PI}{Performance Index}
	\acro{RQ}{Research Question}
	\acro{RSVP}{Rapid Serial Visual Representation}
	\acro{RSME}{Rating Scale of Mental Effort}
	\acro{SVM}{Support Vector Machine}
	\acro{TCT}{Task Completion Time}
	\acro{UI}{User Interface}
	\acro{VR}{Virtual Reality}
	\acro{WoZ}{Wizard of Oz}
\end{acronym}

	\cleardoublepage
\makeatother

\mainmatter


\part{Introduction and Foundations\label{part:introduction}}

\chapter{Introduction}
\label{ch:introduction}

\epigraph{\textit{Cogito, ergo sum.}}{René Descartes}
The world has undergone a substantial change in using computing systems for information intake. Since the beginning of mankind, the speed at which information is delivered and consumed has increased dramatically and keeps accelerating with the increasing availability of past information platforms and the proliferation of modern media. Local transfer systems such as speaking and writing have evolved into a global information network with the advent of the World Wide Web. Thus, extensive knowledge has become available for the public through different modalities. Buckminster coined this as the \textit{knowledge doubling curve}~\cite{fuller1981critical}.

The distribution of information has started to form a society that obtains their knowledge from other sharing peers. The amount of available information is expected to rise further with ubiquitous tools that proliferate into our daily routine. The consistent availability of information requires constant attention and strains the cognitive demand of users. Various stimuli compete for the attention of people and it is up to the user which ones are eligible to be processed, requiring strategies to decide on how much of the available cognitive resources should be spent on specific information channels. Miksch and Schulz elaborate this aspect in their work \textit{Disconnect to Reconnect}, where they found that the reduced use of digital technologies is associated with improvements in \textit{performance}, \textit{self-control}, \textit{well-being}, and \textit{maintenance of real-life relationships}~\cite{8944615}.

Additionally, information has become more accessible with the proliferation of mobile ubiquitous computing devices. This includes smartphones, smartwatches, notebooks, and smart eye-wear technologies. These devices contain rich capabilities to provide output in mobile scenarios. This enforced a shift on how humans process information, since content can be consumed on-the-go or in any other stationary setting. For example, popular applications include reading (\textit{e.g.}, Kindle), auditory skimming (\textit{e.g.}, Blinkist), language learning (\textit{e.g.}, Duolingo), or activity on social media platforms (\textit{e.g.}, Instagram).

However, with the establishment and an increasing number of apps that compete for cognitive resources, full attention cannot be paid to multiple stimuli that demand the same perception channels~\cite{salvucci2010multitasking}. This results in mental fatigue, difficulties with concentration, or decline in attention. This ongoing trend is expected to continue with an increasing number of devices, which negatively impacts the operation of user interfaces in short and long-term usage. Cognitive overload due to complicated user interface operations or extended multitasking as well as cognitive underload because of boredom is the result. While user interfaces in personal home environments can be disposed of, safety and time-critical environments depend on reliable operators that require few cognitive resources for operation. The \textit{Hawaii False Missile Incident} in 2018~\cite{barbash2018hawaii}, in which a false missile alert was issued due to a series of operation errors, shows the need for interfaces that are aware of the user's cognitive states to avoid far-reaching interaction mistakes.

Since cognitive resources change throughout the day, users find themselves at times where they can accomplish challenging tasks as well as times where they are unable to concentrate. Systems that implicitly assess cognitive states already exist in driving contexts where they evaluate eye movements to infer the driver's fatigue level. However, these are restricted to specific application scenarios. Mobile and stationary applications do not take the current task difficulty and the user's cognitive demand into account, and therefore, rely on explicit user actions (\textit{e.g.}, user intervention) or trigger interruptions (\textit{e.g.}, notifications) at times where users are focused, thus interrupting the workflow.

This thesis investigates how ubiquitous technologies can be used to sense the current level of cognitive demand of users while they interact with user interfaces. Users seldom express their mental demands explicitly, hence implicit detection methods for cognitive workload provide the potential for continuous monitoring instead of selective sampling. Ubiquitous technologies become equipped with rich sensing techniques that can determine cognitive and physical states. Human bodies emit physiological signals that mediate mental states, such as stress level or current level of physical activity. Advances in physiological sensing enable the quantification of the concurrent cognitive workload levels. Therefore, this thesis focuses on how systems achieve awareness about the user's perceived workload and explore applications that provide cognitive augmentation. This includes the design of assistive technologies that mitigate or foster cognitive workload (\textit{i.e.}, investigating \textit{how} to assist), evaluation of physiological sensing modalities that provide real-time metrics for cognitive workload (\textit{i.e.}, creating methods \textit{when} to assist), and the presentation of applications that benefit from these insights (\textit{i.e.}, in \textit{which} application scenarios to assist). This \textbf{workload-awareness} is envisioned as an extension of the context-aware computing paradigm and enables the development of novel application scenarios.

Workload-awareness allows applications to match the user's current available cognitive resources with the complexity of the presentation of a user interface. By estimating their available cognitive resources, opportune moments for cognitive augmentation can be selected and matched with the preferred exhibited mental demand. By matching task requirements and cognition, peak performance levels are achieved while avoiding frustration and boredom. The scheduling of tasks towards a currently perceived or expected cognitive workload level can be arranged by modifying the schedule, adapting the presentation of the current task to the cognitive context, provide visualizations that inform the course of cognitive demand throughout the day, and enable designers to evaluate their user interfaces regarding the placed cognitive demand.

To build workload-aware systems that use physiological sensing as a marker for cognitive effort, we initially focus on the phenomena of cognitive capabilities and effective perception of information. At its core lies the theory of cognition, a crucial factor for perceiving, processing, memorizing, and responding to information. This thesis presents a series of studies that investigate the design requirements of systems that utilize cognitive workload and leverage physiological sensing to detect cognitive workload as a strategic metric for user interface adaptations and evaluation. We explore physiological characteristics in combination with machine learning techniques to reliably detect changes in cognitive demand across individuals. To complement, we provide use cases and implementations that quantify and use workload in real-time.

We develop tools that can be applied by researchers and engineers to create workload-aware interfaces that take the cognitive context into account. Therefore, we present algorithms and methods to detect the current level of mental demand, which includes workload-awareness as an additional variable within context-aware computing. We create and evaluate concrete workload-aware systems: the concepts developed in this thesis include the use of an in-situ \ac{AR} interface that supports persons with cognitive impairments during daily tasks, visualizations that include direct representation of how the brain is affected by cognitive processing, and the assessment of text reading efficiency. Taking cognitive capacities into account, interface designers receive novel tools to engage with users, implicitly evaluate interfaces, and control the information flow.

\section{Vision: Workload-Aware Interfaces}

Sensing cognitive workload has become relevant in different scenarios, such as workplaces or in educational settings, to provide self-reflection or content-based adaptation. This requires ubiquitous methods to record and analyze user actions. Since we are integrating technologies in our daily lives, clothes, environments, or even our bodies, seamless physiological data collection and analysis can be carried out. We believe that such technologies can support people relative to workload measurements. In the following, we outline our vision of workload-aware systems that facilitate user interaction throughout the day.

The goal of technology is to integrate seamlessly into the environment and support users~\cite{weiser1993ubiquitous}. Thus, the more devices know about physical and physiological constraints or states, the better they can provide contextual assistance. A fully integrated piece of technology knows \textit{when} to approach users and \textit{how} to provide adaptive support with a \textit{suitable} adaptivity level to lower or increase cognitive workload. This process includes multiple steps. First, user states have to be detected. Physical and physiological data is collected and analyzed in real-time to estimate the current level of cognitive workload. Second, the user needs to be approached at the right time to avoid a mismatch between user expectancy and system. The final step copes with the content modification that assesses environmental factors suitable for displaying the newly adapted content. Reinforcement-based assessment algorithms determine if an improvement in physical and physiological measures has taken place. If not, the agent can modify spatial and temporal factors to optimize the user experience. Repeating this process enables the agent to optimize workload-related parameters for the individual user. We define workload-aware systems in the following:

\begin{center}
\begin{GrayBox}
\textbf{Definition \textit{Workload-Aware System}}
\newline
\center{
\parbox{0.98\textwidth}{
\emph{Workload-Aware systems provide interfaces that enable implicit and explicit sensing of cognitive workload. Workload-aware interfaces integrate cognitive workload as a dimension into the paradigm of context-aware computing which can be taken into account for user interface evaluation, workload reflection, or the creation of adaptive user interfaces.}
}}
\end{GrayBox}
\end{center}

A workload-aware system needs information about the users' current actions as well as tasks that have to be accomplished to fulfill the definition. Cognitive default states (\textit{i.e.}, resting or engaged states) need to be monitored across time to detect deviations. Furthermore, the system must be aware of its environment to provide support at the right place and at the right time. The idea of context-aware computing is to use several extrinsic and intrinsic factors to provide support for users just-in-time. Workload-aware systems focus on cognitive workload to augment the user's mental processing. Those systems can sense highs and lows in cognitive workload throughout the day and can enhance cognition. Systems that are aware of the users' current workload throughout the day can support and enhance the cognitive capabilities in various ways:

\subsection*{User Interface Adaptation}
Being aware of the current cognitive capacities of the user can be used as a parameter to influence the representation of a task. An adaptive user interface can provide more complex task representations to engage the user and increase the overall task success rate if sufficient cognitive capacities are left or untouched. Providing adaptive in-situ assistance during daily tasks such as cooking and at manual assembly workplaces can provide cognitive alleviation. Switching assistance modalities in real-time according to the current cognitive workload enables users to effectively engage and solve tasks in a shorter amount of time. At the same time, a system can detect boredom, reduce the provided assistance, and increase the commitment of users' to the current task.

\subsection*{Evaluation Tool}
Subjective measures, such as questionnaires, interviews, or think-aloud protocols are heavily employed to gain insights into the usability of user interfaces. For example, questionnaires have established as a measure for workload where cognitive demand represents a subset for quantification~\cite{HART1988139, pauzie2008method}. However, these methods are prone to subjective perception, require participants to memorize their experience regarding workload, and do not allow real-time insights into cognitive processing during an experiment. For example, knowledge about the current cognitive state allows experimenters and user interface designers to find distracting or unclear elements. A workload-aware system that points this out can be used by the experimenter afterward to explore difficulties in the \ac{UI} design in post hoc interviews.

\subsection*{Reflection}
Reflecting states of cognitive workload is the initial step to self-improvement. By making cognitive states spatially and temporally accessible, users can reflect on their performance throughout the day to schedule demanding tasks at suitable time slots. Hence, productive phases can be identified and exploited. On the other hand, tasks that do not require cognitive attention can be scheduled to time slots where the user is aware of having low cognitive capacities. For example, a user or a system can decide when it is the best time to whether to read a complex scientific article or when to wash the dishes. Synchronizing task difficulty and cognitive workload have the potential to reduce frustration as well as boredom. Instead of communicating cognitive workload only with the user, it can be shared with other persons and agents. This enables collaborative scenarios, in which multi-agents schedule tasks, assign tasks automatically to other persons, match collaboration partners, or communicate the perceived cognitive workload with the environment. This provides other persons and the environment with the ability to reflect on the cognitive workload of other individuals.

\subsection*{Individual Assistance}
Workload-aware systems may become individual assistants who use recent advances in \ac{AI} to create models from cognitive and behavioral patterns. These models are beneficial when it comes to schedule tasks and appointments according to the required cognitive performance. Such devices weave altogether with the required performance and upcoming tasks. By sharing these models, others can synchronize their capacities in workload to optimize the outcomes. The fluctuation of workload depends on multiple parameters, such as nutrition, quality of sleep, workout routine, personal wellbeing, upcoming daily chores, or incoming information consumption. For example, a meeting can be scheduled when a set of parameters that influence workload are optimized among all participants. Teaching technology how cognitive workload affects the priority of tasks throughout the day goes beyond simple context-aware applications. Workload-aware interfaces ubiquitously provide cognitive alleviation by matching chores to cognitive workload and structure tasks in a way to optimize the overall performance. This has the potential to help users to be efficient at their tasks, reduce boredom as well as frustration, and increase well-being. Going one step further, workload-aware interfaces can autonomously assess the difficulty of a task while being aware of the user's available cognitive resources. These tasks can then be matched to the current cognitive fitness to ease its processing, hence providing a workload-aware task schedule.

\subsection*{Interruption Management}
Interruptions in physical as well as virtual environments drain attention and reduce the capability to remain focus on a task. Workload-aware systems become an interface between a user and other individuals to communicate the current state of workload. Users may not be interrupted by notifications or inconvenient office visits. For example, a workload-aware system can mediate the current cognitive state using lights or heat on the door handle. Furthermore, senders of messages may be notified that the current message is delayed and should be forwarded only under important circumstances. Flow states are fostered due to times in which high focus is required, but also for situations in which interruptions are inappropriate. For example, disruptions during highly focused conversations between individuals can be proactively prevented and delayed to a more suitable time slot. Workload-aware systems can, therefore, serve as a potential protection mechanism for periods that require high focus.

\section{Research Questions}
Three pivotal aspects have to be considered before integrating workload-awareness into ubiquitous computing technologies: the users and their context, available workload sensing modalities, and the application. Table~\ref{tab:rq} provides an overview of the \ac{RQ} that are addressed within the scope of this thesis.

\begin{table}[h!]
\resizebox{\textwidth}{!}{%
\begin{tabular}{lll}
\toprule
\textbf{RQ} & \textbf{Research Questions} & \textbf{Chapter}\\
\midrule
\rowcolor{Gray}
&\textbf{Human Cognition}&\\
RQ1 & What are the design requirements for systems that provide opportune cognitive augmentation?& \ref{ch:smart_kitchen_requirements}\\
RQ2 & What are suitable in-situ feedback modalities for cognitive support?& \ref{ch:support_modalities}\\
\midrule
\rowcolor{Gray}
&\textbf{Sensing Cognitive Workload}&\\
RQ3 & Does electroencephalography provide measures of cognitive workload for user interface evaluation? & \ref{ch:workloadbyeeg}\\
RQ4 & Do eye gaze metrics enable the classification of cognitive workload states? & \ref{ch:smooth_pursuit} \& \ref{ch:pupil_dilation}\\
\midrule
\rowcolor{Gray}
&\textbf{Applications}&\\
RQ5 & Does in-situ feedback provide cognitive support during a cooking task?& \ref{ch:smart_kitchen_deployment}\\
RQ6 & How can complex physiological signals be visualized to be utilized by non-expert users? & \ref{ch:visualization_reflection}\\
RQ7 & How can RSVP reading parameters be selected based on cognitive workload? & \ref{ch:workload_information_consumption}\\
\bottomrule
\end{tabular}}
\caption[Research Questions]{Overview of the investigated research questions. The first part investigated opportune moments and suitable feedback modalities for workload-aware environments. The second part looked at physiological modalities for workload quantification. Finally, the third part combines both concepts and presents workload-aware applications.}
\label{tab:rq}
\end{table}

Overflowing users with information and support-related stimuli may reduce the capability to process further information. The result can be boredom, frustration, or excessive workload. Therefore, it is important to understand \textbf{human cognition} and the design requirements of workload-aware interfaces before augmenting workload. To identify opportune moments regarding workload assistance, we employ qualitative inquiries to derive design implications for workload-aware systems (\textbf{RQ1}). Here, we observe and conduct semi-structured interviews with tenants as well as the labor of sheltered living facilities. This specific user group is used to report the need for cognitive assistance during all facets of their daily life. Hence, it allows us to sketch and weigh a sequence of daily life tasks according to their cognitive demand, leading to design guidelines for future workload-aware systems. Upon deriving these design guidelines, we investigate which output modalities are suitable to communicate assistance using tactile, auditory, and visual feedback (\textbf{RQ2}). This part describes \textit{how} cognitive assistance can be provided. 

Ubiquitous awareness of cognitive states requires applications that are \textbf{sensing cognitive workload} to provide relevant insights in real-time. First, we leverage an established method to estimate the level of cognitive processing using \ac{EEG} measures (\textbf{RQ3}). Second, we investigate how eye activity reflects cognitive workload as a contactless alternative to \ac{EEG}. We exploit smooth pursuit as specific eye movement which requires participants to lock on a target and closely follow moving objects. Furthermore, we investigate how daily math tasks influence the fluctuation of the pupil size. Both approaches provide a robust real-time classification (\textbf{RQ4}). This part investigates sensing modalities that reveal \textit{when} cognitive assistance should be delivered.

The \textbf{application} section represents the final part of this thesis. Addressing workload-awareness as a promising dimension for future context-aware applications, we focus on applications that aim to amplify or improve the overall user experience. In a user study, we show how cognitive assistance is provided using in-situ assistance that comprises auditory and visual feedback (\textbf{RQ5}) and investigates how cognitive processes can be visualized to the individual user. Individuals gain the possibility to review the level of workload during specific tasks or certain points of time. Furthermore, current brain areas that are responsible for cognitive processing can be reflected in real-time to raise self-awareness (\textbf{RQ6}). Finally, we assess \ac{EEG} as a tool for evaluating information processing parameters that influence the space and time of \ac{RSVP} presentations. We show how \ac{EEG} can be used as an implicit workload measure from which adaptive systems and user interface designers benefit when designing interfaces that represent textual information (\textbf{RQ7}). Finally, this part is concerned with use cases \textit{which} benefit from cognitive augmentation.

\section{Challenges and Contribution}
The extensive consumption of information is increasingly proliferating throughout society through several modalities, such as mobile or wearable devices. At the same time, these interfaces require our full attention while our time and resources are limited. With workload-aware systems that are conscious regarding the user's cognitive capabilities, content can be mapped suitable to the mental state. Thus, we investigate how technologies can help us to unload cognitive workload and enable more efficient interaction with the user. Thereby, we put the focus within this thesis on the following three challenges:

\begin{enumerate}
\item Users, although aware of their mental capacities, do not expect the user interface to react to the current level of workload. With individual cognitive capabilities changing throughout the day, such systems have to employ cognitive assistance \textit{just-in-time} when needed. When succeeding, such assistance can result in higher efficiency and focus due to mental offloading. However, this process has to be seamless. The cognitive augmentation and ``disaugmentation'' through computing systems without making the user aware of it is a pressing challenge. Furthermore, new upcoming tasks can be scheduled depending on the currently available cognitive capacities. Two decisive parameters are commonly manipulated within the domain of adaptive user interfaces: the presentation modality, such as auditory, visual, or tactile feedback, and the respective representation of the modality such as a tactile pattern or specific auditory alert. However, how the prior mentioned parameters are adjusted depends (a) on the personal preferences of the user and (b) on the task the user is currently working on. A system that chooses sub-optimal values for these parameters may frustrate the user due to placing additional cognitive workload on it or provoke boredom through excessive assistance. Finding the sweet spot that provides an engaging user experience is a key challenge for workload-aware systems.

\item Ubiquitous technologies -- which end users can acquire nowadays -- become more and more context-aware. Today computing devices and their accompanying sensors are gaining computation power while being reduced in their size. As an example, devices such as fitness trackers support health goals, yet operate without regarding the cognitive state of the user. Technologies that quantify cognitive workload as an additional dimension provide the possibility to adapt the current user interface, give user interface designer insights into their design, and enable task scheduling that is mapped between task difficulty and cognitive capacity. The challenge is to exploit technologies that extract diurnal measures in real-time that predict the currently perceived workload so that user interfaces can operate accordingly.

\item Finally, little research has been conducted on manipulating the self-awareness and reflection of cognitive demand throughout the day. Stationary, mobile, and wearable technologies can sense user activities and behavior constantly and may reflect cognitive states to the user. This, combined with a recording of conducted activities throughout the day, allows users to optimize their schedule to match their availability of cognitive resources to a suitable task. Moreover, recent advances in deep learning pave the way for automatic scheduling of tasks depending on the required cognitive demand. Smart assistants can assess that information to provide implicit assistance in task scheduling for cognitive assistance.
\end{enumerate}

In this thesis, we tackle the aforementioned challenges by applying theories of cognitive psychology and physiological sensing on technology to provide cognitive assistance. This includes working memory theories, the utilization of ubiquitous sensing, constant availability of physiological data, and applying recent advances in \ac{AI}. More specifically, we set the goal of

\begin{enumerate}
\item objective real-time quantification of mental workload to identify the need for cognitive assistance during the use of technology as an additional dimension within the paradigm of context-aware computing.
\item deciding and identifying opportune moments that require technologies to deliver content that is tailored to the users' remaining mental capacities.
\item visualizing cognitive demand with the intent to provide user reflection or deeper insights for user interface designers that cope with the creation of mentally gentle or demanding \acp{UI}.
\end{enumerate}

\begin{figure}
 \includegraphics[width=\columnwidth]{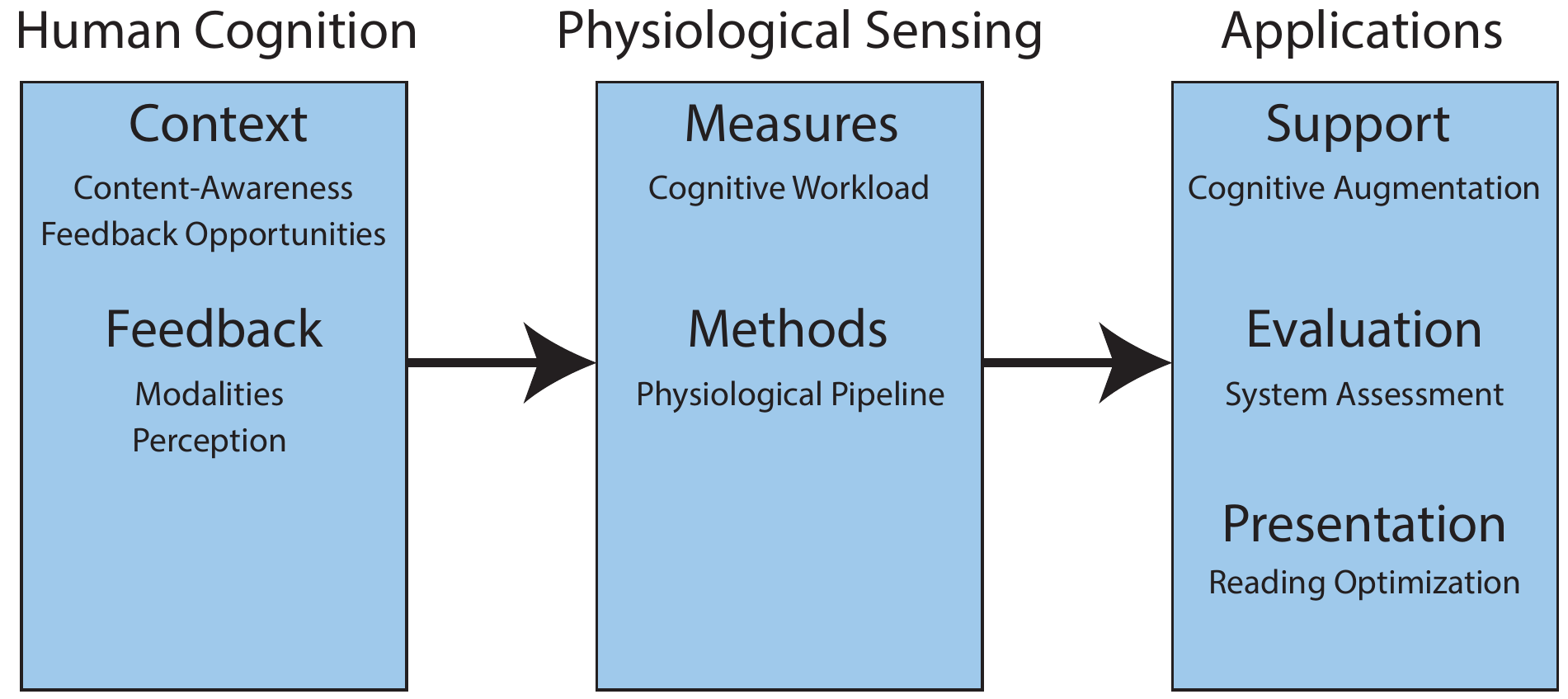}
 \caption[Structure of the thesis]{Overview of the contributions and challenges tackled during the course of this thesis. Design implications and the efficiency of feedback modalities are investigated in the \textit{Human Cognition} section. We then apply AI-based \textit{Physiological Sensing} to objectively measure cognitive workload. Finally, \textit{Applications} are sketched which benefit from workload-awareness.}
 \label{fig:contributions_and_challenges}
\end{figure}

This thesis focuses on end-user evaluations; in particular workload augmentation and the application layer within the domain of \ac{HCI}, cognitive psychology, and \ac{AI} including \ac{ML} as well as \ac{DL}. A human-centered approach is selected by conducting qualitative inquiries a well as lab and in-situ studies, and feedback for future workload-aware systems is collected and evaluated within technological and algorithmic probes. Table~\ref{tab:researchprobes} provides a summary of the conducted research inquiries while Figure~\ref{fig:contributions_and_challenges} illustrates the structure of the thesis. The following scientific contributions address the aforementioned goals:

\newcolumntype{Z}{>{\raggedright\let\newline\\\arraybackslash}X}
\begin{table*}
\centering
\resizebox{0.92\columnwidth}{!}{
\begin{tabularx}{\textwidth}{Z p{3cm} p{8.0cm} Z}
\toprule
 \multicolumn{3}{l}{\large \textbf{\textsc{Research Inquiries}}} \\
\midrule
& \textbf{Inquiry} & \textbf{Description} & \textbf{Chapter}
\\ \midrule

\parbox[t]{2mm}{
\vspace{1.5em} \multirow{2}{*}{\rotatebox[origin=c]{90}{\textbf{Human Cognition}}}
}
&

\raisebox{-\totalheight+.5em}{\includegraphics[width=\linewidth]{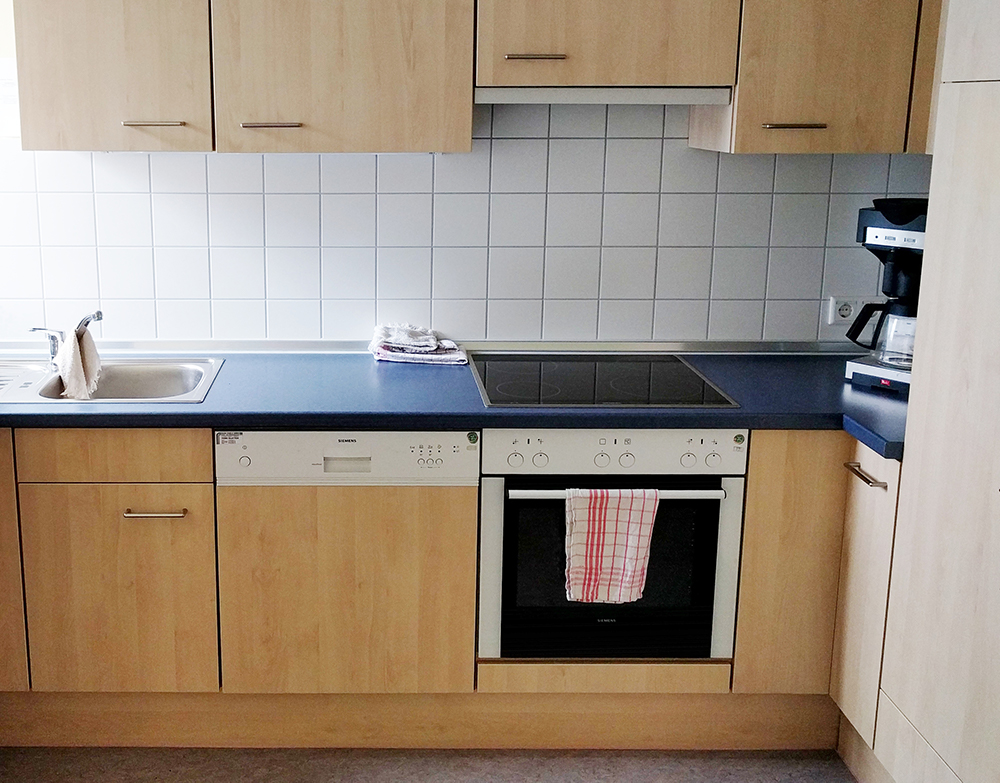}} &
We investigated the design implications of cognitive assistance through qualitative inquiries in a sheltered living facility and derive five design implications for the design of workload-aware systems.&
Chapter~\ref{ch:smart_kitchen_requirements} \\

& \raisebox{-\totalheight+.5em}{\includegraphics[width=\linewidth]{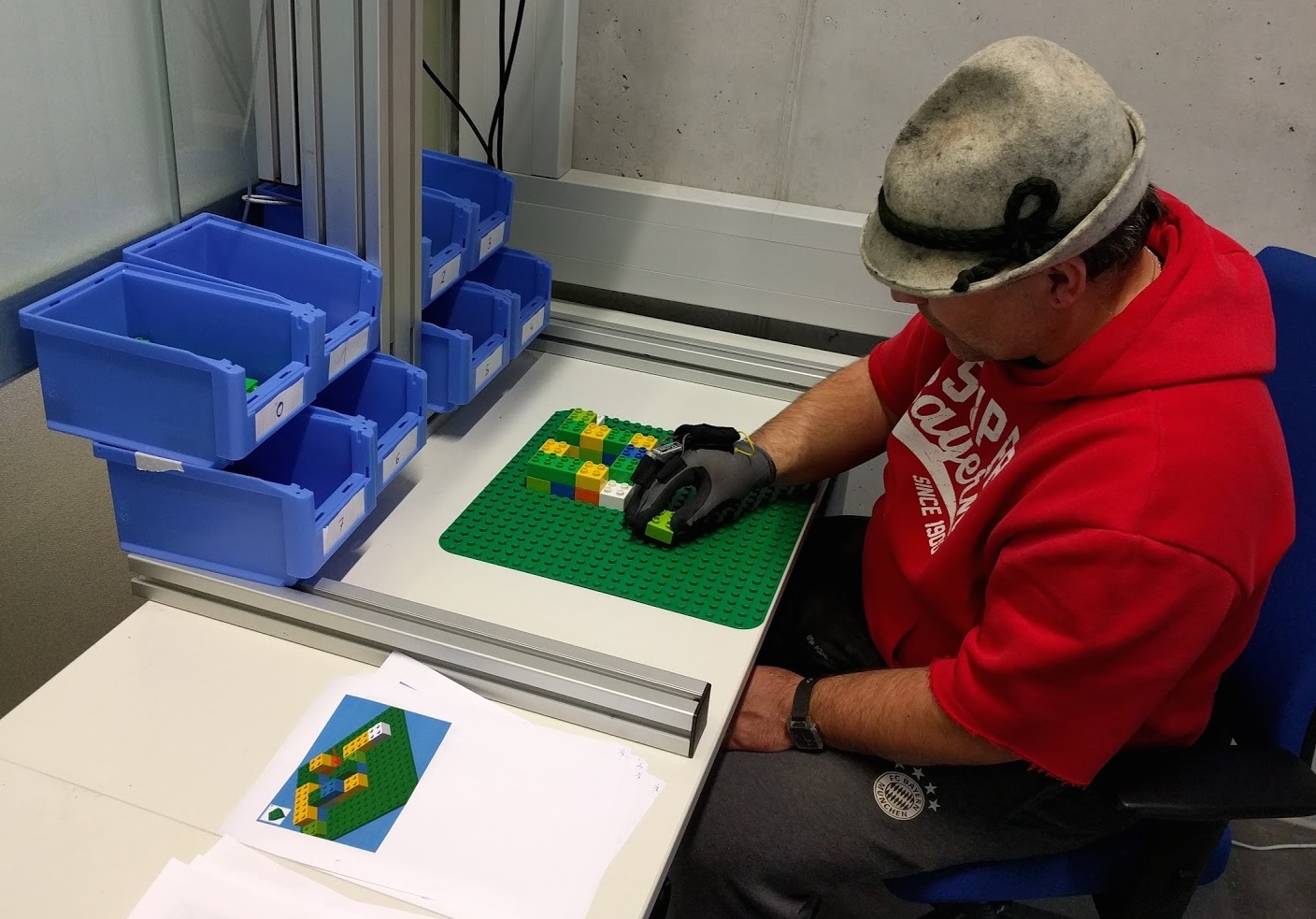}} &
To evaluate the impact of different stimuli on cognitive workload, we performed a study with three different output modalities to communicate errors in an assembly scenario.&
Chapter~\ref{ch:support_modalities} \\

\midrule

& \raisebox{-\totalheight+.5em}{\includegraphics[width=\linewidth]{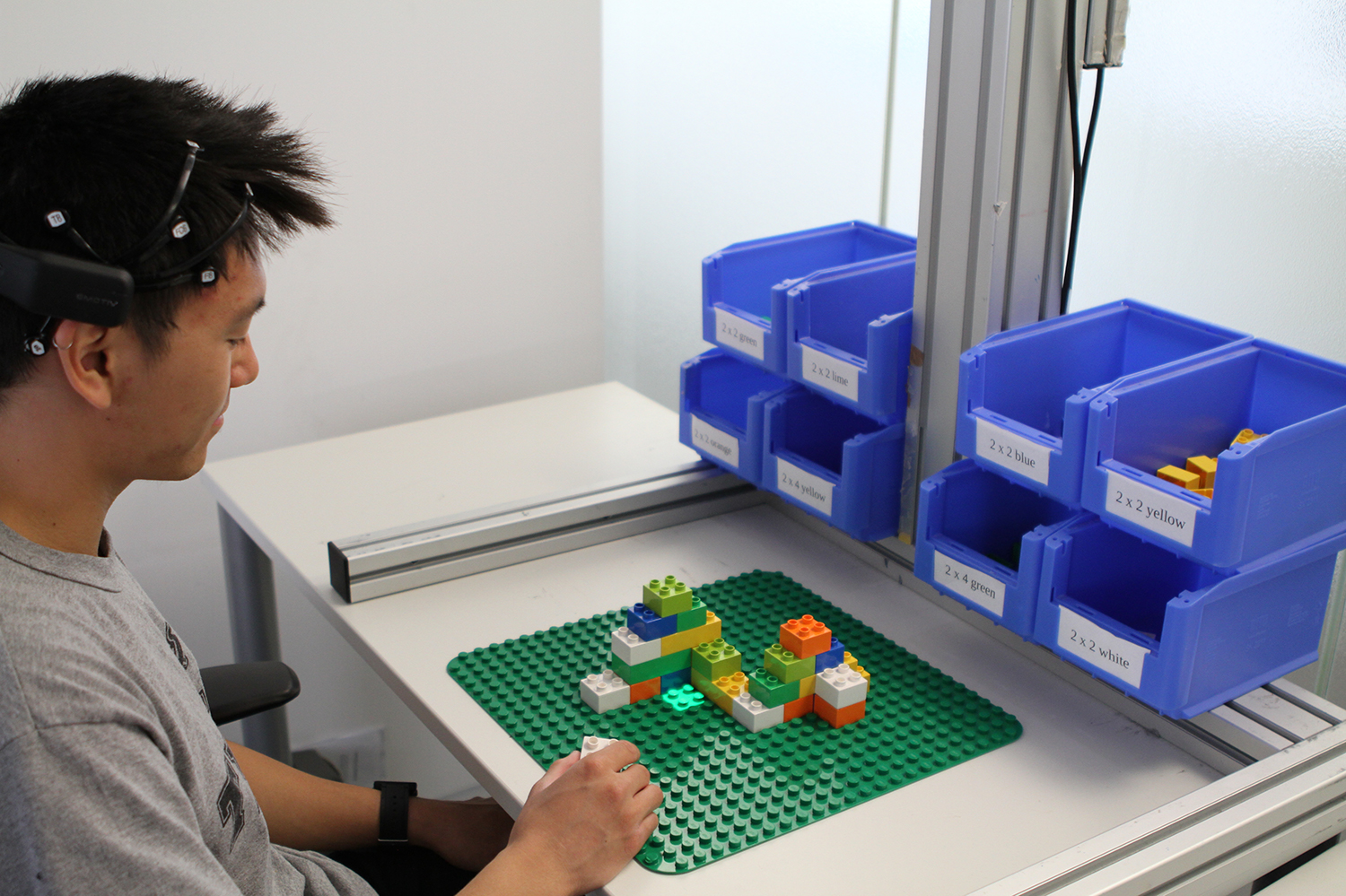}} &
We present the use of \ac{EEG} features to assess the workload induced by two different instruction modalities in a manual assembly scenario and present a \ac{EEG}-based evaluation pipeline.&
Chapter~\ref{ch:workloadbyeeg} \\

\parbox[t]{2mm}{\vspace{-3em}\multirow{2}{*}{\rotatebox[origin=c]{90}{\textbf{Sensing Cognitive Workload}}}}

& \raisebox{-\totalheight+.5em}{\includegraphics[width=\linewidth]{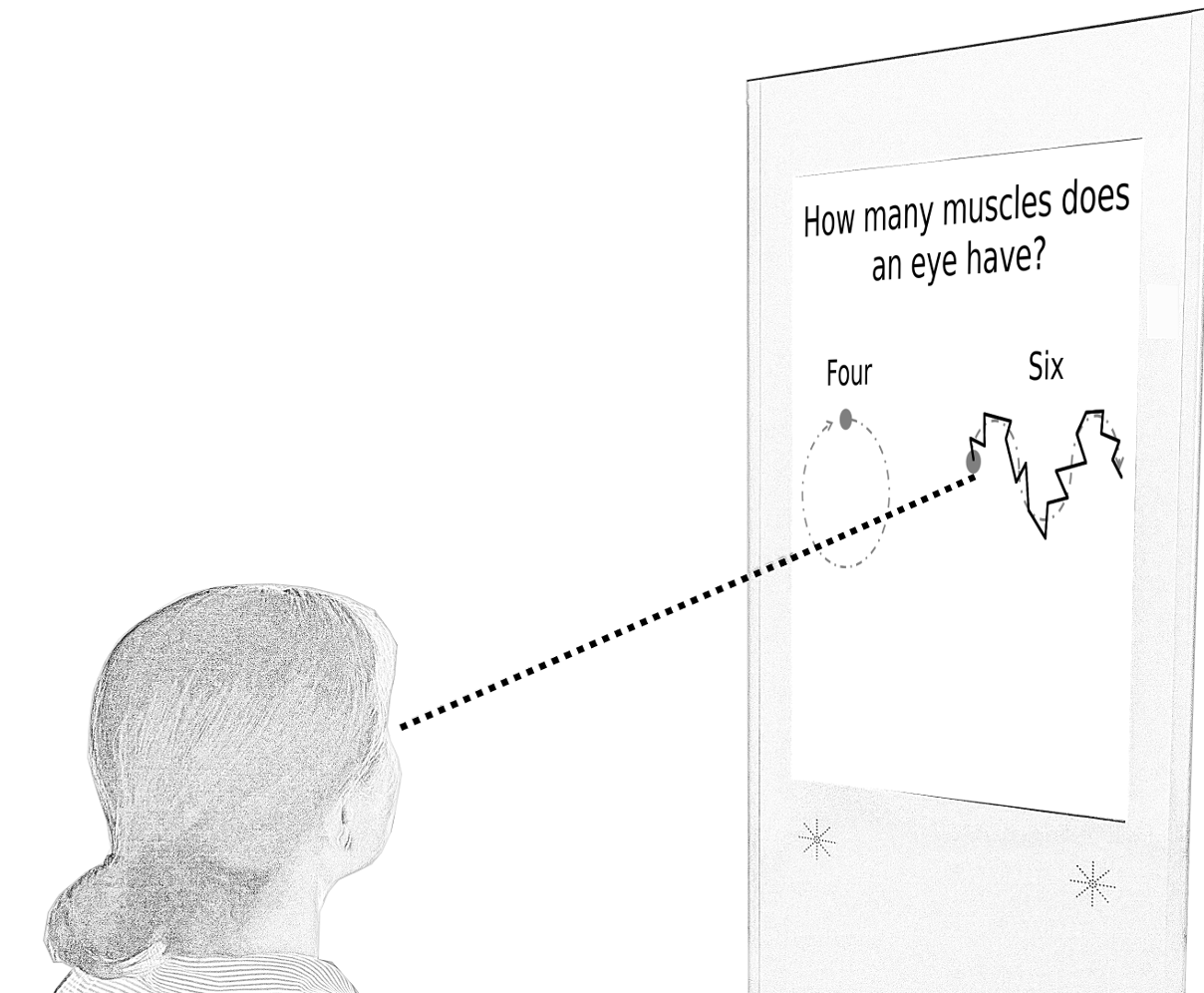}} &
We exploit the performance of smooth pursuit eye movements as an indicator of cognitive workload. Furthermore, we explore the classification performance for potential workload-aware systems.&
Chapter~\ref{ch:smooth_pursuit}\\

& \raisebox{-\totalheight+.5em}{\includegraphics[width=\linewidth]{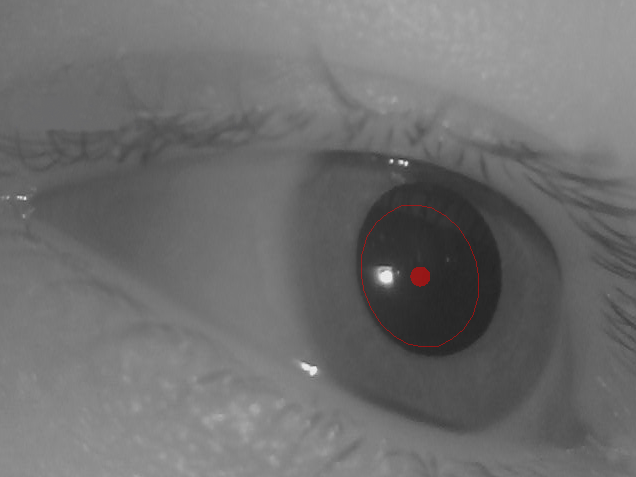}} &
Changes in the pupil diameter during different cognitive workload states. We evaluate the performance of cognitive workload classification in an adaptive math task scenario. &
Chapter~\ref{ch:pupil_dilation}\\

\midrule

& \raisebox{-\totalheight+.5em}{\includegraphics[width=\linewidth]{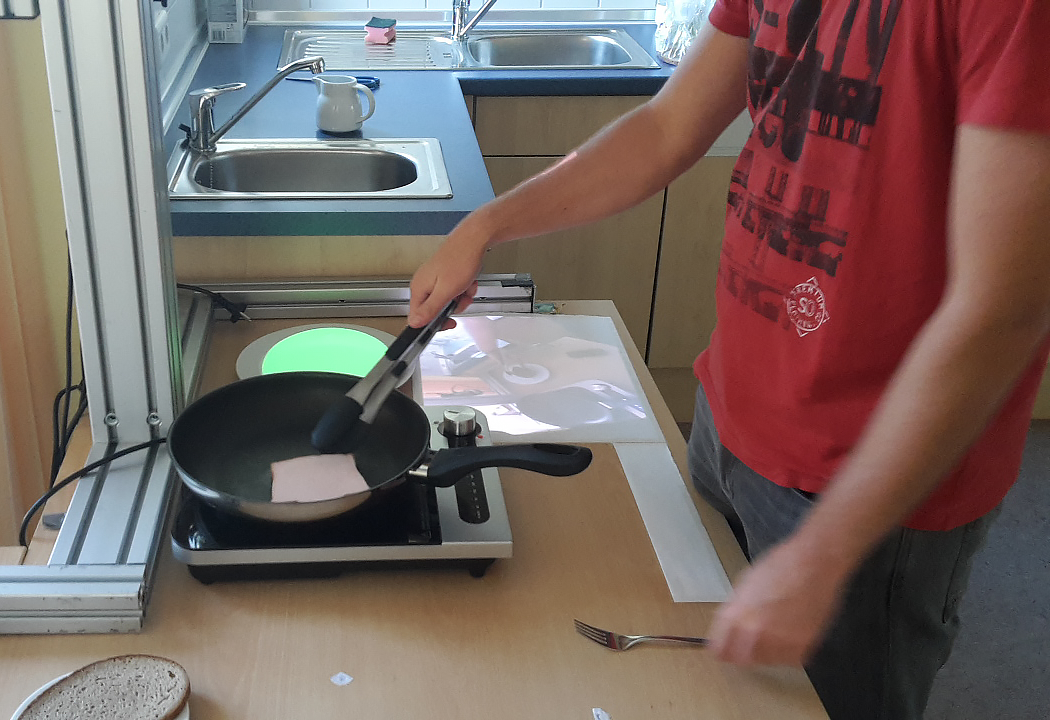}} &
We evaluate an assistive system based on the previous Chaptersthrough a long-term study in a sheltered living facility.&
Chapter~\ref{ch:smart_kitchen_deployment} \\

\parbox[t]{2mm}{\vspace{1em}\multirow{2}{*}{\rotatebox[origin=c]{90}{\textbf{Applications}}}}

& \raisebox{-\totalheight+.5em}{\includegraphics[width=\linewidth]{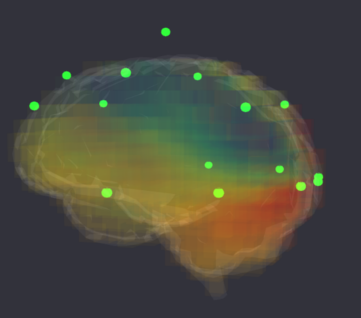}} &
We present the implementation of a brain visualization that displays dipoles from which electrical activity is generated. We show how this approach can be combined with a neurofeedback loop.&
Chapter~\ref{ch:visualization_reflection}\\

& \raisebox{-\totalheight+.5em}{\includegraphics[width=\linewidth]{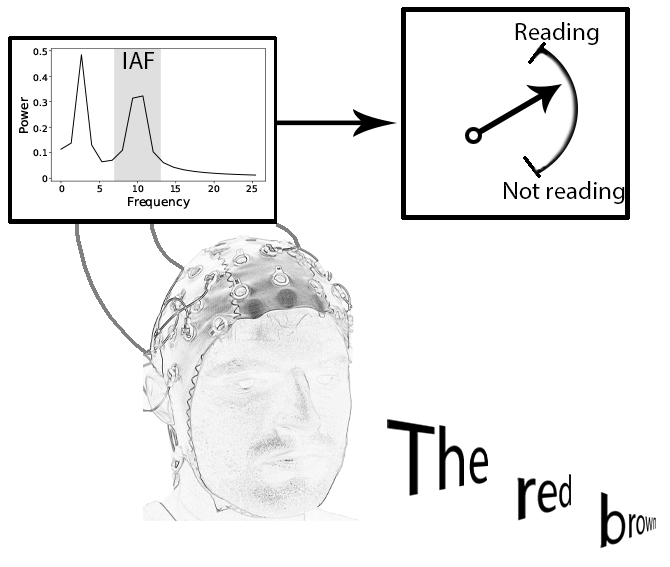}} &
We present a study which investigates the influence of different \ac{RSVP} reading parameters on information and knowledge acquisition. We derive a predictive model that utilizes \ac{EEG} that forecasts the suitable reading parameters.&
Chapter~\ref{ch:workload_information_consumption}\\
\bottomrule
\end{tabularx}}
\caption[Research inquiries within the scope of this thesis.]{Research inquiries developed within the scope of this thesis. Each research implementation is described in detail in a dedicated Chapter or section within this thesis.}
\label{tab:researchprobes}
\normalsize
\end{table*}

\subsubsection{Design Requirements for Workload-Aware Interfaces}
The design requirements and modalities for technologies that adapt during opportune moments were evaluated in a field study. Through qualitative inquiries, design guidelines were derived to match the expectation of the users. Specifically, households that accommodate people with cognitive impairments that require cognitive assistance were targeted ~\cite{Kosch:2018:SKP:3173574.3173845}. Supervised cooking emerged as an important communal activity where cognitive assistance is frequently required but not always available due to shortages of specially trained staff. If not mentally supported, tenants tend to make mistakes that may cause injuries or leave their current task ``as it is'' including the risk of unexpected consequences (\textit{e.g.}, leaving the cooktop on). Our methodology employed observations to create an initial question catalog which was refined in an iterative process. Next, semi-structured interviews were conducted with caretakers of the sheltered living facility to establish design implications for workload-aware interfaces that support tenants during the cooking process. Regarding the preferred feedback modalities of computerized assistance, an additional study revealed significant preferences for auditory and visual feedback during opportune moments for assistance.

\subsubsection{Using Physiological Sensing to Assess Cognitive Workload}
We build upon previous theories of cognitive psychology and the field of \ac{HCI} to quantify the demand of user interfaces on end-users. An objective real-time assessment of cognitive workload is important since questionnaires or think-aloud protocols, the traditional way of workload assessment, are prone to subjective perception, interfere with the experimental workflow, and require participants to remember their experience. We exploit physiological sensing as an alternative to traditional workload measures for two reasons: First, \ac{UI} designer and experimenter receive in-situ feedback about their novel system while participants use it. Secondly, users benefit from self-reflection and \ac{UI} adaptations that prevent boredom or frustration due to \ac{UI} difficulty adjustments. In this thesis, we evaluate the concepts of eye tracking~\cite{Kosch:2018:LME:3170427.3188643, Kosch:2018:YET:3173574.3174010} and \ac{EEG}~\cite{Kosch:2018:ICA:3233739.3229093} as direct workload measurements that can be employed in real-time.

\subsubsection{Identifying Opportune Moments for Cognitive Assistance}
Based on in-situ observations, we identified how user interfaces can adapt to the current cognitive state by evaluating visual, auditory, and tactile feedback. First, we reviewed the aforementioned modalities in a user study~\cite{Kosch:2016:CTA:2982142.2982157} and found a preference for visual and auditory feedback. We thus implemented a system using the preferred stimuli in an in-situ study to determine the feasibility of providing adaptive assistance suited to the current level of cognitive demand~\cite{kosch2019thedigital}.

\subsubsection{Tools and Frameworks for Building Workload-Aware Systems}
Tools and frameworks are essential for the development of context-aware systems. Such tools can support the proliferation of a new dimension in the paradigm of context-aware computing. We present visualizations that support user interface designers to understand how their interfaces affected cognitive processes on a cortical level. Additionally, users can self reflect on their past performance~\cite{kosch2016brain}. In this context, we publish the tools and source code we have implemented. This is complemented by the collected data sets that can be used to innovate or extend existing research.

\section{Ethics}
All studies reported in this thesis were conducted in accordance with the declaration of Helsinki~\cite{world2001world}. Furthermore, we followed the aspects of the \ac{ELSI}~\cite{myskja2014we}. This includes the submission of a detailed study plan to individual project partners that were involved in the related study. Within the context of studies that require proactive involvement and acquisition of participants, consent for data collection was given according to the \ac{GDPR}. Personal data, such as demographic data, photos, and videos, were stored securely to prevent access by third parties. No data was collected by any probes before the agreement of the involved users. The studies conducted in Chapter~\ref{ch:smart_kitchen_requirements}, \ref{ch:support_modalities}, and \ref{ch:smart_kitchen_deployment} included tenants with cognitive impairments from a sheltered living facility. Thereby, the study procedure followed an additional examination process of the \ac{BMBF}\footnote{\url{www.bmbf.de/en/index.html} - last access \urldate} and the \ac{GWW}.

\subsection{Consent}
Written and informed consent was received, when possible, by the participants themselves. Participants were informed about their rights and compensation within the study. To ensure proper consent from participants with cognitive impairments, we submitted our study plan to the \ac{BMBF} and the corresponding living facility \ac{GWW}. Upon gaining approval from both institutions, the tenants of the sheltered living facility were informed about the course of study and were asked for their potential participation. This was achieved by asking participants directly for consent. If this was not possible due to the level of cognitive impairment, the associated legal guardians were asked to provide written consent for the participant. Regardless of the aforementioned procedure, participants were always informed about the course of study, collected data, and their rights to abort the study or request a deletion of their data.

\subsection{Physiological, Behavioral, and Task-Related Data}
During our research, we collected a rich amount of individual physiological, behavioral, and task-related data from participants. This was required to feed the data into \ac{ML} and \ac{DL} algorithms that learn or derive patterns. However, participants might behave differently during an experiment to hide their regular behavior due to privacy concerns. We designed our experimental procedure transparently by showing what can be inferred from physiological, behavioral, and task-related data. Furthermore, participants were introduced into the anonymization process to provide evidence about the intractability of their identity. When employing systems in the wild, a ``\textit{Privacy by Design}''~\cite{10.1007/3-540-45427-6_23} paradigm was employed to ensure privacy in protected places.

\subsection{User-Awareness}
When technology tracks mental states, as well as behavioral data in private settings, the challenge of workload-aware interfaces, is to provide seamless interaction while communicating with the user how certain decisions were made. Users may be unaware of certain triggers that cause an unexpected system behavior based on their cognitive state or behavior.

\section{Research Context}
This thesis presents work that was carried out in the context of different scientific institutes and research partners for four years. The presented research was carried out over two years in the Human-Computer Interaction and Cognitive Systems group at the Institute for Visualization and Interactive Systems in Stuttgart as well as in the Human-Centered Ubiquitous Media group at the Ludwig Maximilian University of Munich. Both groups are supervised by Prof. Dr. Albrecht Schmidt. Several publications resulted in a variety of collaborations.

\subsection{Project \textit{KoBeLU}}
The major part of this thesis was conducted in the context of the project \ac{KoBeLU} (Grant No. 16SV7599K) funded by the \ac{BMBF} within the project call ``Erfahrbares Lernen''. Including the Ludwig Maximilian University of Munich, eight project partners (AUDI AG, MAHLE GmbH, International Center for Ethics in Research (IZEW), \ac{GWW}. Offenburg University of Applied Sciences (HSO), Stiefel Eurocart GmbH, User Interface Design UID) took part in the project. KoBeLU aimed to develop and evaluate assistive systems within the learning contexts. Specifically, the influence of in-situ assistance was subject to studies for different stakeholders in various learning environments. By integrating assistive technologies, this 3-year project (Aug. 2016 - Nov. 2019) researched the degree of cognitive assistance that is provided in real-time. The collaboration between research and application resulted in conjoint publications~\cite{funk2016automatic, kosch2016exploring, kosch2020emotion, Kosch:2016:CTA:2982142.2982157, kosch2019thedigital, Kosch:2018:SKP:3173574.3173845, nowak2020what}.

\subsection{Project \textit{be-greifen}}
Within the project call ``Erfahrbares Lernen'', collaborations were done with the project ``be-greifen'' (Grant No. 16SV7527). This is reflected in several publications with the research associate Pascal Knierim~\cite{knierim2016ubibeam, knierim2018flyables,knierim2020opportunities, knierim2018challenges, knierim2020altering, knierim2017tactile, kosch2018dronectrl, kosch2017chances}. The work there was related to cognitive assistance using \acp{HMD}.

\subsection{Project \textit{AMPLIFY}}
Several projects were supported by the European Union's Horizon 2020 Programme under the ERCEA AMPLIFY (Grant No. 683008). This is reflected in publications with the research associates Jakob Karolus~\cite{karolus2020hit, karolus2018emguitar, liang2018on} and Matthias Hoppe~\cite{hoppe2019dronos, hoppe2018vrhapticdrones, hoppe2019are, hoppe2020dont, wozniak2020madrone}

\subsection{University Collaborations}
Starting in 2017, the workshop series \ac{DAEM} was launched at the conference PETRA in cooperation with the Ostwestfalen-Lippe University of Applied Sciences. The workshop is centered around the development and evaluation of assistive technologies in manufacturing environments. This includes the evaluation of technologies that support and foster the cognitive abilities of workers, resulting in a design space for industrial assistive computing~\cite{buettner2017design}. To date, the workshop was renewed for a fourth iteration in the year 2020.

Furthermore, this thesis presents work concerning people with cognitive impairments. Since this represents a marginalized group, traditional qualitative methods are difficult to apply. To overcome this, we consulted the expert Erin Brady for this research project as she provides the necessary research qualifications. The conducted project is concerned about the behavior of persons with cognitive impairments during communal activities, particularly cooking. This enabled us to derive design implications for assistive kitchen technologies that provide mental assistance for people with cognitive impairments~\cite{Kosch:2018:SKP:3173574.3173845} and point out the potential abuse of assistive technologies in working environments~\cite{kosch2019keep}.

\section{Thesis Outline}
This thesis comprises \textbf{eleven chapters} and is divided into \textbf{six parts}. The structure of this thesis closely follows the depiction of Table~\ref{tab:researchprobes}. The first part denotes background knowledge and differentiates existing work from the novel contributions presented in this thesis. The second part provides a series of studies that investigate the design requirements and implications of workload-aware systems. The third part introduces studies about physiological sensing modalities that enable in-situ measures of workload states. The fourth part presents how the aforementioned studies provide cognitive assistance and enable users to reflect on workload states. Finally, the fifth part concludes with guidelines and provides an outlook for workload-aware systems in the domain of computational interaction. The sixth and last part includes the bibliography and listings.

\subsection{Part \ref{part:introduction}: Introduction and Foundations}
\subsubsection{Chapter \ref{ch:introduction} -- Introduction} The first Chapter describes the motivation and vision for workload-aware interfaces, presents use cases, states the context in which this thesis was conducted, states collaborations, and summarizes challenges as well as opportunities which were faced throughout our research.

\subsubsection{Chapter \ref{ch:background} -- Background and Foundations} The second Chapter introduces key concepts of human cognition and the concept of cognitive workload. We present state-of-the-art measures that facilitate explicit and implicit measures for cognitive workload. This is followed by the presentation of recent developments of ubiquitous assistive technologies that support workers and persons with cognitive impairments. Ethical considerations are discussed since sensor-dependent systems need to be viewed critically from a data collection perspective.

\subsection{Part \ref{part:human_cognition}: Human Cognition}
\subsubsection{Chapter \ref{ch:smart_kitchen_requirements} -- Opportune Cognitive Augmentation} Detecting suitable situations for cognitive augmentation is crucial to avoid context-aware mismatches. Therefore, we first focus on the disclosure of opportune moments for cognitive assistance. We assess the characteristics of diurnal chores that benefit from cognitive support in a case study with persons with cognitive impairments.

\subsubsection{Chapter \ref{ch:support_modalities} -- Feedback Modalities}
In this chapter, we evaluate the efficiency of feedback modalities after detecting situations that require assistance. We conducted a lab study that compared different visual, auditory, and tactile feedback patterns.

\subsection{Part \ref{part:sensing_cognitive_workload}: Sensing Cognitive Workload}
\subsubsection{Chapter \ref{ch:workloadbyeeg} -- Mobile Electroencephalography}
Cortical activity provides markers for cognitive demand and exhaustion in real-time. In a lab study, we evaluated the cognitive demand of two different user interfaces by analyzing \ac{EEG} data. We found that \ac{EEG} is a reliable complementary measure for cognitive workload and proposes an evaluation pipeline that supports researchers and designers in their assessment of user interfaces regarding the placed mental demand. Machine learning techniques are applied to show potential real-time adaptations of future workload-aware visualizations.

\subsubsection{Chapter \ref{ch:smooth_pursuit} -- Smooth Pursuits}
Human eye gaze behavior is affected under different levels of cognitive workload. We investigate changes in smooth pursuit eye movements under different levels of cognitive workload. Applying machine learning techniques reveals a reliable distinction for different cognitive workload levels for interpersonal and intrapersonal classification.

\subsubsection{Chapter \ref{ch:pupil_dilation} -- Pupil Dilation}
The pupil is an eye component that is heavily affected by changes in cognitive demand. We show how differences in pupil dilation can be utilized to build reliable on-the-fly workload-aware systems, and how eye-tracking data can be used for the physiological prototyping of workload-aware systems.

\subsection{Part \ref{part:applications}: Applications}
\subsubsection{Chapter \ref{ch:smart_kitchen_deployment} -- Design Pipeline Evaluation}
We evaluate the results and the design pipeline from Chapter \ref{ch:workloadbyeeg} in a case study at a sheltered living facility. Using an exemplary assistive system that incorporates the aforementioned results, tenants with cognitive impairments reacted favorably to the provided cognitive augmentation and were unusually independent throughout the task.

\subsubsection{Chapter \ref{ch:visualization_reflection} -- Reflection}
Communicating changes in cognitive demand help to understand how diurnal tasks can be mapped to available resources that vary throughout the day. We propose the implementation of a real-time visualization which we evaluate in the context of a neurofeedback loop.

\subsubsection{Chapter \ref{ch:workload_information_consumption} -- Workload-Aware Reading Interfaces}
This chapter describes the concept of an application scenario for workload-aware adaptive reading scenarios. By evaluating \ac{EEG}, we propose strategies to adjust \ac{RSVP} reading parameters depending on the current level of perceived workload. We propose predictive models that adjust the reading speed regarding the cortical activity.

\subsection{Part \ref{part:guidelines_and_conclusion}: Conclusion}

\subsubsection{Chapter \ref{ch:conclusion} -- Conclusion and Future Work}
In this chapter, we summarize and conclude the findings of this thesis and revisit the research questions stated in the beginning. We reflect on the presented research approach, discuss future work, and present implications for workload-aware systems.



\chapter{Background and Foundations}
\label{ch:background}
\epigraph{\textit{My problem is that I have been persecuted by an integer. For seven years this number has followed me around, has intruded in my most private data, and has assaulted me from the pages of our most public journals.}}{George A. Miller}

The research presented in this work is based on the field of ubiquitous computing, assistive technologies, and physiological sensing. To create and design workload-aware interfaces, we apply theoretical concepts from psychology, cognitive load theory, and current practices of employing sensing modalities in assistive technologies. This includes ethical considerations that arise with interfaces that can sense physiological states ubiquitously. Here, we provide readers with the necessary knowledge that deepens the foundations and selected methods for the studies presented in the upcoming chapters.

\section{Ubiquitous Computing}
The term \textit{Ubiquitous Computing} describes the era in which computers started to "weave themselves into the fabric of everyday life until they are indistinguishable from it"~\cite{weiser1993ubiquitous}. This included a deviation from the many-to-one relationship, where many persons relied on the use of one computer. With the introduction of the \textit{MITS Altair 8800} personal computer in the year 1975, computers became available for broad masses of people. In the following years, computers became smaller and their capabilities in sensing capabilities and computational power increased. Nowadays, the distribution of computers has changed into a many-to-one relationship, where many computational units are owned or operated by a single person. These can sometimes be not distinguishable as they are seamlessly integrated into the environment: adaptive light switches, slide doors, or voice assistants are just a few examples of technologies that facilitate our life~\cite{Weiser1997}.

These technologies accomplish tasks for us which would usually drain our cognitive attention. They provide mental alleviation and allow us to schedule cognitive resources to other tasks that require human action. However, in emergency cases, ubiquitous devices might compete for our attention to take action, so we should be made aware of our cognitive states to provide selective assistance. In the following, we summarize previous research on cognitive workload theory.

\subsection{Perception of Workload}
Workload, regardless of mental or physical nature, is an important component throughout the perception of diurnal activities. The body-mind interaction is believed to be affected when strained or relaxed over a certain time span~\cite{10.2307/2254848}. Hence, the workload is ubiquitously present and often associated with the accomplishment of daily tasks. Reducing workload to support users in daily tasks is an established way to increase a person’s overall life quality. For example, robots and machines reduce physical workload by overtaking tasks that are dangerous or exhaustive for humans.

Cognitive workload is one type of workload that exhausts mental resources and that is attributed to individual properties. Past research aimed to reduce cognitive workload that arises through everyday tasks, such as digital calendars, task scheduling apps, or notification management systems~\cite{Sellberg2014}. Providing too much cognitive support might lead to boredom or loss of important information that comes along with a context-aware mismatch. No presence of cognitive support might lead to frustration due to failing in accomplishing the present task. Both aspects are detrimental for the user in terms of performance, user experience, and task performance which might elicit dangerous situations~\cite{GALY2012269}. Finding the ``sweet spot'' for mental demand is necessary to keep the user in flow states, where flow states are an essential part of maintaining the users' engagement and task efficiency~\cite{csikszentmihalyi1997flow, csikszentmihalyi2013flow}. We elaborate on the term cognitive workload and the relevant semantics for this thesis in the following.

\section{Cognitive Workload}
We process information by perceiving events in our environments. In this case, cognitive workload describes any information processing load placed on cognitive mechanisms. It is often associated with the level of measurable effort that is mobilized to accomplish a task. Even more often, cognitive workload is falsely used interchangeably with the terms \textit{stress}, \textit{fatigue}, or \textit{physical workload}~\cite{hancock2000stress}. In the following section, we provide an overview of the semantics of the term \textit{cognitive workload} and its theories which are deemed relevant for the scope of this thesis.

\subsection{Working Memory}
As humans perceive information in their environment, they are obliged to store relevant ``\textit{chunks of information}'' in their short-term memory for further processing. This temporary storage, known as working memory~\cite{BADDELEY197447}, utilizes the five senses sight, touch, taste, smell, and hearing, to encode information. It is assumed that cognitively demanding tasks can only be efficiently accomplished with a sufficient amount of working memory. Thereby, the number of objects that can be stored simultaneously is limited to \textit{seven} plus or minus two chunks~\cite{cowan2010magical}. Without sufficient capabilities to store information temporarily, cognitively demanding tasks might be completed with less efficiency. How large the capacity of working memory is and how it can be quantified is highly disputed in the field of Psychology. In common parlance, it is said that ``seven chunks plus/minus two'' can be memorized by an individual. Miller argues how this number might have emerged as the status quo in research~\cite{miller1956magical}. Further research has been conducted in this regard, indicating that the number seven is the limiting factor that should be considered as critical when relying on temporary storage~\cite{SAATY2003233}. Thereby, large inconsistencies in a set of observed events provide a better memory performance.

\subsection{Cognitive Workload Theory}
Previous work has inquired into different theories on how cognitive workload and mental models work. The term ``cognitive workload theory'' has been designed to optimize knowledge intake and boost learning performance during information processing. It is assumed that working memory is limited in dealing with the perception through our five senses. Thereby the processing of visual, auditory, or tactile information is limited by the number of ``chunks'' an individual can remember and mentally process at the same time. The overall aim of cognitive workload theories is to provide guidelines that minimize the demands upon working memory~\cite{sweller1998cognitive}. In the context of \ac{HCI}, designers could reduce the number of visual chunks a user has to remember (\textit{e.g.}, menu items), or utilize the Gestalt laws~\cite{hartmann1935gestalt} to group similar chunks as means to reduce the effort for item searching. John Sweller~\cite{sweller1988cognitive} presented a cognitive workload theory which divides workload, learning, and problem-solving into three components: \textit{Intrinsic Workload}, \textit{Germane Workload}, and \textit{Extraneous Workload}, that describe factors that contribute to the demand for working memory.

\textit{Intrinsic Workload} describes the inherent complexity of a task. \textit{Intrinsic Workload} is not trivial to manipulate since the intrinsic task complexity is associated with the necessary individual cognitive demand to process and act upon specific information. If possible, \textit{Intrinsic Workload} should be kept at a minimum to avoid unnecessary internal processing of users. However, this is considered difficult because the task itself needs to be modified to manipulate the complexity~\cite{zbrodoff2005everyone}.

\textit{Germane Workload} represents the effort to process patterns within a task. Realizing schemes that help to solve a task may increase task engagement and foster learning. Hence, maximizing the share of \textit{Germane Workload} is emphasized as a crucial factor when designing engaging user interfaces.

\textit{Extraneous Workload} is manipulated by the task representation that can be perceived with human senses. For instance, a well-designed visualization enhances the interpretation of data far better than bad design. Relative to the other workload types, \textit{Extraneous Workload} is trivial to manipulate since the task representation can be exchanged through the appearance of the problem (\textit{e.g.}, visually or auditory). \textit{Extraneous Workload} should be minimized to avoid unnecessary allocations of cognitive resources required to resolve the intrinsic task complexity. Figure \ref{fig:sweller_overview} illustrates the components of Sweller's cognitive workload theory.

\tikzstyle{block} = [rectangle, draw, rounded corners, text width=8em, text centered,minimum height=4em]
\tikzstyle{sub_block} = [rectangle, draw, rounded corners, text width=4em, text centered,minimum height=4em]
\tikzstyle{arrow} = [draw, thick, -latex', shorten >=0pt]

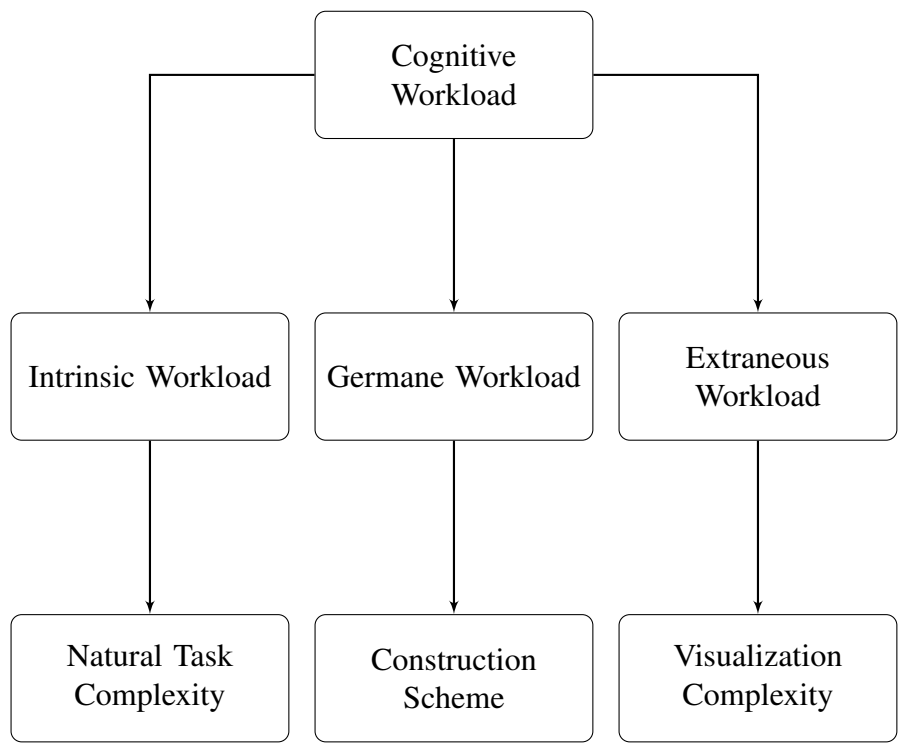
\begin{figure}[!h]
\centering
\begin{tikzpicture}[node distance=4cm]
\draw ++(0,0) node [block] (cognitive_workload) {Cognitive Workload};
\draw ++(0,0) node [block, below of=cognitive_workload] (germane_workload) {Germane Workload};
\draw ++(0,0) node [block, left of=germane_workload] (intrinsic_workload) {Intrinsic Workload};
\draw ++(0,0) node [block, right of=germane_workload] (extraneous_workload) {Extraneous Workload};
\draw ++(0,0) node [block, below of=germane_workload] (schema_construction) {Construction Scheme};
\draw ++(0, 0) node [block, below of=intrinsic_workload] (natural_complexity) {Natural Task Complexity};
\draw ++(0, 0) node [block, below of=extraneous_workload] (interaction_modalities) {Visualization Complexity};
\path [arrow] (cognitive_workload) -- (germane_workload);
\path [arrow] (cognitive_workload) -| (intrinsic_workload);
\path [arrow] (cognitive_workload) -| (extraneous_workload);
\path [arrow] (intrinsic_workload) -- (natural_complexity);
\path [arrow] (germane_workload) -- (schema_construction);
\path [arrow] (extraneous_workload) -- (interaction_modalities);
\end{tikzpicture}
\caption[Graphical representation of the cognitive workload theory model by John Sweller]{Graphical representation of the cognitive workload theory by Sweller \cite{sweller1988cognitive}. \textit{Intrinsic}, \textit{germane}, and \textit{extraneous} workload represent form the overall experienced mental demand. While \textit{intrinsic} and \textit{germane} workload cannot be easily changed due to the inherent task complexity and tools provided to the user, \textit{extraneous} workload can be manipulated by the task representation.}
\label{fig:sweller_overview}
\end{figure}

\subsection{Cognitive Workload Assessments}
The assessment of the cognitive workload that is placed by user interfaces is a factor that has been viewed as crucial by designers. Different measures, ranging from direct subjective assessments to implicit physiological measures, have been employed to sense cognitive workload. Although being far from exhaustive, the following section informs about current practices of workload measures that have gained popularity in \ac{HCI} research. These focus primarily on self-rated measures, brain-sensing, eye tracking, and physiological measurement modalities.

\subsubsection{Self-Rated Measures}
\begin{figure*}[t!]
\centering
\subfloat[][]{
\includegraphics[height=0.48\columnwidth]{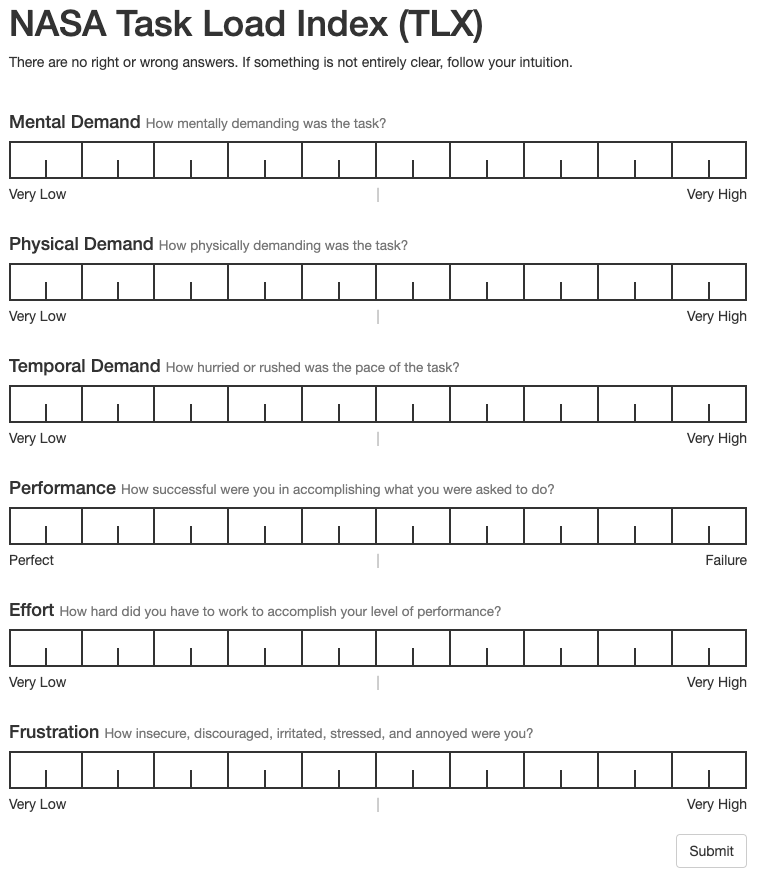}
\label{fig:nasa_tlx_questionnaire}}
\subfloat[][]{
\includegraphics[height=0.48\columnwidth]{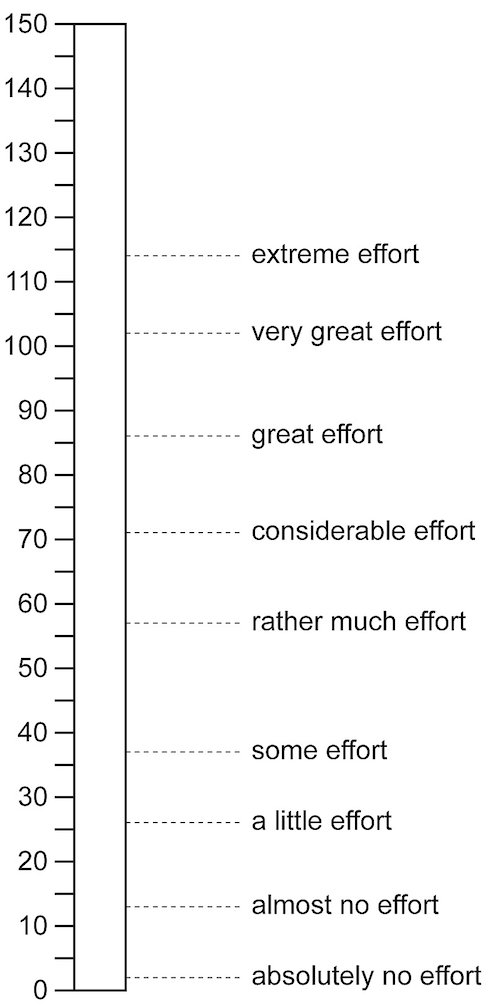}
\label{fig:rsme_questionnaire}}
\subfloat[][]{
\includegraphics[height=0.48\columnwidth]{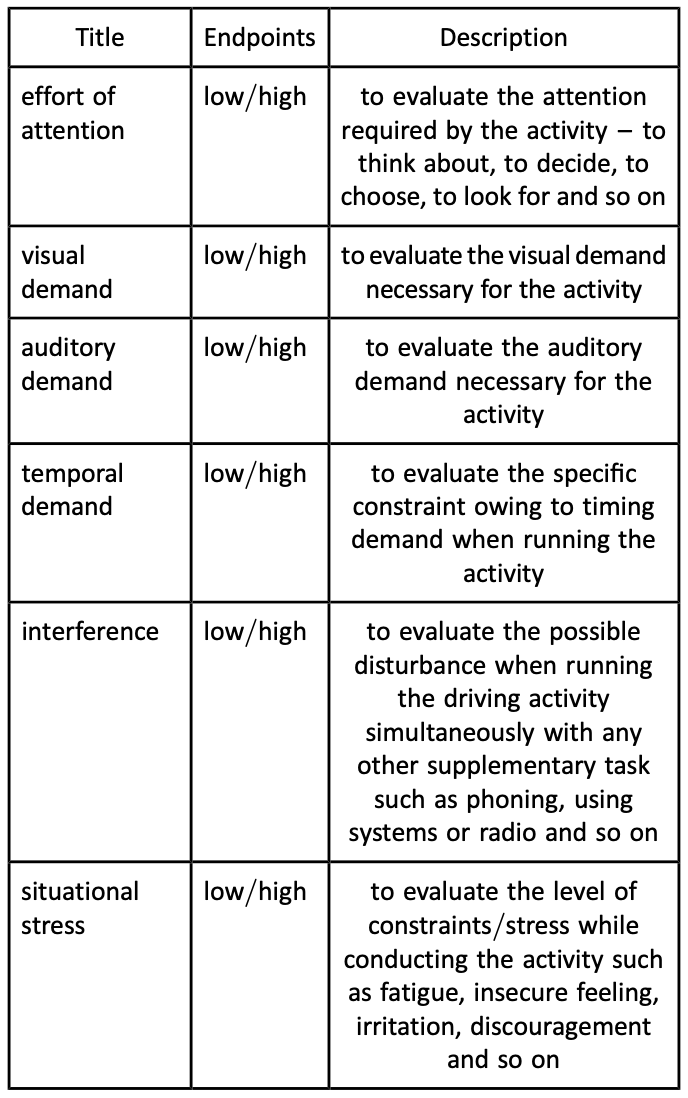}
\label{fig:dali_questionnaire}}
\caption[Examples of questionnaires that are frequently used to evaluate cognitive workload]{Example of questionnaires which measure cognitive workload. \textbf{(a):} NASA Task Load Index, \textbf{(b):} Rating Scale of Mental Effort, and \textbf{(c):} Driver Activity Load Index.}
\label{fig:sample_questionnaires}
\vspace{0.4cm}
\end{figure*}

Asking users in-situ or post hoc about their cognitive strains during interaction is a common way to evaluate user interfaces. Subjective inquiries benefit from being an uncomplicated measurement that can be used without the need for complicated hardware or foreknowledge. Popular questionnaires which are used among researchers are, for example, the \ac{NASA-TLX}~\cite{hart2006nasa, HART1988139}, \ac{RSME}~\cite{WIDYANTI201370, zijlstra1985construction}, or \ac{DALI}~\cite{iet:/content/journals/10.1049/iet-its_20080023} (see Figure~\ref{fig:sample_questionnaires}). Questionnaires reign among the most used measures for cognitive workload since they are easy to employ and be applied during or after experiments. However, they face critical trade-offs: They are susceptible to the subjective perception of participants, they place additional cognitive workload on the participants by requiring them to remember their experience, and they lack concrete real-time assessments.

\subsubsection{Electroencephalography}
The human brain is the information and control unit of a human being. Thereby, approximately 100 billion neurons underlie human cognition~\cite{herculano-houzel_human_2009}. Electrical activity in several brain regions can be measured by placing electrodes on the scalp. Neurons communicate by exchanging electrical activity using neurotransmitters, a chemical transferred between neurons. In this work, we focus on measuring this electrical activity via an \ac{EEG} headset. \ac{EEG} is commonly leveraged in clinical application and yields a non-invasive method to estimate cortical activity~\cite{mason_general_2003, nicolas-alonso_brain_2012, wolpaw_braincomputer_2002}. Electrical potentials between 1\,$\mu v$ and 100\,$\mu v$ (microvolts) are measured by placing conductive electrodes on a scalp. An additional electrode serves as a reference electrode, which can be placed on the earlobe or scalp~\cite{malmivuo_bioelectromagnetism:_1995}. The measured \ac{EEG} potentials allow the processing and extraction of different features. Machine learning has been used on these features to retrieve insights into cognitive processes~\cite{1741-2552-15-3-031005, 1741-2552-4-2-R01}. This approach allows to discriminate between cognitive states~\cite{frey_review_2013, Grimes:2008:FPC:1357054.1357187, Lee:2006:ULE:1166253.1166268} and enables \ac{EEG} as a modality for user experience evaluation~\cite{Frey:2016:FEE:2858036.2858525}. For example, changes in electrical potentials are observed by analyzing frequency bands. Previous work found a drop in frequencies of alpha (8 - 12 Hz) and an increase of theta (4 - 8 Hz)~\cite{Klimesch1993, Kosch:2018:ICA:3233739.3229093} when subjects have to raise mental capacities. An alternative approach is the assessment of \acp{ERP} to infer mental workload~\cite{1741-2552-9-4-045008, 1741-2552-13-2-026019}. To complement this, real-time brain visualizations enable deeper insights into sequences of neuronal activity~\cite{kosch2016brain}.

\begin{figure*}[t!]
\centering
\subfloat[][]{
\includegraphics[height=0.4\columnwidth]{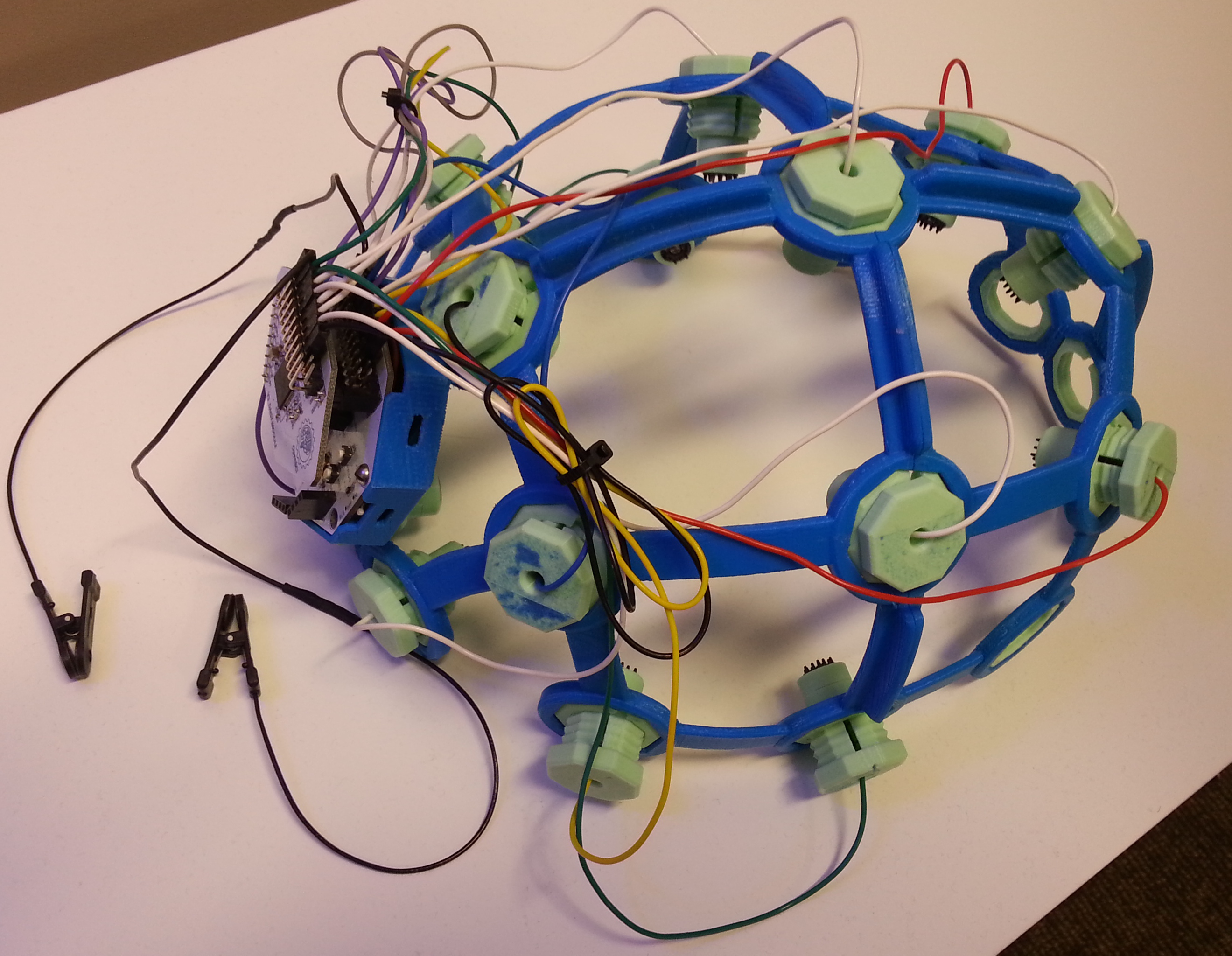}
\label{fig:openbci_headset}}
\subfloat[][]{
\includegraphics[height=0.4\columnwidth]{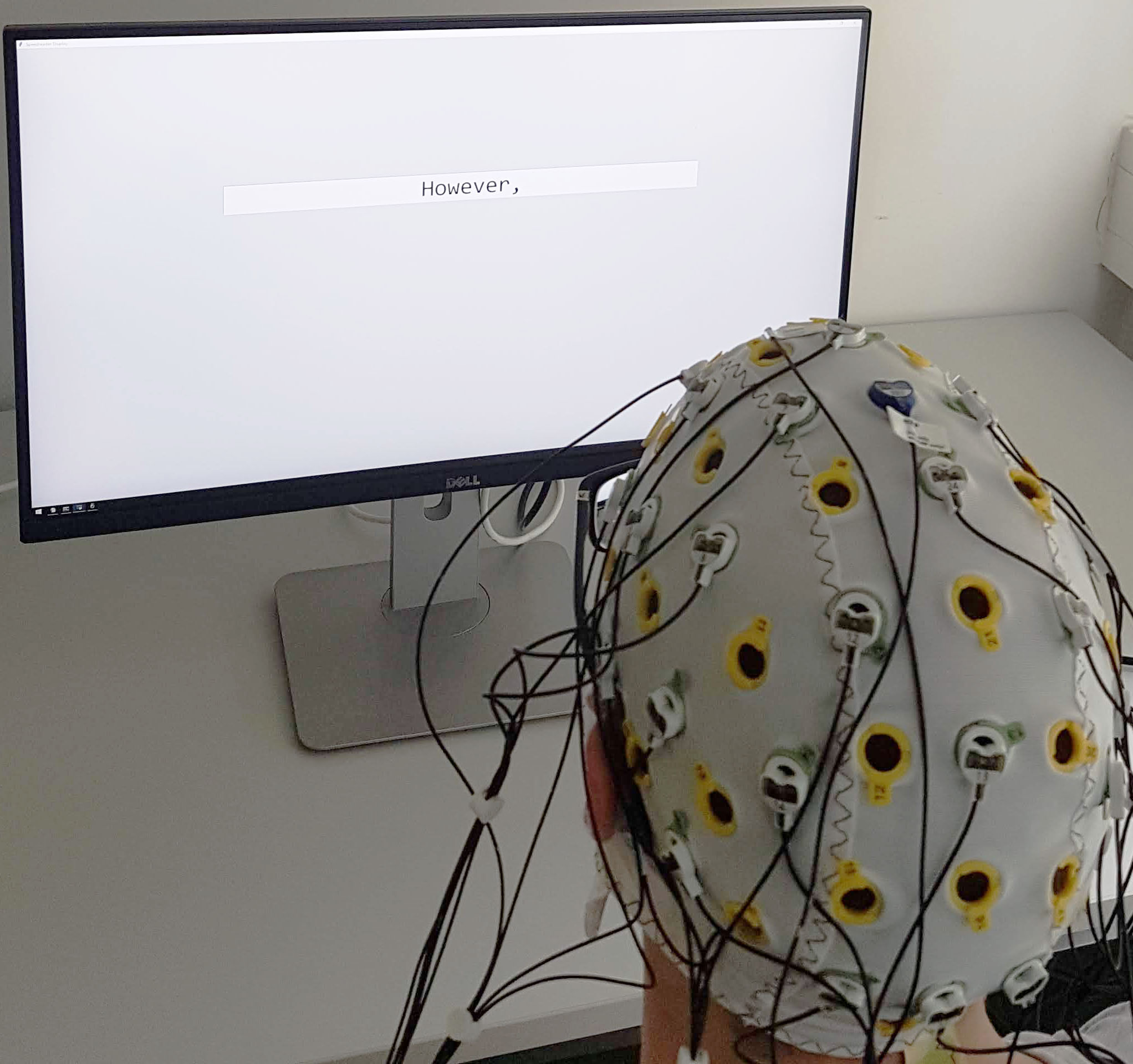}
\label{fig:64_electrodes}}
\caption[Depiction of two different \ac{EEG} headsets]{Different \ac{EEG} setups. \textbf{(a):} Affordable off-the-shelf solutions can be achieved using the OpenBCI with a customized 3D printed headset. \textbf{(b):} More expensive devices enable integrated impedance checks and a dense electrode placement with up to 64 electrodes.}
\label{fig:eeg_headsets}
\vspace{0.4cm}
\end{figure*}

However, \ac{EEG} measurements are prone to noise. Head movements, muscle contractions, eye movements, and even eye blinks cause changes in the electrical field on the scalp. Researchers are concerned about this and invest great effort to reduce the number of measurement artifacts~\cite{Cutmore1999129, friedman_facial_1991, korshakov2010line, NIKULIN20111528}. Since artifacts cannot be completely avoided, the use of \ac{EEG} often requires a controlled environment comprising minimal body movements of the user. Such conditions are often impractical for end-users due to their high experimental control. However, recent technical advances ameliorate these disadvantages~\cite{mullen2015real} that allow artifact corrections for \ac{EEG} during mobile settings~\cite{blum_eeg_2017}. The barrier \ac{EEG} in real-world settings has been lowered by making \ac{EEG} headsets accessible to the consumer market. These are usually priced between \$249\footnote{\url{www.choosemuse.com} - last access \urldate} and \$1600\footnote{\url{www.futurehealth.org/bm_at1.htm} - last access \urldate}. More affordable open source solutions can be acquired within the OpenEEG Project\footnote{\url{www.openeeg.sourceforge.net} - last access \urldate} (see Figure~\ref{fig:eeg_headsets}).

\subsubsection{Frequencies}
Different oscillations in \ac{EEG} signals are attributed to several semantics of cognitive states. The six bandwidths delta (1 \ac{Hz} - 3 \ac{Hz}), theta (4 \ac{Hz} - 7 \ac{Hz}), alpha (8 \ac{Hz} - 12 \ac{Hz}), lower beta (13 \ac{Hz} - 20 \ac{Hz}), upper beta (21 \ac{Hz} - 30 \ac{Hz}), gamma (31 \ac{Hz} - 100 \ac{Hz}) power are usually investigated when analyzing frequencies. Changes in alpha and theta frequencies are correlated with the mental demand placed on working memory as well as increase or decrease in task engagement~\cite{gevins1997high}. The alpha power is associated with changes in brain resting states. If the brain is resting, a modulation of alpha waves is achieved which decreases when users perform tasks that require their memory. Theta oscillations correlate with changes in task engagement at the frontal anterior cingulate cortex~\cite{BUSH2000215}. Thereby, an increase in theta power is associated with higher task engagement while lower theta power is an indicator of low task engagement. Since it is unknown if correlations between the demand of working memory (\textit{i.e.}, alpha power) and task engagement (\textit{i.e.}, theta power) might elicit flow states, researchers use the theta-alpha ratio as a coherent metric for cognitive demand.

\subsubsection{Event-Related Potentials}
\acp{ERP} are specific amplitudes that characteristically appear after users perceive a stimulus. Different types of amplitudes can be measured depending on the type of provided stimulus. For example, if a stimulus does not match the expectation of a user, a negative peak after approximately $400\,ms$ is measured in the \ac{EEG} signal. Another example is the processing of stimuli that require the users' attention which typically resemble a positive peak after $300\,ms$~\cite{brouwer2012estimating}. Smaller \ac{ERP} amplitudes are expected when users pay less attention to a specific stimulus. Tasks that require large amounts of working memory are more difficult to process, thus resulting in smaller \ac{ERP} amplitudes. Hence, \acp{ERP} can be measured to assess the mental vigilance towards auditory cues~\cite{10.1145/3173574.3174046} or to detect vocabulary gaps~\cite{schneegass2020braincode, 10.1007/978-3-030-29387-1_17}.

\subsubsection{Eye Tracking}
Eye tracking is a method to follow the gaze of a person. The gaze trail, a sequence of gazed objects, or the pupil diameter can be measured without great effort. Eye tracking has become an interesting topic for the \ac{HCI} community since it provided the possibility to computationally link attention and objects of interest with visual perception. The \textit{Mind-Eye Hypothesis} states that gazed objects are an indicator of visual attention~\cite{just1980theory}. This is not necessarily true: objects can be gazed without paying visual attention to them. However, eye tracking represents an interesting tool to learn about human attention, focus, cognition, and potential modality for advanced user inputs~\cite{Duchowski2002}.

\begin{figure*}[t!]
\centering
\subfloat[][]{
\includegraphics[height=0.31\columnwidth]{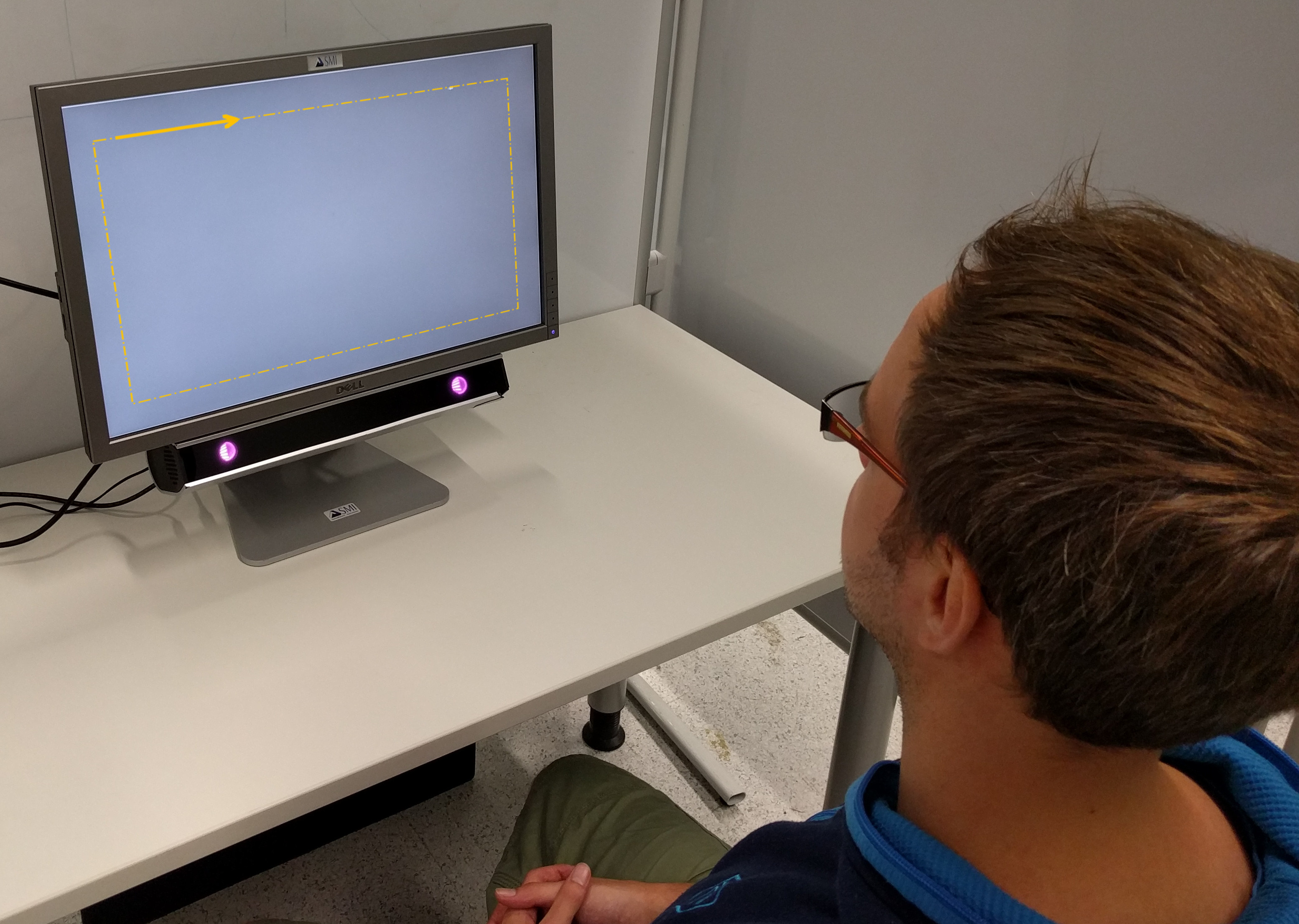}
\label{fig:stationary_eye_tracker}}
\subfloat[][]{
\includegraphics[height=0.31\columnwidth]{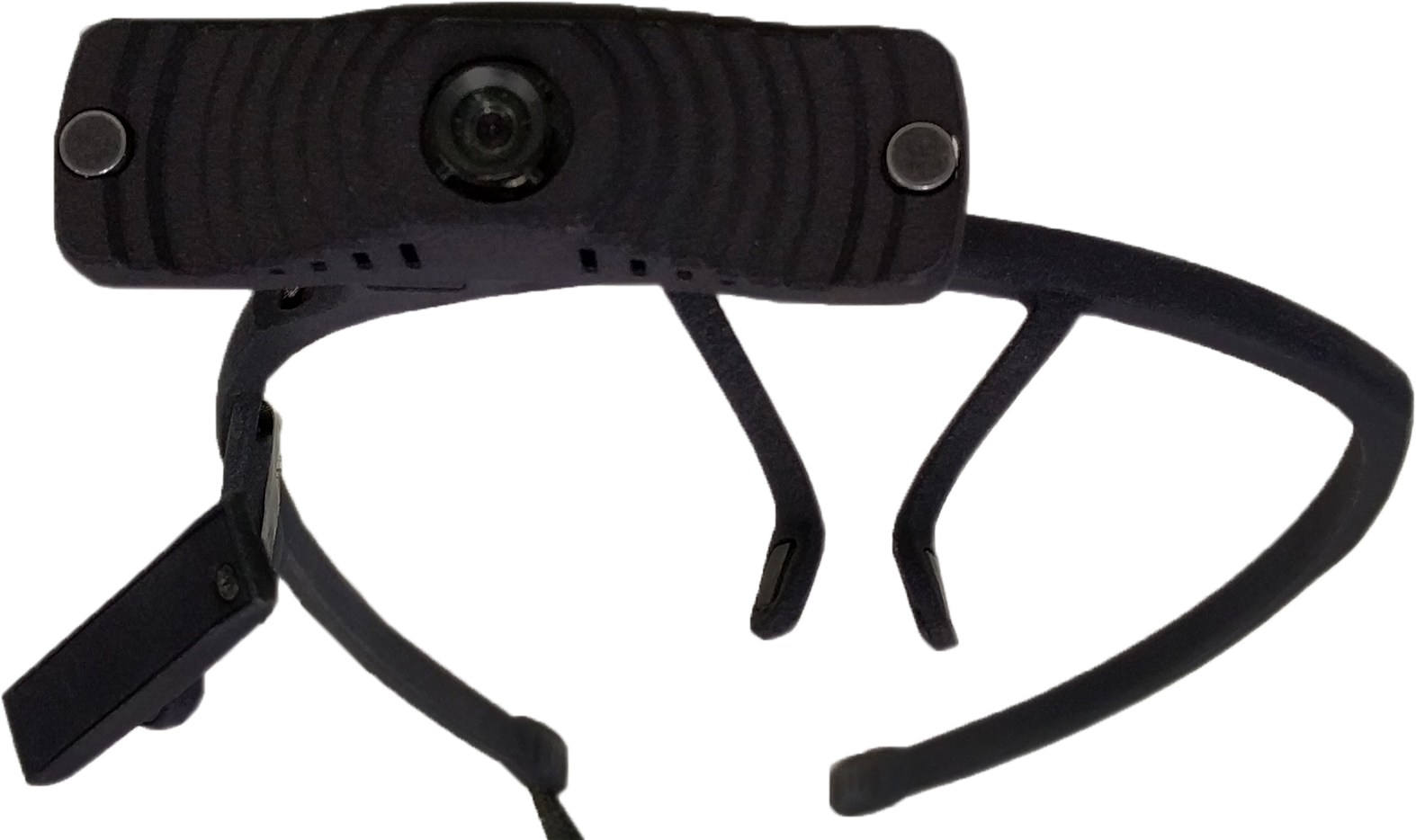}
\label{fig:mobile_eye_tracker}}
\caption[Stationary and mobile eye trackers]{\textbf{(a):} Stationary eye tracker which emits infrared light. This light reflects on the cornea which is captured by the device to estimate the gaze points. \textbf{(b):} Head-worn mobile eye tracker. An infrared camera captures the reflection of infrared light and maps the gaze to a world camera.}
\label{fig:eye_tracker_types}
\vspace{0.4cm}
\end{figure*}

Eye trackers are available in stationary and mobile settings. Stationary eye trackers are traditionally mounted below a display and emit infrared light which is reflected on the eye (see Figure~\ref{fig:stationary_eye_tracker}). This reflection, known as corneal reflection, is captured by the eye tracker to estimate the gaze point on the screen~\cite{poole2006eye}. Additional variables that can be captured include the pupil diameter, gaze point differences between the left or the right eye, and environmental lighting conditions. In contrast, mobile eye trackers resemble goggle-like devices that are worn by users (see Figure~\ref{fig:mobile_eye_tracker}). Instead of putting the eye tracker below a monitor, the infrared emitter can be worn like glasses. Similar to stationary eye tracking, the corneal reflection is captured and the gaze is mapped to the currently perceived view. Mobile eye tracking is not limited to a physical screen and can be used to track the eye gaze throughout mobile scenarios, where the eye gaze is mapped on a recorded video feed~\cite{5586690}. Alternative approaches to mobile eye-tracking include the evaluation of \ac{EOG}. \ac{EOG} describes electrical activity which is created by muscle contractions through eye movements. \ac{EOG} goggles employ electrodes that have contact with the skin and serve as a measurement unit for eye-based activity~\cite{Bulling:2008:YET:1409635.1409647}. The current eye gaze and movement directions can be evaluated by analyzing the emitted electrical activity. The incorporation of mobile eye-tracking has recently taken place in \ac{AR}~\cite{vanderMeulen:2017:WMA:3132272.3132278} or \ac{VR}~\cite{Duchowski:2000:BET:355017.355031} applications and is likely to become a popular usability tool.

We elaborate on well-known eye movements and metrics in the following. Although being far from exhaustive, we focus on metrics that are relevant for the understanding of this thesis, where a comprehensive review is provided by Duchowski~\cite{duchowski2007eye}. \textbf{Fixations} are pauses over regions which attract the attention of users~\cite{Salvucci:2000:IFS:355017.355028}. Thereby, the visual gaze remains over the region for the duration of the fixation. \textbf{Saccades} describe rapid eye movements between fixations~\cite{Salvucci:2000:IFS:355017.355028}. Saccades are very fast and are considered to be among the fastest movements a human body can produce. \textbf{Smooth Pursuits} are smooth eye movements that occur when closely following a moving target on a screen~\cite{vidal2013pursuits}. Calibrating an eye tracker, which is a challenging procedure that disrupts the user experience, is not necessary to detect smooth pursuits. On the contrary, smooth pursuits can be used to calibrate eye trackers~\cite{khamis2016textpursuits}. Finally, the \textbf{Pupil Diameter} can be measured as an indicator for cognitive workload~\cite{pfleging2016model, Gollan:2016:DEC:2968219.2968550}. However, pupil dilation is easily affected by external noise factors such as changing lighting conditions. Therefore, careful interpretations have to be derived when using pupil dilation as a measure for cognitive states.

\subsubsection{Other Physiological Measures for Cognitive Workload}
Alternative measures to sense cognitive workload have emerged besides the use of questionnaires \ac{EEG} and eye tracking. \textbf{\ac{fNIRS}} is a non-invasive modality that uses near-infrared range light between 650\,$nm$ and 1000\,$nm$ to measure concentration changes in \ac{HbO} and \ac{HbR} in the brain~\cite{VILLRINGER1997435}. The concentration levels can be estimated by placing light emitters on the scalp that measure the returning light from an infrared round trip~\cite{HONG201587}. This method is popular for \ac{BCI} applications since \ac{fNIRS} is easy to set up and process~\cite{10.3389/fnhum.2015.00003}. However, \ac{fNIRS} is known to have significant latencies since it takes between five to ten seconds until changes in oxygen are visible in the bloodstream~\cite{doi:10.1111/psyp.12648}. Another non-invasive measure for brain activity is \textbf{\ac{fMRI}}. \ac{fMRI} uses magnetic resonance imaging to analyze blood flow levels~\cite{engel1994fmri}. In contrast to the presented non-invasive methods, invasive methods require electrodes to be placed below the scalp. Methods, such as \textbf{\ac{ECoG}}, require electrodes to be placed below the skull to capture electrical signals~\cite{1642767}. \textbf{\ac{HR}} measures the heartbeat where changes in \ac{HR} and \ac{HRV} are frequently used as an assessment for stress and cognitive workload~\cite{RajendraAcharya2006}. \textbf{\ac{EMG}} measures electrical activity emitted from muscle contractions by body movements. Changes in muscular micro-movements are an indicator of cognitive workload~\cite{baranauskas_towards_2007}. \textbf{\ac{EDA}} describes the activity of sweat glands as an indicator of stress and workload~\cite{boucsein2012electrodermal, Kosch:2019:YSR:3319499.3328230}. Cognitive workload can have an impact on \textbf{haptic interaction}, such as touch or pen input. Changes in force or interaction patterns can be measured when users experience cognitive workload~\cite{Yu:2011:CLE:1943403.1943481}. Cognitive workload is also known to influence \textbf{Speech} regarding lexical density~\cite{Khawaja:2010:ULC:1719970.1720024}, tempo~\cite{Chen:2016:SFD:2993148.2993149}, pauses~\cite{Khawaja:2008:TBY:1517744.1517814}, or voice pitch~\cite{VUKOVIC2019116}. \textbf{Body Temperature} changes have been correlated with cognitive load~\cite{Schaule:2018:ECW:3200905.3191764}. Finally, direct measures of cortisol levels have been used to measure cognitive workload~\cite{Cinaz:2013:MMW:2434601.2434680}. However, cortisol levels require laboratory analysis and are not suitable for real-time analysis.

\section{Assistive Technologies and Systems}
Assistive technologies started as devices that aided persons with disabilities, such as voice recognition systems, braille, hearing aids, wheelchairs, or screen readers. This basic idea of using technologies to enhance human capabilities has attained popularity for the general population. Such assistive systems have become popular tools in home and work settings. Speakers, phones, and TVs have been enriched by technologies that enable them to communicate with users and provide information just-in-time. Assistive systems evolved from the need for explicit interaction (\textit{e.g.}, mobile phones) to devices which require subtle interaction (\textit{e.g.}, speech recognition). Cognitive assistance through in-situ feedback enables the ubiquitous availability of information that has to be memorized for a short amount of time. In particular, assistive systems have proliferated for two stakeholders: people working at production lines with frequently changing assembly batches, and therefore constant changes in assembly instructions, have cognitively benefited from visual in-situ assistance~\cite{funk2017working}. Furthermore, persons with cognitive impairments have shown positive effects by the use of assistive systems in home and work environments~\cite{funk2015using}. We summarize relevant research about the role of assistive systems for both user groups, focusing on lowering cognitive effort, fostering collaboration among peers, and providing assistance that is necessary to accomplish daily tasks. Thereby, we use the term \textit{assistive technology} and \textit{assistive system} synonymously to describe a technical artifact that supports individuals physically and mentally regardless of the presence of a cognitive impairment.

\subsection{Assistive Systems at Workplaces}
Small batch sizes are pivotal to the proliferation of assistive technologies at production lines due to the individual demand for goods. Instead of producing the same part over and over again, workers at production lines have to memorize variations of assembly instructions. The constant memorization of assembly instructions strains the working memory~\cite{7140647, Kosch:2018:ICA:3233739.3229093} which eventually results in a larger number of errors and a decrease in manufacturing quality. Especially new workers that have to learn novel variations of assembly instructions are confronted with a steep learning curve. This has detrimental consequences for the assembly process: production times, error rates, and cognitive workload temporarily increase as new assembly instructions have to be memorized.

Several assistive systems exist to reduce assembly errors and task completion times by keeping the cognitive effort low. McCalla et al.~\cite{mccalla1997peer} presented PHelpS, a system that can convey tasks to other colleagues who are willing to help. PHelpS consists of a repository that algorithmically matches colleagues that are suitable and eager to provide support. Klinker et al.~\cite{Echtler2004} incorporated \ac{AR} into a welding gun, using projection-based and HMD-based \ac{AR} to support workers during their welding tasks. Another approach for assistive computing was presented by Gulevich et al.~\cite{Gurevich:2012:TVA:2207676.2207763}. Their project \textit{TeleAdvisor} uses a camera-projector-system to provide learners with real-time instructions in a TV setup scenario. Gauglitz et al.~\cite{Gauglitz:2012:IPE:2371574.2371610} presented the use of a hand-held device for mobile \ac{AR} assistance. \ac{AR} instructions can be annotated to operate a Boeing 737. Funk et al.~\cite{funk2016motioneap} presented motionEAP, a camera-projection system to assist with in-situ projections. Several studies found that in-situ projections provide a higher assembly efficiency and cognitive alleviation when projections are adapted to the current assembly performance~\cite{funk2017working, funk2016interactive}. However, how to adapt to the individual perception and skill set remains an open question.

\subsection{Assistance for Persons with Cognitive Impairments}
\label{lab:mental_assistance}
Past work has recognized the potential of mental augmentation for persons with cognitive impairments. Today, approximately 20\% of the world's population lives with some level of cognitive impairment~\cite{world2011world}. This trend is expected to increase with an aging society. Sheltered work and living organizations, as well as researchers, have recognized this and begun to evaluate the efficiency of assistive systems.

Levine et al.~\cite{levine1992people} conducted an early review of how persons with cognitive impairments perceive several assistive systems. They present a model for the use of assistive technologies for people with different disabilities. Comprehensive literature reviews which present past work about the usability of assistive systems for persons with cognitive impairments have been presented in the past~\cite{scherer2005assistive}, showing that mental support through assistive technologies is considered as a cognitive prosthesis for persons with cognitive impairments. Funk and Schmidt~\cite{7140647} exploited the potential of several assistive systems for cognitive assistance at workplaces. A series of studies revealed that assistive systems contributed to lower task completion times, fewer errors, and enhanced engagement during an assembly at production lines~\cite{funk2015using,Funk:2015:UIP:2700648.2809853, korn2013potentials}.

Different types of cognitive impairments raise individual requirements for assistive technologies. For example, dementia-related impairments such as Alzheimer's or Parkinson's disease require prospective aids that amplify memory~\cite{BHARUCHA200988}. Using such memory aids helps affected persons to concentrate on the current task or remember important actions that lead to their accomplishment. This has to be implemented carefully: novel technologies are less frequently adopted by individuals with cognitive impairments and are more likely to be abandoned when direct advantages or rewards are not immediately visible~\cite{korn2015design, doi:10.3109/09638289609165907}. Assistive systems are often considered as complicated and likely to cause difficulties in the process of their adoption~\cite{Dawe:2006:DSS:1124772.1124943}. Hence, assistive systems should be designed in a user-centered design process that involves the relevant stakeholders accordingly. The type of cognitive impairments that were present in our studies were trisomy 21, autism, dementia, and general impairments that were legally defined as cognitive disabilities. Such impairments can, for example, influence how information is processed and retained, hence hindering independent living.

\subsection{Ethical Considerations}
\label{lab:ethical_considerations}


Ethical issues mays arise with the collection and use of personal data through the mundane attunement of users, as the use of assistive systems is considered an authorization to implicitly use personal data. With this, ethics regarding participatory \ac{HCI} research that includes persons with cognitive impairments has emerged as a relevant aspect~\cite{10.1145/3313831.3376627}. Recent research in philosophy, as well as ethics, discussed the consequences of systems that can ubiquitously collect data and adjust their services~\cite{Waycott:2015:EEH:2702613.2702655, Waycott:2016:EEH:2851581.2856498}. Research ethics follow the principles of ``autonomy'', ``justice'', ``benevolence'', and ``non-maleficence'', which are cornerstones of ethical theorizing~\cite{beauchamp2001principles}. We highlight the relevant ethical considerations \textit{empowerment}, \textit{autonomy}, \textit{informed consent}, \textit{privacy}, and \textit{social interaction} in the following that have been considered in the previous research.

\textit{Empowerment} is considered as an intuitive aspect for assistive computing: through assistive systems, people are physically and cognitively supported to live autonomously and accomplish tasks with less or without cognitive effort. This is achieved by technology compensating for certain forms of physical or cognitive demand. For instance, an assistive system could function as an externalized memory for a person with a memory weakness. The assistive system is assumed to work as cognitive support. But from a philosophical point of view, this raises interesting questions about the boundaries of the mind and also the ethical consideration of whether assistive systems could be paternalizing their users by nudging them into certain behavioral patterns~\cite{doi:10.1111/1467-8284.00096}.

There are further reasons to be hesitant about subscribing to \textit{empowerment}. For instance, the underlying concept of cognitive disabilities presupposes a biomedical understanding of support, according to which a person has a situational or permanent impairment that is physically or cognitively compensated by assistive technologies~\cite{Boorse2010, hacking1990taming}. For persons with cognitive impairments, such a model tends to overlook the role of social environments in shaping or perpetuating disabilities, as well as the different forms of discrimination persons with cognitive impairments are confronted with. That notwithstanding, while there might be clear limits to how much empowerment might be achieved through assistive systems, its benefits for helping people with cognitive impairments in leading better lives cannot be denied.

\textit{Autonomy} describes the capacity of a person to exhibit self-governance~\cite{buss2002personal}. What that means is that if a person acts, the acting is motivated by personal reasons. An autonomous action might be contrasted with cases where a person is either coerced or manipulated to act in a certain way. The necessary conditions for being autonomous are much debated in the relevant philosophical literature~\cite{arpaly2002unprincipled, doi:10.1086/426304}, nevertheless it is fair to assume that, to be autonomous, one must possess a certain set of psychological capacities such as the ability to self-reflect. This includes that a person's beliefs and intentions must be coherent and stable over time. In addition to that, an agent must possess higher-order evaluative judgments on what matters to them~\cite{arpaly2002unprincipled, Frankfurt1988}. For instance, if someone is persuaded by a fitness app to go for a run, the relevant action being autonomous hinges on whether running is an activity that the person value. In the context of assistive systems, \textit{autonomy} as an ethical criterion is supposed to ensure that the person and not the technology exhibits control~\cite{doi:10.1086/426304}. There are some challenges for using \textit{autonomy} as an ethical criterion within the context of people with cognitive impairments~\cite{doi:10.1111/hypa.12119}. The conditions for \textit{autonomy} are mentally and physically highly demanding, increasing the importance of \textit{autonomy} as an ethical principle. For example, even if a person does not meet the requirements for full-fledged \textit{autonomy}, they might still have certain beliefs, desires, and values that express what matters to them. Therefore, one ethical criterion for the development of assistive systems or technologies has been that these do not interfere with the basic values and desires of their users.

The problem of cognitive demand also applies to \textit{informed consent} and \textit{privacy}~\cite{Eyalmedethics-2012-100490, miller2010ethics}. Assistive systems can collect fine-grained data on a person and, therefore, might infringe upon their \textit{privacy}. Whereas strict guidelines have been established, governing the collection and storage of personal data, it is still a problem to sensitize users to issues related to informational \textit{privacy}~\cite{nissenbaum2009privacy}. One factor is the high degree of abstraction regarding \textit{privacy} issues, where many users find it difficult to grasp the content of \textit{informed consent} and terms of service documents or to appreciate their relevance~\cite{brey2005freedom, van2008information}. While many decisions concerning welfare are deferred to caretakers, we still consider it as crucial that a person with cognitive impairments has authority over their interests~\cite{rossler2018value}. Thus, we are currently experimenting with developing easy to understand, \textit{informed consent} documents in regards to \textit{privacy} issues that make use of simple language and are accompanied by informational videos. Admittedly, this is still an area where a lot of empirical testing is required~\cite{Barocas:2014:BDE:2684442.2668897, doi:10.1177/2053951717743530}.

We need to consider how assistive systems affect the \textit{social aspects} of the lives of persons with cognitive impairments. While assistive systems have the potential to improve the life quality for persons with cognitive impairments, they should foster social and communal activities~\cite{doi:10.1258/135763307780096195, Coeckelbergh2010, doi:10.1080/14639230410001684387}, which eventually results in better care work through freeing other resources of caretakers. Consequently, it is one of the critical ethical aims to ensure that not merely the technology, but also the social settings of the care- and tenant-relationship are designed in a way which promotes the welfare of users.

\part{Human Cognition\label{part:human_cognition}}



\chapter{Opportune Cognitive Augmentation}
\label{ch:smart_kitchen_requirements}

\epigraph{\textit{Today we serve technology. We need to reverse the machine-centered point of view and turn it into a person-centered point of view: Technology should serve us.}}{Donald A. Norman}







Cognitive workload, although different from the individual constitution and individual amount of required cognitive resources, is ubiquitously present when persons work on chores. The relationship between exerted cognitive resources and the detection of the need for cognitive assistance is an important factor that has to be considered when designing workload-aware interfaces. This poses the following question, assuming that effective workload measures are integrated into the environment: \textit{when} is the right time to intervene with cognitive assistance?

The vision of workload-aware user interfaces requires careful investigations of design requirements and implications for the respective user group. Persons with cognitive impairments, who are living in sheltered living facilities, are a user group that requires cognitive assistance to achieve basic life skills. Such cognitive impairments include trisomy 21, autism, dementia, and general impairments that are legally defined as cognitive disabilities. As persons with cognitive impairments receive ``\textit{handcrafted}'' ubiquitous assistance  (\textit{e.g.} through family members or trained labor) during daily chores, their barrier to express the need for cognitive assistance is lowered. This supports the social intention to support persons with cognitive impairments by researching workload-aware assistive technologies since the need for cognitive assistance is expressed early to family members or caretakers. This chapter focuses on persons with cognitive impairments to explicitly detect when manual mental augmentation through a supervising person is desired: these cues are potential entry points for workload-aware systems. Cognitive impairments can impact someone's ability to complete traditional activities of daily living, such as cooking, bathing, or acquiring groceries~\cite{perneczky2006complex}. Numerous government and volunteer-driven organizations exist to provide specialized training to people with cognitive impairments as they learn these independent living skills. As average lifespans increase, the number of people affected temporarily or permanently by cognitive impairment will continue to rise~\cite{JGS:JGS4752}, and this training will be in higher demand. The world report on disability~\cite{world2011world} approximately assesses 20\% of the human population a cognitive impairment where the numbers are expected to rise.

In continental Europe, one venue for learning independent living skills are sheltered living facilities, which are housing communities where tenants with cognitive impairments live together with supervision from at least one caretaker. In these facilities, 20 to 30 people share living quarters and are coordinated in learning daily living skills by expert staff members and teams of volunteers. The ultimate goal of a sheltered living facility is to teach the inhabitants to live independently with other tenants who help and support each other. The organization supports tenants in moving to a shared apartment, where four to six people help each other without any caretaker supervision. Consequently, the facilities provide training in both the acquisition of life skills and the social organization of a community of tenants.

Communal cooking is one activity that integrates both of these aspects of living in sheltered housing facilities. Tenants learn to complete their kitchen duties to benefit other members of the community. In sheltered housing, instructors provide the group with specialized supervision on safe cooking methods and techniques to cook collaboratively. However, due to worker shortages in this field, individual instructions are rarely possible. Further, assistance in cooking may still be needed by tenants who have recently transitioned to an independent living facility.

Contextualized assistance (\textit{e.g.}, delivered through displays~\cite{funk2015using}, augmented reality~\cite{funk2015comparing}, or context-aware support systems~\cite{7427601}) has shown to be effective in helping people with cognitive impairments perform chores independently. While the workload from caretakers may be alleviated when using digital assistants in communal cooking activities, support and learning can be provided independently from the sheltered living facility. In this chapter, we describe current training practices in sheltered living facilities. Furthermore, we explore potential challenges and benefits of providing \emph{communal} contextual instruction and support for the development of independent living skills. Using contextual observation of communal cooking sessions in sheltered living facilities, we conduct supplementary interviews with staff members and volunteers. We chart the opportunities, trade-offs, and constraints involved in the design of assistive systems for communal kitchens for persons with cognitive impairments. In this chapter, we report a qualitative inquiry of current communal cooking practices in sheltered establishments. Based on the qualitative inquiries, we describe four themes regarding the design opportunities and constraints in a communal kitchen for persons with cognitive impairments. Finally, we conclude with five implications for assistive technologies supporting cooking in sheltered housing. The research question we answer is:

\begin{itemize}
\renewcommand{\labelitemi}{$\rightarrow$}
\item \textbf{RQ1}: What are the design requirements for systems that provide opportune cognitive augmentation?
\end{itemize}

\begin{center}
\begin{GrayBox}
\centering
\parbox{0.98\textwidth}{
\emph{This section is based on the following publication:}
\vspace{3mm}
\publicationsbegin
\item \bibentry{Kosch:2018:SKP:3173574.3173845}
\publicationsend
}
\end{GrayBox}
\end{center}

\section{Related Work}
The work presented in this chapter bases on the use of assistive technologies by persons with cognitive impairments as outlined in Section~\ref{lab:mental_assistance}. This section provides an overview of technologies supporting persons with cognitive impairments during everyday tasks and describes current research which aims to provide contextual assistance in kitchens.

\subsection{Accessibility through Assistive Systems}
Assistive technologies have proliferated into the home environments of people with mental and physical disabilities. Users of assistive technologies are ubiquitously surrounded by technologies that monitor them. This includes smartphones, tablets, ambient computers, or wearable devices, all of which have become increasingly popular. Assistive technologies intend to provide mental assistance for persons with cognitive impairments. Past research has investigated the design of assistive technologies for persons with cognitive impairments and elderly people that provide mental assistance, and to this end, reduce the complexity of the underlying system.

Bouchard et al.~\cite{Bouchard2006} developed a plan recognition framework in smart homes for people suffering from dementia. Using microsensors, the framework analyses previous actions to predict the original intention of the person. Pollack et al.~\cite{Pollack2005} define three goals for assistive technologies when used by older people with cognitive impairments: providing assurance for elderly, use technologies to compensate for impairment, assess the users' status. Furthermore, three different types of assistive systems are defined: assurance, compensation, and assessment systems. Since smart environments need to be equipped with microsensors to measure contextual parameters of their surroundings~\cite{LiJiang2004}, ethical implications for assistance systems in home environments were to be defined~\cite{stip2005environmental}. Assistive technologies provide several benefits, such as workload reduction, socialization, or care delivery. However, the use of such technology has ethical ramifications and considerations (see Section~\ref{lab:ethical_considerations} for a comprehensive elaboration). To enable independent living for the people with cognitive impairments, a smart home system leveraging several sensors, such as infrared motion sensors, microphone arrays, and accelerometers, was used to sense contextual data~\cite{arcelus2007integration}. The collected data was used to train artificial intelligence, which provides adaptive assistance whenever required. However, challenges such as preserving privacy~\cite{courtney2008privacy, courtney2008needing} and acceptance by senior users~\cite{brownsell2000community, demiris2008senior} have to be considered. Mihailidis et al.~\cite{Mihailidis2008} present the concept of an assistive system that supports older adults with dementia during regular daily routine tasks. Specifically, this includes a study where washing hands were trained via an audio-based system.

Within workplaces in industrial areas, augmented reality was used to provide cognitive alleviation for workers with cognitive impairments in workplaces~\cite{Chang:2015:AOV:2700648.2811354} and assembly lines~\cite{funk2015comparing}. They found that by projecting in-situ information about the current assembly step, increased efficiency concerning time and number of errors would be achieved, and concluded that assistive systems at workplaces foster learning by skill transfer while releasing cognitive resources at the same time~\cite{7140647, funk2015using}.
Gamification can be incorporated into assistive systems to maintain or increase motivation~\cite{korn2015design, korn2015towards, Korn:2015:TGI:2774225.2774834}.

Research has also concerned notifying persons about events, such as warning of dangers or sending reminders. In previous research, we~\cite{Kosch:2017:OSF:3123024.3124395, Kosch:2016:CTA:2982142.2982157} investigated how events of interest can be communicated efficiently with persons with cognitive impairments within workplaces. By comparing visual, auditory, and tactile feedback, their findings show that visual in-situ cues are perceived efficiently. A survey exploring how notifications in smart homes can be communicated was carried out by Voit et al.~\cite{Voit2016}. Leveraging an online survey, they investigated suitable devices and locations displaying notifications. Their results showed that smart home-related notifications should be received by mobile devices which would be easily perceivable when worn on the body. Wiehr et al.~\cite{Wiehr2016} define how the increasing number of notifications raises challenges regarding the design, implementation, and psychological factors of users.

\subsection{Assistive Technologies in Kitchens}
Cooking represents an important communal and social activity for people with cognitive impairments. Cooking entails a starting point (\textit{e.g.}, starting to prepare the ingredients), a structured or unstructured processing pipeline (\textit{e.g.}, cooking using a fixed or unstructured procedure), and an endpoint that facilitates the cooking result (\textit{i.e.}, the cooked meal). Previous research has researched smart kitchens extensively as it revealed challenges and opportunities that are generalized to other assistive technologies.

Thereby, smart kitchens have been the focus of Blasco et al.~\cite{Blasco2013}. They developed and assessed smart kitchens for older adults and evaluated the simulation of specific situations, such as making dinner or washing up. Besides assisting in kitchens, calorie and nutrition-aware contextual cooking plans can be provided ~\cite{Chi2008, Chi2007}. By displaying information about nutrition during cooking, healthier ingredients can be chosen by cooks. Hashimoto et al.~\cite{Hashimoto2008} designed algorithms for smart kitchens, which recognize the users' cooking actions and food material. Since recipes play an important role in cooking, Schneider et al.~\cite{Schneider2007} developed a semantic cookbook. The semantic cookbook is a system enabling different parties to share their recipes among their smart kitchens to display it on an output device. As most handwritten recipes are passed down by previous generations, digital recordings can ensure the continued existence of recipes and can be easily shared. Cooking can also represent a method for social communication and interaction. Therefore, Terrenghi et al.~\cite{Terrenghi2007} present the "Living Cookbook". In their work, cooking experiences are recorded to educate others, practice cooking techniques, or share cooking experiences. An evaluation of the "Living Cookbook" shows that its use increased motivation and improved social communication. Hooper et al.~\cite{Hooper:2012:FKT:2370216.2370246} investigates how the new material can be learned within instrumentalized environments to support task-based learning. During their research, the efficiency and design space of learning new languages within regular cooking tasks were evaluated. Bonanni et al.~\cite{Bonanni2005} evaluate several augmented kitchen interfaces regarding their usability. Their studies focus on usability as well as interfaces that are not demanding in terms of attention and cognitive workload. Miyawaki et al.~\cite{Miyawaki:2009:CSS:1630995.1631004} prototyped a kitchen for people with higher brain dysfunction. Olivier et al.~\cite{Olivier:2009:AKD:1579114.1579161} presents a lab-based replication of a smart kitchen, where designers can evaluate novel solutions. Also, Scheible et al.~\cite{Scheible:2016:SME:2991561.2998471} show how social and emotional components can be incorporated into the cooking process.

\subsection{Collaborative Accessibility}
People with disabilities often \emph{co-construct accessibility} in spaces alongside the other disabled or able-bodied peers who inhabit them \cite{branham2015collaborative}. This process of co-construction involves a large amount of collaboration among individuals as they stage environments to be more accessible; divide tasks and responsibilities based on ability, and intervene or request assistance for tasks that are not accessible.

Branham and Kane~\cite{branham2015collaborative} described the collaborative accessibility practices of blind and sighted domestic partners in their home settings. Within these close partnerships, partners can set up plans for managing household tasks, object placement, and requests for assistance. However, this collaboration to make accessible environments can be less successful in work settings, where an intimate understanding of the disability is not possessed by a person with a disability or co-workers~\cite{branham2015invisible}; or in public spaces where accessibility cannot be co-constructed in advance~\cite{williams2014just}.

The study of this collaboration work has been rare among people with cognitive impairments. In our observations in this study, we examined the collaboration between individual tenants with cognitive impairments and the neurotypical\footnote{We use the term ``neurotypical'' here to describe individuals without any measurable level of cognitive impairment. This term is often used to differentiate people with a neurological disorder, like individuals on the Autism spectrum or with cognitive impairments, from individuals without a neurological disorder.} instructors who supervised them; however, we also discuss opportunities for collaboration between multiple tenants with cognitive impairments without supervision.

Overall, previous work has investigated effort into the development, design, and evaluation of assistive systems. However, the design space of smart kitchens for people with cognitive impairments has not been considered yet in related research. Through the execution of qualitative contextual inquiries with caretakers and observations of tenants in sheltered living organizations for people with cognitive disabilities, we close this gap by charting the design space for assistive technologies in smart kitchens for users with cognitive impairments.

\section{Methods}
Sheltered living facilities offer people with cognitive impairments assistance with learning everyday tasks, including cooking. The main goal is to teach elementary skills in a methodical way that can be reapplied in independent living environments. When tenants make progress in applying their skills without caretaker intervention in the facility, they can move to live more independently in houses shared with a small number of other people with cognitive impairments. The research for this project was conducted in collaboration with a sheltered living organization in Germany which operates both a sheltered living facility and independent houses for tenants who have completed their training.

\subsection{Context: Communal Kitchens}
\begin{figure}[t!]
\centering
\includegraphics[width=0.5\columnwidth]{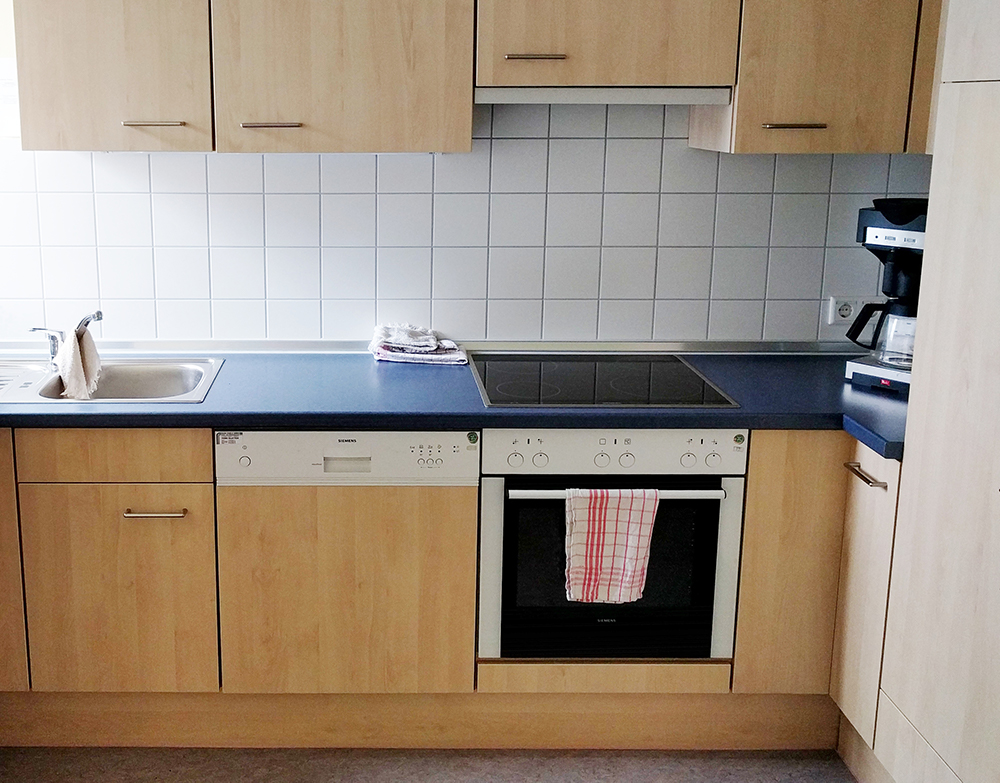}
\caption[A kitchen in a sheltered living facility]{A kitchen in a sheltered living facility, used by the tenants with cognitive impairments and the caretakers. Tenants overtake practiced tasks which include cutting ingredients or laying the table. Dangerous tasks, such as the operation of the oven or stove, are accomplished by the caretakers.}
\label{fig:kitchen}
\end{figure}

The kitchen in the sheltered living facility and independent living homes use regular components necessary to cook a meal, such as an oven, stove, tabletop, and refrigerator. The kitchen from the sheltered living facility is shown in Figure~\ref{fig:kitchen}.

Within the sheltered housing, tenants and caretakers cook dinner every weekday evening. If enough inhabitants stay at the sheltered housing during weekends, both lunch and dinner are cooked. The way supervision is provided depends on whether the inhabitants are living in sheltered facilities or residing in their own independent homes.

Major supervision during cooking is mostly provided by a caretaker within the sheltered housing. A sheltered living facility comprises 20 to 30 inhabitants depending on its size. Groups of four to six inhabitants cook together with at least two caretakers who explain the course of cooking in the beginning, distribute tasks, and make sure that the participants avoid serious injuries. Currently, no standardized assistive technology supporting caretakers is enrolled in such facilities. Dangerous tasks, such as operating the oven, are only performed by caretakers.

Minor supervision is necessary for persons with cognitive impairments living in independent housing. This independent place is managed by the sheltered living organization but spatially separated from the sheltered living facility. Such independent living places comprise between four and six inhabitants living together. This principle enables maintaining social settings where people can help each other. For cooking, the same rules apply for sheltered housing. However, only a single caretaker is present to avoid possible injuries and provide advice on demand. The caretaker also assesses the social setting, cognitive development, and skill progress of the inhabitants.

\subsection{Data Collection}
Data for this study was collected through two observation sessions of communal cooking in a sheltered living facility and an independent living facility in Germany and was supplemented by semi-structured interviews with staff members from those organizations. The tenants who participated in the cooking sessions ranged between 30 and 49 years old. Ethical approval for all components of the study was given by the sheltered living organization and the German Federal Ministry of Education and Research according to institutional guidelines.

We chose to conduct observations rather than interspersing interview questions throughout the task as in a contextual inquiry, due to the level of cognitive impairment among our participants and the time-based nature of the cooking tasks being taught. Cognitive impairment impacts executive functioning, like task planning and memory. Interruptions during a time-based task make it more difficult for a person with a cognitive impairment to hold onto their original intention, and less likely to return to the task after being distracted~\cite{troyer2007memory}. At the direction of the caretakers at the facility, we limited interaction with the tenants during the cooking tasks and used follow-up interviews with staff members to supplement our understanding of the observed data. The contextual inquiry was thus carried out with the caretakers, which have been observed and interviewed during this study~\cite{blandford2016qualitative}. Participants were monetary  compensated through funding from a joint project.

\subsection{Observations}
We conducted two observations in communal kitchens as tenants prepared meals. We observed communal cooking processes among four to eight participants - the number of tenants in the kitchen at one time varied as they arrived and left the kitchen at their discretion, but a total of twelve participants were observed between the two sessions. Two rounds of observation were performed --- one of a \emph{supervised cooking session} in the sheltered living facility, and one of a \emph{unsupervised cooking session} in an independent household managed by the same organization. Each observation session lasted at least one hour. A caretaker was present in each kitchen at all times during our observations but did not intervene in the cooking activities in the unsupervised session.

In the supervised cooking session, ten participants we observed were born with a cognitive impairment limiting their ability to understand and process information. In addition to this, two other participants were affected by light motoric impairments. All participants we observed in the unsupervised cooking session were affected by a cognitive impairment limiting their ability to understand and process information. None of the participants were affected by sensory impairments or dementia.

Our observations were documented through notes and photographs. We took pictures of the cooking environments before the tenants began cooking to limit task disruptions. During the observation, two researchers took detailed field notes on the session. Observations focused on how the cooking process was organized and the interactions between the tenants. In the observation of the structured cooking session, we also observed how the caretakers provided instruction to the tenants and how they directed their attention. We noted how physical artifacts in the kitchens were used, arranged and assigned by the tenants. In each of the two observations, we remained in the kitchen for the entire duration of the cooking process, from before the tenants started arriving in the kitchen to when they left the room after having eaten. We have not interacted with the tenants directly as we wanted to avert confusion and avoid disruptions during the regular cooking procedure.

\subsection{Interviews}
After the observations, we constructed specific semi-structured interview protocols for the caretakers to clarify and refine our understanding of the data collected during the cooking sessions. This enabled us to obtain information that could not be disclosed in the presence of the tenants and have the full attention of the caretaker. Table~\ref{tab:demographics} summarizes the participants' role and working experience.
The primary focus of these interviews was to better understand the work processes and roles assigned involved in the structured cooking sessions, such as the reasons for distributing particular tasks, rules for using appliances, and methods for distributing the workload among tenants. We were also able to inquire further about the role of cooking in the community of tenants in both structured and unstructured settings.

\begin{table}
\centering
\begin{tabular}{c l c l}
\toprule
\textbf{Participant} & \hspace{11mm} \textbf{Role} & \textbf{Age} & \textbf{Experience}\\
\midrule
P1 & Housing officer & 33 & 15 years\\
P2 & Sheltered caretaker & 49 & 30 years \\
P3 & Sheltered caretaker & 20 & 0.5 years \\
P4 & Independent caretaker & 46 & 6 years \\
\bottomrule
\end{tabular}
\caption[Demographic data and working experience of the interviewed participants]{Demographic data and working experience of the interviewed participants.}
\label{tab:demographics}
\vspace{-1em}
\end{table}

The interviews comprised one interview with the housing officer of a sheltered living institution (a person whose full-time job is to manage the facilities), two interviews with volunteer caretakers working at a sheltered housing facility, and one caretaker responsible for an independent group of persons living together. Overall, 5:45 hours of recordings were collected. The mean age ranged from 20 to 49 years (M $= 37$, SD $= 13.29$) and working experience of caretakers ranged from six months to 30 years (M $= 12.88$, SD $= 12.87$). P1, who is a housing officer and responsible for multiple living facilities, provided us with most data about the observed cooking processes as well as general information. The interviews from P2, P3, and P4 were largely used to gain new insights and to confirm the statements provided by the housing officer.

\subsection{Data Analysis}
We conducted a qualitative analysis of our observation notes and interview transcripts to understand the constraints and opportunities for communal cooking experiences for tenants. We used a team consisting of two coders working with the Atlas.ti software package. We conducted an initial open coding of 25\% of the data by both researchers. Afterward, a coding tree was established through iterative meetings. We then coded the rest of the data with the agreed-upon set of codes. A final meeting was conducted where we grouped codes to establish the four emergent themes: \textsc{work organization}, \textsc{community}, \textsc{supervision}, and \textsc{practicalities}.

\section{Findings}
In this section, we present four themes that emerged from the analysis of our data set. We present each theme and our understanding of the constraints, tradeoffs, and opportunities involved. The presented quotes have been transcribed from their original German into English.

\subsection{Work Organization}
Cooking was seen as a key activity in the learning process required for gaining a higher degree of independence. In our observations, both tenants and supervisors gave heavy attention and concern to how tasks were divided within the cooking group. Much of this task division within the supervised cooking session was done by the supervisors in advance, and tasks were created both for the tenants cooking that day and for the supervisor on duty. The supervisors communicated clearly which tasks were intended to be performed by tenants and which were reserved for supervisors:
\begin{quotation}
\textit{"Tasks are usually divided between tenants and supervisors. The parts were carried out until completed; nobody stops suddenly within their task." (P1)}
\end{quotation}
One supervisor showed us how a weekly cooking plan helped organize the communal activities in the group. The cooking duties were distributed every week within a weekly meeting where supervisors and tenants took part. The cooking responsibilities were discussed verbally. Tenants were informed about their responsibilities by the assigned supervisors.

These plans served as a valuable organization and accountability tools on tenants' journey to greater domestic independence. One officer explained that managing schedules and making sure all tenants were involved in cooking served not only practical purposes but also worked as a means of assuring participation in this important activity of daily living on days when tenants were not intrinsically motivated:
\begin{quotation}
\textit{"Many tenants do not feel like they want to cook\ldots 
A cooking plan is created, which forces everyone to cook regularly. Usually, we find out whether there are any important appointments before a cooking schedule is created." (P1)}
\end{quotation}
The supervisors assumed responsibility for developing these plans, leveraging their knowledge of tenants' current ability levels and other commitments to make the plans maximally effective in the limited time available.

Coordination in the kitchen was necessary, as the kitchens constituted the only common rooms in the facilities we visited. We observed tenants constantly entering and exiting the kitchens. Some tenants who were not involved in cooking tasks that day came to socialize or observe the cooking tenants. Other tenants who were assigned to specific cooking duties might leave their task briefly, and return to them after a short walk. The diverse patterns of the tenants within the kitchen at any time meant that supervisors had to engage in constant monitoring of task completion to ensure that coordination was occurring.

The division of labor among stakeholders was emphasized not only in terms of which tenants performed which kitchen duties, but also concerning assigning paid personnel and volunteers to supervision:
\begin{quotation}
\textit{"There is a duty roster for interns and employees. This plan specifies, how many supervisors have to be there during cooking." (P2)}
\end{quotation}
External factors also affected the way kitchen work was organized. For example, drinks provided by an external supplier arrived on a fixed schedule that also needed to be included in the kitchen plans:
\begin{quotation}
\textit{"Drinks are delivered every week on Wednesday. The group buys groceries by themselves every Tuesday and Friday. On Wednesday, we buy groceries together. Usually, one supervisor is associated with each tenant." (P1)}
\end{quotation}
These considerations show a considerable amount of coordination that is required for the supervisors and tenants in the supervised cooking environments. While the meta-level scheduling work was performed solely by the supervisors, individual cooking sessions and supportive tasks were more collaborative efforts between volunteers and tenants.

Overall, we observed the multiple aspects of the division of labor, time management, and logistics, all of which significantly affected the way the cooking was enacted.

\subsection{Community}
The social dynamics of the sheltered living community were important to both tenants and to the supervisors, who saw these social relationships as valuable learning experiences. Aspects of communication and collaboration between tenants and supervisors were often observed and mentioned in our data set. Tenants interacted with each other while cooking and the supervisors reported that due to social expectations, tenants did not eat alone and instead joined communal activities:
\begin{quotation}
\textit{"Nobody would cook together and then decide to eat alone in his room." (P1)}
\end{quotation}
Tenants who were not assigned cooking tasks enjoyed observing the cooking tenants performing their duties and would choose to join the kitchen and monitor the cooks as they worked through their tasks.

While the tenants did not initiate social activities outside of the facility, the supervisors often planned these events as an important training experience for gaining independence:
\begin{quotation}
\textit{"We go to the cinema or do other activities on weekends. But this depends on the initiative of the supervisors." (P1)}
\end{quotation}
These independence-building activities helped to foster community among the tenants and supervisors, and could be useful in preparing tenants to navigate social relationships in the more independent housing options.

Despite this emergent community among the tenants and supervisors, there was potential for conflicts to occur connected with the execution of the kitchen work. Supervisors, in particular, were concerned about these potential conflicts and helping the tenants resolve these conflicts amicably. One supervisor reflected on how some tenants were particularly concerned about the alignment of cutlery on the kitchen table, indicating that tenants with different levels of ability might experience frustration with each other:
\begin{quotation}
\textit{"How exactly cutlery is placed depends on the person. Some execute their tasks very accurately, while others do their tasks in a rudimentary way. We often experience conflicts between inhabitants because of that." (P1)}
\end{quotation}
Similarly, another supervisor was concerned about how critique could be communicated gently, and how it affected tenants differently:
\begin{quotation}
\textit{"Some have problems accepting critique and answer with statements like "This has been always been done this way." (P2)}
\end{quotation}
These interpersonal disagreements are likely to arise in more independent kitchens as well, so supervisors desired more effective ways to resolve these conflicts.

Within the community of tenants, different tenants assumed unique roles in the cooking process. This often was at odds with the goals of the supervisors, who wanted tenants to participate in cooking equally as part of their rehabilitation. Instead, more motivated tenants often tried to monopolize the cooking positions:
\begin{quotation}
\textit{"Our tenants are very motivated to shop for the ingredients. However, the motivation regarding cooking is different per person. Surprisingly, it is always the same persons who volunteer for kitchen service." (P3)}
\end{quotation}
Some tenants were not intrinsically motivated to do the complex work required in the kitchen. In these instances, social factors were often mentioned by supervisors as key to increasing their motivation. Enabling tenants to express approval of each other's actions and their role in the community was seen as extremely valuable to the group dynamics:
\begin{quotation}
\textit{"Approval of other tenants is important; much more important than approval from the supervisors." (P2)}
\end{quotation}

\subsection{Supervision}
The nature of the supervision provided to the tenants was a unique aspect of work in a sheltered housing facility. Supervisors explained that maintaining engagement and limiting frustration for tenants was one of the most important purposes of their work and one of the primary goals of the more structured cooking sessions. They conducted this supervision work both pre-emptively (while doing \textsc{Work Organization}, as described above) and by being available to help or intervene during the tasks themselves:
\begin{quotation}
\textit{"We sort from the beginning inhabitants for specific tasks; for example if it is clear that a particular person should not cut hard things instead of vegetables and salad. They call for help if something is too difficult for them. We have to avoid frustration and confusion if inhabitants are not able to do their task." (P1)}
\end{quotation}
Tenants were able to request help if needed, and the amount of supervision needed varied from tenant to tenant. In general, there was a high emphasis on tenant independence, and the details in the instructions provided to the tenants were left sparse to allow them to develop skills on their own. In their role as supervisors, the staff was most concerned that a given individual would be able to generally complete kitchen actions and that they pay attention to learning precise techniques to a sufficient degree:
\begin{quotation}
\textit{"Most people can cut. We do not complain if something is cut too thick or thin. The most important thing is that it is cut and can be cooked." (P1)}
\end{quotation}
Guidance from supervisors was primarily required when tenants made mistakes or grew confused. When this occurred, both supervisors and other tenants stepped into a supervisory role to help the tenants understand the issues with their work. Feedback from all members of the group was often provided. Tenants who intervened were often concerned about the process implications of an error (\textit{e.g.}, if a mistake would impact the overall meal) or highlighted safety concerns when they noticed unsafe behavior. For supervisors, helping the tenant understand what went wrong in the task execution was the key educational aspect:
\begin{quotation}
\textit{"This [intervention] is very individual. It always depends on the mistake they make. Some do what we tell them and they do not get confused while doing their kitchen tasks and accept the critique." (P2)}
\end{quotation}
Certain complex aspects of cooking required constant verification from the supervisors and fewer opportunities for independence. For these tasks, the supervisors often overstepped their role as supervisors to join the cooking process directly. For example, the spice cupboard required careful instruction and often the supervisors would administer spices without any help:
\begin{quotation}
\textit{"Managing spices is an interesting question: Who dispenses the spices? Our experience shows that adding spices is a fine-motor task requiring experience to not, for instance, over salt a meal. Spicing is mostly done by supervisors." (P1)}
\end{quotation}
However, some tenants had stronger fine motor skills than others and performed spicing in the supervised session with direct instructor supervision.

This intervention was in part based on tenants' skill limitations and was an integrated part of the supervision. Supervisors needed to bear in mind the specific tenants' perception of the cooking experience in each session, and be aware of stimuli that the tenants could not always process:
\begin{quotation}
\textit{"I do not think that persons with cognitive impairments are able to determine when a meal is cooked or not." (P1)}
\end{quotation}
Finally, supervisors recognized that more instructions could be provided to the tenants and they could benefit from more attention to improve their long-term educational outcomes within the kitchens. However, the number of personnel available for communal cooking was limited:
\begin{quotation}
\textit{"It is a communal kitchen. They have their room, but in principle, they should be able to cook for themselves. The process of learning how to cook independently in one's own house should be learned here." (P1)}
\end{quotation}

\subsection{Practicalities}
Our last theme addresses the practical concerns involved in cooking activities. A core concern is maintaining the safety of the tenants, who may enter the facility with little knowledge about safe cooking procedures. As many kitchen tools are potentially dangerous and electrical appliances were in use, supervisors were especially careful to monitor tenants as they performed potentially dangerous actions, but had to make crucial tradeoffs between safety concerns and independence:
\begin{quotation}
\textit{"Knives, blender, everything that can cause potential injuries is dangerous. We try to avoid every chance where injuries could happen. But this is not right since they will not learn how to cut or blend\ldots however, they \emph{have} to learn how to cut or blend [to cook independently]. Therefore, negative experiences are necessary to learn what causes injuries. But our primary goal is to avoid these injuries as much as we can." (P1)}
\end{quotation}
Thus, the staff made risk assessments to determine if the use of particular skills outweighed their danger. Further, some supervisors mentioned that they prioritized the use of devices that offered a greater degree of safety, such as the microwave oven:
\begin{quotation}
\textit{"The microwave offers comfort and security. Food is just put in and the microwave tells them when the food is ready." (P1)}
\end{quotation}
Some tenants were aware of the safety risks of certain tools, and would monitor each other and repeatedly provide feedback when they observed safety violations.

Another facet of the cooking experience that we observed was making sure that the process was hygienic. Hygiene rules were imposed on the facilities by legal regulations, and needed to be followed and enforced strictly by the staff:
\begin{quotation}
\textit{"We provide a hygienic plan. The housekeeping management organizes hygiene schooling. Furthermore, hygiene is taught to the tenants over time, for instance, how to wash hands, disinfect them, and how to wear gloves. Additionally, supervisors know the hygiene-related habits of tenants. This means that the individual hygiene is very diverse." (P1)}
\end{quotation}
The medical condition of the tenants also affected the cooking experience. Distributing medicine was an integrated part of the communal cooking process, and the supervisors distributed pre-mixed dosages of pills in color-coded containers. The choice of the menu was also affected by health concerns:
\begin{quotation}
\textit{"We serve salad as a side dish because of health reasons to ensure enough vitamins in meals. If we ask what they want to eat, they would prefer sausages, fried chicken, or fries all the time." (P1)}
\end{quotation}
Lastly, participants were aware that cooking was connected with other activities, such as setting the table, and different tenants would perform these tasks in parallel with the cooking tasks. While these activities required less help, supervisors still monitored if they were completed and coordinated:
\begin{quotation}
\textit{"The tenants doing kitchen service needs to set up the table accordingly; including drinks and cutlery." (P1)}
\end{quotation}

\section{Implications for Design}
We present the lessons learned in the form of implications for design. These implications present an overview of the design space of assistive systems for cooking in sheltered living facilities. The implications can be used by future designers to assure that possible solutions for the cooking benefit the users with cognitive impairments. These implications are derived from the themes presented above. Although these implications are derived for persons with cognitive deficiencies, we believe that these could be generalized to implement smart kitchens for elderly populations, who often begin to experience cognitive decline through the aging process~\cite{Blasco2013, chan2009smart, courtney2008needing}, as well as assistance systems integrated into other household environments.
\subsection{Support Clear Task Division}
From our inquiries, we learned that tasks have to be intelligently distributed and communicated among tenants.
The goal of the tasks was communicated verbally since most of the tenants were not able to read or write. During our studies, we found that continuous support was needed. Supervisors were reminding tenants from time to time about their current task and kept them motivated. Furthermore, the quality of their work was evaluated and criticized by both supervisors and other tenants when necessary, as we observed in the \textsc{supervision} theme. As a consequence, future smart kitchens should communicate recipes and tasks; \textit{explicitly providing feedback on the component tasks that make up the cooking process}. Furthermore, attention and motivation should be fostered by highlighting the current task to be performed repeatedly. Division of labor should be explicitly supported, giving the supervisor freedom in orchestrating the cooking experience based on their awareness of tenants' diverse capabilities, and should help them maintain an overview of all the activities in the kitchen.
\subsection{Embrace the Group Experience}
Our analysis revealed that maintaining social ties between supervisors and cohabitants was an important factor on each tenant's route to greater independence. In the \textsc{community} theme, we observed how cooking represented an important social activity, where tenants worked together to achieve a communal goal. Managing social activities and minimizing opportunities for conflicts within the group was regarded as a key role for supervisors. Furthermore, some tenants were not motivated to cook, since it required extra effort and paying attention to additional constraints as shown in the \textsc{practicalities} theme, but were more motivated by their role in the larger community of the kitchen. Our research suggests that adaptive motivating elements should be \textit{integrated to maintain and augment the social experience of cooking together}. For example, showing visible achievement metrics or cheering cartoon figures as a token of appreciation for completing a task may foster social interaction. Further, assistance systems should support conflict resolution by accounting for the tenants' unique abilities and personalities.
\subsection{Prioritize High-Safety Instructions}
Safety during cooking was the main concern for supervisors. As persons with cognitive impairments perceive pain differently, the risk of severe injuries is higher than among neurotypical populations \cite{Buffum2007}. In the \textsc{practicalities} and \textsc{supervision} themes, we observed how instruction always prioritized safety features for tasks that posed a possible danger. Consequently, smart kitchen systems for tenants should \textit{only feature displays in safe areas}, where the probability of getting injured is minimized. As dangerous tasks, such as operating the stove, cannot be fully avoided, the system should communicate safety hazards to avoid severe injuries. As we observed, some of the tenants had issues retaining focus; drifting in and out of the kitchen space during the completion of a single task. Future systems should \textit{clearly show dangers that have a temporal aspect}, for instance, communicating that a hot plate cannot be touched until it has cooled down to a safe temperature.
\subsection{Enable Customization for Different Abilities}
In our work, we observed users with cognitive impairments with a diverse set of abilities and impairments. In the \textsc{practicalities} theme, we observed that some users were able to occupy themselves with dispensing spices, measuring water, or performing tasks that required fine motor skills; while these tasks were inaccessible to others. While supervisors strived to find an optimal division of the tasks, managing multiple parallel activities and the complex learning curves of multiple tenants was a very complex task as shown in the \textsc{supervision} theme. Future systems can not only aid in finding optimal ways to divide the tasks involved in preparing a meal but also \textit{adjust the difficulty of the task to support the learning process} of a given tenant. It is important to note that the information provided should foster individual abilities, therefore the system should monitor the learning process and adaptively increase or decrease assistance. Furthermore, once tasks are complete, instant individualized rewards should be available to maintain management and support communication with the supervisor.
\subsection{Provide Opportunities for Explicit Communication}
The \textsc{community} theme showed how users often required confirmation from their peer group or superiors when performing cooking duties. Further, in the \textsc{work organization} and \textsc{practicalities} themes, we observed how the logistics of food preparation and constant movement in the kitchen affected attention and communication between the supervisors and tenants. Consequently, we see opportunities for fostering reporting to supervisors and increased awareness of when tasks are finished or in progress. Future assistance systems for kitchens for persons with cognitive impairments should \textit{explicitly encourage users to communicate with the supervisor} and show the group when tasks are completed. This, however, needs to be done in ways that do not provide distractions that may affect cooking performance.

\section{Discussion}
The considerations we present above focus on designing for assistance in kitchens within sheltered living facilities. In contrast, here we brainstorm future directions for assistive technologies that assist persons with cognitive impairments in unsupervised, independent, or group-based cooking sessions.

\subsection{Accessible Assignment of Complex Tasks}
One of the primary roles of the instructors in the kitchens was dividing tasks among the tenants, monitoring ongoing preparations, and coordinating their work to create a full meal. We see this coordination work as an important space where intelligent technologies could supplement or replace the trained instructors in the future, both to address the staffing shortages that limit the use of sheltered living facilities and to enable individuals with cognitive impairments to train in these skills from their own homes or the homes of family members.

The instructors created long-term plans for cooking tasks which pushed tenants to continue building their cooking skills and to complete cooking tasks even when they were not intrinsically motivated to cook at that particular time. Intelligent systems which assign cooking tasks must balance these complex needs by generating tasks that are incrementally more difficult over time, and by sensing and responding to users' current level of ability and motivation in a given cooking session. These ability levels may change dramatically between sessions, especially among individuals with cognitive impairments caused by brain injury or aging, and an automated system must be able to flexibly adapt after observing a users' competencies or difficulties in the kitchen.

Intelligent systems may also be useful for home environments, where a person with cognitive impairment may wish to cook collaboratively with neurotypical family members. Family members of people with cognitive impairments can be inadvertent perpetuates of disability stigma against their loved ones, viewing them as less capable and giving them simplistic tasks and responsibilities \cite{dobbs2008ethnographic}. These systems could engage in task division between the family members by identifying tasks that are appropriately complex for both the persons with cognitive impairments and neurotypical participants, reducing biases around cognitive impairment and allowing the person with cognitive impairment to maintain their role within the family structure.

\subsection{Examining Collaborative Accessibility}
Prior work has found that norms in environments where people with disabilities are the majority (\textit{e.g.}, the National Federation of the Blind's annual conference) differ from norms in environments where people with disabilities are the minority \cite{easley2016let}. In the independent housing areas of the sheltered living facilities we studied, all members of the cooking process have cognitive impairments, raising critical questions about how technologies can be designed to facilitate collaborative work without impacting the disability-specific norms of the tenants.

In the supervised observations, tenants turned to the instructing caregiver to ask questions or get feedback. This reliance on the instructor may make it difficult for tenants to develop truly independent skills in a supervised setting. As a result, the instructors gradually reduce their involvement until the tenants can demonstrate a level of independence that qualifies them for independent housing. However, the 'independent' housing is anything but independent -- tenants are surrounded by other residents with cognitive impairments who they can turn to for support or instruction. Designing for these \emph{collaborative} domestic settings is rarely studied in accessibility literature or is looked at between disabled and able-bodied partners rather than among members of a single disabled community. Future study of cooking practices among individuals with cognitive impairments can contribute greatly to our understanding of the co-construction of accessibility.

We recognize the importance of including representative users in accessibility research, as argued by \cite{sears2012representing}, and strove to accurately represent the cooking practices of the tenants with cognitive impairments we observed. However, due to the level of cognitive impairment among the tenants, and the facility's desire to minimize researchers' direct intervention into the cooking process, we could not perform a typical contextual inquiry with the tenants. Instead, we combined our observations with interview data from the facility staff, who may have their own biases in their interpretations of the tenants' actions and interactions \cite{dobbs2008ethnographic}. While other HCI work studying users with cognitive impairments also relies on caregiver stakeholders as informants (\textit{e.g.}, \cite{Dawe:2006:DSS:1124772.1124943}), we see this as a major limitation of our current work. As we continue this thread of research, we intend to develop new methods to facilitate working directly with tenants with cognitive impairments, perhaps drawing from prior work which leverages participatory design methods \cite{frauenberger2016designing}.

\subsection{Study Limitations}
Our contextual observations and evaluation were affected by other constraints. The study and data collection were conducted focusing on West European living standards. The sheltered living facility model is unique to this context. Compared to other locations, different living habits regarding persons with cognitive impairments may be present. Additionally, both of our observations were conducted in different sections of the same sheltered living organization, whereas cooking procedures in other facilities may be executed differently. Communication of complex or novel information can confuse, while simplified visualizations may neglect to foster cognitive abilities and should be considered in a follow-up study.


\section{Study Conclusion}
We investigated the design space of smart kitchens for persons with cognitive impairments living in sheltered housing facilities. Within qualitative contextual inquiries, we obtained data by observing persons with cognitive impairments during a cooking session. Furthermore, interviews were conducted to explore current design gaps of assistive technologies in-depth. We derived the four relevant themes \textsc{work organization}, \textsc{community}, \textsc{supervision}, and \textsc{practicalities}, which emerge as important factors, through a thematic analysis of the interviews. We conclude with five design implications which should be considered when designing smart kitchens for persons with cognitive impairments: clear communication of tasks, fostering the group experience, prioritizing safety, providing rewards, and enabling contextual adaptivity.

\section{Chapter Summary}
In this chapter, we conducted a qualitative inquiry to reveal opportune moments and potential interventions for cognitive assistance through workload-aware interfaces. Through the combination of observations and interviews with caretakers of a sheltered living facility, we derived four themes as well as five design implications for the design of assistive technologies that augment cognitive capabilities. Persons with cognitive impairments were studied as a target group since they express the need for cognitive assistance explicitly to their supervisors. In a first step to understanding how tenants with cognitive impairments process and act using the available information, we conducted an observation to gather information during a communal cooking session. Using the results from the observations, we created a question catalog that addressed aspects of cognitive augmentation through assistive technologies.

Our findings show how persons with cognitive impairments collaborate during cooking with other tenants as well as the supervisor. We find that social collaborations, especially the social ties between tenant and supervisor, play an important role. For example, caretakers repeatedly reminded the tenants to be careful during dangerous or mentally demanding tasks. Situations that may cause injuries are therefore one particular moment that has the potential to benefit from cognitive assistance. We further found that communal settings require a clear task division. It is difficult for persons with cognitive impairments to manage task division by themselves without clear instructions from a caretaker. Assistive technologies can use this circumstance to deliver in-situ cognitive assistance. Our interviews revealed that the caretakers were compelled by the vision of constant cognitive augmentation. Although delivered through ambient visualizations, assistive technologies should be attentive during the whole cooking procedure and deliver assistance whenever needed. Hence, \textbf{RQ1} is answered by the presentation of the associated design implications and the aforementioned situations that benefit from cognitive assistance.

\chapter{Feedback Modalities}
\label{ch:support_modalities}

\epigraph{\textit{For every user action, there should be an interface feedback. For frequent and minor actions, the response can be modest, whereas, for infrequent and major actions, the response should be more substantial.}}{Ben Shneiderman}

So far we have investigated user-centric characteristics when encountering situations in a cooking scenario that require cognitive assistance. This resulted in five design implications for workload-aware assistive systems. While these results deepen our understanding to recognize opportune moments for cognitive assistance, the feedback modalities of such systems have to be designed carefully. Cognitively exerting tasks tend to demand the full attention, alertness, and vigilance of a person. Unsuitable feedback modalities may break the attention span and effort that was put into the task. Hence, it can take between ten to 15 minutes before users can return to their original task after being disrupted~\cite{Iqbal:2007:DRC:1240624.1240730}. Contrary, feedback modalities that demand the attention span frequently have a detrimental effect on the user's task performance. In this chapter, we compare modalities that utilize the visual, auditory, or tactile sensory channel to deliver feedback. For example, visual and auditory feedback can be employed in collaborative work settings while tactile feedback can be used to notify users while preserving privacy. Hence, feedback modalities may be perceived differently by each individual and for each environmental situation. The previous chapter investigated tenants from sheltered living facilities as a target user group for systems that provide cognitive assistance. The roles, as well as social ties between tenants with cognitive impairments and the supervising caretakers, have been investigated to receive insights for the future design of workload-aware interfaces. Thereby, the tenants have shown affinities with technologies that foster communication (\textit{e.g.}, instant messaging or phone calls) throughout their tasks (\textit{e.g.}, cooking).

\begin{figure*}[t!]
    \centering
    \subfloat[][]{
    \includegraphics[height=0.35\columnwidth]{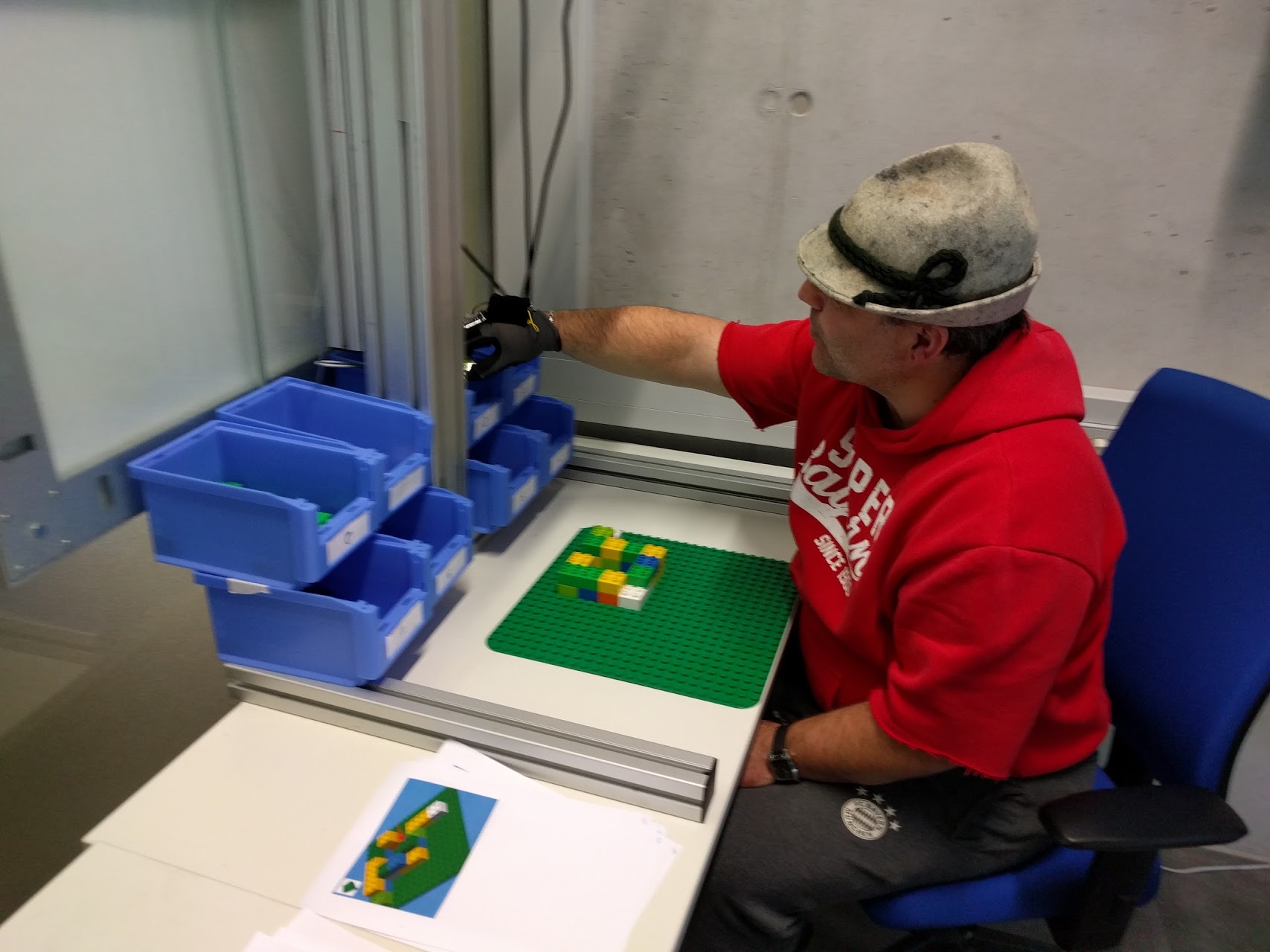}
    \label{fig:splash2}}
    \subfloat[][]{
    \includegraphics[height=0.35\columnwidth]{figures/splash}
    \label{fig:splash1}}
    \caption[Person assembling with tactile feedback]{A participant is wearing an augmented glove providing tactile error feedback during assembly tasks.}
    \label{fig:splash_feedback}
\end{figure*}

The ``United Nations Convention on the Rights of Persons with Disabilities''~\cite{un2008rights} describes how to ensure and protect fundamental rights of people with disabilities. This comprises the inclusion in daily life tasks, such as work and public leisure activities. Sheltered working has emerged to include persons with cognitive impairments in working life. Sheltered work organizations employ workers with cognitive impairments in a supervised work setting. This establishes social ties with peers and provides an income for personal use. Procedural assembly tasks, such as manual assembly at production lines, are common tasks conducted by persons with cognitive impairments. The assembly instructions can be broken down into a small task complexity to fit the mental capacities of workers with cognitive impairments. Traditionally, workers with cognitive impairments rely on a human instructor who supports them during assembly tasks. The instructor is often responsible for multiple workers which makes it difficult to supervise and control each assembly step of each worker for errors.

The use of assistive systems to provide assembly instructions has proved to reduce the number of assembly errors while decreasing the overall assembly time~\cite{funk2015using}. Since errors can not be completely avoided, assistive systems can detect assembly errors and provide appropriate feedback. However, error feedback is perceived different on an individual level. The type of intrusiveness, demand for attention, and privacy aspects are factors that have to be considered~\cite{funk2016haptic}. Previous approaches conclude that a combination of haptic and visual feedback is a suitable way of presenting error messages through assistive technologies. As these results were collected with participants that were not affected by cognitive impairments, it is unclear if which error feedback is suitable for workers with cognitive impairments. In this chapter, we close this gap by presenting a study with 16 workers with cognitive impairments comparing the three most common modalities for providing error feedback, including tactile, auditory, and visual feedback (see Figure~\ref{fig:splash_feedback}). We present the results of a user study that investigates the aforementioned feedback modalities and we provide implications when designing feedback modalities in working environments. This chapter sets the boundaries of communication modalities for workload-aware interfaces:




\begin{itemize}
 \renewcommand{\labelitemi}{$\rightarrow$}
 \item \textbf{RQ2}: What are suitable in-situ feedback modalities for cognitive support?
\end{itemize}

\begin{center}
 \begin{GrayBox}
 \centering
 \parbox{0.98\textwidth}{
 \emph{This section is based on the following publication:}
 \vspace{3mm}
 \publicationsbegin 
 \item \bibentry{Kosch:2016:CTA:2982142.2982157}
 \publicationsend
 }
 \end{GrayBox}
\end{center}

\section{Related Work}
The related work presented in this section includes past evaluations of feedback modalities and developments in assistive technologies for workplaces.


\subsection{Feedback Modalities}


Different feedback modalities have been used for providing information in different scenarios. The most important categories are tactile, auditory, and visual feedback. Tactile feedback is for example used by Bial et al.~\cite{bial2011enhancing} by using a glove that is equipped with vibrating motors. Results show that tactile feedback can be used for navigation tasks. They use the tactile feedback to provide information for motorcyclists during driving tasks. Considering auditory feedback, Rauterberg and Styger~\cite{rauterberg1994positive} proposed adding additional auditory feedback to traditional visual feedback for assembly tasks. Visual and auditory feedback has been provided at the same time while managing computer-numeric-controlled centers. Their study suggests that combining auditory and visual feedback leads to a more positive mood and improves the participant's performance. The multimodal representation of feedback could lead to a high mental demand for participants with cognitive impairments. Plain visual feedback is for example used by Funk et al.~\cite{funk2015comparing} when comparing different visual approaches for providing feedback for workers with cognitive impairments at manual assembly workplaces. In their study, they compare video-based, pictorial and contour instructions. The results of the study suggest that visual contour instructions are perceived well among all \ac{PI} groups of workers with cognitive impairments. Moreover, Cuvo et al.~\cite{cuvo1992promoting} use textual feedback while instructing persons with mild cognitive impairments. In their study, they found that performance feedback is crucial.

Other fields already experimented with combinations of haptic, auditory, and visual feedback. Akamatsu et al.~\cite{akamatsu1995comparison} compared the three feedback types in a mouse pointing task. Their study revealed interesting design implications although no difference in \ac{TCT} was found. Furthermore, Richard et al.~\cite{richard1994comparison} and Petzold et al.~\cite{petzold2004study} compared the three feedback types when manipulating objects and assembling in virtual environments. Visual feedback was delivered on a screen, auditory feedback was provided by headphones and tactile feedback was triggered using pneumatic micro-cylinders, which applied pressure on the fingertips in a glove. Richard et al. found that both haptic and auditory feedback improve the workers' performance in manipulating virtual objects while Petzold et al. found that the performance is increased using additional haptic feedback.

\subsection{Assistive Systems for Workplaces}

In 1991, Pierre Wellner~\cite{wellner1991digitaldesk} suggested using a camera-projector system to provide additional digital information for regular physical objects. In his prototype, Wellner combined a digital and a physical workspace for enabling to use the best features of both spaces in one physical workspace. Later Pinhanez~\cite{pinhanez2001everywhere} was using a camera-projector system combined with a mirror for turning arbitrary surfaces into digital displays. Using Pinhanez's system, nearly every surface can become a display that shows information. Since then, systems using camera projector systems were deployed in different scenarios to provide cognitive assistance. For example, R{\"u}ther et al.~\cite{ruther2013assistance} provide assistance in sterile environments, L{\"o}chtefeld et al.~\cite{lochtefeld2010shelftorchlight} augment a shopping scenario, and Butz et al.~\cite{butz2004searchlight} used it to support searching tasks. In 2008, Bannat et al.~\cite{bannat2008towards} used a similar setup for providing assembly instructions at manual assembly workplaces. Their system uses a camera to detect the position of picking bins and a projector, to provide pictorial instruction for the assembly process. Similarly, B{\"u}ttner et al.~\cite{buttner2015extending} uses in-situ projection at manual assembly workplaces by providing pictorial instructions. They found that using in-situ projection is faster and leads to fewer errors compared to displaying instructions on smart glasses~\cite{buttner2016using}. Further, Korn et al.~\cite{korn2014context, korn2013potentials} used in-situ projection in combination with gamification approaches for motivating and instructing workers with cognitive impairments during assembly tasks. Their study with workers with cognitive impairments revealed great potential for using gamification and in-situ instructions for instructing and motivating workers at assembly workplaces. However, they did not find a statistically significant difference. Further, a comprehensive summary of assistive systems for supporting workers at the workplace is provided by Korn et al.~\cite{korn2015assistive}.

Recently, Funk et al.~\cite{funk2015using} used in-situ projection to provide instructions during assembly tasks at the workplace. Their results show that with increasing complexity, using in-situ instructions is leading to significantly fewer assembly errors and a significantly faster assembly time compared to pictorial instructions. In their system, they were providing red error feedback that illuminates the current picking bin when picking the wrong part. However, no scientific analysis of the error feedback was conducted. Therefore, Funk et al.~\cite{funk2016haptic} further investigated the effects of haptic error feedback, auditory error feedback, and visual error feedback by conducting a study with students. Their results reveal that a combination of haptic and visual feedback might be the best way to communicate errors while performing assembly tasks. However, this was only tested with students in a lab study.

Previous work suggested the use of haptic feedback for communicating errors during assembly tasks. This is considered a privacy-presuming way of communicating errors at workplaces~\cite{funk2016haptic, petzold2004study}. We argue that assistive systems will have a great impact on manual production lines and envision the inclusion of people with cognitive impairments in the work life~\cite{behrendt2015ethical,sauer2010assistive}. Thus, we are interested in how concepts of error feedback influence the efficiency of workers with cognitive impairments. Further, we investigate which error feedback modalities are preferred by the workers themselves.

\begin{figure*}[t!]
 \centering
 \subfloat[][]{
 \includegraphics[height=0.19\columnwidth]{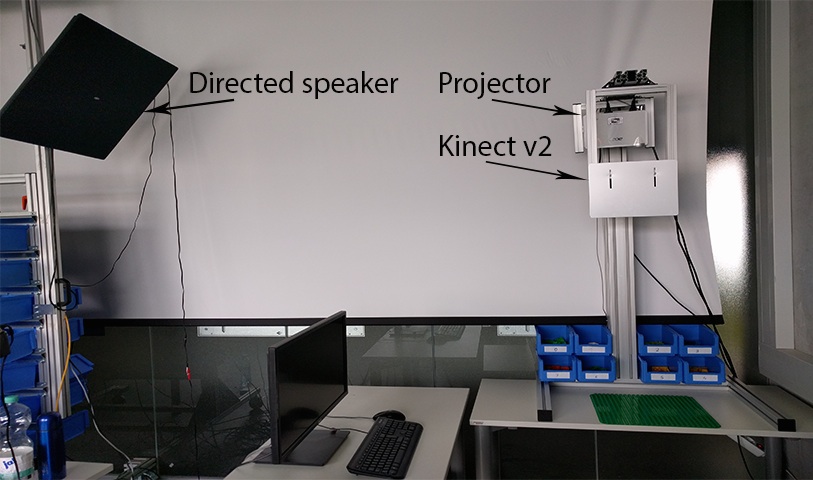}
 \label{fig:system}}
 \subfloat[][]{
 \includegraphics[height=0.19\columnwidth]{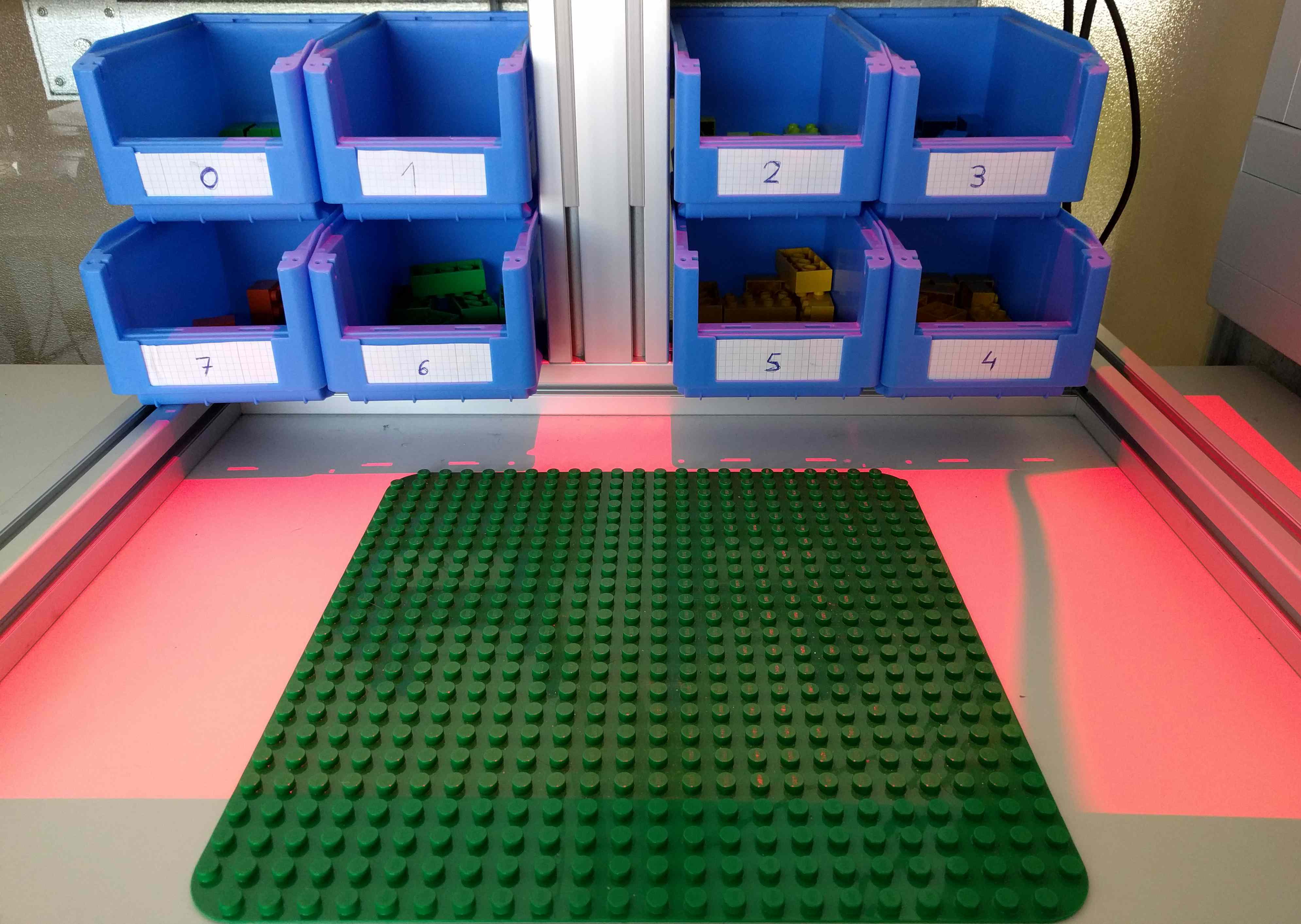}
 \label{fig:redlight}}
 \subfloat[][]{
 \includegraphics[height=0.19\columnwidth]{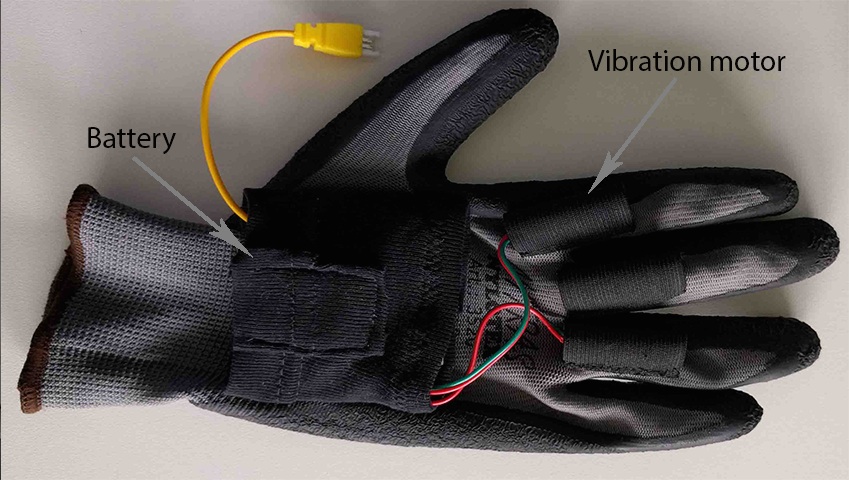}
 \label{fig:glove}}
 \caption[Feedback modalities used in the study]{\textbf{(a):} The system uses a projector and a directed speaker for providing visual and auditory error feedback. Further, a Kinect v2 observes the picking of items from bins. \textbf{(b):} Red light providing visual error feedback. In case of an error, the whole working area is highlighted. \textbf{(c):} A glove equipped with vibration motors is further used to provide tactile error feedback.}
 \label{fig:studysetup}
 \vspace{0.4cm}
\end{figure*}

\section{System}
To incorporate the use of tactile, auditory, and visual error feedback in an assistive system for workplaces, we extended the system presented by Funk et al.~\cite{funk2015using}. The system consists of modular components that are designed for providing different modalities of error feedback. The main system, which tracks assembly steps and provides feedback to the assembly area, is constructed out of multiple aluminum profiles. The profiles can be used to mount different hardware on top of the workplace. The system uses a top-mounted Kinect v2~\footnote{\url{https://developer.microsoft.com/en-us/windows/kinect } - last access \urldate} to detect picking steps. Therefore, the system can distinguish between correct and incorrect picks.

We placed the system on a table and placed a height-adjustable chair in front of it. Therefore the workers can work while sitting at a comfortable height. The system is constructed in a way that the assembly area is $70\,cm$ wide and $49\,cm$ high. This is enough space for placing instructions and assembling Lego Duplo constructions. The system further features $2\times4$ picking bins which are filled with Lego Duplo bricks. Each picking bin is filled with one type of Lego Duplo bricks that is unique in either color or shape. In the middle of the system, there is a Lego Duplo plate firmly taped to the work area (see Figure~\ref{fig:redlight}). The system triggers an error when either a picking error was made, or an assembly error was made by the participants. Picking errors are detected automatically by the Kinect v2 if the user places his or her hand in a wrong bin. For detecting assembly errors, the system uses a wizard of oz approach where a study assistant observes the assembly process and uses a wireless presenter for triggering an error. In the case of an assembly error or picking error, an error message is triggered. Depending on the condition, our system is capable of presenting error feedback using the following three modalities: \\

\textbf{Visual} error feedback is provided using a projector (see Figure~\ref{fig:system}), which is mounted at the top of the assistive system. If an error occurs, a red light illuminates the whole work area (see Figure~\ref{fig:redlight}). While previous approaches only highlight the incorrect picking bin or the incorrect assembly position if an error was made, we decided to illuminate the entire work area. Therefore, the error message is harder to miss. \\

\textbf{Auditory} error feedback is provided using the Holosonics Audio Spotlight 24i\footnote{\url{www.holosonics.com/15-products} - last access \urldate} (AS24i) speaker. The speaker uses ultrasonic waves to prevent the sound from diffusing. Therefore, it is only noticeable by the person sitting at the workplace and therefore retains the user's privacy. As an error sound, we are using a deep error tone exactly as used by Funk et al.~\cite{funk2016haptic}. In case an error is made, the error sound is played by the speaker (see Figure~\ref{fig:system}). \\

\textbf{Tactile} error feedback is provided using a glove equipped with two vibration motors (see Figure~\ref{fig:glove}). We decided to choose standard safety gloves, which are mostly used by workers while performing assembly tasks. The motors are placed on the index finger and the ring finger. Our glove uses an ESP8266 microcontroller and a battery to receive the error trigger messages via WiFi. In comparison to Funk et al.~\cite{funk2016haptic}, our glove can, therefore, be used without using a wire connected to a computer. This wireless feature makes the gloveless obstructive while performing assembly tasks. In case an error is made, the glove uses an alternating vibration pattern that activates each motor twice directly after each other for 0.3 seconds each. This results in a vibration time of 1.2 seconds per error. A summary of the design and duration of every error feedback modality can be found in Table~\ref{tab:feedbackdesign}.

\begin{table}[h!]
 \begin{center}
 \begin{tabular}{ l l l }
 \toprule
 \textbf{Stimulus} & \textbf{Feedback Design} & \textbf{Duration in ms} \\
 \midrule
 Visual & Projecting red light & 2500 \\
 Auditory & Playing deep error sound & 2000 \\
 Haptic & Vibration in worker gloves & 1200 \\
 \bottomrule
 \end{tabular}
 \caption[Details of the feedback stimuli]{Feedback design and duration of each stimulus used in the study. The duration is depicted in milliseconds (ms).}
 \label{tab:feedbackdesign}
 \end{center}
 \vspace{-0.5cm}
\end{table}

\begin{figure*}[t!]
\centering
 \subfloat[][]{
 \includegraphics[height=0.19\columnwidth]{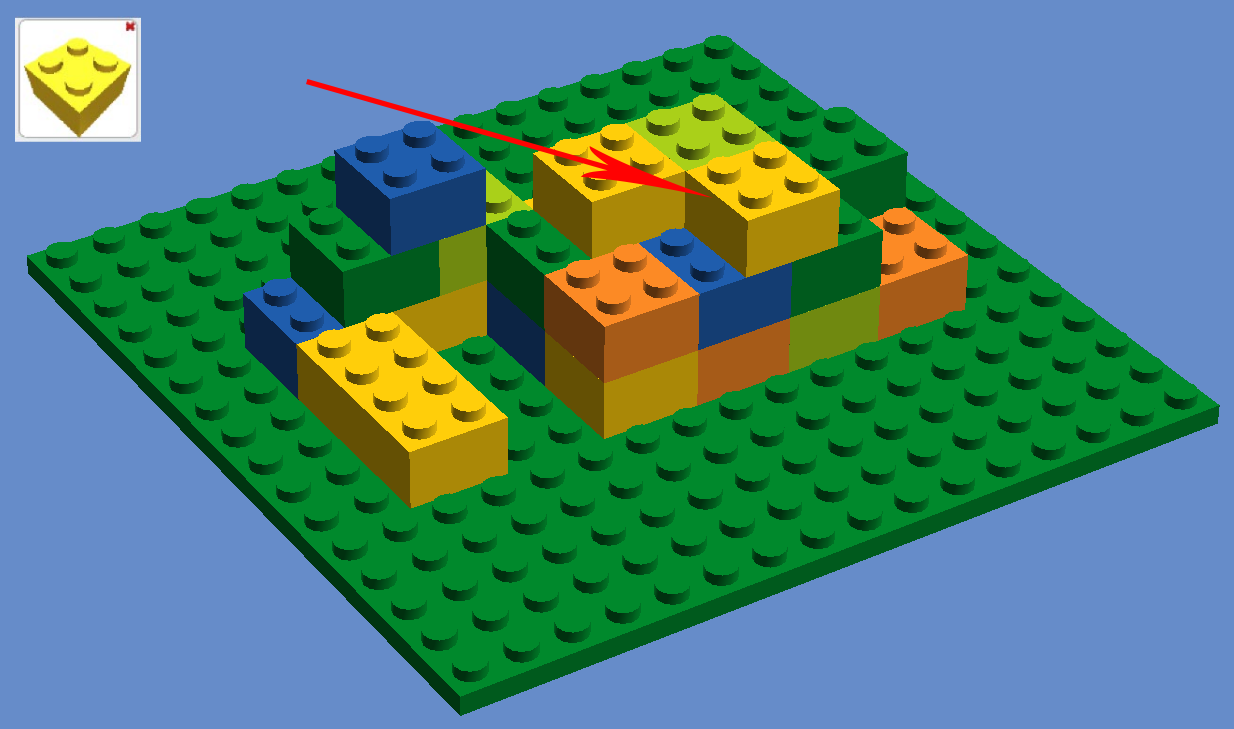}
 \label{fig:taskA}}
 \subfloat[][]{
 \includegraphics[height=0.19\columnwidth]{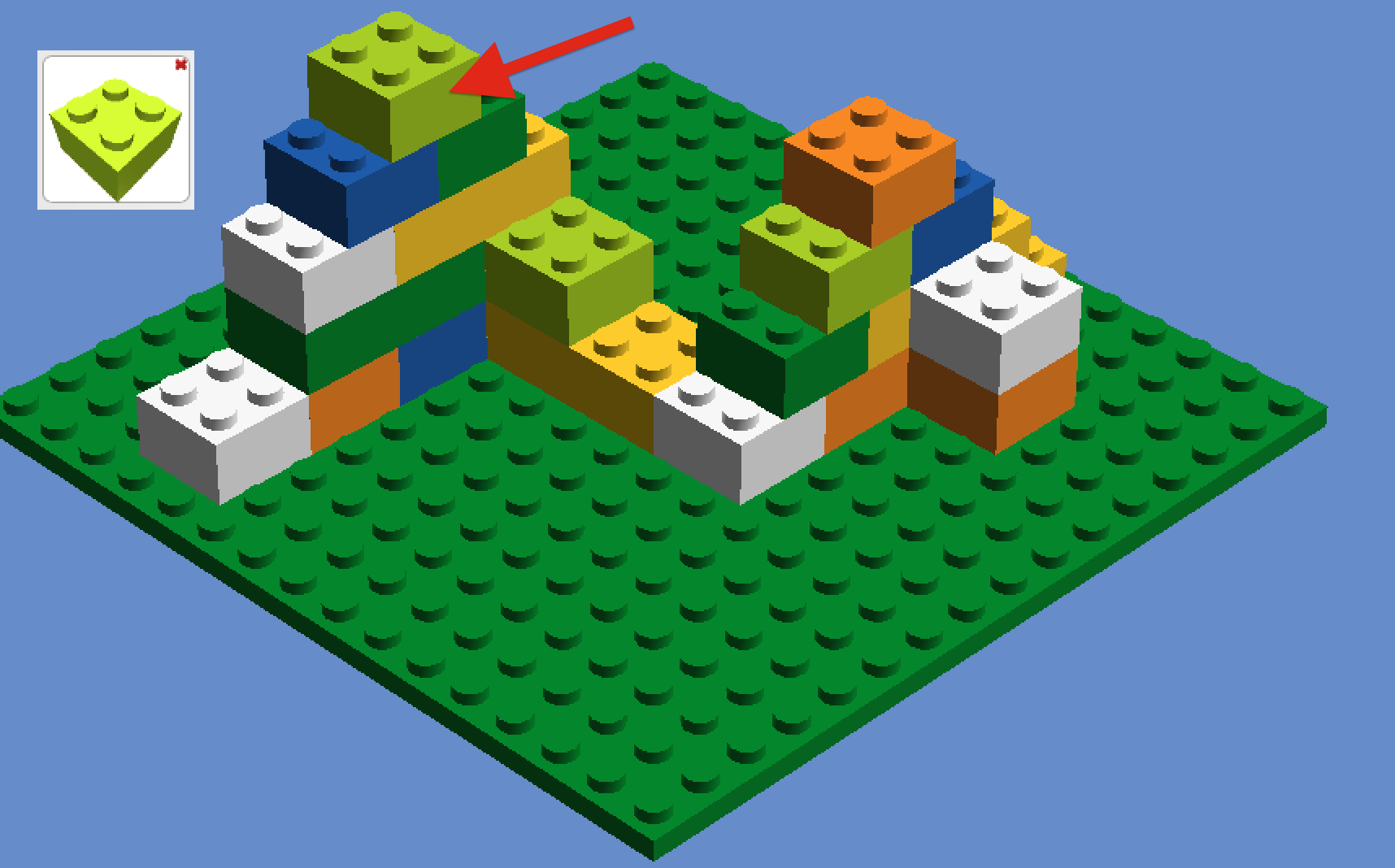}
 \label{fig:taskB}}
 \subfloat[][]{
 \includegraphics[height=0.19\columnwidth]{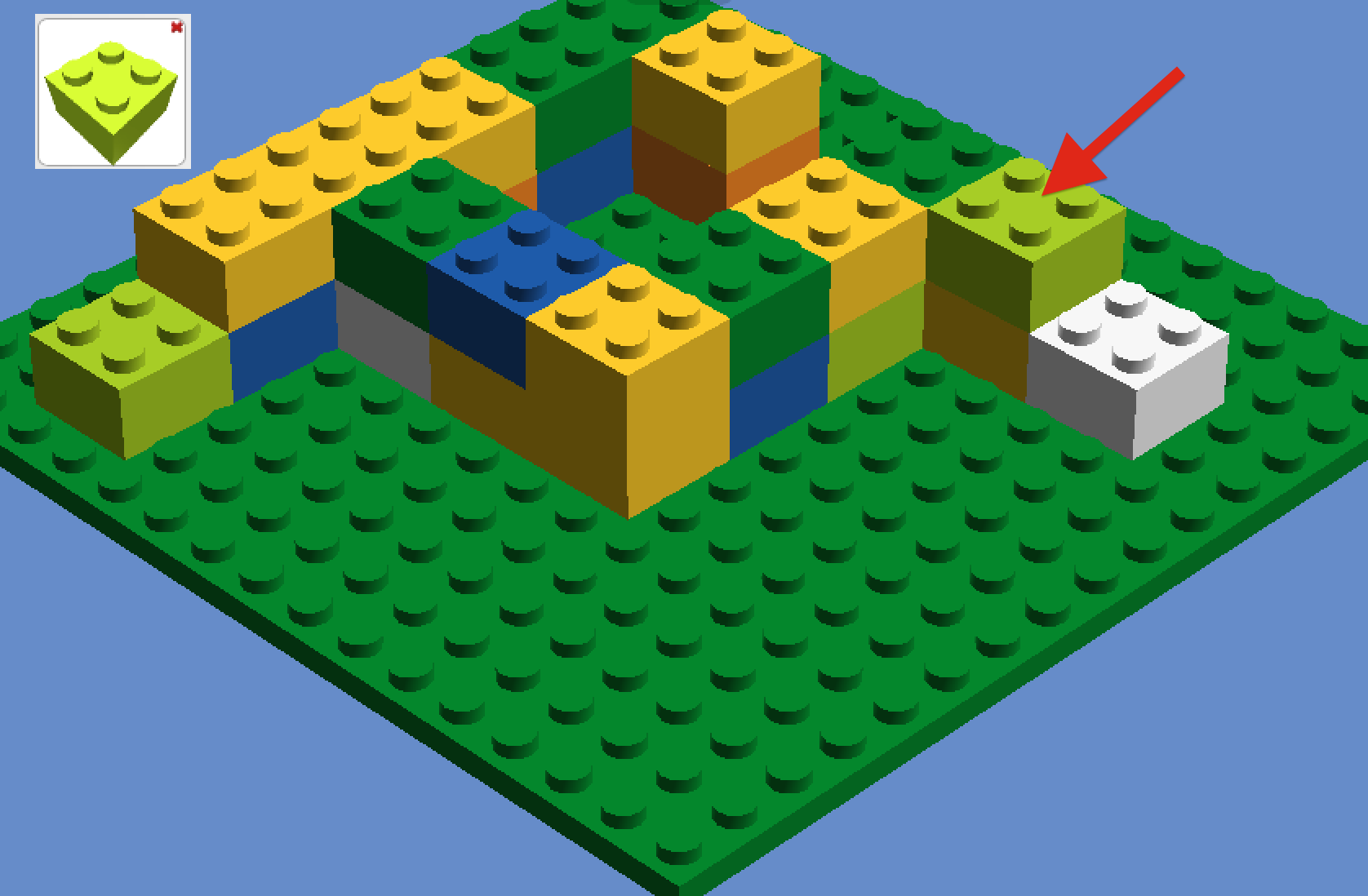}
 \label{fig:taskC}}
 \caption[Assembly instructions used in the feedback study]{The assembly tasks used in the study. We used three different assembly tasks with equal complexity. The images depict the final step of the pictorial instructions.}
 \label{fig:instruction}
\end{figure*}

\section{Evaluation}
We conducted a user study with workers with cognitive impairments to evaluate the usefulness and impact of different error feedback modalities. This section describes the user setup for the study, explains the procedure, and reports the results of the quantitative measures. Ethical approval for this study was given by the employee organization of our partner sheltered work organization and by the German Federal Ministry for Economic Affairs and Energy.


\subsection{Design}
To find the most suitable error feedback modality for communicating errors to workers with cognitive impairments, we designed our experiment following a repeated measures design with the used error feedback modality as the only independent variable. As dependent variables, we measure the \ac{TCT}, the number of assembly errors, and the number of picking errors. We counterbalanced the order of the feedback modalities according to the Balanced Latin Square. Additionally, we collect qualitative feedback through semi-structured interviews and observations. We decided not to use a baseline condition that measures the time and errors without any error feedback as related approaches showed that error feedback is beneficial for workers with cognitive impairments. They usually ask their socio-educational instructors for feedback about the assembly~\cite{funk2015comparing,funk2015using}.

\subsection{Apparatus}
For our experiment, we use the system providing tactile, auditory, and visual error feedback described in the previous section. We configured it that only one feedback modality is active per condition.
As previous work suggested, using Lego Duplo tasks provides a good abstraction of assembly tasks, which enables changing the complexity of tasks without changing the task itself~\cite{funk2015benchmark, funk2015using, tang2003comparative}. Therefore we decided to use a Lego Duplo assembly task. As we designed the experiment to have three conditions, we created three unique Lego Duplo assembly tasks consisting of 24 bricks per task\footnote{We provide the created assembly instructions to other researchers for reproducing the study:\\\url{www.github.com/hcum/comparing-tactile-auditory-and-visual-assembly-error-feedback} - last access \urldate}. The tasks are mainly inspired by the tasks that were used by Funk et al.~\cite{funk2015using}. Therefore, we used their 24 bricks task, increased their twelve bricks task by twelve more bricks, and used their 48 brick task and stopped at 24 bricks. The instructions in their final assembly state are depicted in Figure \ref{fig:instruction}. We printed the instructions on a single-sided A4 sheet of paper in a way that one assembly step is printed on one sheet of paper. The upper left corner of the instruction depicts the brick that has to be picked from one of the eight picking bins. The assembly position is depicted in a way that the brick to assemble is shown in its final position. Further, a red arrow highlights the assembly position that it can be found immediately. Further, the instructions were designed in a way that the same brick does not have to be picked twice directly after the first occurrence.

\subsection{Procedure}
In preparation for the study, we asked for written consent from either the participants or their legal guardians before the study. As the study was conducted in our laboratory, which was a new environment for every participant, we initially made all participants familiar with our laboratory environment. Accompanied by their regular socio-educational instructors, we initially explained the assistive system and the three modalities that are used in the study. At first, we informed the participants that their participation in this study is voluntary. We told them that they should inform us whenever they felt unwell or uncomfortable, as we would immediately abort the study in this case. Afterward, we explained the intention of the study and why the tasks they perform are relevant. After explaining the course of the study, we made the participants familiar with the paper assembly instructions, \textit{i.e.} which part to pick and where to assemble it. Once the participants felt confident using the paper assembly instructions, we introduced the error feedback modality for the current condition and explained what we count as an error and what the system will do if an error occurs. As the participants felt that they understood both error feedback and the paper assembly instructions, we began with the study and started measuring the \ac{TCT}. During the study, three researchers were present at the scene. The first researcher triggered the assembly error feedback in case an assembly error was made as a \ac{WoZ}. Picking errors were detected using a Kinect v2, which triggered the error feedback automatically. The second researcher was measuring the \ac{TCT} and counted the picking errors and assembly errors that were made for each condition. The third researcher was observing the worker reacting to the error feedback and taking subjective notes based on the observation. The error feedback was displayed immediately while performing the assembly when an assembly or picking error was triggered. After the assembly was completed, we asked the participant for their opinion about the used error feedback. Then we repeated the procedure for the other two remaining conditions. At the end of the third condition, we asked each participant for a subjective rating which error feedback was perceived the best by her or him and we asked them why they preferred or disliked the feedback.

\subsection{Participants}
We invited 16 participants for our user study. The participants were aged from 34 to 53 years (M=$40.33$, SD=$6.36$) and were employees of a sheltered work organization working in manual manufacturing. All participants worked on manual assembly tasks daily. None of the participants were familiar with our system, the used task, or the error feedback modalities used in our study.
We invited the participants according to their \ac{PI} in a way that they represent the population of the sheltered work organization. Therefore, we used the \ac{PI} of the sheltered work organization to categorize their capabilities. The \ac{PI} is a percentage ranging from 0\% to 100\%, which indicates how capable a worker with cognitive impairments is of performing a work task. Inspired by previous work~\cite{funk2015comparing,funk2015using}, we divided the population to belong to one of three \ac{PI} groups: 5-15\%, 20-35\% and above 40\%. Accordingly, we invited 5 participants belonging to each of the three \ac{PI} groups. The study took approximately 40 minutes per participant. Participants were monetary  compensated through funding from a joint project.

\subsection{Results}
During our study, four participants belonging to the 5-15\%~\ac{PI} group aborted the study as they did not want to wear the glove anymore. Therefore, we excluded these 4 participants from the quantitative evaluation.

We statistically compared the \ac{TCT}, the number of assembly errors, and the number of picking errors, between the error feedback modalities using one-way repeated measures \ac{ANOVA}. Mauchly's test showed that the sphericity assumption was violated for the number of assembly errors ($\chi^{2}$(2) = 6.852, p $=$ .033) and the number of picking errors ($\chi^{2}$(2) = 9.201, p $=$ .010). Therefore, we used the Greenhouse-Geisser correction to adjust the degrees of freedom ($\epsilon$ = .668 for the number of assembly errors and $\epsilon$ = .624 for the number of picking errors). Further, we used a Bonferroni correction for all the posthoc tests.

\begin{figure*}[t!]
 \centering
 \subfloat[][]{
 \includegraphics[width=0.48\columnwidth]{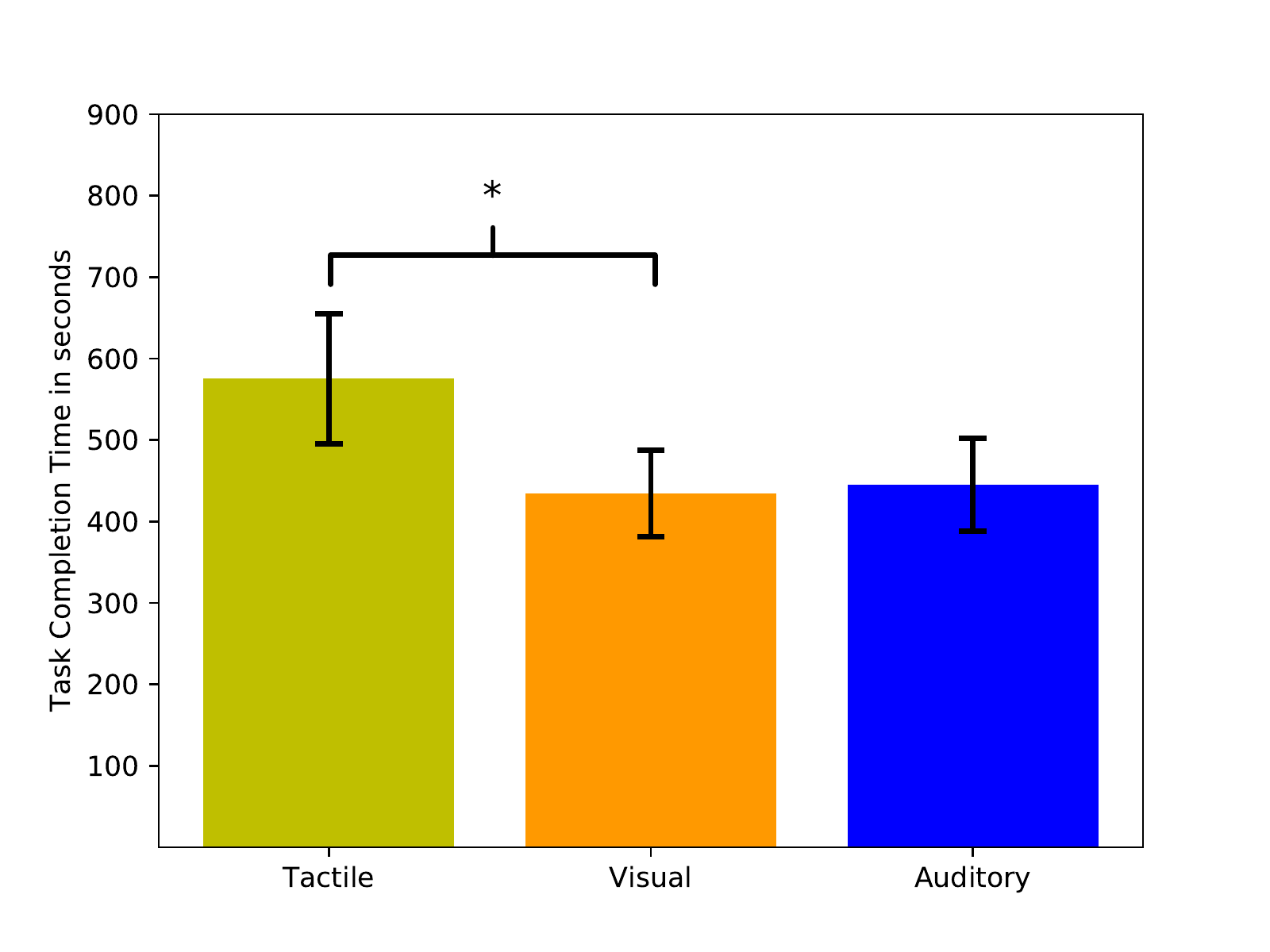}
 \label{fig:tct}}
 \subfloat[][]{
 \includegraphics[width=0.48\columnwidth]{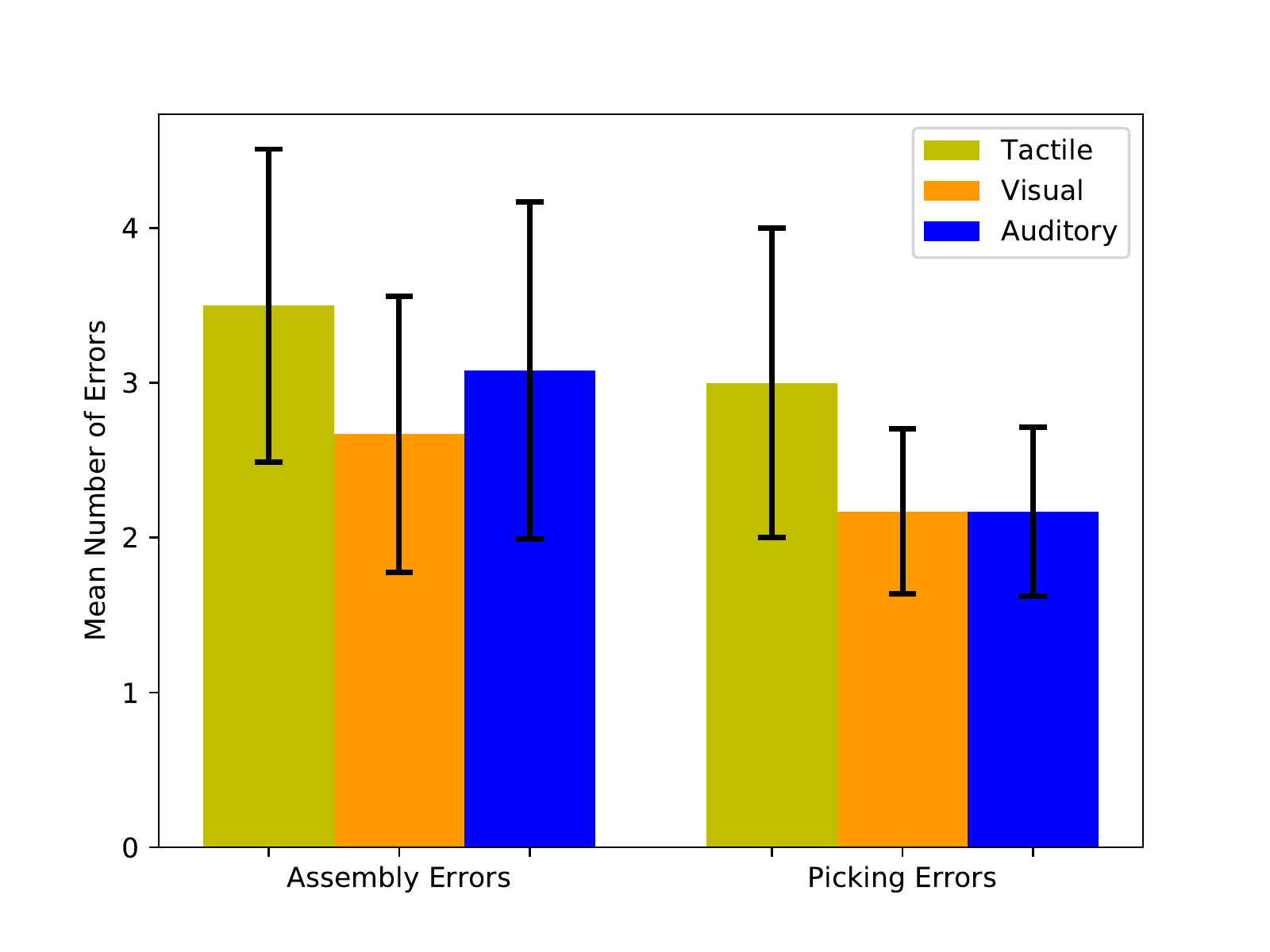}
 \label{fig:er_combined}}
 \caption[Task completion time and number of errors measured in the feedback study]{\textbf{(a):} An overview of the average Task Completion Time to assemble the Lego Duplo construction using the different error feedback modalities. \textbf{(b):} The average number of errors that were made using the different error feedback modalities for both assembly errors and picking errors. The error bars depict the standard error. The brackets indicate significant results.}
 \label{fig:tct_er_combined}
 \end{figure*}


Considering the \ac{TCT}, the participants assembled the Lego Duplo construction fastest using the visual error feedback (M = $434.42$~sec, SD  =$185.27$~sec), followed by the auditory error feedback (M = $445.08$~sec, SD = $197.40$~sec), and the tactile error feedback (M = $575.42$~sec, SD  =$276.71$~sec). As the Shapiro-Wilk test did not show a non-normal distribution, we use a parametric one-way \ac{ANOVA}. The one-way repeated measures \ac{ANOVA} showed a statistically significant difference in the \ac{TCT} between the feedback modalities $F(2,22)=4.634$, $p=.021$. The posthoc tests reveal a significant difference between the visual and tactile feedback. The effect size estimate shows a large effect ($\eta^{2} = .296$). Figure~\ref{fig:tct} shows a graphical overview of the results.

Analyzing the number of assembly errors that were made by the participants using the different error feedback modalities, the visual error feedback resulted in the fewest assembly errors (M = $2.67$, SD = $3.09$), followed by the auditory error feedback (M = $3.08$, SD = $3.77$), and the tactile error feedback (M = $3.5$, SD = $3.5$). As a Shapiro-Wilk test showed a non-normal distribution for all three modalities (all $p<.05$), we use a non-parametric Friedman test.
Accordingly, the Friedman test did not reveal a significant difference between the error feedback modalities ($p>.05$). The results are depicted in Figure~\ref{fig:er_combined}.

For the number of picking errors that were made using the different error feedback modalities, the visual error feedback (M = $2.17$, SD = $1.85$) and the auditory error feedback (M = $2.17$, SD = $1.89$) lead to the same number of picking errors. Using the tactile error feedback, the participants made the most picking errors (M = $3.0$, SD = $3.46$). 
As a Shapiro-Wilk test showed a non-normal distribution for all three modalities (all $p<.05$), we use a non-parametric Friedman test. However, the Friedman test could not reveal a significant difference ($p>.05$). The results are depicted in Figure~\ref{fig:er_combined}.



\section{Qualitative Observation}
Considering the qualitative observational feedback that was collected during the user study, two researchers continuously observed the participants during the study. The first researcher observed the interaction of the participants with different error feedback modalities and asked questions after the condition was finished in a semi-structured way. The second researcher was performing an observational study for subjectively analyzing the effect of the different error feedback modalities on the participants.

When asked about the Lego Duplo task itself, a participant stated that \emph{``with an increasing number of steps the task gets more complicated''} (P1). Another participant pointed out that the used task has several error sources as \emph{``[he] had difficulties in distinguishing between the two different types of green bricks''} (P2). During the study, we subjectively observed that ten of our twelve participants had difficulties to recognize bricks, whereas most of them stated that they did not have problems differentiating between the different colors. 

We also asked the participants about how easy it was to perceive the different feedback modalities. One participant liked the visual feedback as it is \emph{``directly in the field of view''} (P2), others said that the visual feedback is \emph{``easy to see''} (P4).
The participants were unsure about the tactile feedback as \emph{``the vibration is rather unpleasant during the work task''} (P2) and \emph{``[it] distracted me in my workflow''} (P4). Some participants (P3, P8) also did not feel or pay attention to the tactile stimulus. Further, four participants had to abort the study after the tactile feedback condition. Two of these participants stated that they did not feel well wearing a glove, the other two were uncomfortable or scared of the tactile vibration.
Considering the auditory feedback a participant stated that the \emph{``sound was very easy to perceive''} (P2). On the other hand, three participants (P3, P8, P14) stated that they were distracted by the auditory error feedback as it \emph{``scared [them] when it triggered''}.

Considering the privacy implications of the different error feedback modalities, some participants stated that auditory feedback would be \emph{``not that good because others can hear when [I] made an error''} (P2, P13). On the other hand considering the visual error feedback many participants told us that \emph{``[they] don't care if others could see that they made an error''} (P2, P4, P5, P12, P14). When we asked why the participants responded that \emph{``usually the supervisors are watching the work steps as quality control and are telling us when an error occurred''}. (P4) One participant was concerned that his supervisor would not be able to see anymore when an error was made as only he could perceive the tactile feedback (P16).

One participant (P7) stated after the study that he perceived working with the system using any error feedback as very easy as he could completely relax and rely on the system because it would tell him when he makes an error. Two other participants stated that they \emph{``could imagine using the visual feedback daily''} (P4, P5).

\begin{figure}[!t]
\includegraphics[width=\columnwidth]{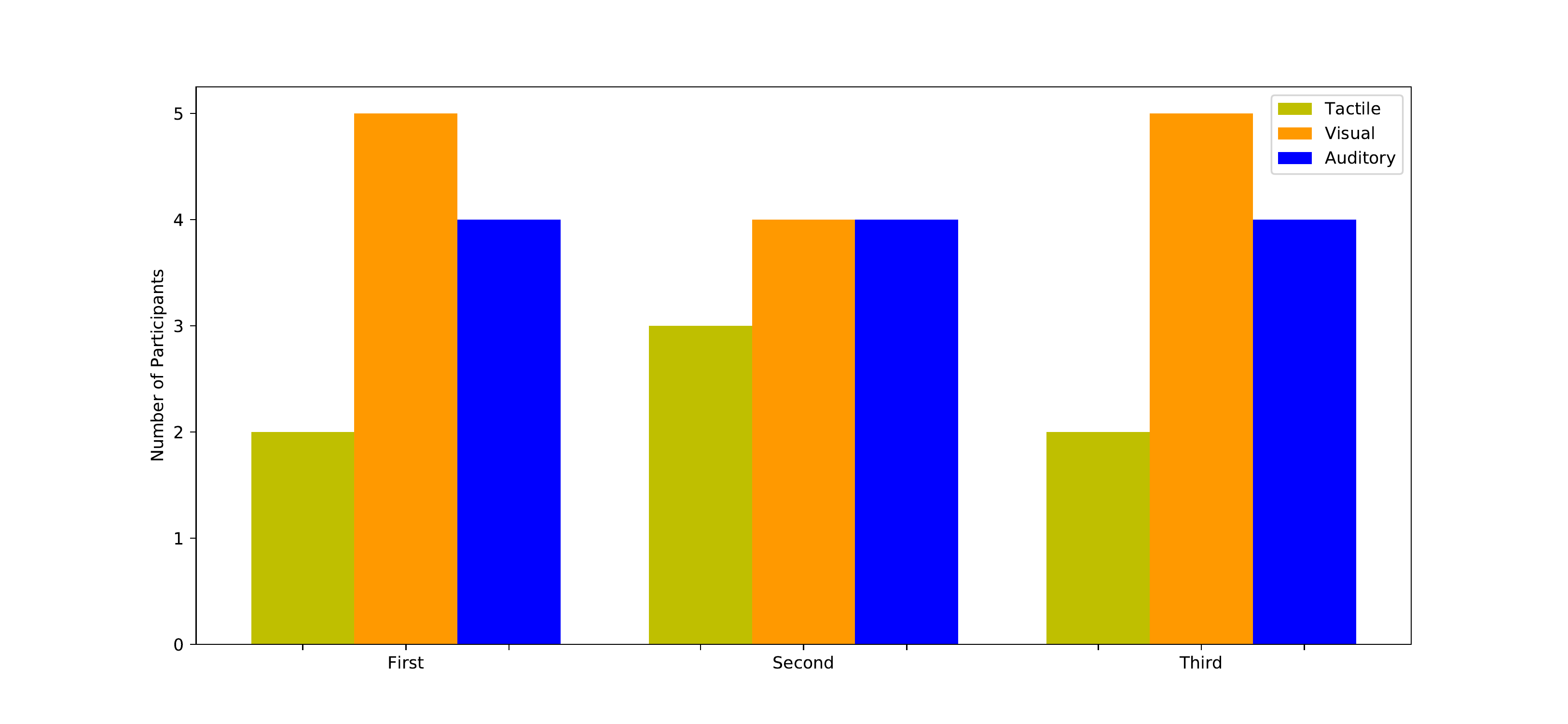}
\caption[Subjective rankings by participants for each feedback modality]{The subjective ranking of the feedback modalities ranked by the participants.}
\label{fig:subjective_rank}
\end{figure}

At the end of the semi-structured interview, we asked the participants to rank the feedback modalities according to which modality they liked best and to also consider which error feedback they would use daily. One of our 12 participants who finished the study did not want to rank the error feedback modalities. The results reveal that participants' subjective impression of the visual error feedback was best ($5\times$ first, $4\times$ second, and $2\times$ third) followed by the auditory feedback ($4\times$ first, $4\times$ second, and $3\times$ third). The tactile feedback was perceived as the worst ($2\times$ first, $3\times$ second, and $6\times$ third). The results are also depicted in Figure~\ref{fig:subjective_rank}.

In addition to the interviews, five participants made comments on their performance or formulated their thoughts during the assembly task. In most cases, these participants showed a higher TCT than other participants. 
Another four participants needed to be guided verbally during their tasks and reacted positively to the verbal help. 

\section{Discussion and Implications}
The results of our study reveal that both the number of assembly errors and the number of picking errors are not significantly different between the error feedback modalities. We argue, that the number of errors is not different between the error feedback modalities as the error feedback occurs after the participant made the error. Thus, we could not measure that the used feedback modality is influencing the participant in the way they are working, \textit{e.g.} paying more attention to a correct picking or a correct assembly of the Lego Duplo construction or being insecure and making more errors due to an error feedback modality. However, the \ac{TCT} is statistically different between tactile error feedback and visual error feedback. We assume that this difference in the \ac{TCT} is caused by a faster error recovery after an error was made. The visual error feedback was perceived significantly faster than the error feedback provided by the tactile glove.

The qualitative observations and the answers we got from the participants through semi-structured interviews after using the different error feedback modalities tend towards using visual error feedback. Although some participants did not care about using error feedback that preserves their privacy at the workplace, auditory error feedback was perceived as distracting. Considering the tactile feedback using the vibrating glove, some participants were able to perceive the tactile feedback while others did not react to the tactile feedback at all. 

While using privacy presuming error feedback mechanisms might be a good decision in general~\cite{funk2016haptic}, the implications for using these error feedback mechanisms for workers are different. Through interviews, we found that errors that are made at the assembly workplace are considered non-private information, which is acceptable to communicate to supervisors. Compared to related work~\cite{funk2016haptic}, which analyzed the error feedback for workers with no disabilities, the design implications are different as error privacy has a higher priority.

\section{Study Conclusion}
We compared visual, auditory, and tactile error-feedback for workers with cognitive impairments at manual assembly workplaces. In a user study, we obtained data from twelve participants to estimate the best feedback modality in terms of task completion time, measured errors, and qualitative feedback. We found a significant difference between visual and tactile error feedback regarding the task completion time, but could not find a significant effect between picking errors and assembly errors between each error feedback modality. We did not observe that altering the error feedback modality had an impact on the way the participants worked on assembly tasks. Therefore, the number of assembly errors throughout all error feedback modalities were not significantly different. However, we measured a significantly faster task completion time using visual error feedback compared to using tactile error feedback. Using visual error feedback, the participants were able to recover faster from errors than using tactile error feedback.


\section{Chapter Summary}
In this chapter, we investigated the efficiency of visual, auditory, and tactile feedback modalities during a Lego Duplo assembly task. Awareness about the current assembly and work context enables us to tailor feedback modalities for collaborative (\textit{i.e.}, using visual or auditory feedback) or privacy-protective settings (\textit{i.e.}, using tactile feedback). The presented study investigated the task completion time, number of errors, and user acceptance by using the three feedback modalities. We find a strong preference for visual feedback regarding acceptance and assembly efficiency followed by auditory and tactile feedback (\textbf{RQ2}). The study shows that the employed feedback modality plays an important role that has to be adjusted to the contextual and individual needs. While tactile feedback showed the lowest preference and performance, it was more preferred when focusing on privacy-related aspects. This trade-off between efficiency and privacy has to be considered during the development of workload-aware systems and interfaces.

\part{Sensing Cognitive Workload\label{part:sensing_cognitive_workload}}










\chapter{Mobile Electroencephalography}
\label{ch:workloadbyeeg}

\epigraph{\textit{Humans read minds. We constantly try to understand what our fellow humans are thinking and feeling -- and how they are going to act.}}{Hank Greely}

In Chapter~\ref{ch:smart_kitchen_requirements} and Chapter~\ref{ch:support_modalities} we outlined the design requirements of workload-aware interfaces. Our results show that, regardless of the current chore, people seek assistance when their short-term memory is demanded frequently. The qualitative inquiry presented in Chapter~\ref{ch:smart_kitchen_requirements} revealed that cognitive offloading is desired during repetitive tasks that are subject to small changes. This requires us to update the learned task while avoiding careless mistakes from the old execution pattern, and incorporate respective changes in task execution routines. Manual assembly at production lines is an example of such routine work. Modern manufacturing processes are increasingly defined by smaller lot sizes of bespoke designs. Gone are the days that require workers to act purely on rote learning. Instead, novel assembly instructions must frequently be committed to memory as soon as new designs and components enter into the production pipeline~\cite{7140647}. Thus, assembly is increasingly defined by its cognitive instead of physical demands. Manual assembly places workload on cognitive processes that underlie executive working memory~\cite{baddeley1974working, gevins2003neurophysiological}. Given that cognitive workload impacts the individual task performance~\cite{engle1999working,HOLLENDER20101278}, it follows that profit margins may suffer as a result of reduced production throughput and error-prone manufacturing.

Assembly instruction systems have been introduced to ameliorate this. Such assistance consists of instructions on printed paper, external displays, or \textit{in-situ} instructions projected directly on the assembly workplace. Such systems are designed to simultaneously reduce cognitive demand while optimizing performance. Indeed, previous research has suggested that design manipulations of the visual representation and modality of assembly instructions can have a significant influence on self-reported measures of cognitive workload~\cite{7140647, Kosch:2017:OSF:3123024.3124395}. Strong arguments have been made for the use of \textit{in-situ} instructions that can be integrated ``just-in-time''~\cite{Buttner:2015:EDS:2786567.2794342, funk2016interactive}. Furthermore, visual \textit{in-situ} instructions are attributed to a significant increase in production efficiency~\cite{funk2015pick, Kosch:2016:CTA:2982142.2982157}.

\begin{figure}
\centering
\subfloat[][]{
\includegraphics[width=0.5\columnwidth]{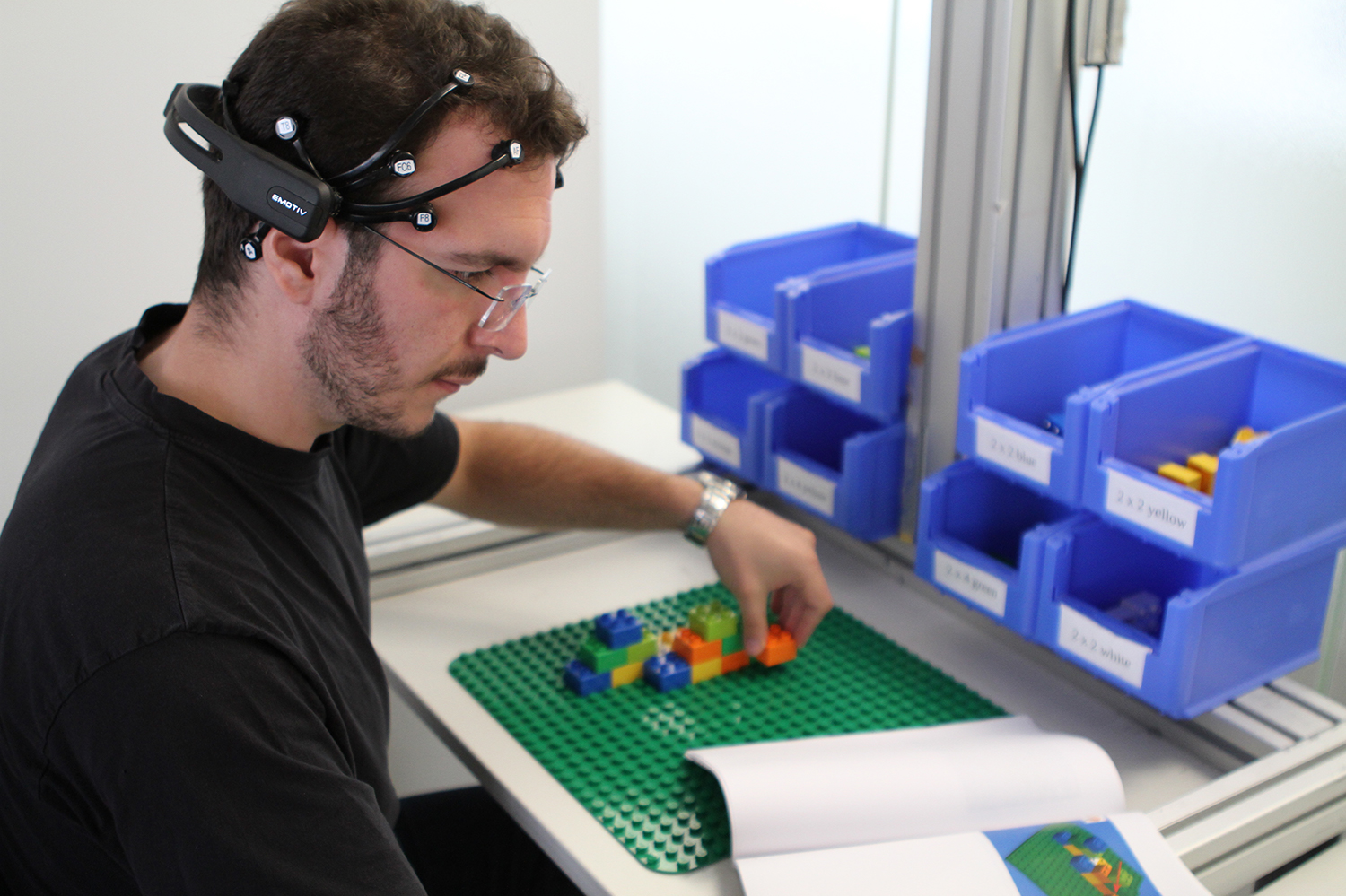}
\label{fig:teaser_paper}}
\subfloat[][]{
\includegraphics[width=0.5\columnwidth]{figures/IMG_0431-2.JPG}
\label{fig:teaser_projection}}
\caption[Paper-based and projection based Lego Duplo assembly while recording electroencephalography]{\textbf{(a)} Paper-based and \textbf{(b)} projected \textit{in-situ} instructions at a manual assembly workplace. An EEG headset measures the level of working memory. This provides objective insights about the working memory load that is placed during manual assembly.}
\label{fig:splash_eeg}
\end{figure}

Novel user interfaces -- for instance, assistive instruction systems -- are often evaluated for their improvements regarding the performance or subjective workload to justify their implementation over the \textit{status quo}. Performance is typically assessed in terms of error rate and assembly completion time, while cognitive workload is often assessed using self-rated measures or semi-structured interviews~\cite{funk2017working}. Commonly used self-rated measures include the NASA-TLX~\cite{HART1988139} or the simpler RSME~\cite{brookhuis2002assessment, zijlstra1993efficiency}. However, these metrics are susceptible to individual differences in subjective reporting. For instance, extroverted confident individuals might be less likely to indicate workload to the same degree as introspective modest individuals~\cite{kaptchuk2003effect}. Furthermore, novel assistive systems and visualizations might introduce a novelty effect. This means, that participants provide self-reported scores that are influenced by the excitement over a new technology~\cite{wells2010effect}. Additionally, subjective reports can only be performed after test completion and rely on cognitive processes, namely working memory. Such methods are limited in their objective and real-time assessment of cognitive workload when evaluating assistive technologies.

Therefore, physiological methods are increasingly employed as a means to estimate mental states, such as cognitive workload~\cite{Haapalainen:2010:PMA:1864349.1864395, Kosch:2018:LME:3170427.3188643, Kosch:2018:YET:3173574.3174010}. For example, \acp{BCI}~\cite{CGF:CGF1928, Vallabhaneni2005} have been used successfully to assess the complexity of presented information in a variety of scenarios~\cite{CGF:CGF1928, kosch2016brain} (see Figure~\ref{fig:splash_eeg}). A \ac{BCI} records brain activity in real-time and computationally derives estimates of a targeted mental state to either provide feedback about the physiological state of the user or enable device control through neuronal activity. Increasingly, advances in brain-recording technology and neuroscience findings contribute towards a vision of ubiquitous \ac{BCI} deployment in everyday places of work and play~\cite{bulling2014cognition, Kitamura:2003:THB:642611.642684, Velichkovsky:1996:NTW:238386.238619}. In the current context, a \ac{BCI} for cognitive workload sensing could be used in collaborative work settings by allocating tasks to workers who are less fatigued~\cite{Fraser:2017:WAD:3025453.3026036, funk2017working} or provide adaptive assistance whenever necessary~\cite{funk2015stop}. However, for this future to be realized, it is necessary to ensure that recorded estimates are robust and valid for its given setting.

\acp{BCI} may record brain activity via \ac{EEG}, which has been preferred for its lightweight and mobile hardware, high temporal resolution, and non-invasive nature~\cite{Grimes:2008:FPC:1357054.1357187}. EEG measures neuronal signals from scalp electrodes, relative to a reference electrode~\cite{fisch1999fisch, gloor1969hans}. Previous research has utilized EEG as a measure for cognitive workload to evaluate task difficulties in real-time~\cite{gevins1998monitoring, gevins1997high, STIPACEK2003193}. To some extent, it is possible to rely on EEG to classify cognitive workload in real-time, particularly those that involve working memory processes~\cite{Grimes:2008:FPC:1357054.1357187}. This makes EEG to a viable alternative or complementary metric for mental workload measures.

In this chapter, we present a method to assess cognitive workload using mobile electroencephalography. \ac{EEG} facilitates the measurement of cortical activity which represents markers for cognitive effort~\cite{gevins1998monitoring, gevins1997high}. We present a study that compares two assistive systems for manual assembly which have shown significant differences in subjective perception. This is complemented by the presentation of a design pipeline that suggests the use of \ac{EEG} as a complementary measure alongside subjective assessments. Henceforth, the evaluation of \ac{EEG} cognitive workload measures yields the potential for real-time adjustments of user interfaces. We seek to answer the following research question:

\begin{itemize}
\renewcommand{\labelitemi}{$\rightarrow$}
\item \textbf{RQ3}: Does electroencephalography provide measures of cognitive workload for user interface evaluation?
\end{itemize}


\begin{center}
\begin{GrayBox}
\centering
\parbox{0.98\textwidth}{
\emph{This section is based on the following publications:}
\vspace{3mm}
\publicationsbegin
\item \bibentry{Kosch:2018:ICA:3233739.3229093}
\item \bibentry{10.3389/conf.fnhum.2018.227.00114}
\publicationsend
}
\end{GrayBox}
\end{center}

\section{Background}
Previous research has investigated effort in examining \ac{EEG} data while subjects perform specific tasks. This section summarizes relevant research regarding the use of \ac{EEG} as a non-invasive method to quantify cognition and presents assistive systems that adapt to the measured brain activity.

\subsection{Quantifying Cognition using \ac{EEG}}
Much research has been performed to define and quantify cognitive workload. Sweller et al. \cite{sweller1988cognitive, sweller1994cognitive} defined three key components of cognitive workload comprising intrinsic, germane, and extraneous cognitive workload. Intrinsic workload describes the inherent complexity of the task itself, and can therefore not be manipulated by external sources. Germane workload describes the cognitive effort subjects need to comprehend and process new information. Extraneous workload describes the cognitive demand to understand and process the visual representation of the underlying information. Experiments often evaluate extraneous workload by influencing the visualization of information. Intrinsic and germane workload is not easy to manipulate since they are task-related or depend on individual mental capabilities.

More precisely, cognitive workload defines the mental effort being used in working memory, which is responsible for fast information processing and limited within its capacities \cite{baddeley1974working}. Gevins et al.~\cite{gevins1997high} invested effort to quantify the occurrence and amount of working memory using \ac{EEG}. They found a decrease in alpha frequencies and an increase in theta frequencies during cognitively demanding tasks. Jensen et al.~\cite{jensen} replicated their results using a memory demanding task. Scharinger et al.~\cite{scharinger2017comparison} also showed a variety of other tasks that correlate with increased working memory. However, working memory strongly depends on individual attributes, such as age or neuronal health state~\cite{Klimesch1993}.

Machine learning became popular to classify mental states using \ac{EEG} data~\cite{Lotte2014, 1741-2552-15-3-031005, 1741-2552-4-2-R01} and has been used to estimate workload in different contexts~\cite{muhl2014eeg}. Lee et al.~\cite{Lee:2006:ULE:1166253.1166268} collected \ac{EEG} data of participants to classify cognitive demanding tasks according to their difficulty. Participants were asked to do different cognitive tasks, including mental rotation and mental arithmetic. Instead of using previously recorded data associated with similar tasks~\cite{anderson1996classification, 64464}, data was collected from scratch. Using machine learning, a classification accuracy of $92.4$\% was achieved. However, the experiment has not focused on classifying working memory itself. An approach to classifying working memory was undertaken by Grimes et al.~\cite{Grimes:2008:FPC:1357054.1357187}. Participants performed a \textit{N}-back task, which is a popular task to demand short-term memory~\cite{mehler2011agelab, HBM:HBM20131}. Their study included four different difficulties, where their results yielded a classification accuracy of $99$\% for binary classification and an accuracy of $88$\% for classification between four classes of difficulty. Nonetheless, a medical \ac{EEG} headset was utilized for their experiment, which is hard to deploy on the fly. Furthermore, an artificial task was used to measure the level of working memory instead of using a real-world scenario.

\subsection{Evaluating and Adapting User Interfaces}
Besides using \ac{EEG} for quantifying and classifying working memory researchers have employed this measurement to enable the development of cognition-aware systems. Zander et al.~\cite{zander} investigated how passive brain activity measurements can be leveraged to handle take-over tasks during autonomous driving. Prinzel et al.~\cite{LawrenceJ.Prinzel:2001:EAE:886462} researched how task allocation can be carried out efficiently using \ac{EEG}-related metrics. Finally, El-Komy et al.~\cite{elkomy2017abbas} used physiological sensing devices in an assembly environment comprising additional artificial tasks to measure the workers' stress levels. However, only results from the emotion and arousal scores retrieved by the integrated BCI interface were reported.

To the best of our knowledge, no prior work concerning the evaluation of \ac{EEG} to derive cognitive workload during manual assembly and with different instruction systems has been done. In this work, we close this gap by utilizing standardized manual assembly tasks with two different instruction systems. We record and compare \ac{EEG} data in a repeated measure experimental design to evaluate \ac{EEG} as a valid assessment tool for cognitive workload induced by assistive technologies during manual assembly.

\begin{figure}
\centering
\subfloat[][]{
\includegraphics[width=0.33\columnwidth]{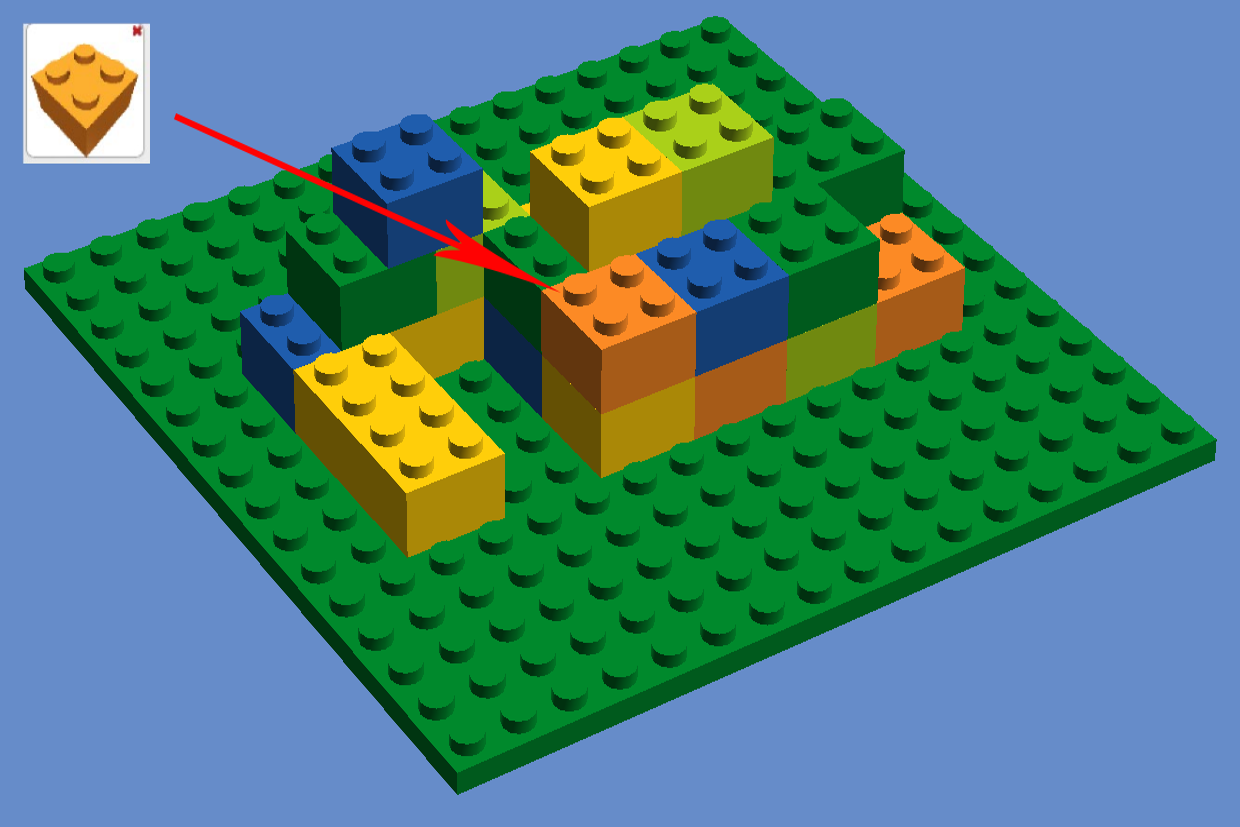}
\label{fig:lego}}
\subfloat[][]{
\includegraphics[width=0.33\columnwidth]{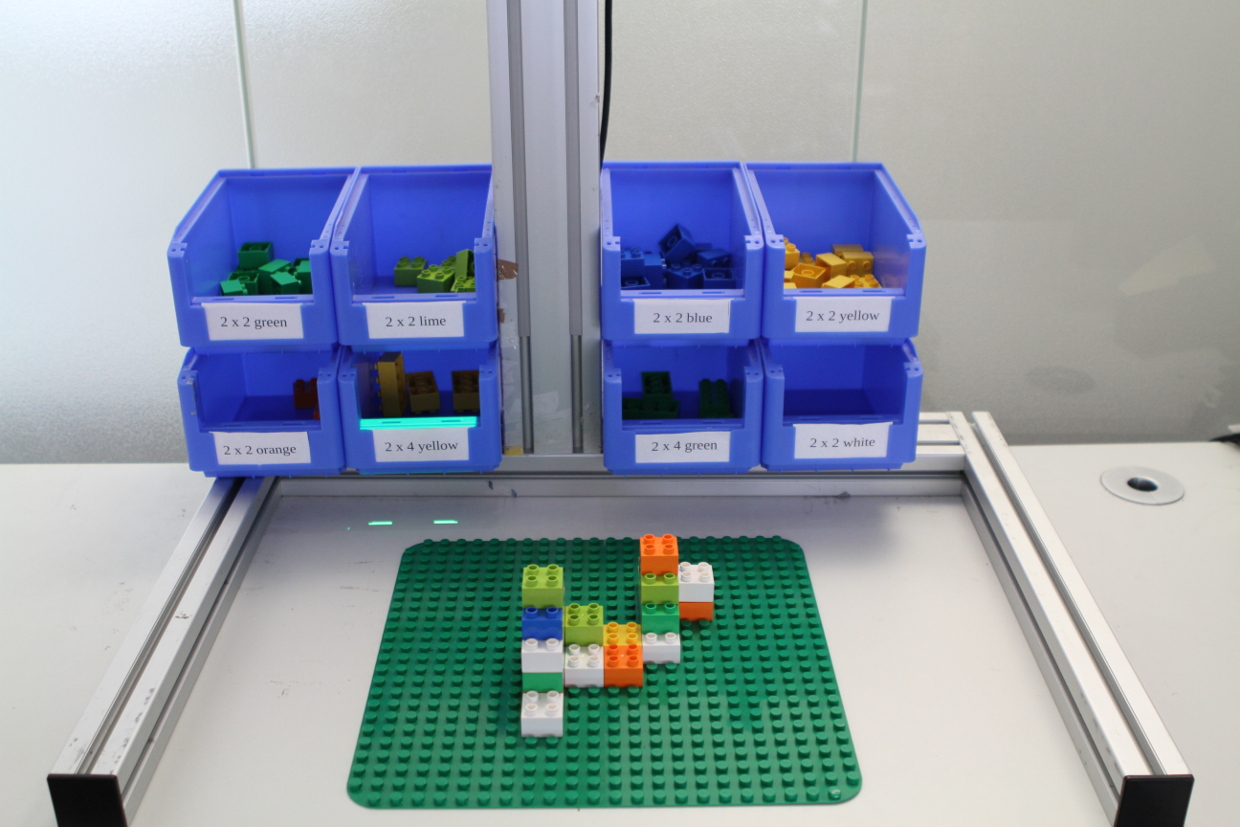}
\label{fig:projected_box}}
\subfloat[][]{
\includegraphics[width=0.33\columnwidth]{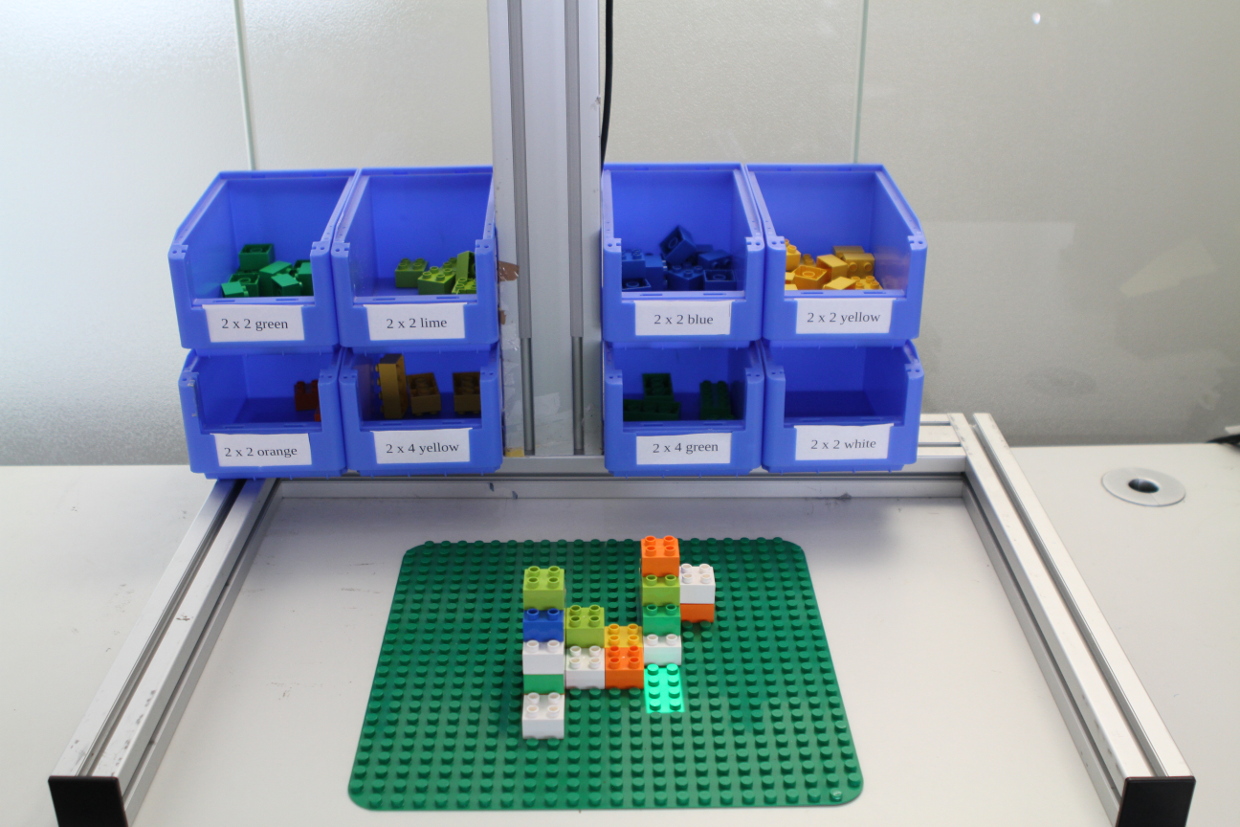}
\label{fig:projected_plate}}
\caption[Used Lego Duplo instructions for the assembly]{\textbf{(a):} Snippet from the used Lego Duplo paper instruction. The next brick can be seen in the upper left corner. A red arrow denotes the final assembly position of the brick. \textbf{(b):} Highlighted box using \textit{in-situ} projections. \textbf{(c):} Projection of the target position of a selected brick.}
\label{fig:lego_dummy2}
\end{figure}

\section{Assembly Instruction Systems}
We identified two different instruction systems from related research, which are different regarding subjective perception~\cite{funk2016interactive}. Within both instruction systems, Lego Duplo bricks are assembled. This task resembles a full replacement for a real assembly task and enables to change the complexity of the task without changing the task itself~\cite{funk2016interactive, Kosch:2016:CTA:2982142.2982157}.
\subsection{Assembly Instruction Visualizations}
Informed by related work, two instruction visualizations are identified which differ in overall interpretation complexity~\cite{funk2016interactive}. We have chosen paper instructions, as they represent the current state of the art when it comes to transferring assembly instructions in manual assembly lines~\cite{funk2015benchmark}. This is compared to projected \textit{in-situ} instructions, where assembly instructions are projected in the workplace. We have chosen these two assembly instruction modalities as subjective measures suggest an alleviation of workload for \textit{in-situ} projected instructions compared to printed paper instructions~\cite{funk2016interactive, funk2015pick}. In the following paragraphs, we describe both instruction systems used in the study in detail.
\subsubsection{Paper Instructions}
We printed single-sided instructions on an A4 sheet of paper. Each work step was printed on a single page, such that the position and size of every step were the same. The paper instructions were put together in the correct order using a folder and positioned to the left or right of the users, depending on their handedness. The folded assembly instruction remained the same position relative to the user's position. The instruction shows a brick on the upper left corner which is required to select for completing the current work step. Marked by an arrow, the assembly instruction shows the final position of the brick (see Figure~\ref{fig:lego}).

\subsubsection{In-Situ Instructions}
We compare the presented paper instructions with \textit{in-situ} projections displaying assembly instructions. We use a similar system as shown by Funk et al.~\cite{Funk:2015:UIP:2700648.2809853}. A projector mounted above the work table displays the next assembly step on the workspace. A Kinect v2\footnote{\url{https://developer.microsoft.com/en-us/windows/kinect} - last access \urldate} validates each work step of the Lego Duplo construction. This includes the verification of correct item selections from bins by observing the hand movements of the participant (see Figure~\ref{fig:projected_box}) and assembly steps by comparing the (see Figure~\ref{fig:projected_plate}). The next work step is displayed when the current work step is performed correctly. The system is waiting until the current work step is carried out correctly and does not proceed if the user makes an error.

\section{Manipulating Cognitive Workload}
We manipulate working memory to assess the validity of our setup. We use a visual \textit{N}-back task~\cite{mehler2011agelab} with two levels of task difficulty (\textit{N}$=0$ and \textit{N}$=2$) to induce cognitive workload, namely executive working memory. In the \textit{N}-back task a series of numbers is presented (\textit{i.e.}, numbers). Each symbol appears in a fixed position. Upon symbol representation, participants have to decide if the current symbol is equal to the symbol shown \textit{N} steps ago. Participants have to keep a sequence of \textit{N} symbols constantly in their memory, decide if a match occurred, and then update the sequence in their memory.

For example, the \textit{0}-back task requires participants to compare each displayed symbol with the first one seen in the series. Since the currently displayed number matches with the first one in the series during the \textit{0}-back task, no memory updates are required. However, the task difficulty can be manipulated by changing \textit{N}~\cite{gevins1998monitoring, Grimes:2008:FPC:1357054.1357187, HBM:HBM20131}. For \textit{N}$=2$, participants have to memorize the last two symbols in the series while paying attention to matches when the currently displayed symbol is the same as shown two items ago. Table~\ref{tab:nback} shows an example of the \textit{N}-back.

\renewcommand{\arraystretch}{1.2}
\begin{table}
\centering
\begin{tabular}{l|c|c|c|c|c|c|c}
\toprule
\textbf{Displayed number} & 5 & 8 & 3 & 4 & 3 & 9 & 1\\
\hline
\textbf{Press button (\textit{0}-back task)} & 5 & 8 & 3 & 4 & 3 & 9 & 1\\
\textbf{Press button (\textit{1}-back task)} & & 5 & 8 & 3 & 4 & 3 & 9 \\
\textbf{Press button (\textit{2}-back task)} & & & 5 & 8 & 3 & 4 & 3 \\
\textbf{Press button (\textit{3}-back task)} & & & & 5 & 8 & 3 & 4 \\
\bottomrule
\end{tabular}
\caption[Example of the \textit{N}-back task]{Example of the \textit{N}-back task task. Participants have to confirm by a button press, that the currently displayed number matches with the number seen \textit{N} numbers ago.}
\label{tab:nback}
\end{table}

The motoric requirements remain constant across difficulty levels during the \textit{N}-back task. This enables comparisons between different difficulties as working memory load is measured while excluding reactions from external stimuli.

A smartphone app\footnote{\url{www.play.google.com/store/apps/details?id=cz.wie.p.nback} - last access \urldate} is used to display a single matching \textit{N}-back task. Throughout the experiment, we use a Nexus 5X\footnote{\url{www.gsmarena.com/lg_nexus_5x-7556.php} - last access \urldate} with a screen size of 5.2 inches to run the \textit{N}-back trials. The displayed symbols ranged between 0 and 9, which appeared in random order. Each number is displayed in the center of the smartphone screen for one second. Afterward, the screen remains blank for 2.5 seconds before the next number appears. Participants have to press a button on the screen if a match occurs.

\section{\ac{EEG} as Indicator for Workload}
In the following study, we assess the validity of our setup by conducting two \textit{N}-back tasks with different complexities as described before. Afterwards, we conduct two assembly tasks per participant, each with the two previously mentioned assembly instruction systems. The overall hypothesis is that \textit{in-situ} projections induce less cognitive workload than printed paper instructions. This leads to the following hypotheses:
\begin{itemize}
\item \textbf{H1:} Projected \textit{in-situ} instructions will induce \textit{higher alpha power}, relative to printed instructions.
\item \textbf{H2:} Projected \textit{in-situ} instructions will produce \textit{lower scores of subjective workload}, relative to printed instructions.
\item \textbf{H3:} Projected \textit{in-situ} instructions will produce \textit{fewer item selection errors}, relative to printed instructions.
\item \textbf{H4:} Projected \textit{in-situ} instructions will produce \textit{fewer assembly errors}, relative to printed instructions.
\item \textbf{H5:} Projected \textit{in-situ} instructions will produce \textit{faster completion times}, relative to printed instructions.
\end{itemize}

\begin{figure}
\centering
\subfloat[][]{
\includegraphics[width=0.5\columnwidth]{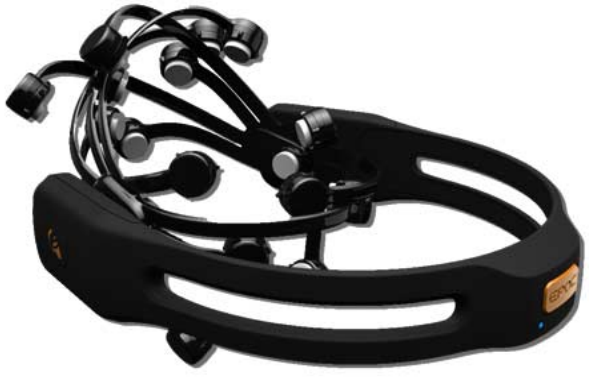}
\label{fig:emotiv_epoc}}
\subfloat[][]{
\includegraphics[width=0.5\columnwidth]{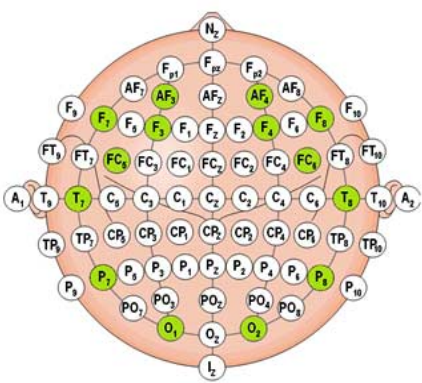}
\label{fig:electrode_layout_epoc}}
\caption[Emotiv Epoc EEG headset and its electrode layout]{\textbf{(a):} The Emotiv Epoc wireless EEG headset featuring 16 electrodes including two reference electrodes. \textbf{(b):} Electrode placement layout of the 14 measurement electrodes~\cite{cybertech}.}
\label{fig:hardware}
\end{figure}

\subsection{Methodology and Measures}
We used the Emotiv Epoc as brain-sensing device during the whole experiment (see Figure~\ref{fig:emotiv_epoc} and Figure~\ref{fig:electrode_layout_epoc}). Since the alpha band varies between participants, we conducted an eyes opened and eyes closed task to estimate the individual alpha band~\cite{BARRY20072765}. The overall duration of this trial was one minute. The participants started with their eyes opened and were verbally instructed to close their eyes after 30 seconds to provoke a sudden increase in alpha power~\cite{BARRY20072765}. The participants kept their eyes closed for another 30 seconds. We use this peak as a reference point for extracting alpha power for later analysis. Furthermore, we took $\rpm 2$ Hz around the peak frequency as a measure for the alpha band.

We continue then with the \textit{N}-back task to verify the validity of our setup and accuracy of our \ac{EEG} measures. Participants start with a \textit{0}-back task to be induced with low workload, followed by a \textit{2}-back task to be induced with high workload. The 2-back induces sufficient complexity to make the differences in working memory between resting and \textit{N}-back task visible in \ac{EEG} data~\cite{brouwer2012estimating}. A NASA-TLX questionnaire is filled out after each \textit{N}-back condition to collect subjectively perceived workload.

We begin with the assembly task afterward. Inspired by several reference tasks proposed by previous research~\cite{funk2015benchmark}, we use a Lego Duplo task to evaluate the paper and projected \textit{in-situ} instructions in terms of measured workload. We prepared two different assembly instructions, where each of them is modeled as paper instruction and projection as shown in Figure~\ref{fig:lego_dummy2}. For the assembly tasks, we used a repeated-measures experimental design with the instruction visualization as a single factor including the levels paper and \textit{in-situ}. We counterbalanced the order of conditions according to the balanced Latin square. As identified in the previous work~\cite{Blattgerste:2017:CCA:3056540.3056547,buettner2017design, funk2016interactive, Sand2016}, we measure the number of errors and the task completion time per trial. The number of errors was divided into item selection and assembly errors. An item selection error is counted whenever participants put their hands into a box where incorrect bricks reside. An assembly error is counted when a brick is assembled in a wrong position. To reduce the number of head movements, we seated the participants before the assembly experiment at a comfortable distance to the assembly setup, so that boxes and the assembly plate can be reached with minimal effort. Participants filled out a NASA-TLX questionnaire after each trial to provide their subjectively perceived workload during the last assembly condition. Ultimately, we asked the participants verbally about their preference between both assembly instruction systems and noted their answers for later analysis.

\subsection{Procedure}
Participants signed a consent form and provided their demographics after we explained the course of research to them. We put the BCI on the participant's head and ensured good connectivity between the scalp and all electrodes. We explained to the participants what \ac{EEG} signals and noisy artifacts are. We asked participants to keep their head as still as possible and to avoid unnecessary eye blinks.

Participants started with a one-minute baseline resting task. Participants kept their eyes open for 30 seconds. After the first 30 seconds elapsed, participants were instructed to close their eyes for another 30 seconds. The study continued with two \textit{N}-back tasks, each comprising 20 numbers. The total runtime of each \textit{N}-back was one minute and ten seconds. The experiment started with the \textit{0}-back task. Participants had to press the match button every time they saw a number since the currently displayed numbers refer to the same number. Participants continued with the \textit{2}-back task. The participants had to press the match button whenever the currently displayed number was equal to the number shown two numbers back. We recorded \ac{EEG} data during all tasks. Participants were asked to fill out a NASA-TLX questionnaire after each \textit{N}-back task.

After the verification procedure, participants started with the assembly procedure. They began with either paper instructions or projected \textit{in-situ} instructions based on the order of the balanced Latin square. Additionally, we shuffled the order of the Lego Duplo instructions itself between the conditions. During the assembly, we recorded raw \ac{EEG} data, counted the number of errors separated into item selection or assembly errors, and the time used for assembly. During all conditions, participants were instructed to keep their head still and avoid unnecessary eye blinks. After every assembly, participants were asked to fill out a NASA-TLX questionnaire to assess the perceived the workload during the usage of the given instruction system. Qualitative comments about the preference of users regarding the instruction systems were collected in the end.

\subsection{Data Processing}
We apply the following data processing procedure: Data is filtered using a spatio-spectral decomposition method~\cite{NIKULIN20111528} with filter thresholds between 0.5 Hz and 20 Hz to remove unwanted frequencies caused by eye blinks or head movements. We average all 14 channels by calculating the element-wise mean of the signals. We remove the first and last four seconds of the signal to avoid unwanted artifacts caused by the beginning and end of the trial~\cite{Lee:2006:ULE:1166253.1166268}. We divide the signal into one-second slices with an overlap of half a second. Instead of extracting the alpha band between 8 Hz and 12 Hz, we determine the maximum peak during the eyes opened and eyes closed task for each individual~\cite{FINK2005252, Klimesch1993}. The power spectra around the maximum peak ($\rpm 2$ Hz) is averaged and used as individual alpha power.

\subsection{Results}
We recruited twelve participants over our university mailing lists (8 male, 4 female). All participants were students and had a normal or corrected-to-normal vision. None of the participants were affected by neurological disorders. The mean age was 23 years (SD = $2.22$). Participants were compensated with 10 Euro for their participation.

\begin{figure}
\centering
\subfloat[][]{
\includegraphics[height=0.36\columnwidth]{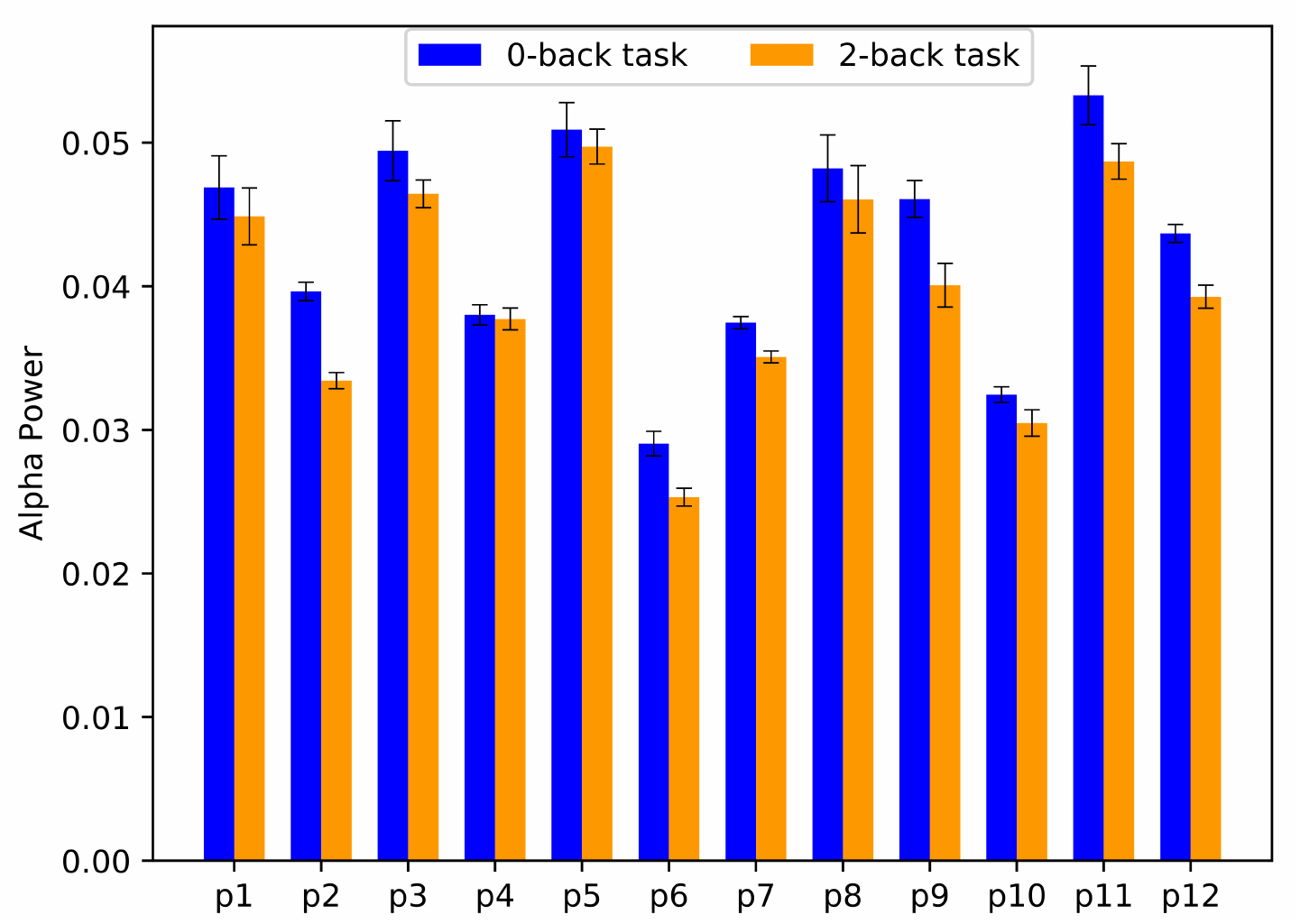}
\label{fig:baseline_ch5}}
\subfloat[][]{
\includegraphics[height=0.365\columnwidth]{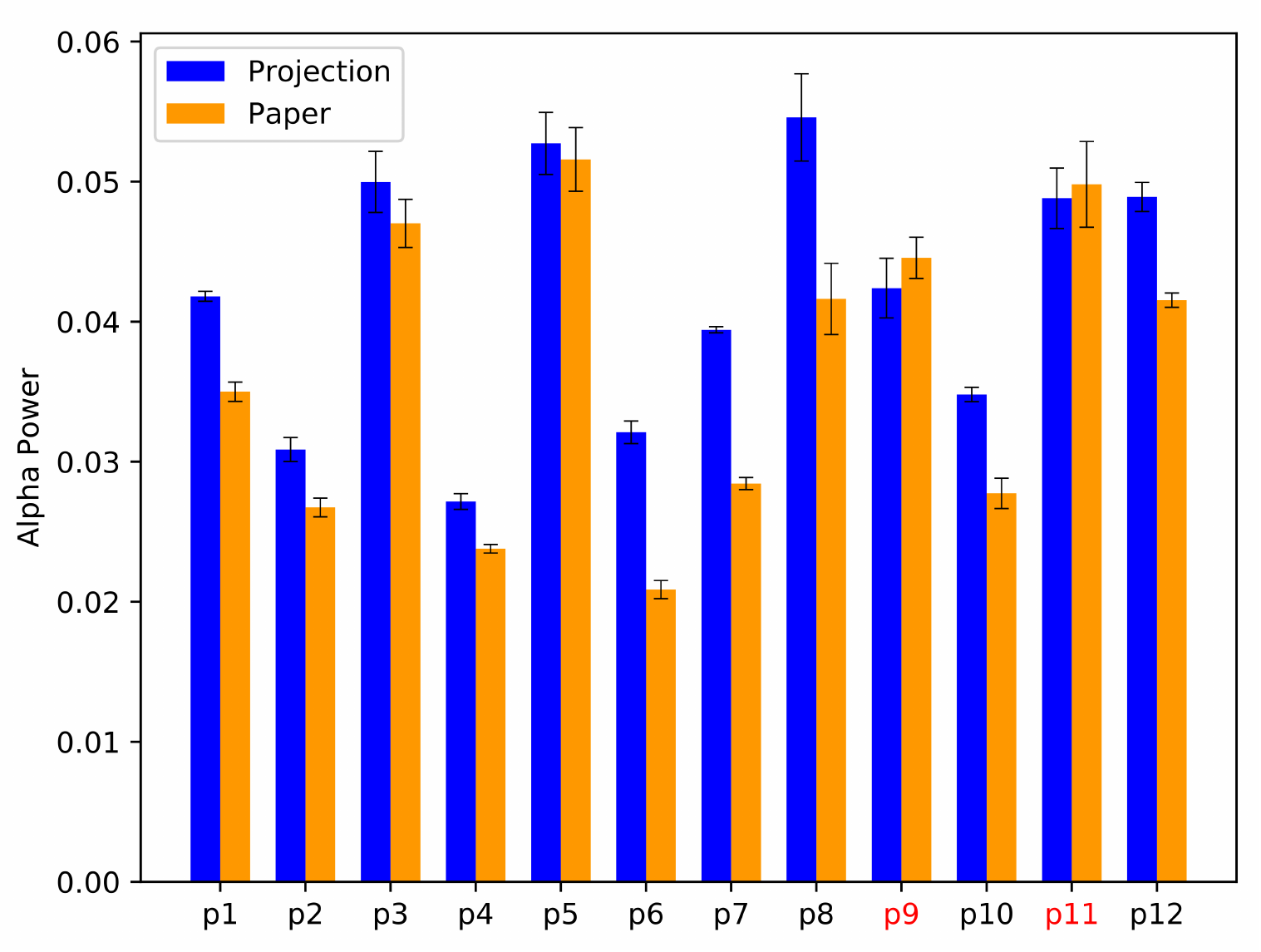}
\label{fig:conditions}}
\caption[IAF power for each condition]{\textbf{(a):} IAF power for working memory load, \textit{i.e.}, \textit{0}-back, and \textit{2}-back. The IAF power is higher for lower working memory load. The error bars depict the standard error of the sample mean. \textbf{(b):} Participant-wise comparison of IAF power when using \textit{in-situ} projections and paper instructions. Except p9 and p11, the use of \textit{in-situ} instructions results in higher IAF power compared to paper instructions. The error bars depict the standard error of the sample mean.}
\label{fig:alpha_overall}
\end{figure}

\subsubsection{Executive Working Memory Load and \ac{EEG}}
We identified alpha desynchronization as features that corresponded with an increased load in executive working memory. The alpha band is known to vary across individuals~\cite{doppelmayr1998individual}. Therefore, we determined an individual alpha frequency (IAF) bandwidth for each participant based on their peak frequency value given their baseline \ac{EEG}. We defined $\rpm2$ Hz around the peak frequency as the individual alpha band. Overall, the mean IAF band ranged between $6.5$ Hz and $10.5$ Hz (SD = $2.14$).

The mean power of the IAF power for each participant was submitted to a two-tailed paired-samples t-test for the factor \textit{executive working memory load} consisting of the \textit{0}-back and \textit{2}-back task. A Shapiro-Wilk test confirmed that the data was normally distributed. The results reveal a significant difference, $t(11)=5.90,~p<0.001,~d=1.70$, between the \textit{0}-back and \textit{2}-back task. Figure~\ref{fig:baseline_ch5} shows the direct comparison of alpha activation between both conditions. To summarize, our chosen \ac{EEG} feature was sensitive to load in an executive working memory task, namely, there was less power in the IAF band when there was a higher load in this task. Thus, we report a large effect size in support of H1.

\subsubsection{Assembly Performance and Alpha Power}
The IAF power of each participant was submitted to a two-tailed paired-samples t-test for the factor \textit{haptic assembly instructions} consisting of paper and \textit{in-situ} instructions after a Shapiro-Wilk test confirmed that the data was normally distributed. The results reveal a significant difference, $t(11)=3.86,~p=0.003,~d=1.12$. Figure~\ref{fig:conditions} shows the mean alpha activation per condition and participant. Similar to our findings with the executive working memory load, we find that IAF power is generally lower for paper instructions compared to when participants experienced projected \textit{in-situ} instructions instead---note that this was not true for p9 and p11 at the individual level. Thus, we report a large effect size in support of H2. More importantly, we show that the IAF power responds for lower executive working memory load as it does for our projected \textit{in-situ} instructions.

\begin{figure}
\centering
\subfloat[][]{
\includegraphics[width=0.5\columnwidth]{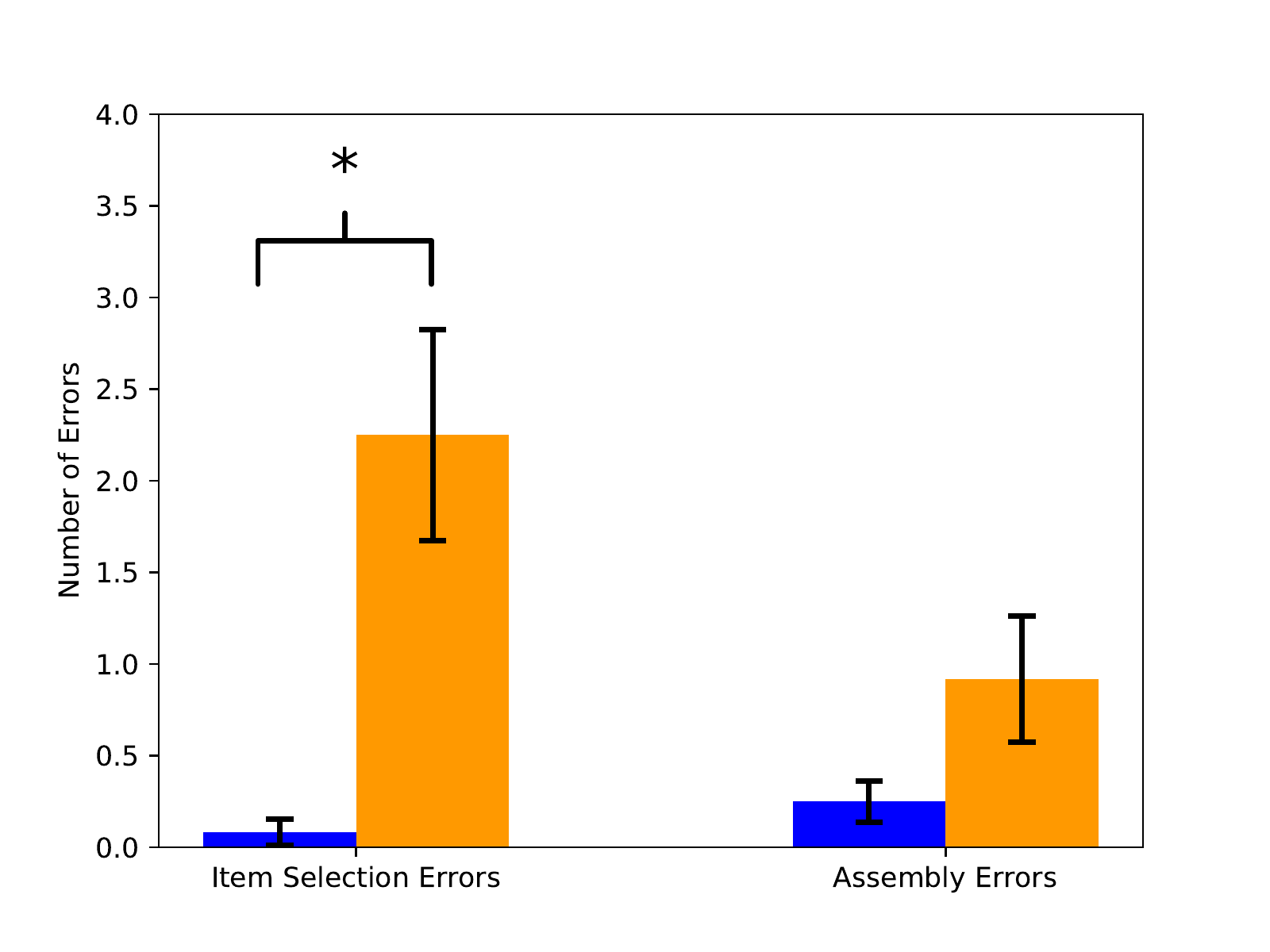}
\label{fig:errors}}
\subfloat[][]{
\includegraphics[width=0.45\columnwidth]{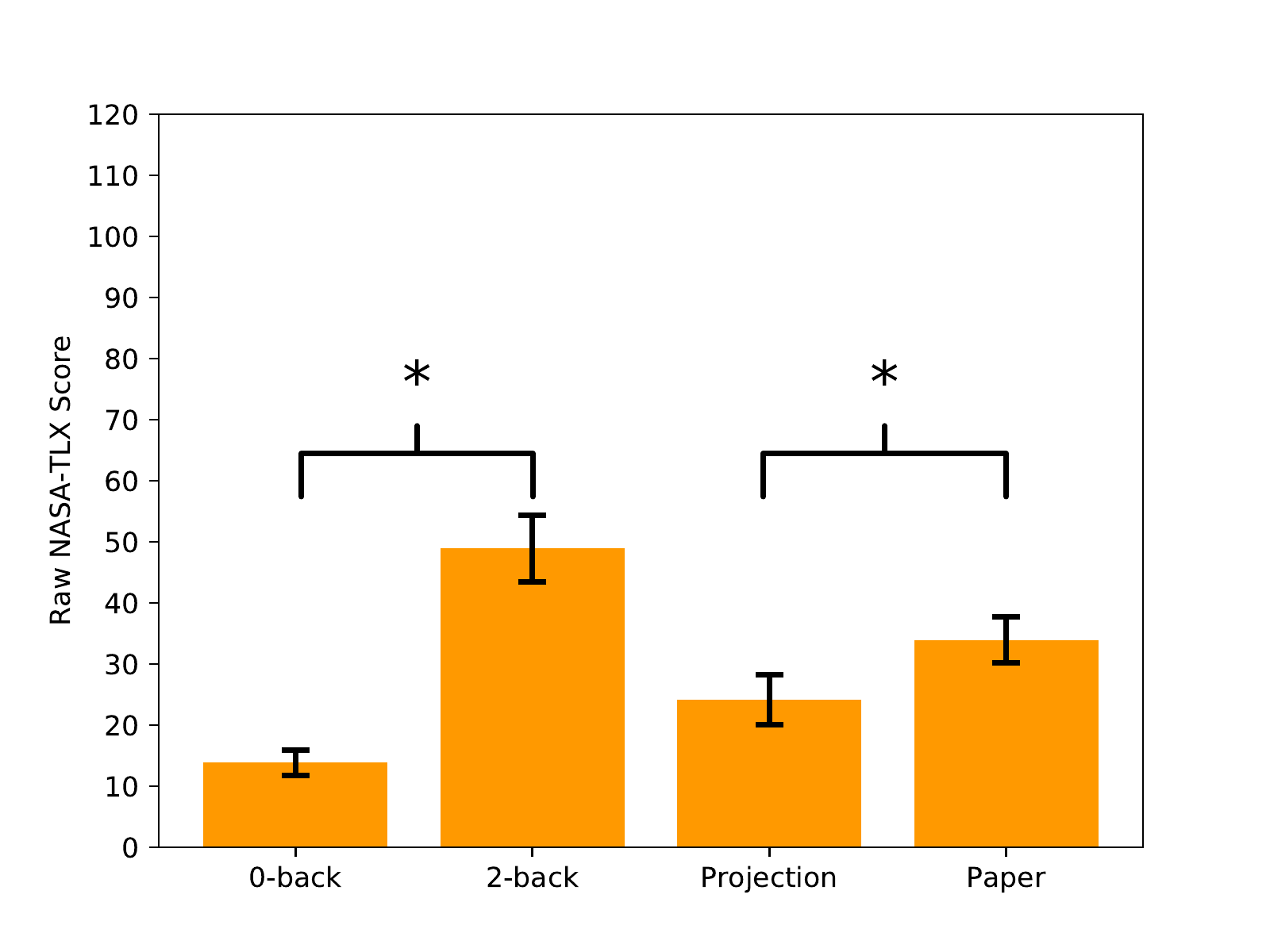}
\label{fig:nasa_tlx}}
\caption[Number of item selection errors and raw NASA-TLX scores]{\textbf{(a):} Mean item selection and assembly errors. The error bars indicate the standard error. Brackets indicate significant differences. \textbf{(b):} Mean raw NASA-TLX scores per condition. The error bars depict the standard error. Brackets indicate significant differences.}
\label{fig:errors_nasa_tlx}
\end{figure}

Additionally, we statistically compare the number of errors made by the participants during assembly as well as the time they required to finish the assembly between the different conditions. We classify errors into item selection errors and assembly errors. Overall, participants did in average $2.25$ (SD = $2.301$) item selection errors during the paper instruction condition and $0.083$ (SD = $0.289$) errors during the \textit{in-situ} condition. $0.917$ (SD = $1.379$) assembly errors were performed when using paper instructions and $0.25$ (SD = $0.452$) during the \textit{in-situ} condition. A Shapiro-Wilk test did not show a normal distribution for the item selection and assembly errors. Thus, we have conducted a Wilcoxon signed-rank test. A significant difference was found for the number of item selection errors, $p = 0.009$. However, no significant difference was found for the number of assembly errors, $p = 0.188$. Figure \ref{fig:errors} compares the number of item selection and assembly errors between both instruction systems.


The task completion time averages to $217.08$ seconds (SD = $40.31$) for paper instructions and $124$ seconds (SD = $13$) for projected \textit{in-situ} instructions. The task completion time shows a significant effect between both conditions regarding task completion time, $t(11) = -8.82$, $p = 0.001$, $d = -2.55$.

\subsubsection{Qualitative Assessment}
We statistically analyze the subjectively perceived workload using the collected NASA-TLX questionnaires. We average the scores received from the six Likert scales per condition to calculate the mean raw NASA-TLX scores. The averaged NASA-TLX score amounts to $13.83$ (SD = $7.34$) for the \textit{0}-back task, $48.92$ (SD = $19.01$) for the \textit{2}-back task, $33.92$ (SD = $13$) for the assembly task using paper instructions, and $24.17$ (SD = $14.26$) during assembly using projected \textit{in-situ} projections.

A Shapiro-Wilk test confirmed that the data is normally distributed. A t-test shows a statistical difference in NASA-TLX scores between the \textit{0}-back and \textit{2}-back condition, $t(11) = -6.41$, $p = 0.001$, $d = -1.85$. Therefore, more workload was subjectively perceived in the \textit{2}-back task than in the \textit{0}-back task. A significant difference between paper and \textit{in-situ} instructions, $t(11) = -3.86$, $p = 0.003$, $d=-1.12$, was found. Therefore, lower alpha activation was measured during the \textit{in-situ} conditions compared to the printed paper instruction conditions. Figure~\ref{fig:nasa_tlx} shows the mean NASA-TLX scores.


Additionally, we collected comments from the participants by asking for their preference regarding the assembly in the instruction system. Most participants provided us with positive feedback. For example, one participant mentioned that \emph{"[\ldots] projected instructions were easier and faster to understand than paper instructions. The light was good guidance."} (P1, similar to P2, P3, P7). Another participant mentioned that \emph{"[he] felt that I was faster using projections since I did not have to flip the paper. Having both hands free enhanced the overall assembly."} (P5). Finally, one participant mentioned that \emph{"It was easier to follow the light than to follow the paper instructions itself. However, I felt like a robot during assembly."} (P4).

However, some participants stated that \emph{"[\ldots] the higher assembly speed using projections was stressful. Paper instructions provided a short relieve in workload when flipping the page."} (P9 similar as P11, P12) or that they were \emph{"[\ldots] unused to it, but [I] could familiarize with it after some time."} (P3, similar to P6, P9).

It is interesting to note that P9 and P11 perceived the assembly as stressful due to its fast pace. Both participants also show less alpha activation during the projected \textit{in-situ} condition than in the paper condition (see Figure~\ref{fig:conditions}). Instructions were provided immediately by the \textit{in-situ} instruction system, which was perceived as stressful two participants (P9, P11). Paper instructions provided cognitive alleviation when flipping the page to mentally prepare for the next work step. Previous research also supports a positive correlation between stress and cognitive workload~\cite{doi:10.1080/10253890600965773, SCHOOFS2008643}. The discomfort of the provided assembly instruction system can also be elicited from our results.


\subsection{Assessing Cognitive Workload in Real-Time}
Our results show a significant difference in alpha activation between paper and projected \textit{in-situ} instructions. Using the collected data, we analyze the real-time applicability of \ac{EEG} for cognitive workload assessment. We achieved this by calculating the alpha fluctuation over time for both conditions.

\begin{figure}
\includegraphics[width=\columnwidth]{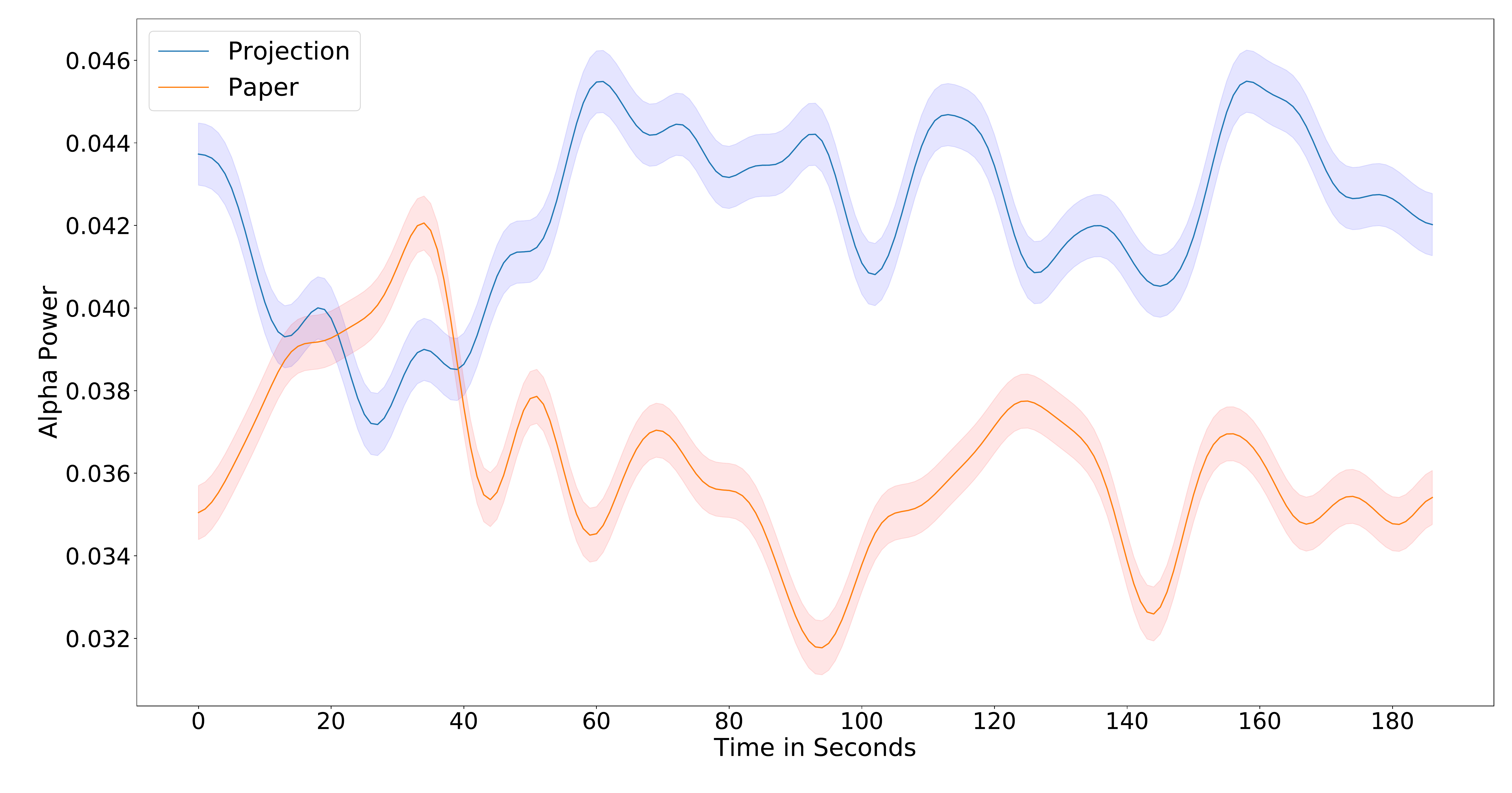}
\vspace{-5mm}
\caption[Mean IAF power fluctuation of all participants across time]{Mean IAF power fluctuation of all participants across time. Alpha power increases when using projected \textit{in-situ} instructions and decreases for paper instructions. The shadowed area describes the standard deviation of each data point per participant.}
\label{fig:alpha_fluctuation}
\end{figure}

Similar to the previous analysis steps, we preprocess the data using the spatio-spectral decomposition method~\cite{NIKULIN20111528} with filter thresholds between 0.5 Hz and 20 Hz and calculate the mean of all channels. We remove the first and last four seconds of the signal to remove unwanted response artifacts. We calculate the individual alpha band per participant and averaged the alpha power overall participants for both conditions. Figure~\ref{fig:alpha_fluctuation} shows the alpha fluctuation for both conditions.

For the paper instruction, there is an increase followed by a decrease of alpha activation over time for the paper instructions. Workload starts to differentiate with time when information from paper instructions have to be held continuously in the short-term memory. This could be a reason for the decrease of alpha power with time.

Projected \textit{in-situ} instructions show a decrease followed by an increase in alpha power. We assume, that the novelty of the system for the participants is responsible for the decrease in alpha power at the beginning of the condition. Alpha power increases when the participants become familiar with the system and less information has to be kept in the short-term memory. A Shapiro-Wilk test showed a non-normal distribution. A Wilcoxon signed-rank test resulted in a statistical difference between both conditions, $p = 0.001$.

\section{Discussion}
We evaluate the viability of a commercially available EEG device for estimating the mental workload of two different instruction systems to augment a manual assembly task. We discuss the validity of our hypotheses and present a framework to evaluate interactive assistive technologies based on \ac{EEG} measures.

\subsection{Validating the \ac{EEG} Setup prior to the Experiment}
To validate the correctness of our \ac{EEG} setup and measures, we manipulated the difficulty of a \textit{N}-back task to vary working memory load and investigated its impact on \ac{EEG} measurements. Specifically, we focused on alpha power in an individually determined frequency bandwidth, proximal to 10 Hz. Previous research has consistently demonstrated lower alpha power in participants who experience high working memory load compared to low working memory load~\cite{gevins1997high}. Thus, we expected lower alpha power during the \textit{2}-back task compared to a \textit{0}-back task. Indeed, we found lower alpha power for higher task difficulty and higher alpha power for lower task difficulty during the \textit{N}-back conditions. In other words, participants experienced less cognitive workload during the \textit{0}-back task compared to the \textit{2}-back task. This supports the applicability of using commercial devices to infer cognitive workload and converges with the results found in previous work~\cite{Lee:2006:ULE:1166253.1166268, Grimes:2008:FPC:1357054.1357187}.

Before evaluating novel interactive assistive systems using \ac{EEG}, our findings suggest to conduct an eyes opened and eyes closed task to elicit the individual's alpha peak. Afterward, a \textit{N}-back task with $N = 0$ and $N > 1$ should be conducted to infer the validity of \ac{EEG} measures in terms of lower alpha activation for higher task complexities. The individual alpha peak is elicited when participants close their eyes~\cite{Klimesch1993} enabling an individual IAF analysis for upcoming \ac{EEG} trials.



\subsection{Evaluating Assembly Instruction Systems}
Statistical comparisons of alpha power confirmed that projected \textit{in-situ} instructions generated significantly higher alpha power than the use of printed paper instructions. In other words, the projected \textit{in-situ} instructions induced less cognitive workload. Thus, H1 is statistically supported. This finding that is based on \ac{EEG} measurements agrees with the subjective self-reporting measures based on NASA-TLX questionnaires~\cite{funk2017working}. A significant difference in subjectively perceived workload was found between paper and projected \textit{in-situ} instructions (H2). In general, our results support the idea that the alpha-band frequency power of \ac{EEG} measurements is a valid metric for estimating the levels of cognitive workload induced by a specific instruction system or visualization. Altogether, this agrees with the motivation of \textit{in-situ} projections, which is to reduce cognitive load by providing situated information at the appropriate times.

Regarding assembly performance, we can partially confirm the outcomes of previous research~\cite{funk2016interactive}. We found a significant difference in the number of item selection errors. We confirm that fewer item selection errors were observed using projected in-situ instructions compared to printed paper instructions (H3). However, no significant difference between both instructions systems was found on assembly errors. Therefore, we cannot confirm H4. Assembling with projected \textit{in-situ} instructions takes significantly less time than printed paper instructions. This agrees with previous research~\cite{funk2016interactive} and supports our final hypothesis (H5).


Our results encourage to measure and evaluate \ac{EEG} to assess the mental demand of assistive technologies in manual assembly processes. However, the experiment has taken place under controlled conditions in lab environments where participants were restricted to a reduced number of eye blinks and head movements. We recommend the evaluation of \ac{EEG} to evaluate assistive technologies in controlled environments before they will be deployed in real-world scenarios or for further scientific research inferring the mental demand during usage. This way, novelty biases can be detected by comparing correlations between \ac{EEG} measures and self-rated assessments. System architects benefit from these evaluation steps as the objectively measured workload can be considered into the design pipeline.

\subsection{\ac{EEG} as Real-Time Evaluation Tool}
\label{sec:eeg_real_time}
Additionally, we investigated the applicability of \ac{EEG} as a real-time evaluation tool for detecting working memory. Previous research used medical \ac{EEG} systems to derive the current level of executive working memory in real-time~\cite{doi:10.1207/s15327590ijhc1702, 4650550, MULLER200882}. We show that differences in cognitive workload are measurable for two different instruction visualization systems using a mobile consumer \ac{EEG} headset.

Figure~\ref{fig:alpha_fluctuation} shows, how the level of alpha power changes with time during assembly. The alpha power for paper instructions decreases with time as the current bin for an item selection has to be recalled with every assembly step. Furthermore, the final position of the brick has to be recognized and placed correctly according to the paper instruction.

Projected \textit{in-situ} instructions show the opposite effect. After a decrease in alpha power, which could be attributed to a lack of familiarity with the system, an increase in alpha power is observed. This can be due to the alleviation of working memory load since projected instructions eliminate the need to maintain a set of manual assembly instructions in working memory. Projected instructions are updated by each step of manual assembly and at the appropriate time steps. Thus, unlike with the use of paper instructions, it is no longer necessary to maintain and recall instructions from working memory.

\begin{figure}
\includegraphics[width=\columnwidth]{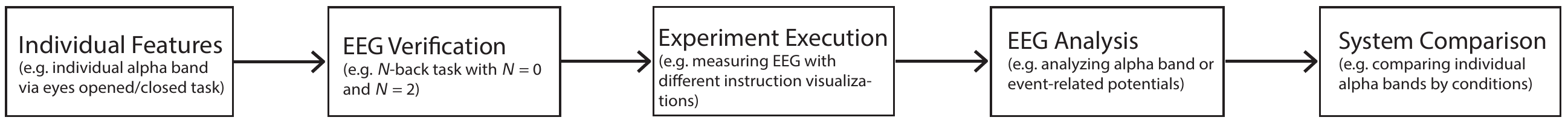}
\vspace{-5mm}
\caption[Physiological design pipeline for user interface evaluation]{Experimental protocol which enables system designers to evaluate mental demanding elements in their user interface design on a cognitive level.}
\label{fig:design_pipeline}
\end{figure}

The stability in alpha fluctuation for both instruction systems support the use of commercial \ac{EEG} devices for real-time workload estimation. This agrees with the subjective perception of workload through NASA-TLX questionnaires and verbal feedback provided by the participants. A real-time system for estimating cognitive load could benefit use-case scenarios such as in evaluating user interfaces, assessment of workload in safety-critical tasks, or real-time adaptation of user interfaces suited to the current level of cognitive workload. By detecting constant low or high alpha fluctuations the provided assistance can be adjusted depending on the measured workload levels. Therefore, we present an experimental protocol that evaluates the cognitive demand assistive systems require during runtime (see Figure~\ref{fig:design_pipeline}). Designers can use this protocol as an assessment template to find distracting or cognitive demanding user interface elements.

Novel engineered assistive technologies can benefit from real-time insights into the mental resources by an operator. This enables user interface designers to test visualization adaptations for different measures of cognitive workload. However, the question of how to provide user interface adaptation for different levels of cognitive workload remains and depends highly on the use case scenario in which assistive technologies are deployed.

\subsection{Other \ac{EEG} Metrics for Evaluating Mental Workload}
We relied on an established \ac{EEG} metric (\textit{i.e.}, alpha power) to confirm previous claims that projected \textit{in-situ} instructions can reduce working memory. Frequency domain measures of cognitive load are well-established and proved to be viable \cite{gevins2003neurophysiological,gevins1998monitoring}. However, frequency-based measures often lack the discrimination of functional interpretations, afforded by time-domain measures. Event-related potentials (ERPs)~\cite{pfurtscheller1999event} refer to the time-domain waveform that is contingent upon the occurrence of a critical event such as the presentation of a stimulus.

Unlike frequency domain measures, the time-varying \ac{EEG} activity of subsequent voltage deflections can be individually attributed to functional mechanisms that underlie information processing. For example, an early negative deflection in the ERP waveform around 100--200 ms is often associated with stimulus detection while a later positive deflection between 250--600 ms is associated with stimulus recognition and working memory updating \cite{Parasuraman1980}. This allows for stronger functional discriminability in the type of cognitive work that is experienced by the user, compared to the all-encompassing term 'mental workload'. Signal processing and classification approaches have been proposed to utilize ERP features for mental workload classification \cite{brouwer2012estimating,kathner2014effects}.

The use of ERPs for estimating mental workload is limited by the fact that ERPs are event-related. This means that ERPs can only be extracted if the event that triggers them is known beforehand. These are typically the items presented in \textit{N}-back tasks. Unfortunately, not all interfaces that are of interest will consist of a \textit{N}-back component. One approach is to introduce task-irrelevant probes for ERPs, such as an environmentally sound. More recent research has shown that involuntary ERPs to task-irrelevant stimuli vary in their amplitudes depending on the cognitive demands of the primary task that they are engaged in \cite{burns2015use,miller2011novel,scheer2016steering}. Application cases include estimating the workload of users playing Tetris across different difficulty levels for estimating the immersion of users in high-fidelity driving simulators.

\section{Study Limitations}
The study was affected by certain limitations. The study was conducted in a controlled laboratory setting, whereby participants were instructed to blink as little as possible and to avoid unnecessary head movements to minimize artifacts in the \ac{EEG} signal. Users are unlikely to abide by such instructions in mobile real-world settings. Nonetheless, we point out that the manual assembly task is one that involves substantial activity. Therefore, our \ac{EEG} recordings were likely to have contained a significant degree of motoric noise and continued to be robust for our current estimation of cognitive load. Cortical activity related to the planning of motor actions can also result in lower frequency power in a bandwidth that overlaps with alpha (\textit{i.e.}, mu-power: 8 Hz to 12 Hz~\cite{doi:10.1080/17470910903395767}). Therefore, the current results could have been affected by motor planning during item selection and placement within the assembly task, rather than cognitive load per se. Nonetheless, we assume that our results are valid for cognitive load and not a motoric activity, given that we validated our \ac{EEG} measure with a corresponding \textit{N}-back task. Furthermore, we extracted the individual alpha band, which is not necessarily located in the mu frequency range.

\section{Study Conclusion}
In this study, we investigated if \ac{EEG} is a suitable measurement modality for cognitive workload. In a case study leveraging an assembly scenario, we find that projected \textit{in-situ} instructions reduce cognitive workload compared to traditional paper instructions. We employed a commercial \ac{EEG} headset to derive direct measurements of neural activity that varied by working memory load in a \textit{N}-back task, namely alpha power~\cite{Klimesch1993}. This same measurement (\textit{i.e.}, alpha power) was larger for projected \textit{in-situ} instructions than the traditional approach of paper instructions, demonstrating that cognitive load was lower when a projected \textit{in-situ} system was employed. To date, only subjective questionnaire estimates have been collected. The current work demonstrates the viability of using a commercial \ac{EEG} device for evaluation purposes, even in a setting that involves a large amount of user activity (\textit{i.e.}, manual assembly). The collected data set is hosted on Github\footnote{\url{www.github.com/hcum/identifying-cognitive-assistance} - last access \urldate}.

\section{Chapter Conclusion}
In this chapter, we investigated if \ac{EEG} represents a suitable real-time metric for cognitive workload. We analyzed alpha oscillations to quantify the current demand of working memory load, a cognitive component that describes the available capacities of the short-term memory. Contrary to post hoc assessments, \ac{EEG} provides a real-time measure for cognitive effort. Using a manual assembly experiment as a case study, we found that cortical activity is a reliable indicator to measure the cognitive demand placed by user interfaces. Based on our results, we provide an evaluation pipeline that suggests the use of \ac{EEG} for user interface evaluation. Reflecting on our results, we conclude that \ac{EEG} is a suitable real-time measure for cognitive workload that answers according to the research question (\textbf{RQ3}). The presented pipeline (see Section~\ref{sec:eeg_real_time}) provides an initial cornerstone for the real-time evaluation and intervention of \ac{EEG}-based workload-aware interfaces. Therefore, we envision the use and development of novel pipelines with existing or alternative physiological modalities that build on the presented approach.




\chapter{Smooth Pursuits}
\label{ch:smooth_pursuit}

\epigraph{\textit{The eyes are the mirror of the soul and reflect everything that seems to be hidden; and like a mirror, they also reflect the person looking into them.}}{Paulo Coelho}

The previous chapter investigated the feasibility of \ac{EEG} signals as a measure for cognitive workload. The analysis of cortical activity has proved useful to detect changes in cognitive demand in real-time. Thereby, \ac{EEG} provides the capability to derive the mental demand directly from the brain. However, \ac{EEG} relies on a correct setup of the headset that has to be kept steady to avoid noise in the signal. Other factors, such as head movements and eye movements, may influence the signal as well. Therefore, alternative modalities have been juxtaposed to overcome the limitations that \ac{EEG} measures pose.

As such, eye tracking has emerged as both inferences for cognitive states as well as an interaction modality. Eye gaze-based interactive systems hold a lot of promise for cognition-aware interaction~\cite{bulling2014cognition}. The human eye represents the central organ which enables visual processing. Complemented by previous research, eye gaze as an input for interactive systems has been extensively explored~\cite{duchowski2002breadth}. One specific eye movements, namely smooth pursuit eye movements, have been utilized as an alternative eye movement input modality for interactive systems~\cite{hutchinson1989human, vidal2013pursuits}. In contrast to \ac{EEG}, smooth pursuit has the advantage to be a contactless measure. Eye movements can be recorded using a remote eye tracker, neglecting the need for hardware that has to be attached to the user. Furthermore, smooth pursuit eye movements cannot be pretended since they require locking onto a moving target~\cite{velloso2016ambigaze}. This robustness against false positives is another benefit of using smooth pursuit eye movement as an interaction technique.

\begin{figure}
    \includegraphics[width=\columnwidth]{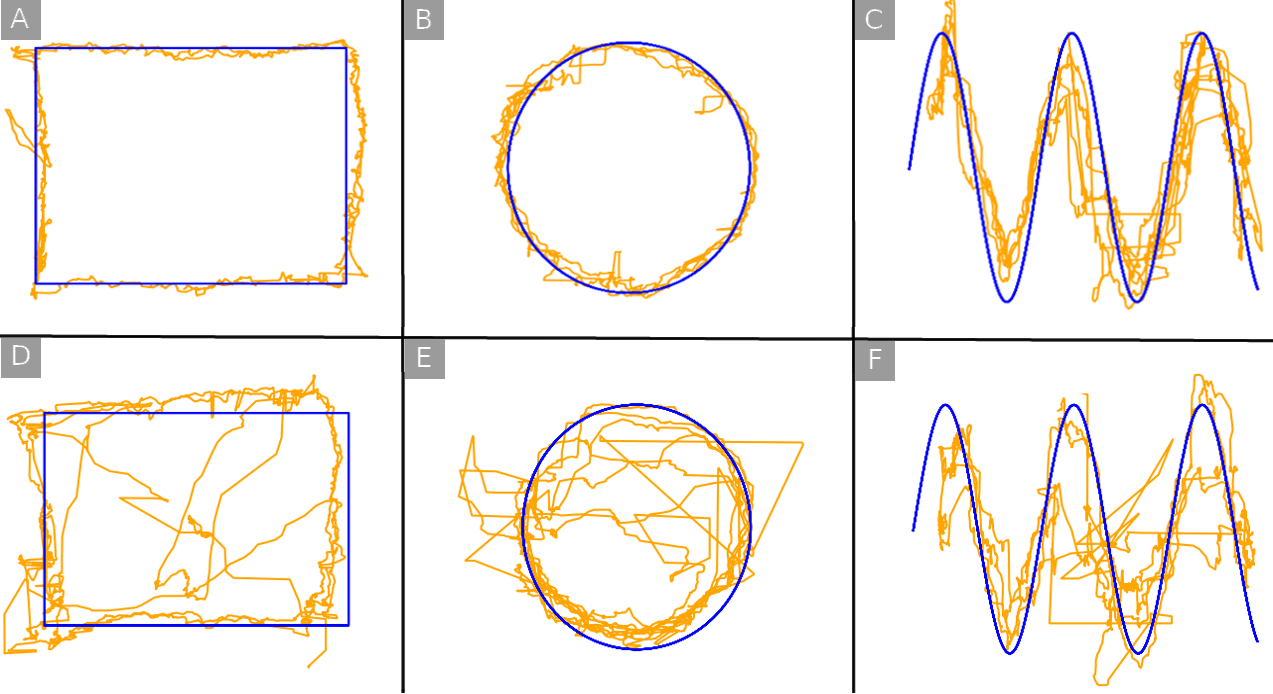}
        \vspace{-1em}
    \caption[Example of smooth pursuit recordings for different workload levels]{Smooth pursuit recordings under different levels of cognitive workload for three trajectories. The blue line shows the displayed trajectory, while the orange line visualizes the gaze path. (A), (B), and (C) show the gaze path during low cognitive workload phases, while (D), (E), and (F) shows the gaze path during states of high workload.}
    \vspace{-1em}
    \label{fig:teaser_intro}
\end{figure}

Research has shown that the behavior of the eye is strongly affected by psychological~\cite{purves2001types, schenck1987rapid} and psychophysiological states of the human body~\cite{sparks2002brainstem, zihl1983selective}. One such state is cognitive workload, which has a remarkable impact on eye movements~\cite{barnes2008cognitive, JUST1976441, stuyven2000effect}. Figure~\ref{fig:teaser_intro} shows how cognitive workload affects smooth pursuit eye movements. Driving in a stressful context or performing multitasking during cognitively demanding tasks under time pressure are just two examples. Previous measures for cognitive workload facilitate questionnaires, such as the NASA-TLX~\cite{doi:10.1177/154193120605000909, HART1988139} or the Driver Activity Load Index (DALI)~\cite{pauzie2008method}, which are often used in \ac{HCI} to evaluate interfaces. Yet these questionnaires are prone to the interpretation of the questions by the user and only allow for an assessment at the end of the task, hence providing rather coarse-grained insights. At the same time, eye-tracking data can be used to assess cognitive workload~\cite{Dingler:2016:CSM:2968219.2968565}. In particular, previous work showed that under controlled conditions, pupil dilation provides an estimate of cognitive load~\cite{peysakhovich2016study, pfleging2016model}. However, pupil dilation is highly sensitive to light conditions, which change in outdoor environments, and hence cannot easily be applied in many ubiquitous computing settings. 


In this chapter, we propose an approach that exploits smooth pursuit eye movements to assess cognitive workload. We hypothesize that working memory, a component of cognitive workload, measurably affects smooth pursuit eye movements~\cite{BADDELEY197447}. Henceforth, we define smooth pursuit describes eye movements that occur as the eyes closely follow a moving object. These movements are evaluated in several studies used for user interface element selection~\cite{esteves2015orbits, velloso2016ambigaze} or intuitive eye tracker calibration~\cite{khamis2016textpursuits, vidal2013pursuits}. By conducting a user study, we found that smooth pursuit can be used for a contactless assessment of cognitive workload with an accuracy of up to $99.5\%$. Thereby, the need for body-worn sensors is obviated. This chapter reports on a user study that investigates the impact of cognitive workload on smooth pursuit eye movements. Based on the insights from the user study we build a classifier to approximate the level of cognitive workload using gaze differences during smooth pursuit eye movements. We discuss how these findings can be used during the evaluations that require the assessment of cognitive workload as well as for the design of cognition-aware interactive systems. This chapter investigates smooth pursuit eye movements as part of the following research question:

\begin{itemize}
    \renewcommand{\labelitemi}{$\rightarrow$}
   \item \textbf{RQ4}: Do eye gaze metrics enable the classification of cognitive workload states?
\end{itemize}

\begin{center}
    \begin{GrayBox}
        \centering
        \parbox{0.98\textwidth}{
            \emph{This section is based on the following publication:}
            \vspace{3mm}
            \publicationsbegin    
            \item \bibentry{Kosch:2018:YET:3173574.3174010}
            \publicationsend
        }
    \end{GrayBox}
\end{center}

\section{Related Work}
We present related work concerning existing research on using smooth pursuit eye movements for interaction and the influence of cognitive workload on eye movements.

\subsection{Interacting with Smooth Pursuit}
Using eye gaze as input for ubiquitous interactive systems has been extensively explored. Duchowski~\cite{duchowski2007eye,duchowski2002breadth} performed a literature survey about current eye-tracking applications and technologies. Focusing on usability, related research shows how eye tracking technologies evolved until they became usable in human-computer interaction research~\cite{jacob2003eye, poole2006eye}.
Recent research has addressed the use of eye gaze for selection tasks~\cite{ohno1998features,sibert2000evaluation}, which can be used by physiologically impaired individuals to perform input on computers~\cite{hutchinson1989human}. Text input via eye input has been researched using dwell time~\cite{bee2008writing, majaranta2002twenty}, off-screen targets~\cite{isokoski2000text}, or eye movements along the y-axis on a display~\cite{tuisku2008now}. The usage of such methods was also evaluated in real-world scenarios~\cite{de2007evaluation,zhang2013sideways}. Furthermore, eye movements and eye gestures can be used to interact with devices.

Recently, smooth pursuit eye movement~\cite{hutchinson1989human,vidal2013pursuits} has been used as an alternative gaze-based input modality to interactive systems. This interaction technique overcomes the need for precise calibration and training of the user before interaction and further supports ubiquitous interaction. Several researchers have investigated how smooth pursuits can be used intuitively. People cannot pretend to perform smooth pursuit movements since they require locking onto a moving target~\cite{velloso2016ambigaze}. This reduces the likeliness of detecting false positives during an interaction. 

Another approach to leverage smooth pursuit as an input modality has been researched by Esteves et al.~\cite{esteves2015orbits}. They used a mobile eye tracker to enable hands-free interaction with a smartwatch by showing moving dots on the smartwatch display. Results show robustness against false positives regarding input, no need for calibrating an eye tracker, and efficient hands-free interaction. Using smooth pursuit to interact with physical real-world devices also showed promising results. Velloso et al. \cite{velloso2016ambigaze} developed and evaluated an object-driven system, which can be operated by only performing smooth pursuit. As soon as a user approaches an object such as a fan or windmill, it begins to present moving targets that trigger actions as soon as selected by smooth pursuit.

Researchers have also used smooth pursuit to calibrate eye trackers. Pfeuffer et al.~\cite{pfeuffer2013pursuit} investigated this approach by using animations on a display to implicitly calibrate an eye tracker. Their results state that a 5-point calibration achieved a higher detection rate and required less calibration time. They concluded that smooth pursuits calibration provides better usability and flexibility for eye tracker calibration on small displays.   

Vidal et al.~\cite{vidal2013pursuits} proposed different applications for smooth pursuit eye interaction, such as selection tasks or password authentication. The implemented applications showed a fast selection and completion time for different tasks. 
Since smooth pursuit can be performed with almost every kind of animated trajectory, more recently Khamis et al.~\cite{khamis2016textpursuits} investigated which trajectories are most suitable for smooth pursuit interaction and calibration when showing text-based content on a display. While showing a question in the top left corner of a screen, they also showed several answers to the question on the same screen displayed in different trajectories. Schenk et al.~\cite{Schenk:2017:GEG:3025453.3025455} used smooth pursuit as an element selection mechanism in desktop settings to avoid the Midas touch problem~\cite{Jacob:1990:YLY:97243.97246}. Similarly, Lohr and Komogortsev~\cite{Lohr:2017:CSP:3027063.3053233} compared smooth pursuit-based input against dwell-based input approaches. A significant faster selection of elements was achieved when using smooth pursuit at the cost of likely unwanted selections.

\subsection{Impact of Cognitive Workload on Eye-based Properties}
Previous research has shown that the behavior of the eye is strongly affected by the psychological~\cite{purves2001types, schenck1987rapid} and psycho-physiological~\cite{sparks2002brainstem,zihl1983selective} state of the human body. One such state is cognitive workload, which has a shown impact on eye movements~\cite{barnes2008cognitive,stuyven2000effect}.

Researchers found a relation between pupillary responses and cognitive demand~\cite{ahern1979pupillary, hess1964pupil}. During a task comprising different task complexities, the pupil diameter of participants was measured as an indicator for cognitive workload.
Results show increasing pupil dilation with increasing task difficulty. Pfleging et al.~\cite{pfleging2016model} created a model to estimate cognitive workload under different lighting conditions based on pupil dilation. The feasibility of measuring pupil dilation using a remote eye tracker has been investigated by Klingner et al.~\cite{klingner2008measuring}. Their findings show a lower accuracy compared to mobile eye trackers due to noise and head movements. Kruger et al. \cite{kruger2013measuring} investigated eye behavior when perceiving a stimulus with and without subtitles using electroencephalography and pupillary measurements. In an experiment with two groups, lower cognitive workload was measured within the group perceiving subtitles than in the group lacking subtitles.

Liang and Lee~\cite{liang2008driver} compared the efficiency of different machine learning algorithms to estimate distractions during driving tasks based on saccadic eye movements. Higher frequencies of saccadic eye movements have been used as an assessment for cognitive workload. Best results were achieved using a \ac{SVM} for machine learning. However, their findings address saccadic eye movements only and do not examine the effects of smooth pursuit eye movements under cognitive workload. 

Benedetto et al. \cite{benedetto2011driver} investigated the correlation between blink duration and visual and cognitive workload of a driver operating a car in a simulated environment. Participants had to perform a Lane Change Test~\cite{mattes2009surrogate} while doing an in-vehicle information system task~\cite{Victor2005167} at the same time. Their findings show a lower blink duration compared to their baseline task than in the cognitively demanding task. Ahlstrom and Friedman-Berg \cite{ahlstrom2006using} also found a significant correlation between shorter blink durations and cognitive workload. 

Stuyven et al. \cite{stuyven2000effect} investigated the impact of cognitive workload on saccadic eye movements. Their findings show an increased occurrence of saccadic eye movements when inducing cognitive workload. Tsai et al.~\cite{tsai2007task} investigated how eyes behave under cognitive workload while performing a paced auditory serial addition task~\cite{gronwall1977paced}, however, their work did not investigate the impact on smooth pursuit eye movements. Recently, Zagermann et al.~\cite{10.1145/2993901.2993908} developed a model and showed concepts to derive cognitive workload from eye behavior, such as saccades, fixations, pupil dilation, and eye blinks. Cognitive workload influences microsaccadic eye movements \cite{siegenthaler2014task}. They found that microsaccades occur more frequently with higher perceived workload during a non-visual task.

The voluntary involvement of smooth pursuit eye movements has been researched by Barnes~\cite{barnes2008cognitive}. He showed that cognitive processes in smooth pursuit eye movements are even involved without voluntary participation. Important factors for following moving stimuli were attention and awareness, which trigger the process of smooth pursuit on a neuro-scientific level. Therefore, smooth pursuit eye movements are voluntary up to a certain degree. In contrast, Collewijn and Tamminga~\cite{collewijn1984human} investigated contexts in which smooth pursuit movements are voluntary. They used different targets and backgrounds to investigate smooth pursuit performances of humans. Contreras et al.~\cite{contreras2011effect} researched the eye-target synchronization performance of people with traumatic brain injury. People suffering from traumatic brain injuries show a worse performance when performing smooth pursuit in terms of deviation points apart from a shown moving object on a display.

Previous work has investigated how smooth pursuit can be leveraged for calibration and interaction. However, to our knowledge, no prior work has proposed to use work to propose using smooth pursuit eye movements as a measurement to derive cognitive workload. This opens new opportunities for researchers and practitioners alike. In particular, using smooth pursuits as a real-time measurement for cognitive workload is valuable in the context of evaluating interactive systems as well as for developing cognitively adaptive systems.

\label{sec:evaluation}
\section{Study}

To understand the impact of cognitive workload on smooth pursuit eye movements, we designed a lab user study where we induced cognitive workload while participants performed smooth pursuit eye movements.

\subsection{Independent Variables}
In our experiment on investigating the impact of cognitive workload on eye movements, we explore the influence of three independent variables when interacting with smooth pursuit systems: (1) trajectory type~\cite{khamis2016textpursuits}, (2) speed of the stimulus~\cite{khamis2015field}, and (3) task difficulty. In the following, we describe these three independent variables in detail.

\begin{figure}
    \includegraphics[width=\columnwidth]{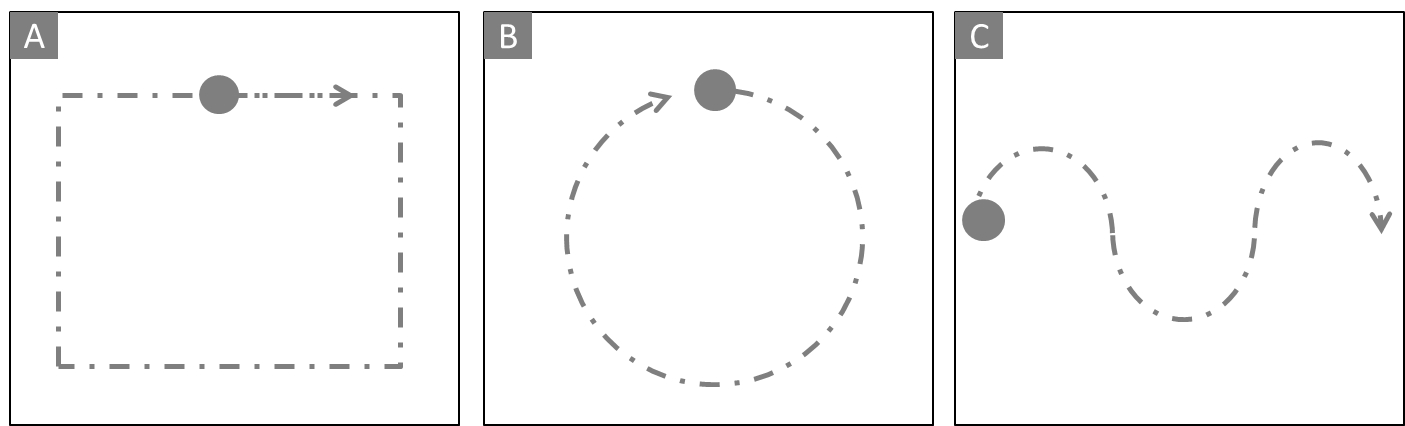}
    \caption[Smooth pursuit trajectories used in the study]{Three different trajectories chosen for evaluation: (A) Rectangular trajectory, (B) Circular trajectory, (C) Sinusoidal trajectory. The size of the moving object in the experiment was 10\,pixels.}
    \label{fig:trajectories}
    \vspace{-1em}
\end{figure}

\subsubsection{Trajectories}
Since smooth pursuit eye movements can be triggered by showing a stimulus moving along a particular trajectory, we implemented a rectangular, a circular, and a sine wave animation to produce this effect. We have chosen these trajectories based on previous research~\cite{khamis2015field,khamis2016textpursuits,vidal2013pursuits}. Furthermore, the chosen trajectories may have a physiological effect paired with the current task difficulty, since the human eye has six muscles responsible for horizontal and vertical movements as well as eye rotations~\cite{leigh2015neurology}. Rectangular trajectories demand muscles on the left and right side of the eye for horizontal movements, while vertical movements demand the upper and lower eye muscles. Circular and sinusoidal eye movements demand four muscles around the eye. Rectangular trajectories require horizontal or vertical movements only, while circular and sinusoidal include diagonal movements as well. In the context of leveraging smooth pursuit as interaction modality, the presented trajectories are used to distinguish between different user inputs. The chosen trajectories are depicted in Figure \ref{fig:trajectories}.

\begin{figure*}
    \centering
        \includegraphics[width=\columnwidth]{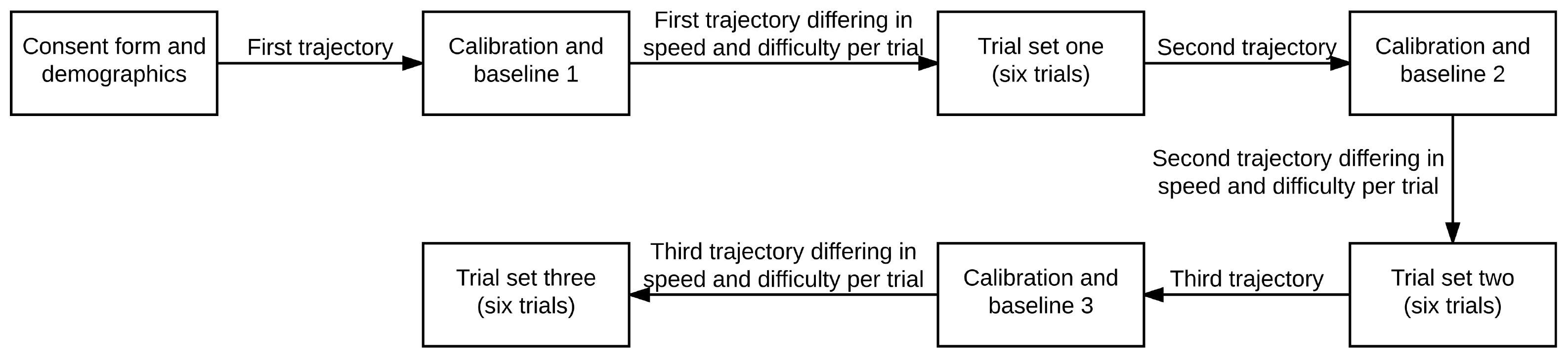}
        \caption[Study procedure]{Study procedure. First, consent approval and demographic data were collected. Afterward, a baseline task was conducted followed by a set of six trials, where each trial differs in trajectory speed and task complexity. This procedure is repeated for every remaining trajectory type.}
        \label{fig:study_procedure}
\end{figure*}

\subsubsection{Speed of Stimulus}
We compare two different speeds at which stimuli are moving based on previous research~\cite{khamis2015field}: $450\,\frac{px}{s}$ (slow\footnote{This corresponds to $17.14^\circ$ per second at a viewing distance of approximately 50 centimeters}) and $650\,\frac{px}{s}$ (fast\footnote{This corresponds to $24.76^\circ$ per second at a viewing distance of approximately 50 centimeters}). Displaying slower or faster animations while experiencing cognitive workload can lead to different performances since eye muscles have to deal with different strains per speed and trajectory~\cite{leigh2015neurology}. Furthermore, trajectory speeds can be used to differentiate user input, since eye movements adapt to different speeds.

\subsubsection{Task Difficulty}
To induce cognitive workload, we use an auditory delayed digit recall \textit{N}-back task from Mehler et al.~\cite{mehler2011agelab} with an English-spoken number set. The \textit{N}-back task is a commonly used task to artificially elicit working memory resources~\cite{BADDELEY197447}, a component of cognitive workload which strains temporal memory capacities and affects secondary task performances negatively~\cite{berka2007eeg,brouwer2012estimating,turner1989working}.

Throughout the study, we use the \textit{N}-back task to manipulate cognitive workload by demanding working memory with different difficulty levels~\cite{kane2007working}. For each trial, participants hear randomized numbers consisting of ten digits between 0 and 9. Hereby, \textit{N} corresponds to the \textit{N}-last digit. After hearing the number, participants have to say out loud the digit they heard \textit{N} digits ago. In our experiment, we use a \textit{1}-back, \textit{2}-back, and \textit{3}-back tasks to induce cognitive workload. By increasing \textit{N}, more digits have to be remembered, hence increasing task difficulty. To collect baseline measures, participants were asked to follow a trajectory without performing an \textit{N}-back task. Similar as in the study presented in Chapter~\ref{ch:workloadbyeeg}, Table \ref{tab:nback2} shows an example of the \textit{N}-back task.

\renewcommand{\arraystretch}{1.2}
\begin{table}[h!]
 \begin{center}
 \begin{tabular}{l|c|c|c|c|c|c|c}
   \toprule
   \textbf{Heard number} & 5 & 8 & 3 & 4 & 3 & 9 & 1\\
   \hline
   \textbf{Number to say (\textit{1}-back task)} & & 5 & 8 & 3 & 4 & 3 & 9 \\
   \hline
   \textbf{Number to say (\textit{2}-back task)} & & & 5 & 8 & 3 & 4 & 3 \\
   \hline
   \textbf{Number to say (\textit{3}-back task)} & & & & 5 & 8 & 3 & 4 \\
   \bottomrule
 \end{tabular}
   \caption{Example of the auditory delayed digit recall \textit{N}-back task. Participants have to remember the \textit{N}-th number back of a spoken number sequence and say the number out loud.}
 \label{tab:nback2}
 \end{center}
 \vspace{-1em}
\end{table}

\subsection{Apparatus}
The study was conducted in a quiet lab with no windows, where lighting conditions were fixed. The setting was spatially divided into an experimenter area and a participant area. A separator divided both areas. While the experimenter controlled the experiment using a laptop, the participant saw an animated trajectory on a 22\,inch screen with a resolution of $1680\,\times\,1050$\,pixels and a refresh rate of 60\,\ac{Hz}. Eye gaze data was collected using a RED250 from SensoMotoric Instruments with a sample rate of 250\,\ac{Hz}. No filter was applied to the captured gaze data, thus raw gaze data was recorded only. We used a Holosonic Audio Spotlight 24i directed speaker to provide the auditory delayed digit recall \textit{N}-back task. 

\subsection{Method and Measures}
We used a repeated-measures design with three independent variables as described in the previous section; namely trajectory (rectangular, circular, sine wave), speed (slow, fast), and cognitive workload (no task, \textit{1}-back task, \textit{2}-back task, \textit{3}-back task). Each experiment consisted of three sessions, where animated trajectories were changed for each session. The animated trajectory consisted of a white dot with a diameter of 10\,pixels. The background was set to gray (RGB: [128,128,128]) to avoid eye exhaustion caused by screen brightness.

Before starting a new session, the eye tracker was calibrated to retrieve gaze points for later analysis of gaze deviations between baseline and smooth pursuit eye movements. Each session began with a 30\,second baseline trial, where the animated trajectory used for the session was shown to participants without inducing cognitive workload. This allowed us to estimate eye movement differences from the displayed trajectory when no cognitive workload was present. We chose a slow speed ($450\,\frac{px}{s}$) to make participants familiar with the displayed trajectory.

Cognitive workload was induced by providing a task difficulty using the auditory \textit{N}-back task described in the prior section while showing the animated trajectory at a certain speed. Task difficulty and speed were counterbalanced during a session according to the Latin square, while each session showed the same trajectory. The order of sessions was counterbalanced participant-wise according to the Latin square. This resulted in seven trials per session, including the baseline task. Running all three sessions, the experiment comprised 21 trials per participant. Overall, the duration of each trial was 25 to 30 seconds, dependent on the length of the spoken number for tasks including task difficulty. After completing a trial, participants were asked to fill out a NASA-TLX questionnaire to assess subjectively perceived cognitive workload. Participants took a 30-second long break afterward. Figure~\ref{fig:study_procedure} shows an illustration of the study procedure.

\subsection{Participants}
We recruited 20 participants (9 female, 11 male), aged between 22 and 34 years (M = $27.5$, SD = $3.13$). Before the experiment, each participant signed a consent form and provided their demographic data. All participants were computer science, students or researchers. All participants had a normal or corrected-to-normal vision. Participants were recruited through university mailing lists. They received sweets and five Euro as compensation. The duration of the study was approximately 30 minutes. 
We explained the purpose of the study and tasks to the participants and informed them they could exit the study at any point. Participants signed informed consent forms and were seated in a comfortable chair, approximately 50 centimeters in front of the display before the experimental setup. Due to technical issues, two participants were excluded from the analysis, as their gaze data was not recorded properly.

\section{Results}
We analyze our data to compare the impact of trajectory type, speed, and task difficulty on smooth pursuit eye movements. We report on quantitative results by comparing measured eye gaze data with the showed trajectory. This is complemented by a subjective analysis through NASA-TLX questionnaires.

\begin{table}
\begin{center}
    \begin{tabular}{lc}
        \toprule
        \textbf{Paired Wilcoxon Signed-Rank Test} & \textbf{Significance}\\
        \midrule
        \begin{tabular}{l|l} Circle Baseline & Circle \textit{1}-back fast \end{tabular} & $p = .002$\\ 

        \begin{tabular}{l|l} Circle Baseline & Circle \textit{2}-back fast \end{tabular} & $p = .002$\\

        \begin{tabular}{l|l} Circle Baseline & Circle \textit{3}-back fast \end{tabular} & $p = .003$\\

        \begin{tabular}{l|l} Circle Baseline & Circle \textit{2}-back slow \end{tabular} & $p = .004$\\ 

        \begin{tabular}{l|l} Circle Baseline & Circle \textit{3}-back slow \end{tabular} & $p = .005$\\

        \begin{tabular}{l|l} Sine Baseline \hspace{1.5mm} & Sine \textit{1}-back fast \end{tabular} & $p < .001$\\ 

        \begin{tabular}{l|l} Sine Baseline \hspace{1.5mm} & Sine \textit{2}-back fast \end{tabular} & $p < .001$\\

        \begin{tabular}{l|l} Sine Baseline \hspace{1.5mm} & Sine \textit{3}-back fast \end{tabular} & $p < .001$\\

        \begin{tabular}{l|l} Sine Baseline \hspace{1.5mm} & Sine \textit{1}-back slow \end{tabular} & $p = .003$\\

        \begin{tabular}{l|l} Sine Baseline \hspace{1.5mm} & Sine \textit{2}-back slow \end{tabular} & $p = .002$\\

        \begin{tabular}{l|l} Sine Baseline \hspace{1.5mm} & Sine \textit{3}-back slow \end{tabular} & $p = .002$\\
        \bottomrule
    \end{tabular}
    \caption[Summary of significant results between the workload levels]{Summary of significant results. Comparisons between other conditions did not result in significant differences.    
}
    \label{tab:significantresults}
    \end{center}
    \vspace{-1em}
\end{table}

\subsection{Smooth Pursuit Differences and Cognitive Workload}
To evaluate the effect of different task difficulties on smooth pursuit performances, we based our analysis on pixel differences between coordinates of the displayed trajectory and measured eye gaze position at the screen. More formally, we calculated the difference between two coordinates $p$ and $q$ using the Euclidean distance with the formula
\begin{equation} \label{eq:euclidean_distance}
         D = \sqrt{\left(p_{t,x} - q_{t,x}\right)^{2} + \left(p_{t,y} -
   q_{t,y}\right)^{2}}
\end{equation}

where $p$ and $q$ depict a two-dimensional vector comprising the baseline coordinates and measured gaze coordinates from participants. We introduce the variable $t$ describing the temporal dependency between eye gaze and displayed stimulus as the used 250 Hz eye tracker might introduce a temporal offset of four milliseconds, which is below the perceptual threshold for interaction.
Differences were normalized with respect to the maximum gaze deviation 
from the shown trajectory. To enable a descriptive analysis, normalized coordinates were averaged for all participants and conditions. Lowest mean eye gaze deviations (M$_{pd}$) were measured for all slow trajectories, where no cognitive workload was induced (Rectangle:~M$_{pd} = 9.14$, SD $= 6.75$, Circle:~M$_{pd} = 9.51$, SD $= 8.03$, Sine:~M$_{pd} = 4.37$, SD $= 1.60$).
Fast rectangular trajectories using a \textit{3}-back task (M$_{pd} = 13.25$, $SD = 11.06$), fast circular trajectories using a \textit{2}-back task (M$_{pd} = 14.54$, SD $= 7.14$) and fast sinusoidal trajectories using a \textit{3}-back task (M$_{pd} = 9.85$, SD $= 3.53$) led to highest mean eye gaze deviations.

Applying a Shapiro-Wilk test on the mean data set showed a non-normal distribution for all conditions (all $p < .05$). A Friedman test showed no significant differences between various levels of cognitive workload and gaze deviations of smooth pursuit eye movements for slow rectangles ($\chi^2(3) = 3.000$, $p = .392$). However, a Friedman test found significant differences within various levels of cognitive workload for fast rectangles ($\chi^2(3) = 11.867$, $p = .008$), slow circles ($\chi^2(3) = 18.867$, $p < .001$), fast circles ($\chi^2(3) = 29.400$, $p < .001$), slow sine waves ($\chi^2(3) = 14.667$, $p = .002$) and fast sine waves ($\chi^2(3) = 30.667$, $p < .001$). We conducted a Wilcoxon signed-rank post hoc test to find significant differences between pairs of task difficulties including baseline trials and normalized gaze deviations after applying a Bonferroni correction (significance level set to $p < .0083$). A summary of significant results can be depicted from Table~\ref{tab:significantresults}. Further, the Cohen's d effect size values of the significant statistical comparisons ranged between $d = 0.66$ and $d = 0.88$.

\begin{figure}
    \centering
    \subfloat[][]{
        \includegraphics[width=0.5\columnwidth]{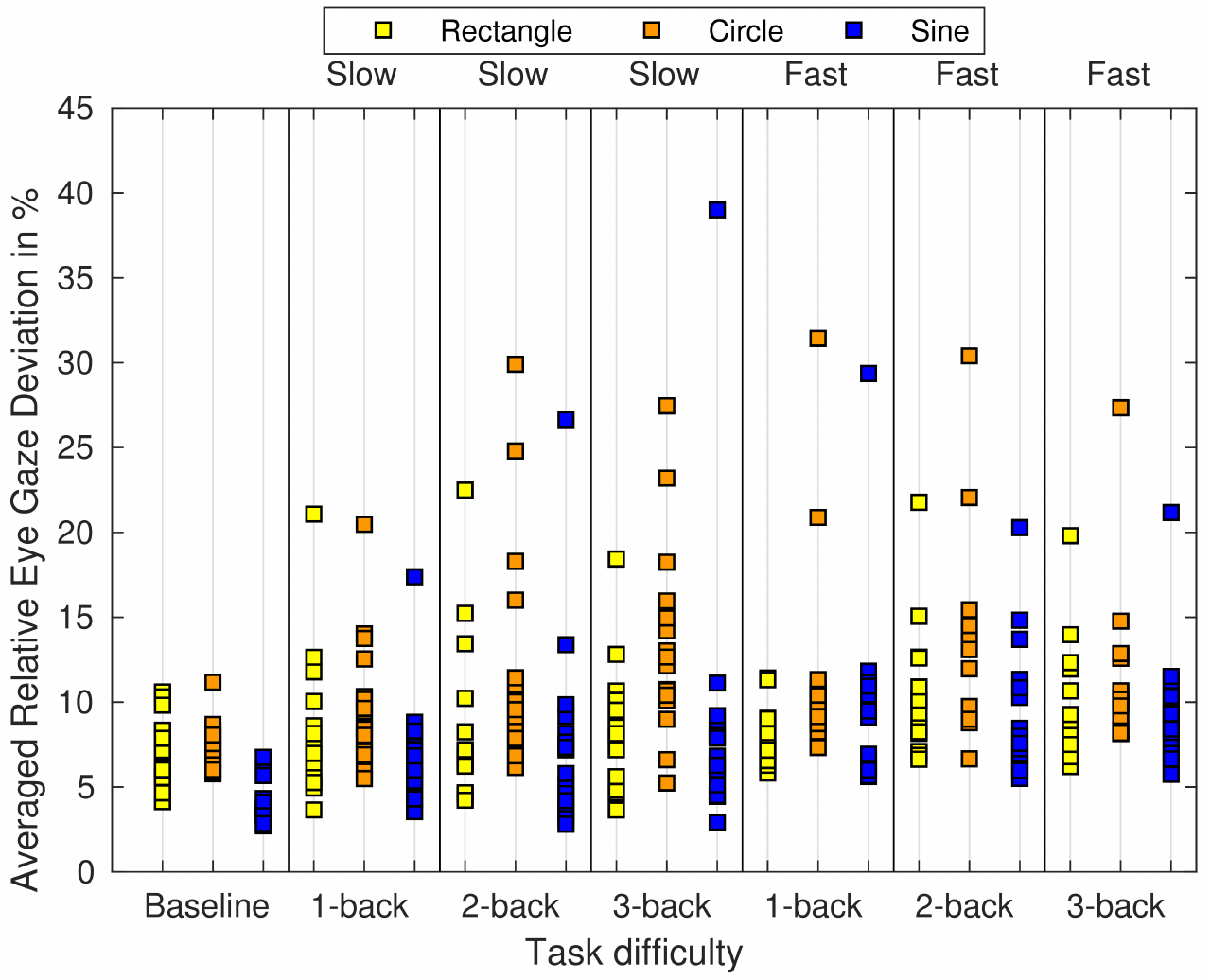}
        \label{fig:errorparticipant}}
    \subfloat[][]{
        \includegraphics[width=0.5\columnwidth]{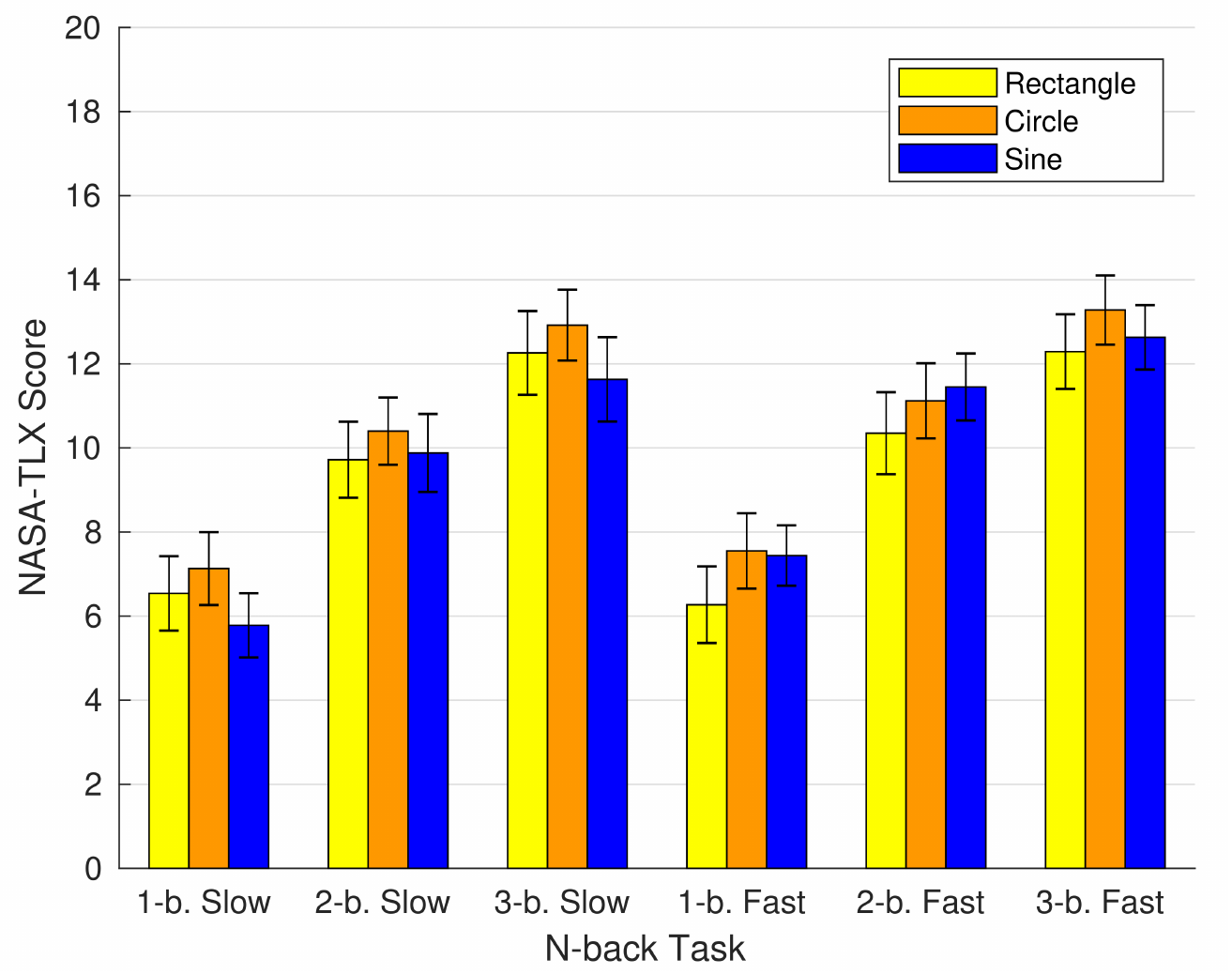}
        \label{fig:nasatlx}}
    \caption[Averaged eye gaze deviations and raw NASA-TLX scores]{\textbf{(a):} Scatter plot of the mean gaze deviation per participant and condition. Each dot represents the mean gaze deviations between the recorded and showed the trajectory of a participant. The baseline measurements show a constant behavior and do not scatter apart from the displayed trajectory. Compared to the baseline tasks, circular and sinusoidal trajectories scatter along the y-axis with increasing task difficulty. \textbf{(b):} Mean NASA-TLX score for different trajectories, task difficulties, and speeds. Increasing task difficulties led to higher NASA-TLX scores. The error bars depict the standard error.}
    \label{fig:errors_tlx}
\end{figure}

To visualize averaged relative differences when performing smooth pursuit eye movements, all averaged relative eye movement differences per participant and condition were plotted. This resulted in one data point per participant and per trial, or 21 data points per participant. The averaged plot is depicted in Figure~\ref{fig:errorparticipant}, where the y-axis depicts the gaze deviations from the shown trajectory in percent. The bottom x-axis is annotated with task difficulties and the top x-axis describes trajectory velocities. The trajectory type is color-coded.

Gaze deviations obtained from baseline trials do not scatter in contrast to gaze deviations measured from trials with task difficulty for circular and sinusoidal trajectories. Furthermore, gaze deviations increase between slow and fast trajectories when raising the task difficulty. However, results show important constraints when assessing cognitive workload from smooth pursuit eye movements. Rectangular trajectories show less eye gaze deviations when increasing task difficulty compared to baseline trials. Thus, evaluating the presence of cognitive workload using rectangular smooth pursuit eye movements is less accurate compared to circular and sinusoidal trajectories. Depending on the setting of a smooth pursuit-driven user interface, rectangular trajectories should be favored when accurate input is required even when the user is impacted by cognitive workload. Circular and sinusoidal trajectories show higher gaze deviations under cognitive workload. Circular and sinusoidal trajectories show clear differences between trials with and without cognitive workload. Such trajectories may be used to determine the existence of cognitive workload.

\subsection{Subjective Analysis of Cognitive Workload}
The lowest NASA-TLX scores per trajectory were subjectively perceived by participants when showing fast rectangles during a \textit{1}-back task (M $= 6.27$, SD $= 4.03$), slow circles during a \textit{1}-back task (M $= 7.13$, SD $= 3.87$), and slow sine waves during a \textit{1}-back task (M $= 5.78$, SD $= 3.41$). Per trajectory, the highest NASA-TLX scores were measured when displaying fast rectangles during a \textit{3}-back task (M $= 12.29$, SD $= 3.97$), fast circles during a \textit{3}-back task (M $= 13.28$, SD $= 3.68$), and fast sine waves during a \textit{3}-back task (M $= 12.63$, SD $= 3.42$). A repeated measures ANOVA showed statistically significant differences between different levels of cognitive workload and NASA-TLX score ($F(2, 321) = 68.503$, $p < .001$). However, no statistically significant differences were found for different speeds ($F(1, 322) = 2.035$, $p = .211$) and displayed trajectories ($F(2, 321) = 21.461$, $p = .722$) compared to NASA-TLX scores. Figure~\ref{fig:nasatlx} illustrates the averaged values from the obtained NASA-TLX scores.

The quantitative analysis shows significant outcomes regarding gaze deviations with different stimuli speeds and \textit{N}-back complexities. To address the impact of different variables on the individual's subjective perception, we conducted a statistical analysis to investigate single NASA-TLX items in correlation to the stimulus speed and \textit{N}-back difficulty. We compare the single NASA-TLX items grouped by the two different speeds. Our results show a significant difference for the \textit{physical demand} ($p = .006$, M$_{slow} = 5.11$, M$_{fast} = 5.69$), \textit{temporal demand} ($p = .003$, M$_{slow} = 7.29$, M$_{fast} = 8.27$), and \textit{effort} ($p = .001$, M$_{slow} = 9.53$, M$_{fast} = 10.29$) scales. However, no significant difference was found in the mental load scale ($p > 0.05$).

To show that different \textit{N}-back difficulties were responsible for higher perceived cognitive workload, we conducted a Wilcoxon signed-rank test to compare the mental NASA-TLX items between the different task complexities. We found significant differences between all three \textit{N}-back difficulties (all $p < .001$) for the \textit{mental demand} scales (M$_{N=1} = 5.84$, M$_{N=2} = 10.56$, M$_{N=3} = 13.84$). Further, we investigate whether the relationship between NASA-TLX scores and gaze deviations was linear using a Pearson correlation. No correlation was found ($.11 < r < .15$, $\forall\,N \in \{1, 2, 3\}$). Thus, it appears that the relationship between the variables is more complex \textit{i.e.} non-linear. This shows the need to understand how smooth pursuit affects subjectively perceived workload.


\section{Predicting Cognitive Workload}
Results from the study showed significantly increased eye movements during cognitive workload for circular and sinusoidal trajectories. The perception of cognitive workload by the individual is supported by subjective ratings of participants using NASA-TLX questionnaires. We train a classifier that predicts cognitive workload from smooth pursuit eye movements. We investigate the performance of person-dependent and person-independent classification for the different experimental conditions.

\begin{table*}
\begin{center}
    \resizebox{\textwidth}{!}{
\begin{tabular}{lcccccccccc}
\toprule
     & \multicolumn{4}{c}{\textbf{Binary Pers.-Indep.}} & \multicolumn{4}{c}{\textbf{Multilabel Pers.-Indep.}} & \multicolumn{2}{c}{\textbf{Multilabel Pers.-Dep.}}\\
    \textbf{Stimulus} & Accuracy & Precision & Recall & F1 & Accuracy & Precision & Recall & F1 & Accuracy & F1 \\
    \midrule
    Rectangle Slow & $75.0\%$ & $37.5\%$ & $50.0\%$ & $42.9\%$ & $40.3\%$ & $26.9\%$ & $40.0\%$ & $30.5\%$ & $48.6\%$ & $52.4\%$\\
    Rectangle Fast & $76.4\%$ & $40.1\%$ & $52.8\%$ & $46.0\%$ & $54.2\%$ & $44.0\%$ & $54.1\%$ & $47.2\%$ & $69.4\%$ & $74.2\%$\\
    Circle Slow & $95.8\%$ & $89.6\%$ & $91.7\%$ & $90.5\%$ & $40.3\%$ & $26.9\%$ & $40.2\%$ & $30.9\%$ & $69.5\%$ & $60.0\%$\\
    Circle Fast & $97.2\%$ & $93.0\%$ & $94.4\%$ & $93.7\%$ & $55.6\%$ & $45.8\%$ & $55.6\%$ & $48.5\%$ & $85.0\%$ & $74.4\%$\\
    Sine Slow & $80.1\%$ & $55.6\%$ & $64.8\%$ & $59.8\%$ & $43.1\%$ & $30.6\%$ & $43.0\%$ & $34.5\%$ & $79.3\%$ & $76.3\%$\\
    Sine Fast & $99.5\%$ & $98.9\%$ & $99.2\%$ & $99.0\%$ & $59.7\%$ & $51.9\%$ & $59.7\%$ & $54.4\%$ & $88.1\%$ & $84.5\%$\\
    \bottomrule
\end{tabular}}\vspace{-1mm}
\caption[Classification performance scores of the eye movement deviation]{Accuracy, precision, recall, and F1 scores of the binary and multi-label person-independent (pers.-indep.) as well as person-dependent (pers.-dep.) classifications. Fast circular and sinusoidal trajectories yield higher classification performances than slower linear trajectories.}
\vspace{-2mm}\label{tab:class_scores}
\end{center}
\end{table*}

\subsection{Attributes, Instances, and Classes}
Data preprocessing was necessary before training a predictive model. We removed the first two seconds per trial to avoid distortions in the signal caused by participants initially searching for the stimulus. We used the collected gaze data and normalized it concerning the coordinates of the shown trajectory. We then calculate the Euclidean distance (see Equation~\ref{eq:euclidean_distance}) between each coordinate of the normalized displayed trajectory and the normalized gaze points recorded from participants. These gaze deviations are defined as the only attribute for the instances we used for classifier training and evaluation later. We smoothed the data using an averaging running window (length 250\,samples) and a hop size of one sample.

The instances used for classifier training and evaluation consisting of a one-dimensional vector, where the normalized gaze deviations represent the only attribute. We define two sets of classes with different class labels. The first class has two labels consisting of \textit{low workload}, referring to trials where no \textit{N}-back task was used, or \textit{high workload}, thus referring to trials including any \textit{N}-back task. We investigate binary classification performances using this class. In contrast, the second class contains the four labels \textit{0-back}, \textit{1-back}, \textit{2-back}, and \textit{3-back}, each referring to the measurements of trials using the corresponding \textit{N}-back task. Overall, we constructed 378 instances, each containing 6000 gaze values, which were used for training and evaluation.

\subsection{Classifier Performance}

An SVM with a linear kernel was used to investigate the prediction performance~\cite{liang2008driver, tsochantaridis2004support, witten2016data}. The previously described instances were used for training and assigned to their appropriate class labels. Within our classifier learning process, we aimed to evaluate the classification efficiency for binary classification, such as detecting low cognitive workload and high cognitive workload. Furthermore, we investigated the classifier performance for determining different levels of cognitive workload concerning the \textit{N}-back task difficulty. We used scikit-learn\footnote{\url{www.scikit-learn.org} - last access \urldate} to train different prediction models.

\subsubsection{Person-Independent Binary Classification}
We performed a person-independent leave-one-person-out classification using two labels; one assigned to low cognitive workload referring to baseline trials and one assigned to high cognitive workload referring to trials comprising a \textit{1}-back, \textit{2}-back, or \textit{3}-back task. A leave-one-person-out classification uses all except one participant for training, while the final participant is used for validating the trained model. The leave-one-person-out classification was carried out for each trajectory and speed separately, where the instances contain the normalized gaze deviations from the shown trajectory per participant and difficulty. The leave-one-person-out classification procedure was repeated for every participant\footnote{Overall 18 runs to use every participant for validation} and the results were averaged. Table~\ref{tab:class_scores} shows the accuracies, precisions, recalls, and F1 scores of the binary classification per trajectory and speed. The classification results favor fast trajectories in combination with changing directions, such as circular or sinusoidal stimuli.

\subsubsection{Person-Independent Multilabel Classification}
To investigate the performance of classifying different levels of cognitive workload, the same leave-one-out validation procedure was conducted for four labels, with each label assigned to instances with their respective task difficulty. Again, the validation was conducted for every participant after using the other participants for training. The results are shown in Table~\ref{tab:class_scores}.

The overall classification efficiency is lower compared to the binary classification accuracy. Different reasons can be responsible for this. First, we are aware that multiple labels may lead to a lower efficiency when the difference between values within the \textit{N}-back conditions is low. Second, we combine multiple participants who may differ in their smooth pursuit behavior individually. The generalized data could, therefore, be biased by individual gaze behavior, which leads to a distorted result. Therefore, we investigate person-dependent training for cognitive workload classification purposes.

\subsubsection{Person-Dependent Multilabel Classification}
We analyzed instances within each person to investigate if higher accuracies could be achieved due to person-dependent properties. Instead of a leave-one-person-out classification, we conducted a leave-one-repetition-out classification for each participant, speed, and trajectory. However, the number of folds had to be set differently, since the animations iterated a different number of times depending on the trajectory and speed\footnote{Each trajectory started and ended at the same position}. Therefore, we adjusted the number of folds of the leave-one-repetition-out classification per trajectory and speed.

The number of folds was set to $k=2$ for slow rectangles, $k=3$ for fast rectangles, $k=5$ for slow circles, $k=7$ for fast circles, $k=2$ for slow sine waves, $k=3$ for fast sine waves. The data per trajectory and speed and participant was randomly partitioned into $k$ folds, where $k-1$ folds were used for training while the last fold was used for evaluation. This procedure was conducted $k$ times per participant and trajectory with the different \textit{N}-back difficulties assigned to the training set. The results were first averaged per participant and then over all participants. Table~\ref{tab:class_scores} shows the results of the cross-validation. Person-dependent classification of multiple cognitive workload levels shows a higher accuracy for fast smooth pursuit movements. Especially fast circular and fast sinusoidal trajectories show better performances compared to rectangular trajectories.

\section{Discussion}
Results from our study show how cognitive workload leads to increased gaze differences of smooth pursuit eye movements during the presence of cognitive workload. We found that faster circular and sinusoidal trajectories led to higher gaze deviations, while rectangular trajectories did not show this effect. We believe, that circular and sinusoidal trajectories required more effort from users since they had to focus more on their task completion due to constantly changing directions. This may be a reason for improved classification results for this kind of trajectories. In contrast, rectangular trajectories can be deployed within smooth pursuit-based interfaces whenever reliable input, independent from the perceived cognitive workload, is required. Depending on the use case, a cognition-aware system designer can decide if reliable input through rectangular trajectories or mental workload estimation by circular and sinusoidal trajectories is desired.

The speed of the trajectories had a quantitative measurable effect on the overall smooth pursuit performances. However, as the statistical investigations of single NASA-TLX scales showed, \textit{mental demand} was not affected compared to \textit{physical} and \textit{temporal demand} as well as \textit{effort} by different speeds. Faster speeds cause therefore a different type of physical workload than cognitive workload which significantly impacts smooth pursuit eye movements. Different trajectory speeds are thus not necessarily responsible for inducing cognitive workload. By comparing different \textit{N}-back difficulties about the NASA-TLX scales, we found a significant difference between all difficulties. This supports, that subjectively perceived cognitive workload was altered by different \textit{N}-back complexities and that both variables manipulate measurable smooth pursuit performances.

The classification results yield higher accuracies for distinguishing between low and high workload levels than for detailed levels of cognitive workload. Furthermore, binary classification can be achieved without the need for person-dependent calibration, while separating different levels of cognitive workload requires a person-dependent calibration regarding cognitive workload.

In a real deployment scenario, binary classification can be used to provide additional help for users when high cognitive workload is classified. This refers to very simple scenarios, where only the estimation of low and high workload is desired. Such places could be public places when, for example, interacting with public displays. In contrast to short interactions in the public, a classifier can be trained for multilevel classifications in private spaces where long-term interaction is conceivable.

Calibrating a classification model with multiple workload levels in the public can result in a cumbersome procedure due to external factors impacting the individual cognitive capacities, such as distracting pedestrians walking by. Our results indicate binary workload classification in the public using a pre-trained classification model, while a classification model with multiple workload levels can be used in private spaces where calibration can be done without any disruptions.

To employ sensing of cognitive workload, smooth pursuit must be elicited. In contrast to smooth pursuit-based user interfaces, other environments require explicit or implicit integration of moving elements. For example, as short waiting times occur as a result of a database query or as a new task is loaded during a user study, feedback on the system status could be presented in a way that fosters smooth pursuit movements. For example, this can be elicited by an animated progress bar. In this way, traditional methods such as the NASA-TLX or DALI questionnaires can be complemented.

Finally, a suitable workload level must be found. Looking back at the classification results of the multi-label model, we achieved reasonable accuracies for person-dependent classification. A cognition-aware system must be able to find the right difficulty for each user, as permanent support by a system may lead to boredom or no support leads to frustration for the user. Keeping the task difficulty at its highest or lowest level might not be favored. However, for deploying binary classification in public use cases minor support might be helpful compared to private settings, where user expectations on cognition-aware computing systems are higher.
\subsection{Limitations}

Our study is prone to certain limitations. We collected gaze data under controlled lab conditions and hence, do not know how our results generalize to other situations, where participants may be distracted for example, by the presence of other people. Still, despite the controlled conditions, participants' gaze behavior may have been influenced by physiological wellbeing, such as lack of sleep. Our calculation of gaze difference is based on 30\,seconds of recording. However, there may be situations in which assessing workload with finer granularity is desirable. Additionally, blink frequency and blink duration were not evaluated during our studies. Another limiting factor was the study execution in a calm lab. Natural distractions in real-world environments could alter our results. Furthermore, we have not investigated the effect of different eye tracking frame rates in our study. Consequently, before assessing cognitive workload through smooth pursuit, eye trackers must be tested for their suitability.

\section{Use Cases for Workload-Aware Systems}
In the following, we present several different use cases to be supported within smooth pursuit scenarios.

\subsection{Support in Safety-Critical Environments}
Smooth pursuit can be utilized to assess cognitive workload during a monitoring task, such as air traffic controller surveying airplane flight processes, to support or warn operators for cognitive exhaustion. Alternatively, workload can be dispatched to a colleague who does not have to cope with high workload during work. Since objects of interest can be visualized using a small moving circle, smooth pursuit eye movements can be triggered this way. As a result, accidents, which occur due to mental overload, distractions, or fatigue, can be avoided. For example, the user interface can be adapted by simplifying displayed content or dispatching a part of the observation task to another colleague. This use case is transferable to other applications, where moving objects occur naturally and require permanent attention of the user. Figure~\ref{fig:teaser_air_tower} shows an example of how smooth pursuit can be used in such situations.

\subsection{Adaptation of Pursuit-based Interactive Systems}
Prior research has introduced many applications that use smooth pursuit primarily for interaction. As we illustrated in our related work section, smooth pursuit interfaces have been used for smartwatch interaction \cite{esteves2015orbits} and interaction with smartphones~\cite{Liu2015}. Using mobile devices enables ubiquitous sensing of cognitive workload in outdoor settings, bypassing the disadvantages of using pupillary measures being prone to lighting conditions. Interacting with large distant displays \cite{vidal2013pursuits}, where touch and gesture interaction have emerged as state of the art to communicate input~\cite{Peltonen:2008:MDT:1357054.1357255, Vogel:2004:IPA:1029632.1029656} can use smooth pursuit to implicitly measure mental states.

\begin{figure}
    \centering
    \subfloat[][]{
        \includegraphics[width=0.5\columnwidth]{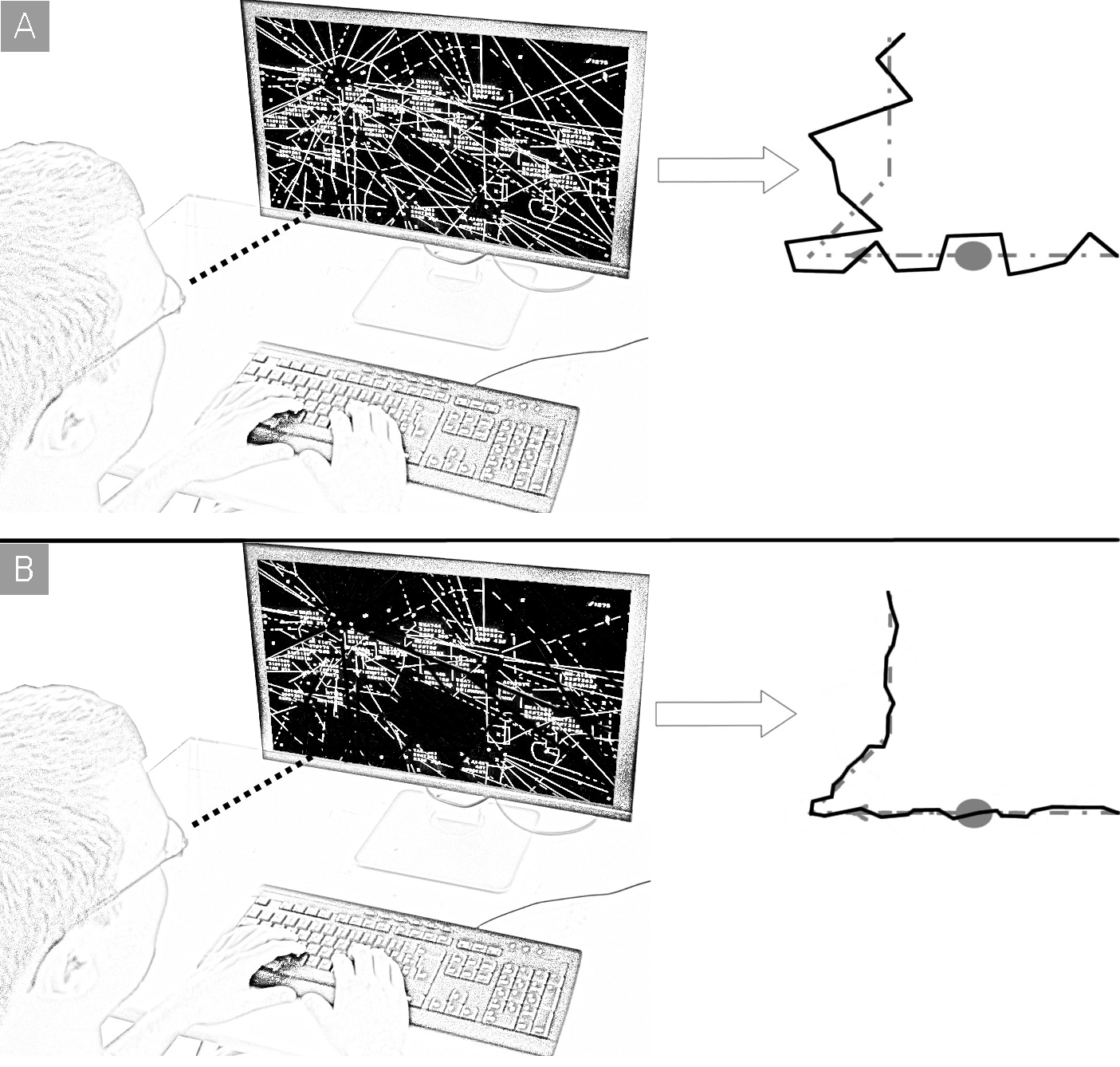}
        \label{fig:teaser_public_display}}
    \subfloat[][]{
        \includegraphics[width=0.5\columnwidth]{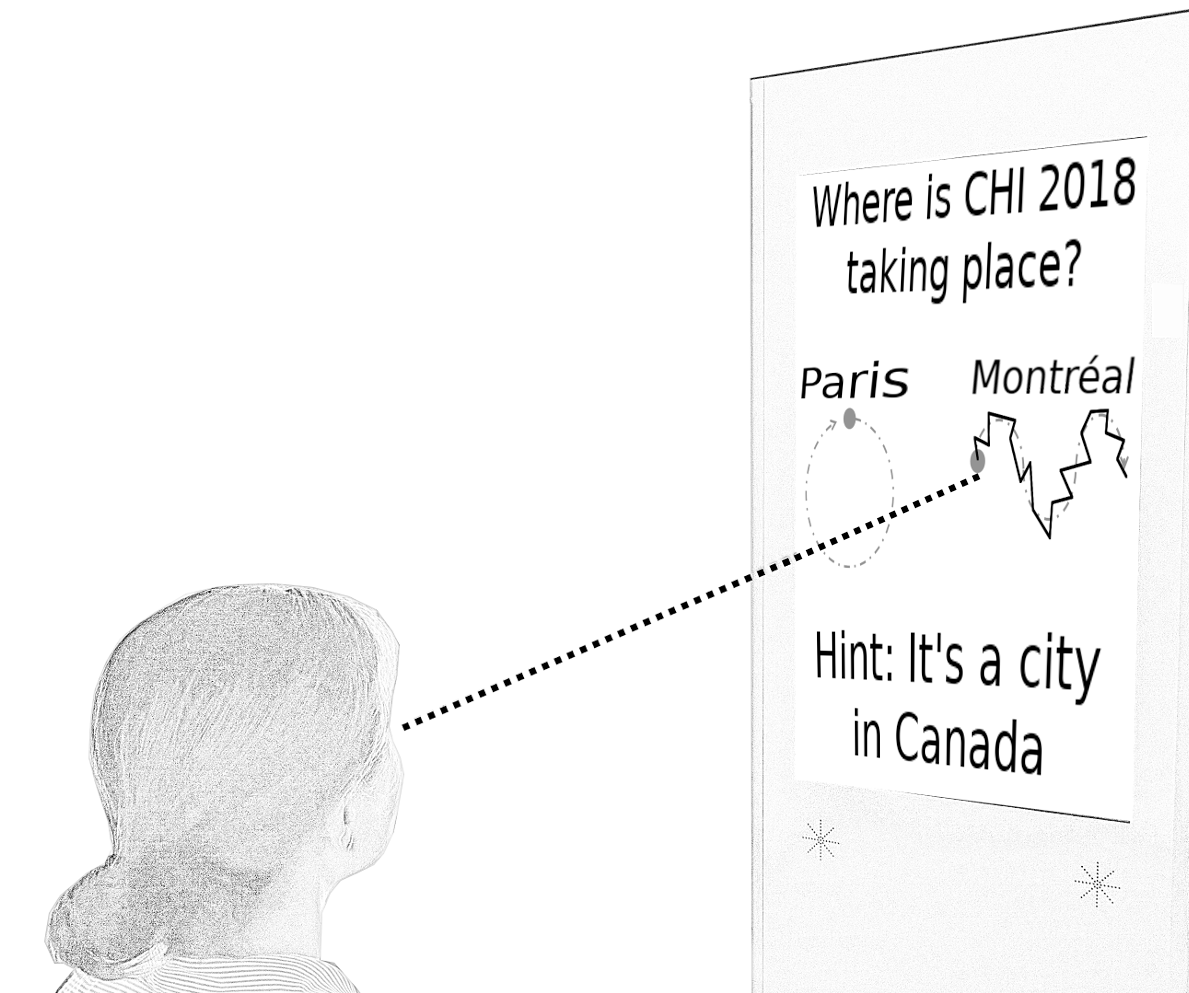}
        \label{fig:teaser_air_tower}}
    \caption[Exemplary use cases for smooth pursuit-based workload assessments]{\textbf{(a):} User working in an air control tower. Moving dots representing airplanes on the screen can cause smooth pursuit eye movements. (A) The system detects high cognitive workload from the user and dispatches some observation tasks to a colleague. (B) Alleviated cognitive workload measured after user interface adaption. \textbf{(b):} User playing a quiz game on a public display. The system infers that the question is inducing high cognitive workload. The system is, therefore, observing if this behavior persists and provides a hint to avoid frustration.}
    \label{fig:teaser_sp}
\end{figure}

Our approach enables implicit contactless assessment of cognitive workload while interacting with these devices. Pursuit-based interactive systems benefit from our classifier to dynamically predict the current cognitive workload level of the user during interaction and adapt to it accordingly. If high task load is identified during an interaction, the user interface or task objective can be modified by an easier one. Figure~\ref{fig:teaser_public_display} shows how the existence of cognitive workload can be estimated to adapt a public display app.

\section{Study Conclusion}
The presented study investigated the influence of cognitive workload on smooth pursuit eye movements using three different trajectories with two different velocities. Using an auditory delayed digit recall \textit{N}-back task to induce cognitive workload, a higher deviation of gaze points from shown trajectories is measured compared to measurements when not inducing cognitive workload. Based on our results, we create a person-independent classifier for estimating binary workload and a person-dependent classifier for distinguishing different levels of cognitive workload. While binary cognitive workload classification can be elicited in the public using smooth pursuit interfaces, private spaces benefit from person-dependent classifier calibration to determine different levels of cognitive workload. The collected data set is available on Github\footnote{\url{www.github.com/hcum/your-eyes-tell} - last access \urldate}.

\section{Chapter Conclusion}
In this chapter, we investigated the characteristics of eye movements during different levels of cognitive workload. Eye movements can be recorded using a remote eye tracker which obviates direct body contact with the user. We show that cognitive workload levels can be classified using specific smooth pursuit eye movements using machine learning techniques. Contrary to \ac{EEG}, no calibration procedure or tedious setup is required. Thus, the real-time analysis of eye movement holds promise as an additional measurement channel for workload-aware computing interfaces.

Having such a measurement modality without the need of body-worn devices goes a step towards real-time mental state estimation in ubiquitous computing environments. User interfaces can then provide intervention mechanisms to relax or help users based on their current context. Our classifier depends on eye gaze only and fits into several application scenarios. It can be deployed in real-world scenarios to estimate the presence of cognitive workload in real-time. Thereby, the assessment can be done without the need for contact or additional bodyworn sensors, which comes closer to Mark Weiser's vision of ubiquitous computing. However, the classification performance varied throughout the displayed trajectories, and for both person-dependent and person-independent cases. Whereas accuracies of up to 99\% can be achieved for binary person-independent classification, person-dependent classification excels at the classification for different \textit{N}-back difficulties. Workload-aware interfaces that are deployed in the public may benefit from cognitive workload estimation using a binary classification whereas it may be acceptable to calibrate for fine-grained workload determination in private spaces. This chapter answers \textbf{RQ4} with the focus of measurable changes in smooth pursuit eye movements, which can be leveraged by workload-aware systems for cognitive workload classification.

\chapter{Pupil Dilation}
\label{ch:pupil_dilation}

\setlength{\epigraphwidth}{6.5cm}
\epigraph{\textit{Sitting in the far horizon\\
The evergreens across the border\\
Can you see a different story\\
With the failure and the glory\\
The truth is in the eyes of the beholder\\
The truth is in her eyes when you hold her
}}
{Andrew Stockdale}

The previous chapter investigated changes in eye movements during different levels of cognitive workload. An \textit{N}-back task was employed to elicit different levels of cognitive workload. Specifically smooth pursuit eye movements were affected and could be successfully classified between and within persons. However, moving elements are required to elicit smooth pursuit movements since they cannot be pretended or simulated by humans~\cite{velloso2016ambigaze}. Alternative metrics for measuring cognitive workload using eye-based properties have found relevance in previous research~\cite{duchowski2002breadth, duchowski2018gaze} including pupil dilation.

\begin{figure}[h!]
    \centering
    \subfloat[][]{
        \includegraphics[width=0.5\columnwidth]{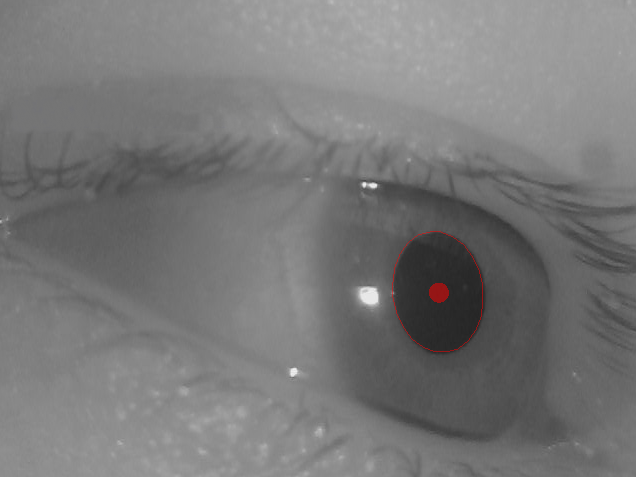}
        \label{fig:pupil_low}}
    \subfloat[][]{
        \includegraphics[width=0.5\columnwidth]{figures/pupil_high_edited.png}
        \label{fig:pupil_high}}
    \caption[Pupil dilation under different levels of workload]{Exemplary impact of cognitive workload on pupil dilation. \textbf{(a):} Measured pupil diameter under low cognitive workload. \textbf{(b):} Measured pupil diameter under high cognitive workload.}
    \label{fig:pupil_compared}
\end{figure}

Pupil dilation has attracted attention in the HCI community as a measure for cognitive workload (see Figure \ref{fig:pupil_compared})~\cite{Duchowski:2018:IPA:3173574.3173856}. Demanding the short-term memory frequently over a timespan causes cognitive effort~\cite{ashcraft2001relationships, ashcraft2007working, heitz2008effects, hess1964pupil, kahneman1966pupil}, a mental process which extends the pupil diameter (see Figure~\ref{fig:pupil_high}). Eye trackers are used to retrieve eye gaze positions which can be used at the same time to evaluate pupillary data in real-time. Using such a metric as an assessment of cognitive workload can be utilized by applications to provide help and assistance for users when high cognitive workload is detected. For example, workload-aware systems can assist in real-time during a cognitively demanding task. Furthermore, the usability of interfaces can be evaluated by conjunctively assessing pupil diameter and eye gaze during user interaction. Although being sensitive to lighting conditions, the evaluation of pupil dilation to infer cognitive workload has been examined by various researchers. Benedetto et al.~\cite{benedetto2011driver} explored the impact of cognitive workload on pupil dilation and eye blink duration. Since pupillary measurements are error-prone to lighting conditions, Pfleging et al. \cite{pfleging2016model} proposed a model for classifying cognitive workload of pupil dilation under different lighting conditions. Kiefer et al.~\cite{kiefer2016measuring} evaluated the pupil diameter under different task difficulties to assess the perceived task complexity. Gollan et al.~\cite{Gollan:2016:DEC:2968219.2968550} examined the assessment of pupil dilation under cognitive workload in real-time. Hess et al. \cite{hess1964pupil} investigated the impact of simple mental processes, such as solving one and two-digit multiplications, on the pupil diameter. Zbrodoff and Logan \cite{zbrodoff2005everyone} found a correlation between higher pupil extensions and increasing math exercise complexity. By using multiplication tasks, a higher pupil diameter was measured when multiplying two-digit numbers than one-digit numbers.

This chapter investigates the performance of a classification model that distinguishes between two different levels of workload. This includes the assessment of pupil dilation to create adaptive workload-aware interfaces that adjust the difficulty in real-time. Thereby, task complexity is set to a suitable difficulty to keep users engaged, thus becoming more difficult or easier depending on current mental workload measurements. This is complemented by presenting and evaluating a proof-of-concept application, which sets its task complexity based on pupillary data to prevent mental underload or overload. Hence, this chapter complements the second part of the research question:


\begin{itemize}
    \renewcommand{\labelitemi}{$\rightarrow$}
   \item \textbf{RQ4}: Do eye gaze metrics enable the classification of cognitive workload states?
\end{itemize}

\begin{center}
    \begin{GrayBox}
        \centering
        \parbox{0.98\textwidth}{
            \emph{This section is based on the following publication:}
            \vspace{3mm}
            \publicationsbegin    
            \item \bibentry{Kosch:2018:LME:3170427.3188643}
            \publicationsend
        }
    \end{GrayBox}
\end{center}

\section{Application}
To examine the usefulness of adapting task complexity based on pupillary data, we constructed a prototype capable of evaluating pupil dilation measurements in real-time.

\subsection{Recording and Feedback Setup}
A mobile eye tracker from Pupil Labs\footnote{\url{www.pupil-labs.com} - last access \urldate} is used to receive pupillary data. Data was updated at 30\,Hertz and processed on the attached computer. An external screen is used to display stimuli. To avoid distractions during the experiment, the setup was divided into two areas using separators, respectively for the experimenter and participant. The experiment was conducted in a room without windows and constant lighting conditions. The stimulus monitor had a resolution of $1920\times1080$ and a screen size of $23$ inches.


%

\subsection{Classifier}
Two baseline trials are conducted in the beginning to obtain ground truth data about pupil dilation during an \textit{easy} and a \textit{difficult} math task, each lasting three minutes. A single complexity was assigned to every trial (\textit{easy}/\textit{difficult}). The trained classifier is then used to adaptively control the complexity of the displayed multiplications in a third trial. The classifier aims to estimate low and high cognitive workload based on the data retrieved from the two baseline tasks. The prototype simulates a cognition-aware system by setting task complexity to \textit{easy} if high mental workload is measured. In contrast, task complexity is set to \textit{difficult} when low cognitive workload is classified from the previous baseline measurements. 

To assess cognitive workload from the users' pupil dilation, we trained individual support vector machines (SVMs) with a linear kernel to infer required changes of task difficulty in real-time. We used the individual pupil dilation as the only feature. We defined two classes, \textit{easy} and \textit{difficult}. The classifier was trained individually for every participant after the two baseline trials. In the last trial, the previously trained classifier is used to predict cognitive workload. The task difficulty is set to the opposite task complexity to force cognitive alleviation or effort. The person-dependent classifier accuracies ranged between 64$\%$ and 99$\%$ (M = $79\%$, SD = $0.16\%$) resulting from a k-fold cross-validation with $k=5$.

\section{Study}
We conducted a pilot study to evaluate the feasibility of changing task complexity based on pupil diameter measurements as an indicator for cognitive workload. The study configuration conforms to the previously described prototype.

\subsection{Cognitive Task}
\begin{figure}
    \centering
    \subfloat[][]{
        \includegraphics[width=0.5\columnwidth]{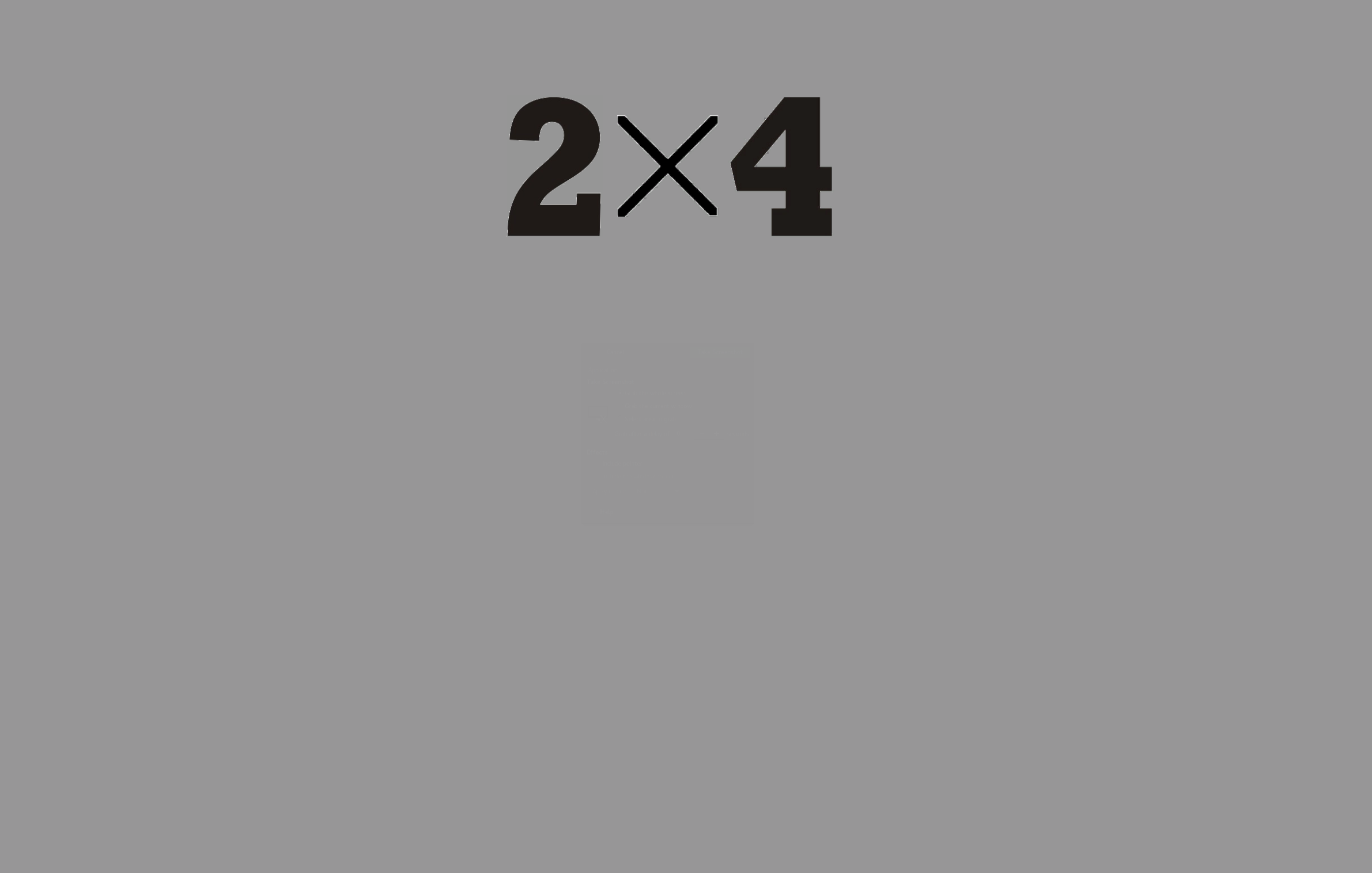}
        \label{fig:math_easy}}
    \subfloat[][]{
        \includegraphics[width=0.5\columnwidth]{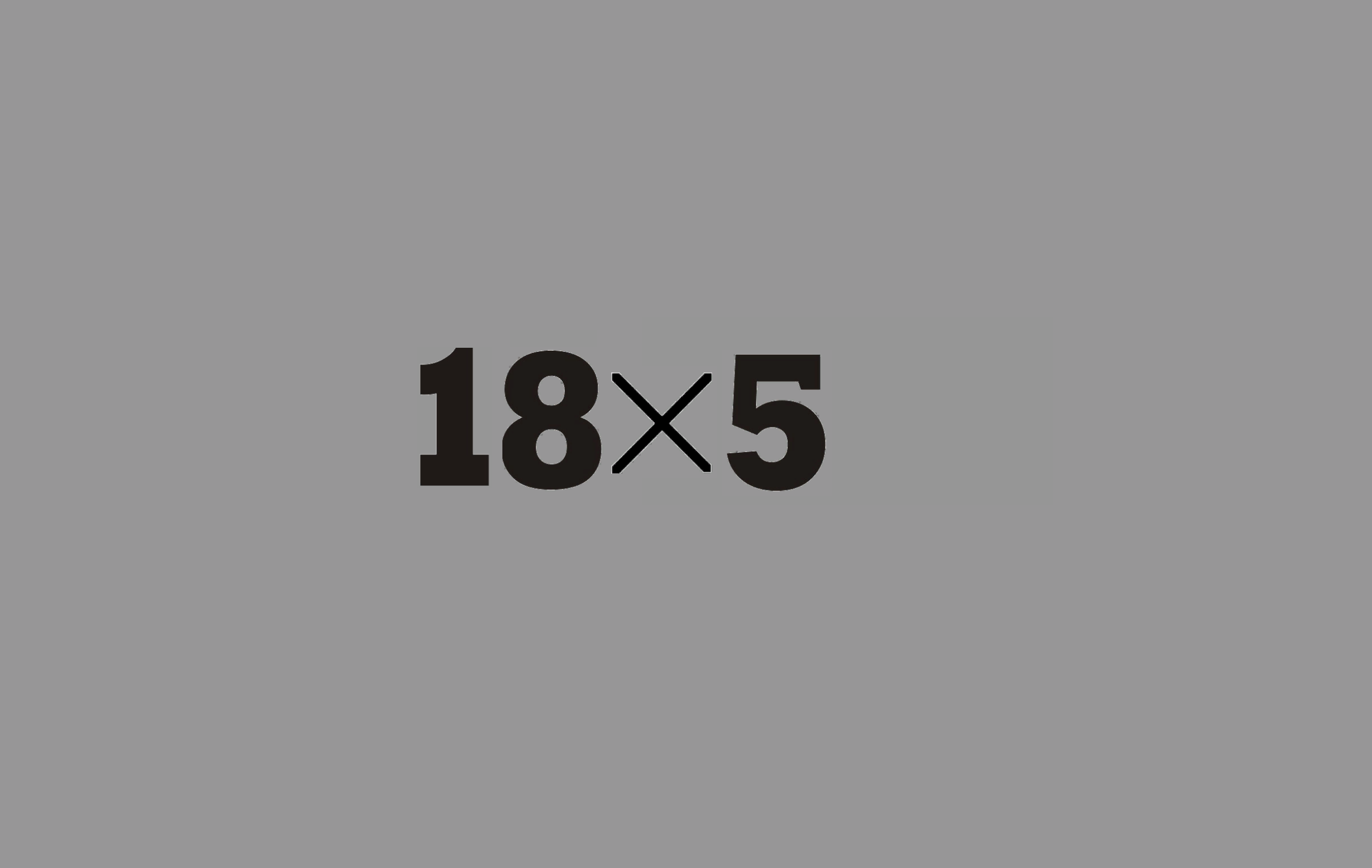}
        \label{fig:math_high}}
    \caption[Math task employed during the experiment]{Multiplication math task with equations moving from top to bottom comprising two different complexities. \textbf{(a):} \textit{Easy} multiplication containing one-digit numbers. \textbf{(b):} \textit{Difficult} multiplication comprising a two-digit number and a one-digit number.}
    \label{fig:mathtask}
\end{figure}

To induce cognitive workload, we use a multiplication math task with two different complexity levels including one-digit and two-digit multiplications. Both complexities lead to different pupil dilations~\cite{ashcraft2001relationships, ashcraft2007working, zbrodoff2005everyone}. To add a time constraint, the multiplication math tasks were moving centered from the top to the bottom of the screen within five seconds. During the time limit, participants were asked to type the correct solution on a keyboard number pad. When the multiplication reached the bottom of the screen, it disappeared and a new multiplication is displayed at the top of the screen. If the user fails to enter a solution during the given time frame, an error is counted. If the solution entered during the time is correct, the multiplication disappears and a new multiplication task is displayed at the top of the screen. The complexity is divided into \textit{easy} and \textit{difficult}. The \textit{easy} condition uses numbers ranging from $0 - 9$, randomly selecting two numbers which have to be multiplied together. In the \textit{difficult} condition, two numbers are randomly chosen, where the first number ranges from $10 - 19$ and the second from $0 - 9$ (see Figure \ref{fig:mathtask}). Numbers are constantly displayed in black font on a gray background (RGB: $[150, 150, 150]$).

\subsection{Participants}
Six participants (4 male, 2 female) took part in the pilot study, ranging from an age between 22 and 34 years (M = $29.17$, SD = $4.41$). All participants were computer science, students or researchers. Participants had a normal or corrected-to-normal vision. The overall duration of the study was approximately 15 minutes. Before the start of the experiment, participants signed an informed consent form and provided their demographic data.

\subsection{Procedure}
First, we explained to the participants the study procedure and familiarized them with the setting. The mobile eye tracker was then calibrated. The study consisted of three trials; two baseline multiplication trials (\textit{easy}/\textit{difficult}) and one \textit{adaptive} multiplication trial. The duration of each task was three minutes. The order of the first two baseline trials was counterbalanced according to the balanced Latin square. The pupillary data from the baseline tasks was used to derive an SVM classifier capable to distinguish between cognitive workload induced by a \textit{easy} or \textit{difficult} task. During the \textit{adaptive} trial, the trained user-dependent classifier is used to adaptively set the complexity of the multiplication task. The complexity is set by classifying measured pupil diameter data in five-second intervals. If the classifier estimates a high diameter difference, the task complexity is set to \textit{easy} to avoid mental overload. In contrast, the task complexity is set to \textit{difficult} if the classifier estimates a low pupil diameter to avoid mental underload. If the classifier determines a complexity change during a multiplication task, the new task complexity is adjusted to the next appearing multiplication. After every trial, participants filled a NASA-TLX \cite{HART1988139} questionnaire to obtain subjectively perceived workload during the trials.

\section{Exploratory Results}
Overall, the \textit{easy} multiplication condition consisted of $618$ displayed calculations, the \textit{difficult} condition $420$ displayed multiplications, and $466$ displayed calculations for the \textit{adaptive} condition. Least errors were measured during \textit{easy} multiplication, comprising $571$ correct ($92$\%) and $47$ wrong ($8$\%) answers. Most errors were made during \textit{difficult} trials with $255$ correct ($61$\%) and $165$ wrong ($39$\%) answers. The \textit{adaptive} condition placed itself in between with $313$ correct ($67$\%) and $153$ wrong ($33$\%) answers.


\begin{figure}
    \centering
    \subfloat[][]{
        \includegraphics[width=0.5\columnwidth]{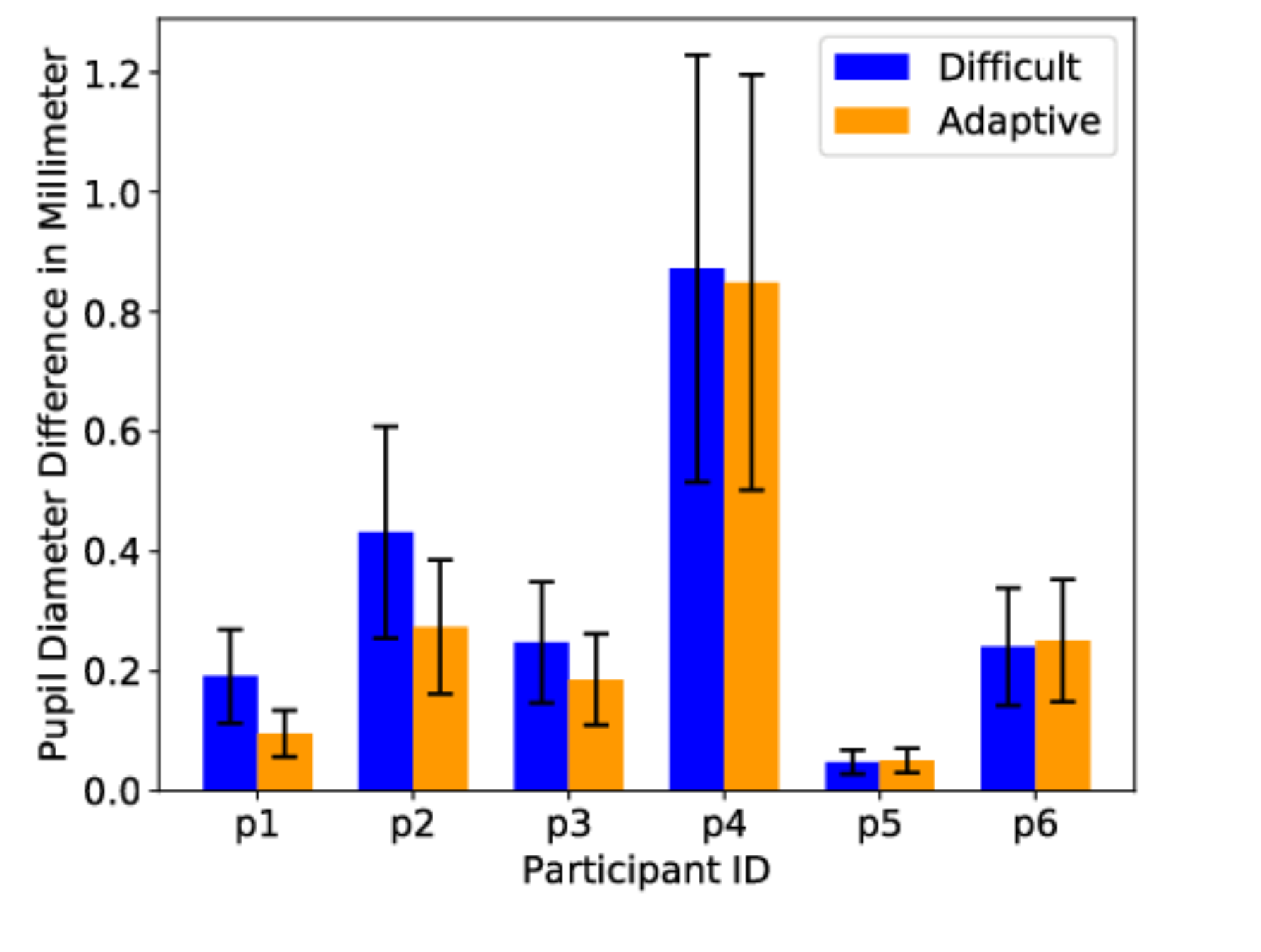}
        \label{fig:averaged_pupil_plot}}
    \subfloat[][]{
        \includegraphics[width=0.5\columnwidth]{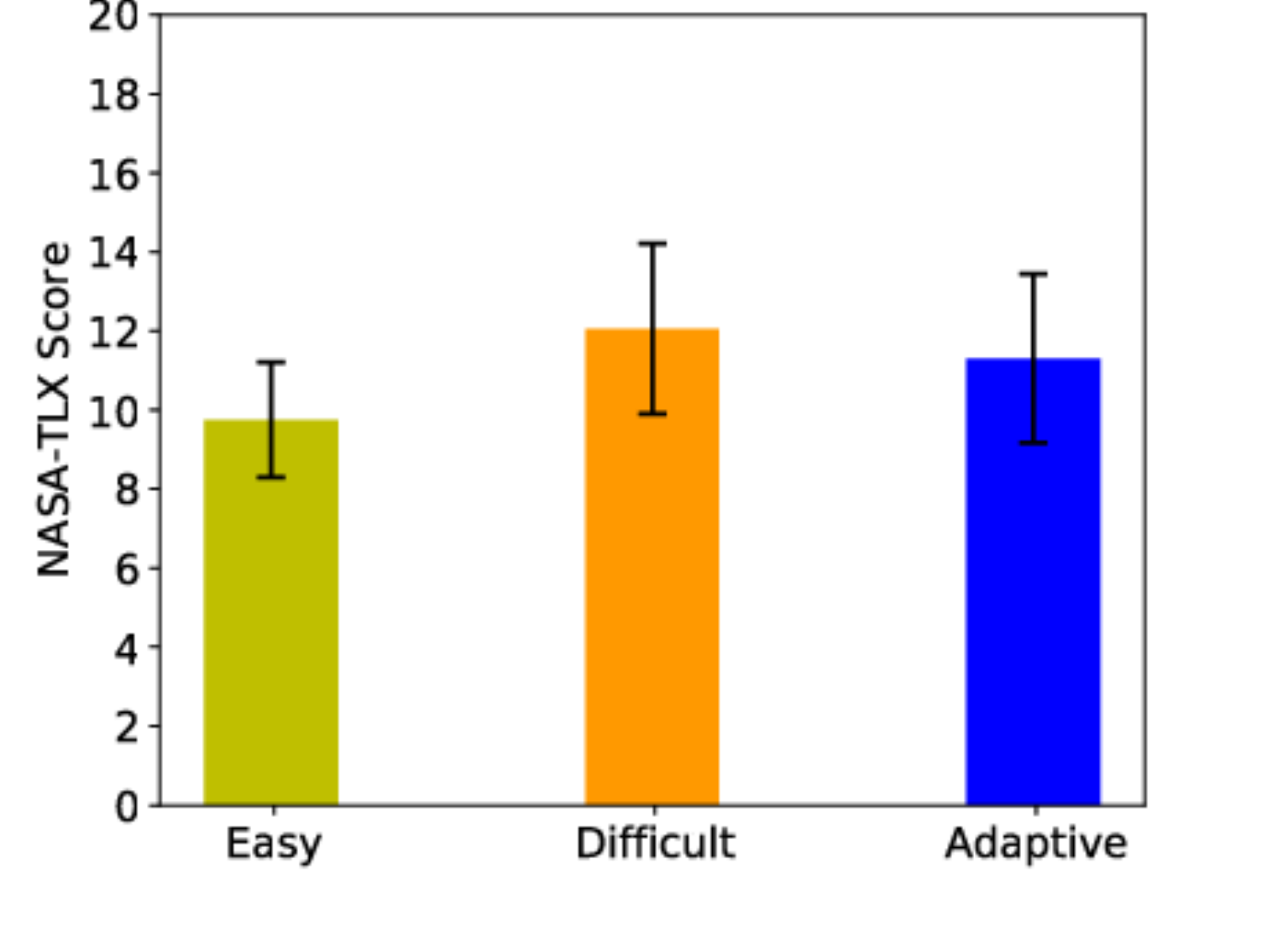}
        \label{fig:nasa_tlx_score}}
    \caption[Pupil diameter differences and raw NASA-TLX scores]{\textbf{(a):} Averaged differences of pupil diameter per participant between \textit{easy} and \textit{difficult} task complexities as well as \textit{easy} and \textit{adaptive} task complexities. The difference between the \textit{easy} and \textit{difficult} task complexity is higher compared to the \textit{adaptive} task complexity except for two participants (p5, p6). The whisker lines depict the standard error. \textbf{(b):} Averaged NASA-TLX scores for the different task difficulties. The whisker lines depict the standard error.}
    \label{fig:results_pupil}
\end{figure}

We investigate individual differences in the pupil dilation measures between the different task complexities. The differences in the \textit{easy} and \textit{difficult} task complexity ranged between $0.05$ and $0.87$ millimeters. The difference between the \textit{easy} and \textit{adaptive} complexity ranged between $0.05$ and $0.85$ millimeters. Figure \ref{fig:averaged_pupil_plot} shows the difference between the averaged pupillary measurements per participant. The mean number of complexity switches during the \textit{adaptive} condition results in $2.6$ complexity changes for a total of $466$ multiplication tasks during the \textit{adaptive} condition. The mean NASA-TLX score (see Figure~\ref{fig:nasa_tlx_score}) reveals lowest subjectively perceived workload during the \textit{easy} condition (M = $9.75$, SD = $1.45$) and highest workload measurements during the \textit{difficult} complexity (M = $12.05$, SD = $2.15$). The \textit{adaptive} condition was rated between both baseline trials (M = $11.30$, SD = $2.14$).

\section{Discussion}
In this study, we explored the impact of adapting the complexity of a cognitive workload inducing task based on the measurement of pupillary changes.

\subsection{Adaptive User Interfaces}
Our findings support the use of pupil dilation to detect mental effort. Assessing mental workload in real-time to change the task complexity has shown its feasibility when participants performed adapted math tasks. The adaptive trial shows how computing systems can make use of the pupil diameter to create engaging user interfaces while avoiding the perception of high or low mental workload.
\subsection{Creating User Specific Models}
The study shows the viability of individually trained classifiers. Since physiological measurements differ among users, person-dependent measurements cannot be avoided. Limited to pupillary data, encouraging results were achieved by allocating short training times to create a user-specific model. This is supported by averaged pupil diameter measurements and subjective feedback provided through NASA-TLX questionnaires.


\section{Study and Chapter Conclusion}
This chapter investigated the assessment of cognitive workload based on pupillary data to change the complexity of a task in real-time. In a preliminary study, we trained person-dependent classifiers through math tasks with different complexities. During an \textit{adaptive} trial, task complexity was changed depending on the classification of pupillary data. The results show the practicability of using pupillary data in controlled environments to evaluate user interface adaptation mechanics and user interface assessment. 

Eye gaze-based properties reflect cognitive states. In this chapter, we investigated the suitability of pupil dilation as an alternative workload measure. A simple classification scheme revealed reliable discrimination between low and high workload states for difficulty alternations, hence assessing pupil dilation as a suitable measure for cognitive workload (\textbf{RQ4}). Performance and subjective reports imply that more engagement and less frustration is fostered when alternating between task difficulties. In contrast to smooth pursuit, dilation-based workload assessments do not require a moving object to be detected. However, while not being affected by moving objects, the pupil size is sensitive to light conditions. The presented experiment leveraged a controlled setup to avoid bias through lighting conditions. A potential solution is the normalization of pupil dilation through in-situ light measurements. The combination of lighting conditions and pupil dilation can provide a decent estimate for the currently perceived workload.

\part{Applications\label{part:applications}}


\chapter{Design Pipeline Evaluation}
\label{ch:smart_kitchen_deployment}

\setlength{\epigraphwidth}{6.0cm}
\epigraph{\textit{For people without disabilities, technology makes things easier. For people with disabilities, technology makes things possible.}}{IBM Training Manual}

So far, we have investigated communication guidelines for workload-aware interfaces and measurement modalities that assess cognitive workload in real-time using physiological measures. This provides an implicit measure to evaluate interactive systems regarding the mental demand placed on the user and adaptive systems that benefit from this insight. In this chapter, we incorporate the results from Chapter~\ref{ch:smart_kitchen_requirements},~\ref{ch:support_modalities}, and~\ref{ch:workloadbyeeg} in a field study. To recapitulate, the previous chapters investigated design implications for assistive technologies and presented an \ac{EEG}-based approach to evaluate these. Persons with cognitive impairments were investigated as a target group as they provide explicit cues when cognitive assistance is needed.

In Chapter~\ref{ch:smart_kitchen_requirements}, cooking was identified as an important communal activity in the context of sheltered living~\cite{Kosch:2018:SKP:3173574.3173845}. The preparation of a meal represents a common task that is achieved by working together. Hence, social skills are conveyed and the awareness of fulfilling duties is invigorated. Cooking transfers a beneficial skill set for future independent living. However, uncommon tasks (\textit{e.g.}, operation of novel kitchen devices) are still perceived as challenging by persons with cognitive impairments. Expert labor collaboratively supervises the cooking process but is scarce due to a worker shortage in the field. Assistive computing systems are therefore more commonly employed in sheltered living facilities.

Our previous research in Chapter~\ref{ch:support_modalities} showed that visual feedback of assistive systems was favored over auditory and tactile feedback by persons with cognitive impairments~\cite{Kosch:2016:CTA:2982142.2982157}. In an assembly scenario, we evaluated the working memory capacities using \ac{EEG} and found that visual in-situ projections decrease working memory strains. In this chapter, we combine the concepts of (a) cooking, (b) visual in-situ feedback, and (c) assistive computing for persons with cognitive impairments within a field study in a realistic context. Specifically, we deployed a projection-based assistive system over two weeks that aims to support a cooking process.

The chapter reports on a user study that investigates cooking performances between \textit{caretaker assistance} and \textit{in-situ assistance}. We then investigate differences in subjective feedback about the individual cooking experience with each feedback modality. Finally, we discuss how a combination of \textit{caretaker assistance} and \textit{in-situ assistance} enables independent cooking for people with cognitive impairments. We believe that the presented concepts translate on other chores and address, therefore, the following research questions:

\begin{itemize}
\renewcommand{\labelitemi}{$\rightarrow$}
\item \textbf{RQ5}: Does in-situ feedback provide cognitive support during a cooking task?
\end{itemize}

\begin{center}
\begin{GrayBox}
\centering
\parbox{0.98\textwidth}{
\emph{This section is based on the following publication:}
\vspace{3mm}
\publicationsbegin
\item \bibentry{kosch2019thedigital}
\publicationsend
}
\end{GrayBox}
\end{center}

\section{Related Work}
The impact of deploying assistive technologies in real-world scenarios has been investigated by past research. We summarize relevant work in the following.

\subsection{Supporting Persons with Cognitive Impairments}
Assistive technologies that focus on support for the elderly and persons with cognitive impairments were the subject of previous research. Assistive systems require design implications and guidelines to coalesce kitchen environments with home environments. For this purpose, Pollack et al.~\cite{Pollack2005} explored how assistive technologies in home environments can be designed to support people with cognitive impairments. This includes a taxonomy for assistive computing systems. Bouchard et al.~\cite{Bouchard2006} developed and evaluated a plan recognition framework for smart homes using microsensors. These sensors made contextual data available~\cite{LiJiang2004} of which people with dementia and caretaker personnel benefited from activity recognition provided by the framework. Furthermore, Arcelus et al.~\cite{arcelus2007integration} used a variety of sensors, such as microphone arrays and accelerometers, to collect contextual information. The collected data was used to train the artificial intelligence that provided individual context-aware assistance. Mihailidis et al.~\cite{Mihailidis2008} found that the integration of training activities into daily life tasks, such as washing hands, reduced the development of dementia. Serious games that simulate regular daily kitchen tasks have shown positive effects on people with cognitive impairments~\cite{10.3389/fnagi.2015.00024}. Ethical aspects need to be considered before using assistive technologies in real-world environments since, depending on the level of impairment, individuals are limited in their ability to give consent~\cite{stip2005environmental}. However, the integration of assistive technologies raises ethical considerations regarding individual autonomy in home environments. This includes privacy concerns~\cite{courtney2008privacy, courtney2008needing} and user acceptance~\cite{brownsell2000community, demiris2008senior}.

Providing \textit{in-situ assistance} at industrial workplaces through spatial augmented reality has shown mental alleviation for people with cognitive impairments~\cite{Chang:2015:AOV:2700648.2811354, funk2015comparing}. Projecting visual in-situ information enhanced the overall assembly efficiency regarding the number of errors and task completion time~\cite{funk2015using}. Furthermore, fewer cognitive resources were utilized~\cite{7140647}. Motivation can be increased or maintained by incorporating gamification into the working environment of persons with cognitive impairments~\cite{korn2015design, korn2015towards}. They found that visual and auditory notifications outperform tactile error alerts concerning usability and understanding.

\subsection{Assistive Technologies in Smart Kitchens}
Integrating assistive technologies in kitchen environments has been the focus of various researchers before. Scheible et al.~\cite{Scheible:2016:SME:2991561.2998471} presented how smart kitchens can be designed to enhance the overall cooking experience. With the availability of microsensors, a huge number of contextual data in kitchen environments can be collected. Thereby, Hashimoto et al.~\cite{Hashimoto2008} analyzed behavioral data to recognize the current cooking action and prepared food materials. Blasco et al.~\cite{Blasco2013} developed and evaluated a smart kitchen system for older adults within real-world scenarios. They found that their participants benefited from cognitive assistance during the cooking sessions. Besides providing support during meal preparation, smart kitchens can provide contextual nutrition-aware information. Based on this, suited recipes can be mediated to ensure the intake of important nutritions~\cite{Chi2007, Chi2008}. Cooking has become a social activity in which recipes are shared and prepared together. Therefore, Schneider et al.~\cite{Schneider2007} present the "Semantic Cookbook" which can capture, share, and exploit cooking experiences semantically. This way recipes can be recorded shared among other smart kitchens and relatives. Since most recipes are passed down to the next generations, Terrenghi et al.~\cite{Terrenghi2007} developed the "Living Cookbook". Cooking experiences were recorded to practice meal preparation techniques and to teach people who are unfamiliar with cooking. Previous research has incorporated language learning tasks into smart kitchens~\cite{Hooper:2012:FKT:2370216.2370246}. Design recommendations were presented for integrating task-based learning within the meal preparation process. Miyawaki et al.~\cite{Miyawaki:2009:CSS:1630995.1631004} developed cooking support for people with higher brain dysfunction. The use of their system during rehabilitation training has shown positive effects. Prototyping new smart kitchen solutions require a robust testing environment. Olivier et al.~\cite{Olivier:2009:AKD:1579114.1579161} presented a prototype setting in which new smart kitchen solutions can be evaluated. Several augmented reality-based applications in smart kitchen environments were evaluated by Bonanni et al.~\cite{Bonanni2005}. This included the assessment of usability, attention, cognitive workload.

Previous work has invested effort into the construction, evaluation, and integration of assistive technologies in home environments. However, the influence of in-situ cooking assistance for persons with cognitive impairments has not been considered yet. To close this gap, we conducted a field study in a sheltered living facility where people with cognitive impairments cooked either with (a) traditional \textit{caretaker assistance} or (b) \textit{in-situ assistance}. We report on differences in cooking performance and subjective feedback between both cooking modalities.

\section{User Study}
Related research has informed how accessibility can be ubiquitously integrated into daily life situations to compensate cognitive impairments. At the same time, assistive technologies have proliferated into home environments that are available for the wider public. However, how assistive technologies impact the behavior of persons with cognitive impairments in sheltered living facilities during cooking has not been explored yet. In the following, we present a study that compares \textit{in-situ assistance} and \textit{caretaker assistance} in terms of cooking performance as well as subjectively perceived feedback. This chapter and the subsequent study aims to answer the following two research questions:
\begin{itemize}
\item \textbf{RQ 1:} How does \textit{in-situ assistance} changes the cooking performance, measured by the overall meal preparation time, compared to \textit{caretaker assistance}?
\item \textbf{RQ 2:} How is \textit{in-situ assistance} subjectively perceived compared to \textit{caretaker assistance}?
\end{itemize}

\begin{figure}
\centering
\includegraphics[width=0.5\columnwidth]{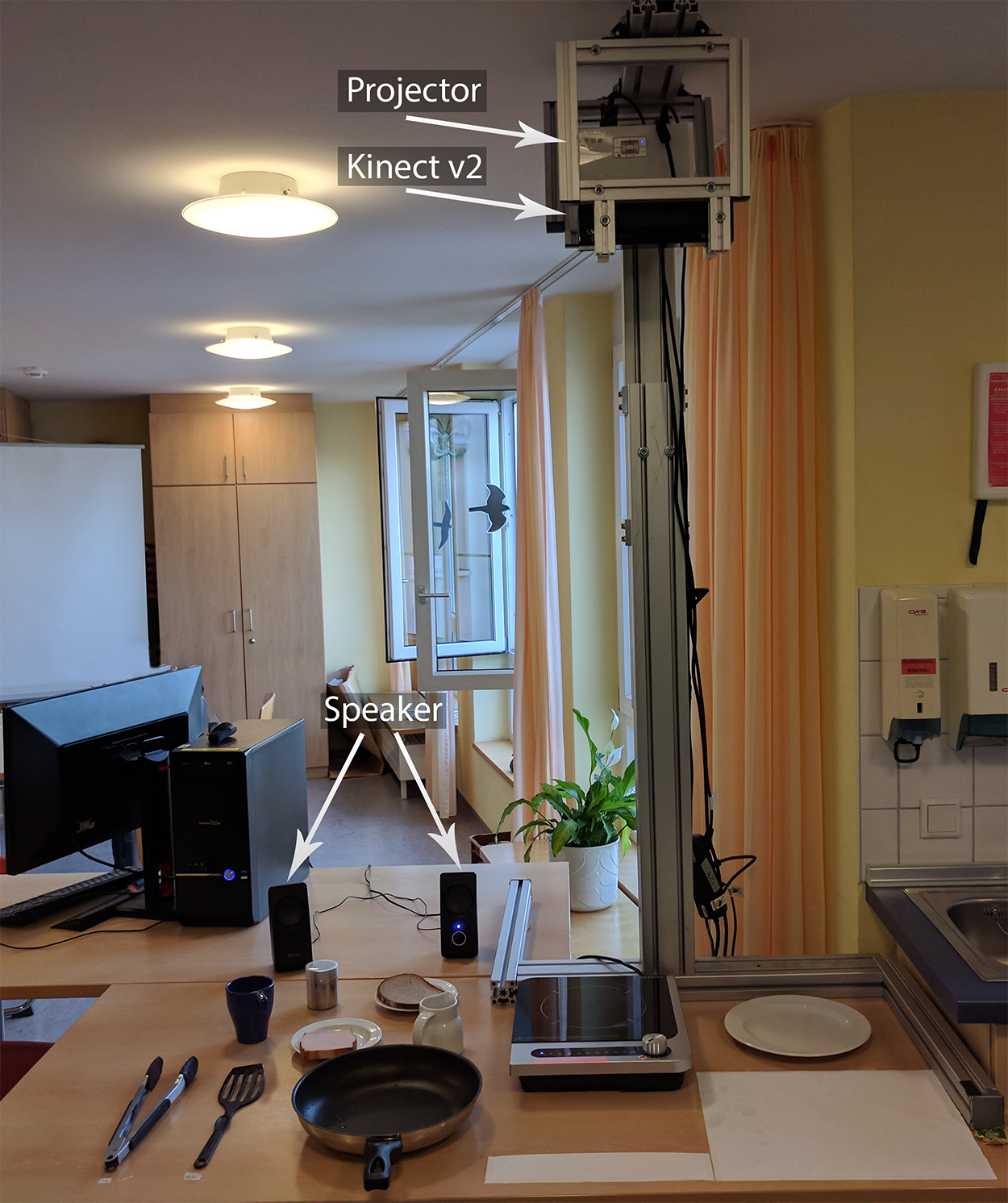}
\caption[Setup of the assistive kitchen system]{Study setup with pre-portioned ingredients. Auditory feedback is provided via external speakers while visual feedback is delivered by projections.}
\label{fig:setup}
\end{figure}

\begin{figure*}[htb]
\centering
\subfloat[][]{
\includegraphics[height=0.20\columnwidth]{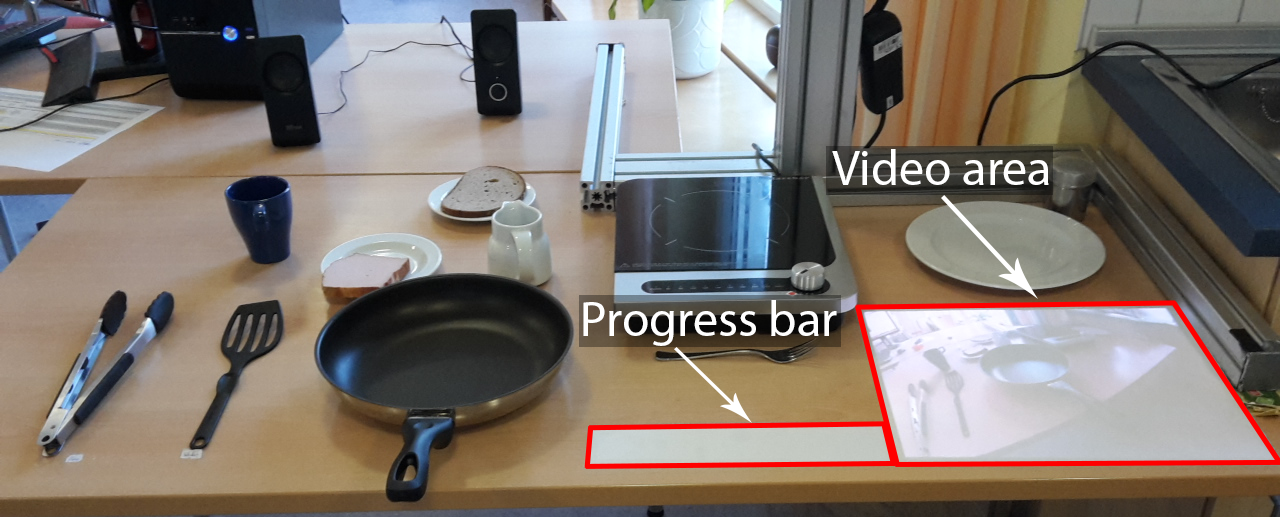}
\label{fig:projection}}
\subfloat[][]{
\includegraphics[height=0.20\columnwidth]{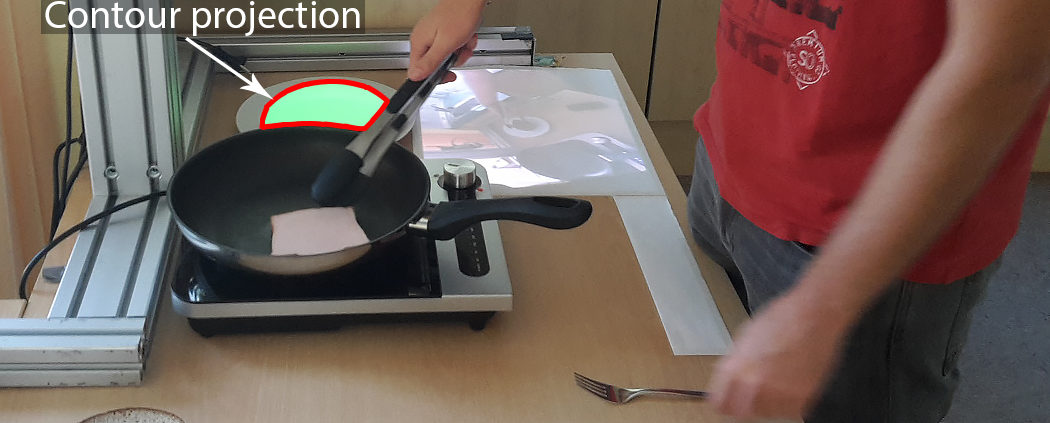}
\label{fig:contour}}
\caption[Output modalities of the assistive system]{Visual output modalities of the cooking instruction system. \textbf{(a):} In-situ projections of video instructions and progress bars were displayed inside the cooking area. The projections were placed on white sheets to enhance the visibility. \textbf{(b):} In-situ projection of video instructions and contours which provide visual cues for grasping or placing objects and food.}
\label{fig:modes}
\end{figure*}

\subsection{Evaluation}
We extended the system of Funk et al.~\cite{funk2015using, funk2017working} by functional modules to provide additional visual and auditory feedback. The main system that tracks cooking steps and provides feedback to the cooking area consists of multiple aluminum profiles assembled. The profiles enable to mount different hardware on top of the cooking area by placing the construction on the cooking area. The system uses a top-mounted Kinect v2\footnote{\url{https://developer.microsoft.com/en-us/windows/kinect} - last access \urldate} which is mounted 1.35 meters above the cooking area to detect finished cooking steps. A projector that is mounted 1.50 meters above the working area presents visual cooking instructions~\cite{funk2015comparing}. Dedicated audio speakers, placed next to the cooking area, provided auditory feedback by playing verbally recorded cooking instructions (see Figure~\ref{fig:setup}). In the following, we describe the cooking feedback modalities which were facilitated in the study.
\subsubsection{Auditory Instructions}
Auditory instructions are the current communication standard for single cooking steps between tenants and caretakers in sheltered living facilities. Most tenants are not able to read and rely on the verbal exchange of instructions. To provide a similar experience through \textit{in-situ assistance}, we recorded single cooking steps with the voice of a caretaker. External audio speakers, set to a suitable volume, were integrated into the cooking environment (see Figure~\ref{fig:setup}). Cooking instructions were played back with each start of a cooking step. We used a Trust 2.0 speaker setup\footnote{\url{www.trust.com/en/product/17595-remo-2-0-speaker-set} - last access \urldate}.
\subsubsection{Video Instructions}
Video instructions were provided by a projector mounted above the cooking environment. The videos were short looped clips, where caretakers showed how to accomplish the next cooking step. The video instructions were projected on a dedicated field right to the cooking area (see Figure~\ref{fig:projection}). Upon step completion and validation through the Kinect v2, the next instruction video was played back. An Acer K330 was used as projector\footnote{\url{www.projectorcentral.com/Acer-K330.htm} - last access \urldate}.
\subsubsection{Contour Projections}
Contour projections were displayed on or around objects which are of interest during the current cooking step. This is accomplished by projecting a green light on different cooking utilities (see Figure~\ref{fig:contour}). Furthermore, a progress bar below the hotplate is projected during waiting times (\textit{e.g.}, when frying ingredients). Figure~\ref{fig:projection} shows the position of the progress bar.

\subsection{Methodology and Measures}
We employ a within-subject design study including the independent variable \textit{assistance modality} consisting of cooking instruction provided by \textit{in-situ assistance} and instructions provided by \textit{caretaker assistance}. The prepared meal was meatloaf with fried eggs and bread. We have chosen this recipe because it was unknown among the participants. Thus, the overall cooking process was unknown to the participants. The runtime of the experiment was ten days (\textit{i.e.}, two weeks without weekends). Two cooking sessions with two different participants were conducted on each day, where one participant cooked with \textit{caretaker assistance} and another participant with \textit{in-situ assistance}. After the first week, the same participants from the last week were invited on the same weekday again with the contrary assistance modality, \textit{i.e.}, participants who used to cook with \textit{caretaker assistance} were either assisted by \textit{in-situ assistance} and participants who were assisted by \textit{in-situ assistance} used \textit{caretaker assistance} instead. By introducing the break of seven days, we reduce the probability that participants remember the cooking procedure when they cook the same meal again using the contrary instruction modality. In other words, five participants cooked with \textit{in-situ assistance} and another five with \textit{caretaker assistance} during the first week. In the second week, participants that started with \textit{in-situ assistance} were invited to cook with \textit{caretaker assistance}. In contrast, participants who cooked with \textit{caretaker assistance} in the first week used \textit{in-situ assistance} in the second week. Note, that all participants in the second week were already aware of the recipe and the cooking procedure from the first week. The required cooking utilities were placed on predefined positions before the experiment. We ensured a similar positioning of cooking utensils for each session.

We measure the meal preparation time for both assistance modalities to investigate temporal differences between both assistance modalities. We subtracted constant waiting times, such as cooking steps that require frying, from the overall task completion time (see Table~\ref{tab:cooking_steps}). By this, individual waiting times between \textit{caretaker assistance} and \textit{in-situ assistance} are removed from the analysis. Overall, four cooking steps facilitated waiting times (\textit{i.e.}, heating oil, frying meatloaf from both sides, and cooking the egg) during one trial. Table~\ref{tab:cooking_steps} shows each cooking step with their accompanied instruction system for \textit{in-situ assistance}. Finally, we conducted semi-structured interviews with the participants about personal preferences in cooking assistance after the experiment.

\subsection{Procedure}
Before the study, we asked for written consent from either the participants or their legal guardians. We conducted the study in a kitchen within a sheltered living facility. We carefully explained the intention of the study to the participants to avoid misunderstandings. Since the system was unfamiliar to the tenants, we made them familiar with the visual and auditory instructions. Furthermore, we showed where the feedback cues were generated to avoid confusion for participants during the study. After being familiar with the system, participants started with the cooking instruction modality according to the balanced Latin square. Before the experiment, ingredients were pre-portioned since tenants affected by motoric disorders could not handle the doses by themselves. The experiment started after ensuring all safety arrangements.

Visual instructions were presented alongside auditory feedback for each step. A voice recorded by a caretaker gave instructions on how to perform the current cooking step. Additionally, a video was projected to the right of the cooking area. These videos had a length between two and four seconds that demonstrated how the current cooking step has to be performed. Furthermore, objects of interest were highlighted by contour projections. A Kinect v2 detected whether cooking steps were performed successfully. If a cooking step was conducted correctly, the visual, auditory, and contour instructions proceeded to the next step. The whole cooking procedure comprised 25 cooking steps including four waiting steps to fry the ingredients (see Table~\ref{tab:cooking_steps}). Since waiting times may occur during frying steps, short cartoons were projected into the video area. This ensured engagement during waiting times since the participants' concentration may be affected and reducing the likeliness of returning to the cooking task \cite{troyer2007memory}. At the same time, a declining progress bar indicates the remaining waiting time (see Figure~\ref{fig:projection}). Notes about the interaction between participants and systems were recorded during the experiment. A caretaker was always present during the cooking sessions to ensure the safety of participants. Participants were lauded at the end of the cooking session.

Afterward, participants participated in a semi-structured interview. The participants were asked about their experience when cooking with \textit{caretaker assistance} or \textit{in-situ assistance}. Furthermore, we asked about suggestions for improvements. The answers were recorded and noted by the experimenters. Overall, the study took approximately 30 minutes including meal preparations and post hoc interviews.

\subsection{Cooking Equipment}
A hotplate and a pan are used to heat the ingredients. We use an induction burner as a hotplate for safety reasons. A spatula and a reacher were used to put and turn ingredients in the pan. Finally, fried ingredients were placed on a plate right to the cooking area (see Figure~\ref{fig:modes}). The whole cooking procedure uses pre-portioned ingredients including a piece of meatloaf, salt, eggs, oil, and a piece of bread.

\begin{table}[t!]
\centering
\begin{tabular}{llccc}
\toprule
Step No. & Cooking Step & Audio & Video & Contour\\
\midrule
1 & Take pan & X &X&\\
2 & Pan on hotplate & X &X&\\
3 & Start hotplate & X &X&X\\
4 & Set heat & X &X&X\\
5 & Take oil & X &X&\\
6 & Put oil into pan & X &X&\\
\rowcolor{lightgray}
7 & Wait until pan is hot & X &&\\
8 & Take reacher & X &X&\\
9 & Take meatloaf & X &X&\\
10 & Put meatloaf into pan & X &X&\\
\rowcolor{lightgray}
11 & Wait until meatloaf is fried & X &&\\
12 & Turn meatloaf & X &X&\\
\rowcolor{lightgray}
13 & Wait until meatloaf is fried & X &&\\
14 & Put meatloaf on plate & X &X&X\\
15 & Put reacher on table & X &X&\\
16 & Take cup with egg & X &X&\\
17 & Put egg into pan & X &X&\\
\rowcolor{lightgray}
18 & Wait until egg is fried & X &&\\
19 & Take spatula & X &X&\\
20 & Put egg on plate & X &X&X\\
21 & Turn hotplate off & X &X&\\
22 & Take bread & X &X&\\
23 & Put bread on plate & X &X&X\\
24 & Put salt on meal & X &X&\\
25 & Completion notice & X &X&\\
\bottomrule
\end{tabular}
\caption[Set of cooking instructions and their respective feedback modality]{Single cooking steps with their assigned instruction systems. The steps 7, 11, 13, and 18 (gray rows) consist of waiting times.}
\label{tab:cooking_steps}
\end{table}

\section{Results}
We analyzed the measures as well as qualitative data collected throughout the study. The results are presented in the following.
\subsection{Participants}
We recruited ten tenants of a sheltered living facility (6 female, 4 male) aged between 24 and 56 years (M $=40.9$, SD $=9.93$). The participants had cognitive impairments that limit their ability to process and understand information. None of the participants were affected by dementia. Five caretakers (3 female, 2 male) aged between 19 and 34 years (M $=26.8$, SD $=7.05$) were involved for the \textit{caretaker assistance} condition. Their individual work experience ranged between several months to twelve years. We constantly assigned each caretaker to two participants. We ensured that the same caretaker was cooking with similar participants during the \textit{caretaker assistance} and observed them during \textit{in-situ assistance}. Participants were monetary compensated through funding from a joint project.

\subsection{Cooking Performance}
We compare the cooking performance between the conditions \textit{caretaker assistance} and \textit{in-situ assistance} regarding the time participants required to prepare a meal. We process the data by subtracting the fixed waiting times from the overall cooking times of \textit{in-situ assistance} and \textit{caretaker assistance} (see Table~\ref{tab:cooking_steps}). \textit{In-situ assistance} comprised a mean waiting time of seven minutes and eight seconds while \textit{caretaker assistance} had a mean waiting time of eight minutes.

\begin{figure*}[htb]
\centering
\subfloat[][]{
\includegraphics[height=0.33\columnwidth]{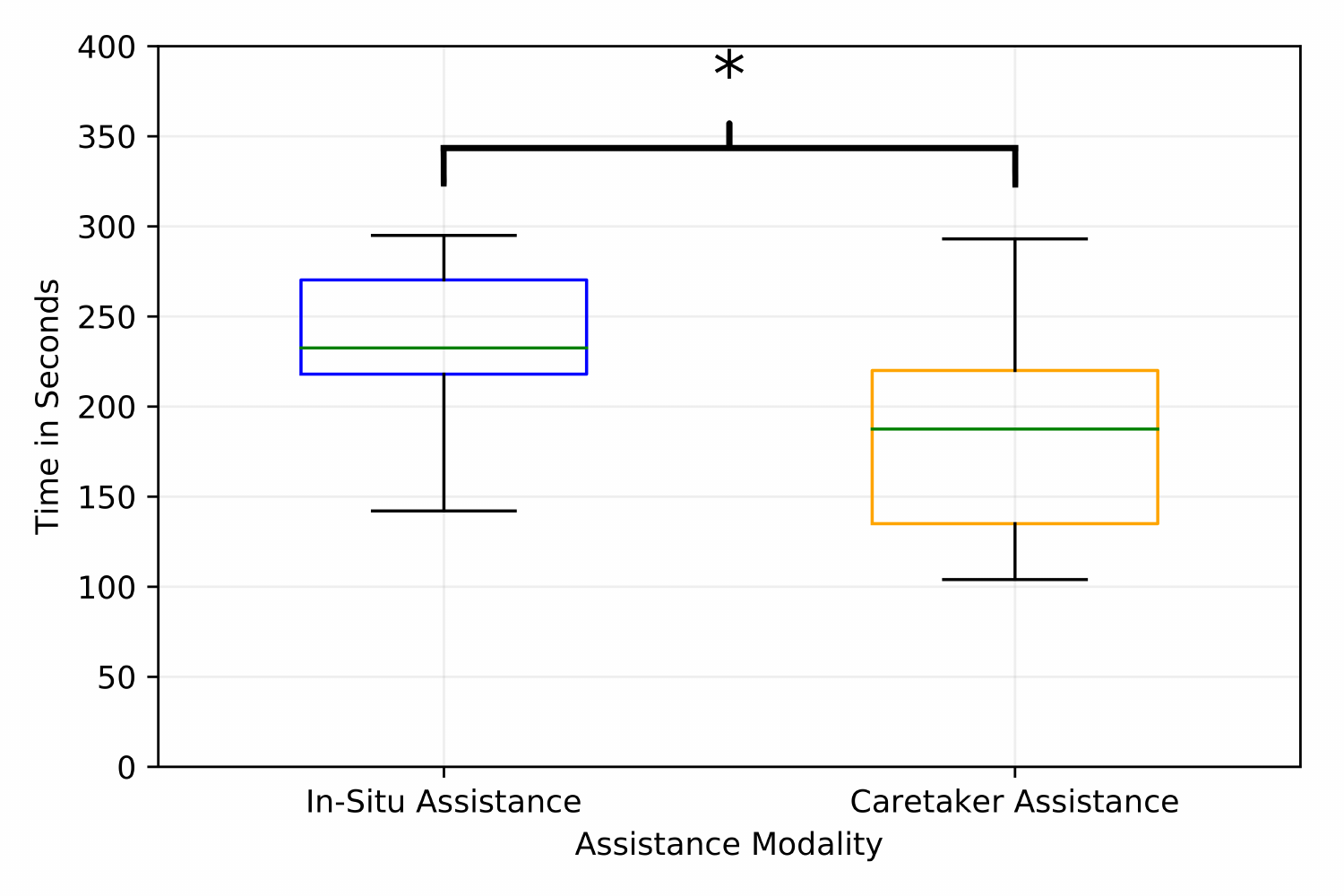}
\label{fig:mean_assistance_system_ch8}}
\subfloat[][]{
\includegraphics[height=0.33\columnwidth]{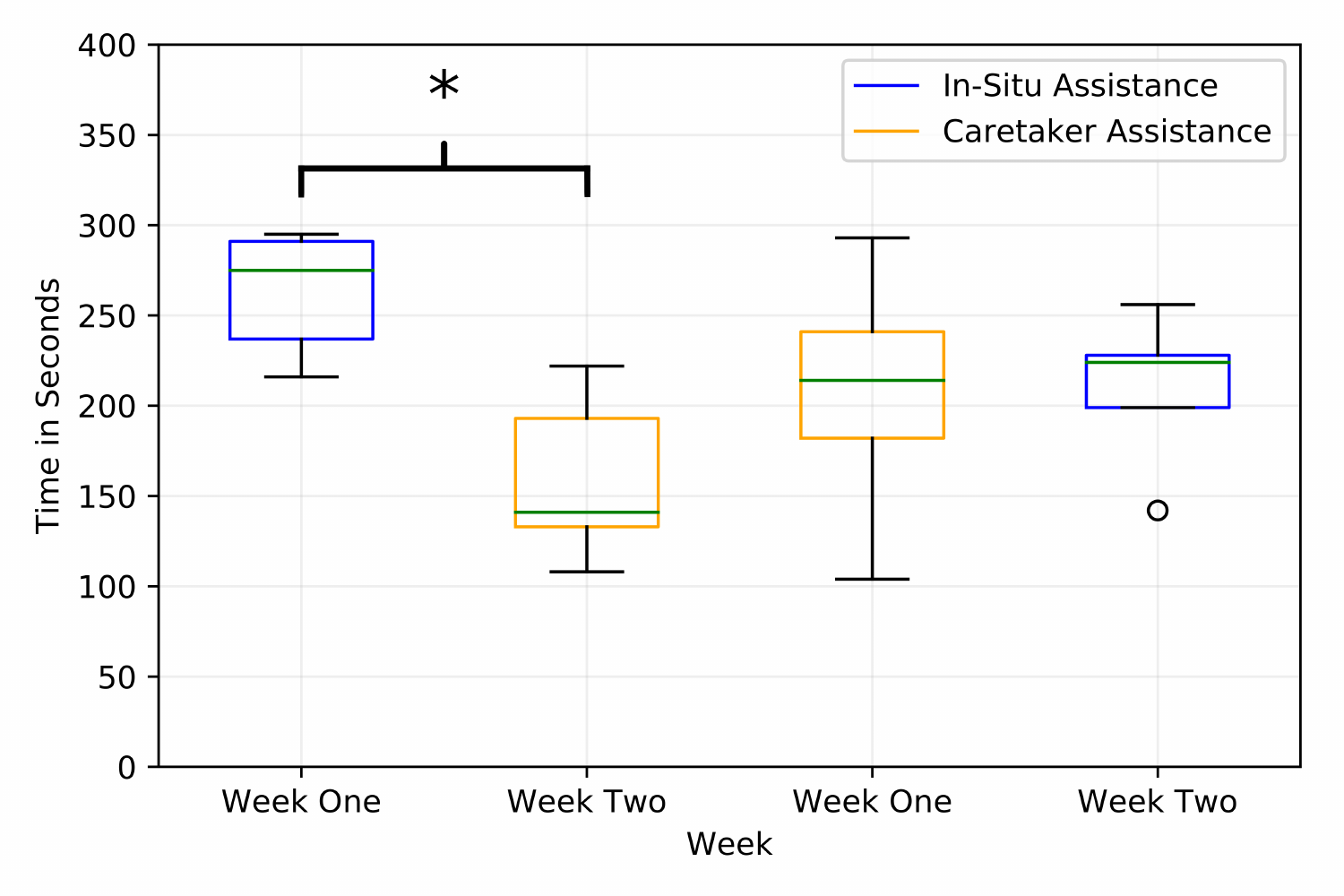}
\label{fig:mean_assistance_system_week_ch8}}
\caption[Mean cooking times between \textit{in-situ assistance} and \textit{caretaker assistance}]{Results of the study. \textbf{(a):} Mean cooking times between \textit{in-situ assistance} and \textit{caretaker assistance} for both weeks. Brackets indicate significant differences. \textbf{(b):} Mean cooking times for \textit{in-situ assistance} and \textit{caretaker assistance} for each week. Participants that started with \textit{in-situ assistance} had higher cooking times in the second week using \textit{caretaker assistance}. Brackets indicate significant differences.}
\label{fig:results_cooking_study}
\end{figure*}

First, we statistically analyze the cooking times between \textit{caretaker assistance} and \textit{in-situ assistance} which were collected during both weeks. A Shapiro-Wilk test did not reveal a deviation from normality, $p > .05$. Thus, we submitted the data to a paired samples t-test and found a significant effect between both assistance modalities, $t(9)=3.628$, $p=.006$, $d=1.147$. Investigating the mean cooking times between both conditions shows longer cooking times for \textit{in-situ assistance} (M $=236.3$, SD $=46.24$) compared to \textit{caretaker assistance} (M $=228.9$, SD $=77.03$). Figure~\ref{fig:mean_assistance_system_ch8} shows the aggregated mean times of the first and second week between both assistance modalities.

Since either five participants cooked with \textit{in-situ assistance} or \textit{caretaker assistance} per week, we separately investigate the cooking times between both conditions for the first and second week. Note that the participants in the second week were aware of the cooking procedure since they already prepared the meal in the first week with the contrary assistance modality. A one-way ANOVA revealed a significant main effect between the cooking times for each week and instruction modality, $F(3, 16)=3.504$, $p=.04$. A Tukey post hoc test revealed a significant effect between the use of \textit{in-situ assistance} in the first week and \textit{caretaker assistance} in the second week, $p=.024$, $d=2.511$. However, we could not find a significant effect when using \textit{caretaker assistance} in week one and \textit{in-situ assistance} in week two. Furthermore, there is no significant effect when using \textit{in-situ assistance} or \textit{caretaker assistance} in the same week. Participants required more time when using \textit{in-situ assistance} in the first week ($M=262.8, SD=34.77$) compared to the second week using \textit{caretaker assistance} ($M=159.4, SD=46.7$). \textit{Caretaker assistance} required less time in the first week (M $=206.8$, SD $=70.42$) compared to \textit{in-situ assistance} in the second week (M $=209.8$, SD $=42.96$). Figure~\ref{fig:mean_assistance_system_week_ch8} shows the separated cooking times of \textit{in-situ assistance} and \textit{caretaker assistance} per week.

\subsection{Subjective Feedback}
We collected qualitative comments from the participants about the experience with traditional \textit{caretaker assistance} and \textit{in-situ assistance}. In the presence of the caretakers, we asked participants if the cooking procedure was pleasant for them. Participants who cooked with \textit{caretaker assistance} and \textit{in-situ assistance} enjoyed cooking. However, one participant stated that ``[\dots] \textit{the videos were not helpful}'' (P4) while others found that the ``[\dots] \textit{videos were funny}'' (P9) and that ``[\dots] \textit{the contour projections were engaging}'' (P7).

Furthermore, we asked participants if they found the visual and auditory feedback helpful. All participants except one (P4) found the projected videos helpful. Four participants stated that they noticed them when they were stuck in the cooking process (P2, P4, P5, P10). The sound was perceived by all participants positively except one participant who stated that ``\textit{The voice was not helpful, the videos helped me more.} '' (P9).

Finally, we asked if the participants could imagine integrating such a system into their daily communal cooking procedure. All participants except one (P8) agreed with an integration of the system into their daily life. However, one participant mentioned the importance of social interaction during cooking activities (P3). One of the participants stated, that the system ``[\dots] \textit{provided more safety when cooking alone}'' (P2). One participant preferred to cook with a caretaker instead of the system (P8).

\subsection{Qualitative Observation}
We observed the interaction between the tenant and the assistive system in the kitchen. We noted, that all participants were able to perform most of the cooking steps by listening to the voice. Video instructions, which were displayed in every cooking step, were perceived when the voice instruction was not understood. The tenants are used to voice-driven instructions since cooking instructions are communicated verbally during the cooking process. A similar observation was found in the perception of the progress bar, which was noticed after they were hinted by the experimenter or caretaker. Four participants required help to understand the system before conducting the first cooking steps. Afterward, the cooking process was guided by the system without interruptions.

\section{Discussion}
We evaluated an assistive cooking system in a sheltered living facility. In the following, we discuss the results of our study.

\subsection{Cooking Performance and Learning Effect}
Overall, we measured longer meal preparation times using \textit{in-situ assistance} compared to \textit{caretaker assistance}. However, we find that the cooking times with \textit{caretaker assistance} in the second week improves significantly when \textit{in-situ assistance} was employed in the first week. Since cooking instructions are usually conveyed verbally, the visual feedback generated by in-situ assistance might complement the cognitive processing of instructions. Therefore, a combination of visual and auditory cooking instructions could result initially in a longer cooking procedure, but can also contribute to an improvement in understanding the cooking process. Furthermore, a learning effect can eventuate as a result of using \textit{in-situ instructions}.
We could not observe this effect when starting with \textit{caretaker assistance} in the first week and continuing with \textit{in-situ assistance} in the second week. This assures the hypothesis that visual feedback can be used as a complementary feedback modality to help tenants to remember or understand complex cooking instructions.

\subsection{Feedback Modalities}
From our observations, we find that most participants could handle simple cooking instructions using auditory feedback. Video and contour projections were considered during waiting times or more complex steps, such as turning ingredients in the pan. These results conform with the results of the Chapter~\ref{ch:smart_kitchen_requirements}, where auditory instructions are preferred to convey and process information. However, most participants considered video and contour instructions useful to understand details although not always needed. We believe that video and contour instructions supplement auditory feedback for rather complex cooking instructions.

\subsection{Subjective Perception}
We asked caretakers and the tenants about their perception of cooking with assistive technologies. The responses were positive by both caretakers and tenants. Participants stated that visual and auditory feedback complement each other in terms of understanding the current cooking step. Since tenants are usually not able to cook without \textit{caretaker assistance}, most of the participants enjoyed their autonomy during the cooking process using \textit{in-situ assistance}. This indicates that suitable feedback content is a critical factor to consider when conducting further research on assistive technologies in private and sheltered living facilities.

\section{Study Conclusion}
We presented a study which compares \textit{caretaker assistance} and \textit{in-situ assistance} during several cooking sessions for tenants with cognitive impairments. We found that \textit{in-situ assistance} requires more meal preparation time in the first week compared to \textit{caretaker assistance}. However, the cooking performance of \textit{caretaker assistance} improves fundamentally when the cooking procedure was conducted with \textit{in-situ assistance} beforehand. Through semi-structured interviews, we find that the visual and auditory feedback provided by \textit{in-situ assistance} is well accepted among tenants. We believe that such assistive technologies can be used in the future to compensate for cognitive impairments. Learning how to accomplish daily life tasks by assistive technologies goes a step towards independent living for tenants with cognitive impairments.

We are aware that our research is prone to several limitations. We did not consider the individual cognitive impairment per participant. Before the study, we ensured that the level of cognitive impairment was similar for each participant. However, assessments to estimate the individual level of cognitive impairment are conducted irregularly. As mental abilities develop further, the assessment may become irrelevant regarding individual impairment. Furthermore, we did not consider the individual level and type of cognitive impairment in our study design. Thus, the results may not be generalizable to people that are affected by other cognitive impairments, such as dementia or motoric disorders. For safety reasons, a caretaker was always present during the study regardless of the cooking assistance modality. Thus, we did not investigate how individual behavior may change when a caretaker is not present. Therefore, it is not clear how people with cognitive impairment interact with assistive technologies in their home environments when caretakers are not nearby. To foster research in this area, we host a reference implementation of the presented workload-aware assistive system on Github\footnote{\url{www.github.com/hcum/kobelu} - last access \urldate}.

\section{Chapter Conclusion}
This chapter presented a case study in which visual in-situ projections were used to guide persons with cognitive impairments through a cooking process. The design of the assistive system followed a user-centric development approach: after recognizing opportune moments for in-situ assistance through qualitative inquiries, we evaluated suitable feedback modalities that were considered during the implementation of an in-situ assistance system. To answer \textbf{RQ5}, we conducted a long-term study over two weeks in a sheltered living facility. The study revealed that no improvements in task completion time and error rate were immediately visible. A potential reason could be the unfamiliarity with the system itself, whereas tenants are used to cooking with their respective caretakers. In contrast, in-situ support was positively perceived and well rated by the participants. Since all participants completed the cooking sessions, we believe that user-tailored assistive technologies will foster autonomy for people with cognitive impairments.


\chapter{Reflection}
\label{ch:visualization_reflection}
\epigraph{\textit{In order to improve the mind, we ought less to learn than to contemplate.}}{René Descartes}

\ac{EEG} has shown to be a feasible tool for workload quantification. In chapter~\ref{ch:workloadbyeeg}, we have conducted a study to develop a pipeline that uses \ac{EEG} as a physiological real-time measure that evaluates the workload placed on users by the respective interface. Such a pipeline enables the development of neurofeedback loops that utilize cognitive workload for feedback mechanisms. However, the brain is a sensitive organ that is highly unexplored. The raw \ac{EEG} signals are difficult for novice users to interpret and process.

In contrast, \acp{BCI} are a category of physiological sensors that are currently coping with these challenges to make their way into the consumer market. Traditionally, they have been used in the medical field to help patients with severe disabilities to gain more control over their lives. \ac{EEG} signals from the brain, gathered through electrodes placed on or inside the scalp, can be interpreted to allow patients to move a wheelchair, a cursor \cite{tanaka2005electroencephalogram, wolpaw1991eeg} or provide feedback to the user in what is called a neurofeedback loop. Neurofeedback training involves interpreting brain signals in real-time and providing the user with feedback about their state through different stimuli, for example, using \ac{VR} environments. \ac{VR} in combination with \acp{BCI} have been researched and used in several applications ranging from treating brain damage \cite{rose2005virtual}, substance abuse \cite{lee2009quantitative} to mindfulness training \cite{anatole2008brain}. Hence, such \ac{BCI} applications abstract the complexity of \ac{EEG} and enable users to utilize brain signals for implicit and explicit interaction purposes.
Using new, light-weight, and less complex consumer BCI devices coupled with a \ac{VR} system allows us to create adaptive virtual environments that can change according to the measured brain signals to increase or decrease activation in certain parts of the brain. This can especially help people with neuropsychiatric diseases (\textit{e.g.} depression) and their doctors to investigate brain activation in real-time during VR-based neurofeedback sessions. The desire to investigate the impact of VR scenes on the brain is high but becomes complicated because of the linearly increasing output of an EEG over time. Whereas the measured brain activity at certain electrodes is what drives the neurofeedback system, the origin of the electrical sources of activation inside the brain is also required to analyze which parts of the brain are affected by the VR scene. The low-resolution electromagnetic tomography algorithm (LORETA) solves this problem for a sufficient number of electrodes in a three-dimensional space \cite{pascual1994low}.

In this chapter, we present a simple and understandable real-time visualization of the mentioned LORETA algorithm. A simulation is executed to show the correctness of the implementation. A study is conducted to show that changes regarding brain activation can be observed in real-time. This is complemented by an evaluation using a \ac{VR}-based neurofeedback loop that reveals the affected brain areas. This chapter showcases the potential that \ac{EEG} holds for neurofeedback platforms for future \ac{UI} adaption systems. Therefore, we investigate the following research question:

\begin{itemize}
\renewcommand{\labelitemi}{$\rightarrow$}
\item \textbf{RQ6}: How can complex physiological signals be visualized to be utilized by non-expert users?
\end{itemize}

\begin{center}
\begin{GrayBox}
\centering
\parbox{0.98\textwidth}{
\emph{This section is based on the following publication:}
\vspace{3mm}
\publicationsbegin
\item \bibentry{kosch2016brain}
\publicationsend
}
\end{GrayBox}
\end{center}

\section{Related Work}
Visualizations of brain data to interpret neuronal activity is an important tool for researchers. In this context, research has been conducted that can be broadly summarized into two categories: (1) the impact of virtual reality on the brain in medical treatments and (2) currently existing brain data visualizations.

Baumgartner et al.~\cite{baumgartner2006neural} have shown that neurophysiological changes can be found when using VR. Increased brain activity at the parietal and frontal lobe could be found. The EEG analysis of the displayed VR scene was carried out using LORETA. However, the output cannot be inspected in real-time and the implementation is not interactive.
Grealy et al.~\cite{grealy1999improving} performed a 4-week intervention program for treating patients with traumatic brain injuries using VR. Significant cognitive increases could be noticed in terms of speech, visual learning and reaction times. Improvements were quantitatively measurable, but the affected brain parts by this training could not be observed at all. Hoffman~\cite{hoffman2004virtual} used functional magnetic resonance imaging (fMRI) to investigate perceived pain when treating burn damages in combination with and without VR. The fMRI has shown that distraction delivered by VR helps to reduce the amount of brain activity in pain-related brain areas. The brain scans presented in the study still lack real-time capabilities to investigate the augmentation and impact of brain activation throughout a VR session. A three-dimensional visualization is presented by the project Glass Brain from the company Neuroscape\footnote{\url{www.neuroscape.ucsf.edu} - last access \urldate}. It enables the visualization of source localization in real-time with interactive features. However, the inner of the visualized brain is hard to investigate since it is filled with a fixed solid color.

Overall, the presented brain visualization solutions lack interactive features, real-time capabilities or suffer performance issues. In contrast, our work shows a simple visual representation of the activities in the brain which allows a human observer to understand the current brain activation in real-time. Currently, we investigate this in two contexts: (1) Providing information to improve medical treatment of brain-related issues, (2) assessment of reactions to visual stimuli in the context of human-computer interaction. In our work, we propose a real-time implementation of LORETA including a three-dimensional representation enriched with interactive features to allow more precise investigations of brain parts. As a proof of concept, we performed a pilot study with an adaptive VR scene which changes according to the measured brain activities to achieve a neurofeedback session. Changes in brain activation are investigated during VR sessions in real-time as well as in post hoc analysis.



\section{System}
The implementation of our algorithm uses neuromore Studio\footnote{\url{www.neuromore.com} - last access \urldate} as base platform. Together it builds a system allowing us to retrieve data measured by a BCI device and processing it with our implementation in real-time. Furthermore, we can create adaptive VR-based neurofeedback sessions based on the measured EEG data\footnote{The visualizations are available in this publication: \bibentry{kosch2016brain}}.

\subsection{Localizing Electrical Source}
We use a multithreaded implementation of the standardized LORETA (sLORETA) algorithm for electrical source localization \cite{pascual2002standardized}. sLORETA allows precise source localization in a three-dimensional space with different settings regarding noise minimization, voxel density and the usage of a head model. The input of the algorithm is the current voltages of extracranial scalp measurements received by the attached electrodes. sLORETA then calculates a density distribution throughout a head and associates each voxel with a magnitude that describes the current density power at the voxel. Voxels are color-coded depending on the calculated magnitude ranging from blue (weak) over green (medium) to red (strong).

\subsection{Visualizing Results}
After retrieving electrode measurements, the algorithm begins processing the data based on the chosen head model and the number of voxels. We have decided to use a spherical head model because it is easier to reproduce and compare results from other research projects. The number of voxels can be set during program runtime, while a higher number of voxels increases the precision of the result at the cost of performance. Interactive features like zoom and rotation are supported to ease the analysis of brain source activation. The computation of the visualization algorithm is GPU accelerated. In contrast to current systems, we can visualize the activated areas in real-time for the observer (doctor, researcher) to directly see the reaction inside the brain.

\subsection{Evaluation}
The implementation is evaluated using the BESA simulator tool\footnote{\url{www.besa.de/downloads/besa-simulator} - last access \urldate} to verify the correctness of the results. The BESA simulator provides the functionality to place an electrical source inside a head. These can be exported as a CSV file, which allows an import into the sLORETA implementation. The evaluation showed that the algorithm is working correctly and can be used in future user studies.


\subsection{Input Stimulus}
Our system uses VR as an input stimulus which enables us to create controllable simulations of real-world situations where the reaction of body and mind can be analyzed. The simulation can be changed based on the measured electrode values to match a VR scene to the cognitive load of a user. Depending on the result of the EEG analysis, the difficulty of tasks or environments can be modified for brain training exercises, for example, to decrease frustration or boredom. Frequent training using VR with suitable parameters can be used to increase overall brain activity. Treatment of neuropsychiatric diseases often aim to increase brain activity in the frontal lobe, since decreased brain activity in these areas often is seen in patients of depression, schizophrenia, substance abuse or obsessive-compulsive disorders \cite{tekin2002frontal}.

\section{Impact of VR}
Virtual reality enables controllable simulations of real-world situations where the reaction of the body and mind can be analyzed. The simulation can be changed to suit the VR scene to the cognitive load of a user. Depending on the result of an EEG the difficulty of tasks can be modified for brain training exercises to decrease frustration or boredom. Therefore it is important to understand what kind of neurophysiological effect different VR tasks have.

VR is often related to video games that provide an immersive experience. Video games contain goals that have to be accomplished by a user. The frontal lobe, which is responsible for motivation-related thinking, is activated during such tasks \cite{anatole2008brain}. Decreased brain activity in this brain area often involves depression, schizophrenia, substance abuse or obsessive-compulsive disorders because of dopamine lack or depletion \cite{tekin2002frontal}. Frequent training using VR with suitable parameters can be used to increase brain activity in the frontal lobe, which may be used to treat neuropsychiatric diseases. Therefore VR is a popular method used in neurofeedback sessions for frontal lobe training.

\section{Study}
A pilot study with eight male participants was conducted to find visible changes of electrical source activation within the brain during a neurofeedback session that used VR to change the stimuli. The average age was $27.12$ years (SD = $5.19$). The participants were healthy and have not participated in a bio- or neurofeedback study before. The study took approximately 40 minutes per participant.

\subsection{Apparatus}
We used an OpenBCI\footnote{\url{www.openbci.com} - last access \urldate} board with 16-channel support as EEG device for the study. OpenBCI and its electrodes are mounted on a 3D printed retainer which can be worn on the head. An Oculus Rift\footnote{\url{www.oculus.com} - last access \urldate} was used to display VR content. The visualization uses 8000 voxels to render the results.

\subsection{Method}
The study is divided into a baseline task, a VR session and a repetition of the baseline task to see visible differences before and after the VR session. The OpenBCI is continuously recording EEG data to enable post hoc analysis. Changes in brain activation are observed in real-time.

The baseline task is conducted to visualize initial source activation. The participant is asked to keep the eyes open for three minutes, followed by three minutes of keeping the eyes closed. Participants were seated in a dimmed room facing a white wall. This is a common task to estimate initial electrical brain source activation \cite{barry2007eeg}.

The VR session displays an infinite tunnel which is traversed by the participant for ten minutes. The traversal speed is influenced by the activation measured by the sum of three electrodes at the frontal lobe. High alpha band magnitudes at these electrodes cause an increasing traversal speed while low voltages slowdown the traversal speed. We told the participant to traverse the tunnel as fast as possible. This training aims to increase brain activation in the frontal lobe. To achieve this, a signal processing algorithm is constructed which focuses on frontal lobe training. The electrodes at the positions F3, F4 and Fz are used to estimate brain activation at the frontal lobe. The measured electrode values are summed up and processed in a \ac{FFT} to estimate frequencies in the range between 8 \ac{Hz} and 13 \ac{Hz}. The estimated frequency is normalized to the interval $[0;1]$. The normalized values are used to set the traversal speed of the VR tunnel, where 0 causes the slowest traversal speed and 1 the fastest traversal speed. In the last step, the baseline task is repeated. This is required to detect differences in brain activation before and after the VR session.


\section{Results}
Results have shown that significant electrical brain source changes could be found within all participants. Different brain activations were observed during the eyes-opened and eyes-closed task. Higher brain activation is noticed during the eyes-closed task. The most active part in the back of the brain which is responsible for visual processing.

Before and after conducting the VR session, a snapshot of brain activation is taken for comparison purposes. The sLORETA visualizations show clear differences regarding brain activation before and after conducting the VR session. The brain shows higher brain activation before the session than the baseline task, which could be the effect of excitement since the participants have not participated in a VR related neurofeedback session before. After conducting the VR session, higher brain activation was measured. Repetition of the baseline task shows higher brain activation than in the first baseline run.

\section{Discussion}
Investigating brain source localization has shown that changes are visible and measurable when using VR as a stimulus. Increased brain activity is observed when participants started closing their eyes. Previous research has shown that this is a natural effect \cite{marx2004eyes}. The back of the brain is permanently active since it is responsible for visual processing. Closing the eyes pushes activity in other brain regions responsible for imagination and multi-sensory actions which explain enhanced electrical source localization during the eyes-closed task.


Considering the effect of VR as the stimulus has shown increased brain activity in different brain regions. The cerebellum and the frontal lobe have shown improvement in brain activation. These regions are responsible for motion sensing and motivation-related thinking. The changes in motion and speed in the tunnel explain brain activation differences in the cerebellum, while the excitement and the desire to speed up the tunnel could explain enhanced activations in the frontal lobe. The repetition of the baseline task has shown sustainable brain activity after the VR session. This is usually not a prolonged effect since permanent brain activation changes require extensive neurofeedback training. Neurofeedback sessions have to be repeated several times for long term lasting effects. The study has shown that real-time brain source localization can be used to understand the effect of stimuli on a neuropsychological level. Improved therapy of neuropsychiatric diseases and tailored enhancements of neurofeedback sessions are possible to increase the treatment efficiency.
\section{Study and Chapter Conclusion}
We presented an interactive three-dimensional real-time implementation of the sLORETA algorithm which calculates a density distribution to estimate the electrical origin of brain source activation. The usage of VR in medical areas has been investigated to estimate neuronal impacts on the brain using EEG devices. Our implementation of sLORETA is used in a user study to find out if it is possible to see visible changes in brain activation during a VR session. The results show that changes in real-time can be observed when using VR scenes as an input stimulus. The outcomes can be used to optimize medical treatment methods or to identify neuropsychiatric diseases. Our visualization proves visible changes regarding brain activation for different stimuli. We see this as a first step to create visualizations that can be interpreted by non-expert users (\textbf{RQ6}). We believe that neurofeedback can be used for the treatment of brain-related diseases like Alzheimer's, epilepsy or attention deficit hyperactivity disorders (ADHD). The use of our brain visualization as an analysis tool by doctors can help to identify upcoming diseases or track the progression of existing diseases. The collected dataset is hosted on Github\footnote{\url{www.github.com/hcum/the-brain-matters} - last access \urldate}.


\chapter{Workload-Aware Reading Interfaces}
\label{ch:workload_information_consumption}

\epigraph{\textit{No entertainment is so cheap as reading, nor any pleasure so lasting.}}{Mary Wortley Montagu}

\begin{figure}
   \centering
   \subfloat[][]{
    \includegraphics[width=0.35\columnwidth]{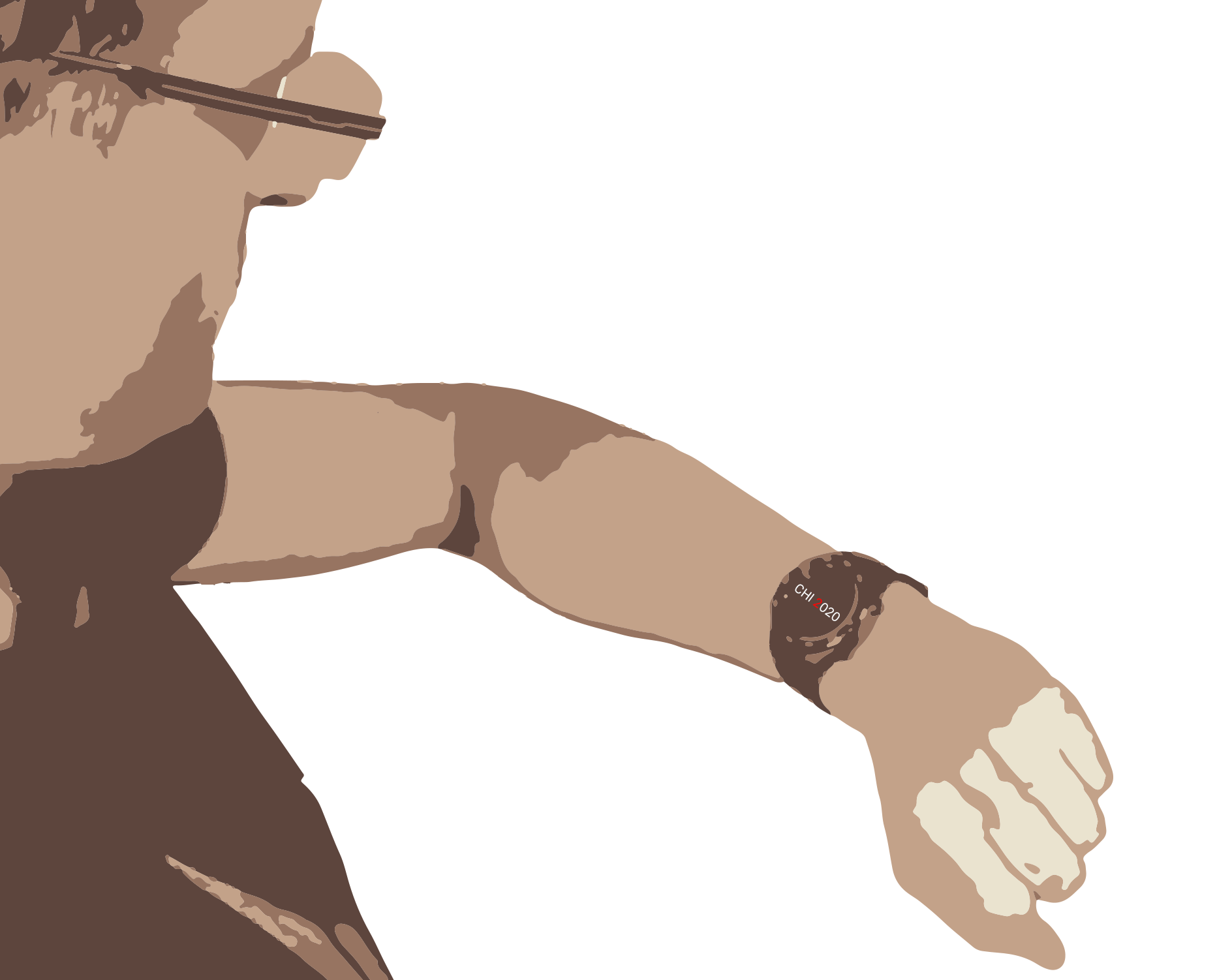}
    \label{fig:teaser_clock}}
    \hspace{3cm}
   \subfloat[][]{
    \includegraphics[width=0.35\columnwidth]{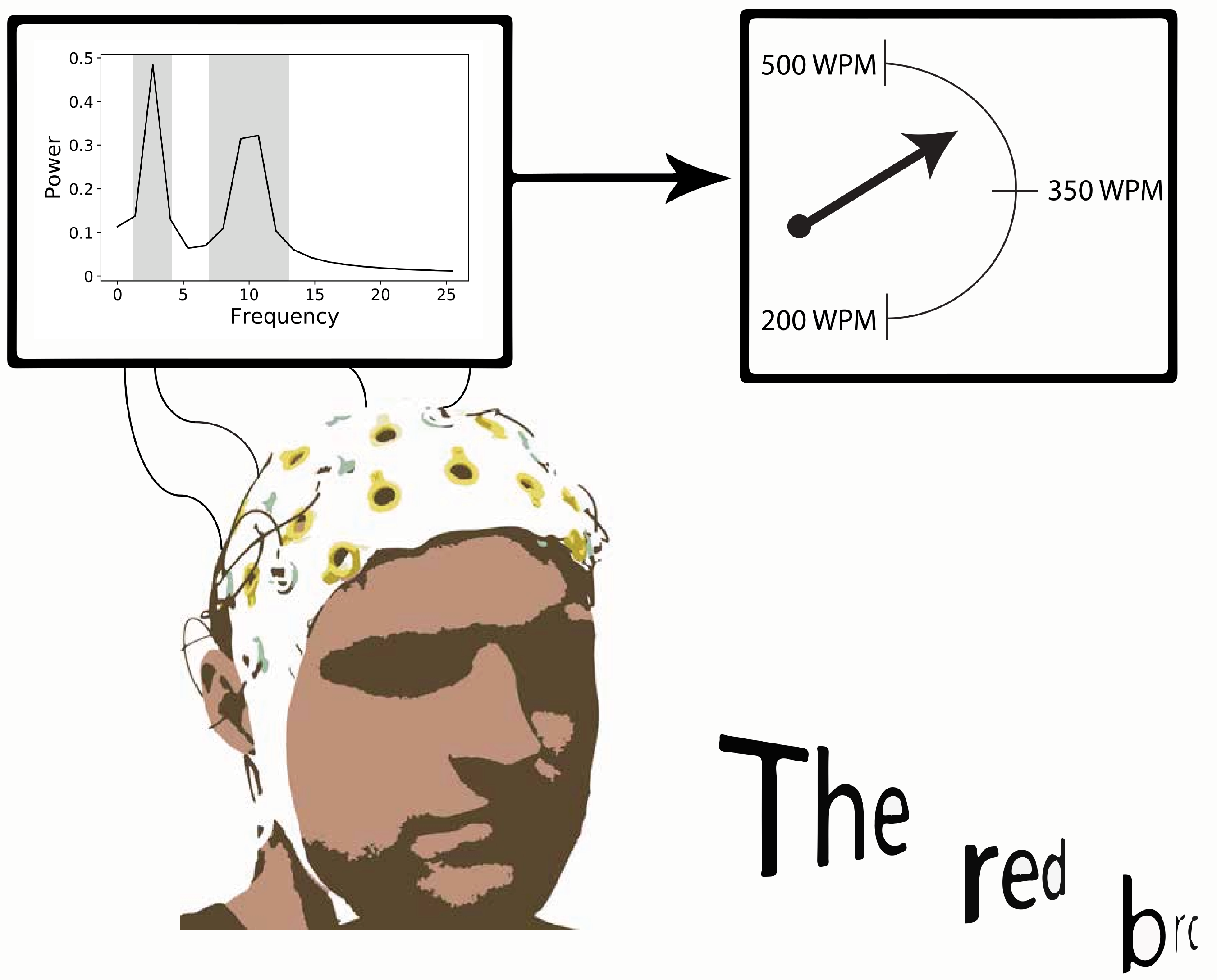}
    \label{fig:teaser_eeg}}
   \caption[Vision of a workload-aware reading interface]{\textbf{(a):} User reading on a device with small screen space. \textbf{(b):} Electroencephalography determines the current level of cognitive workload for different \textit{Text Alignments} and \textit{Presentation Speeds}.}
   \label{fig:reading_interfaces}
\end{figure}

Reading resembles an activity that interprets written text for pleasure or information consumption. It is estimated that approximately \textit{3,500} years B.C. were the first attempts of written communication that took place. Reading and writing itself was reserved for a small population of people. However, with the increase and availability of information daily, novel reading techniques emerged to filter and consume relevant information. One such technique is \ac{RSVP}, a paradigm that displays text in a word-by-word manner to present information. Instead of requiring readers to move their eyes to fixate on individual words whilst reading, readers would fixate on the display center while the words of the targeted text would be presented one after another at a chosen frequency~\cite{Forster1970}. Doing so enables full text to be accurately presented on smartphones~\cite{Georgiev:2012:IUT:2383276.2383324, Oquist:2007:EMS:1329469.1329493}, smartwatches~\cite{Dingler:2016:RGI:2971763.2971794, Dingler:2018:DCG:3173574.3173993}, or smart glasses~\cite{Rzayev:2018:RSG:3173574.3173619}, in spite of the available display area (see Figure~\ref{fig:teaser_clock}) \cite{Guo:2017:SFA:3027063.3053176}. Furthermore, removing the need to move one's eyes and the ability to raise the rate of word presentation above one's normal reading speed allows for clear time-savings, which provide the opportunity to increase the presentation rate for faster text processing. The primary display parameters of most RSVP applications are \textit{Text Alignment} and \textit{Presentation Speed}~\cite{doi:10.1177/154193120104500614}. This is either fixed or adjustable to user preferences. Past research has shown that these two parameters influence perceived workload and text comprehension~\cite{BENEDETTO2015352, castelhano2001optimizing, rubin1992reading}.

The overall aim of \ac{RSVP} is to gain more reading speed by aligning the readers' eye fixation around an Optimal Recognition Point (ORP). Commercial applications, such as \textit{Spritz}, adopted the concept of ORPs where users can adjust various settings. This can reduce cognitive load and result in better comprehension~\cite{kroll1980comprehension, potter1984rapid}, which is highly disputed in the community~\cite{BENEDETTO2015352}. \textit{Presentation Speed} determines the rate of the presented words. Previous research found that \textit{Presentation Speed} is an individual factor that has to be adjusted according to reading ability, text content, and individual preference~\cite{Chen83readingnormal}. Choosing unsuitable \textit{Presentation Speeds} may have negative effects on the overall reading efficiency~\cite{chen2007effects, doi:10.1177/154193120104500614}. For example, setting a high \textit{Presentation Speed} may result in lower text comprehension and text retention since less time is available to process the presented text. In contrast, setting \textit{Presentation Speeds} below the regular reading speed results in a loss of time. Since reading speed is an individual factor, it is difficult for novice users to estimate their optimal \textit{Presentation Speed}. While users can modify the \textit{Presentation Speed} during RSVP to estimate their preferred speed, the adjustment itself places additional workload on users which might compromise their assessment. Hence, the selected \textit{Presentation Speed} remains static during the reading session. Given individual differences and variable user states for and throughout an RSVP reading session of even a single user, it is worthwhile to investigate the viability of developing a real-time measurement that could determine appropriate \textit{Text Alignments} or \textit{Presentation Speeds}.


In this chapter, we investigate the use of \ac{EEG} as a direct metric for information processing whilst reading with RSVP with different \textit{Text Alignments} and \textit{Presentation Speeds} (see Figure~\ref{fig:teaser_eeg}). EEG allows for real-time measurements of a user's brain activity with high temporal resolution across different scalp locations~\cite{nunez2006electric}. While text comprehension and subjectively perceived workload have previously been assessed with \textit{post hoc} questionnaires for evaluating RSVP applications~\cite{Dingler:2015:ASS:2735711.2735796, Rzayev:2018:RSG:3173574.3173619}, we propose that EEG could serve as a direct, implicit, and real-time measurement of cognitive workload during RSVP reading. This overcomes many of the limitations of questionnaires, which include the need for an explicit measurement phase as well as the mitigating influence of the user's subjective self-evaluation (\textit{e.g.}, perceived workload) or their prior knowledge of the read content (\textit{e.g.}, text comprehension). Here, we evaluate if EEG could feasibly be used to predict the individual current gain in reading speed, text comprehension, and subjectively perceived workload. We report on a user study that measures cognitive workload which is raised by the two RSVP parameters \textit{Text Alignment} and \textit{Presentation Speed} using EEG. Thereby, EEG as a direct physiological measure to anticipate the current individual gain in reading speed, subjectively perceived workload, and text comprehension using predictive models from regression analyses. Finally, we discuss how our results contribute to the evaluation of future novel RSVP designs and how our findings can be used to select individual RSVP parameters based on cognitive workload:


\begin{itemize}
\renewcommand{\labelitemi}{$\rightarrow$}
\item \textbf{RQ7}: How can RSVP reading parameters be selected based on cognitive workload?
\end{itemize}

\begin{center}
\begin{GrayBox}
\centering
\parbox{0.98\textwidth}{
\emph{This section is based on the following publications:}
\vspace{3mm}
\publicationsbegin
\item \bibentry{kosch2020one}
\item \bibentry{kosch2019investigating}
\publicationsend
}
\end{GrayBox}
\end{center}

\section{Related Work}
We surveyed related research that investigated the impact of RSVP on the reading experience, performance, working memory, and engagement.

\subsection{Reading using RSVP}
Past researchers investigated how RSVP can be used to increase, control, or manipulate the overall reading speed. While reading with RSVP, words are displayed word-by-word on a fixed space. This is achieved by updating the currently displayed word with a fixed frequency in a single-word-at-a-time notion~\cite{Forster1970, Young1984}.

Using RSVP to read text is believed to increase the overall reading speed. Words are constantly presented in the same visual field of view and thus neglect eye movements, that require time, to jump from word to word~\cite{JUOLA1987499, rayner1998eye}. For example, Rubin and Turano~\cite{rubin1992reading} compared the reading speed and comprehension between the regular reading of a single page letter and previously calibrated RSVP speeds. Their findings show that RSVP speeds up the reading process while maintaining a text comprehension similar to regular reading. However, the presented approach relies on a calibration method that requires users to read the text aloud. Based on the incorrect or correct reading, the RSVP speed was adjusted manually by the experimenter.

Since RSVP presents a word at a time, it reduces display-related space restrictions that enable the representation of longer texts on devices with small screen space. For example, RSVP has been successfully employed on mobile phones~\cite{Oquist:2007:EMS:1329469.1329493}, smartwatches~\cite{Dingler:2016:RGI:2971763.2971794}, and mobile augmented reality devices~\cite{Rzayev:2018:RSG:3173574.3173619}. Since the requirements and context for RSVP designs differentiate between devices, the requirements for RSVP may be different depending on the used device as well as on the user requirements~\cite{Karkkainen:2002:DSD:572020.572052}. For example, users may prefer slower word representation speeds in a mobile scenario than in a static context. In the following, we present related research that investigated relevant RSVP designs and their parameters to provide suitable reading experiences.

\subsection{Design of RSVP Parameters}

RSVP reading operates by presenting single words or short phrases one after another. In contrast to regular text reading, it removes the need for eye movements and controls the rate at which text information is accessed by the reader. Past research was concerned with factors that optimize the overall reading performance.

The centering of text and the rate of its presentation henceforth referred to as \textit{Text Alignment} and \textit{Presentation Speed} respectively, can be argued to critical design parameters. A faster \textit{Presentation Speed} reduces the time it takes to read a document. Meanwhile, effective \textit{Text Alignment} ensures that the presented word or phrase is appropriately fixated by the reader upon presentation without the need for eye movements. Thus, a factor for effective \textit{Text Alignment} is the fixation location of a word or sentence, where a large number of neighboring letters are visible to recognize the word in a short amount of time~\cite{doi:10.1111/j.1467-9817.2005.00266.x}. Some commercial RSVP implementations (\textit{e.g.}, Spritz Inc.\footnote{\url{www.spritzinc.com} - last access \urldate}) adopt an appropriate fixation location, also known as an Optimal Recognition Position (ORP), to eliminate unnecessary saccades and minimize eye movements. For instance, highlighting and centering words around their ORP can reduce the overall word processing time. However, reading with ORPs over longer periods suppresses parafoveal processing, increases subjectively perceived workload, and elicits a higher number of eye blinks which is an indicator for visual fatigue~\cite{BENEDETTO2015352, doi:10.1080/01449290110069400}. Thus, Dingler et al.~\cite{Dingler:2015:ASS:2735711.2735796} explored how alternative ORP representation, such as underlining ORPs instead of coloring them, affect subjectively perceived workload and text comprehension. Although no significant results in subjective workload and text comprehension were found between underlined and colored ORPs, participants elaborated that they adjusted more quickly to colored ORPs instead to underlined ORPs. An alternative approach that utilizes colors in text representations is BeeLine Reader\footnote{\url{www.beelinereader.com} - last access \urldate}. BeeLine Reader integrates color gradients within text lines to save time during return sweeps. Beyond that, Russel and Chaparro~\cite{doi:10.1177/154193120104500614} investigated how different font sizes affected RSVP reading compared to regular reading. They find that font sizes did not significantly influence text comprehension. However, participants stated a preference for font sizes that were readable and did not strain their eyes.

\textit{Presentation Speed} yields another critical RSVP design parameter. An effective \textit{Presentation Speed} has the potential to speed up reading times by accelerating the frequency of displayed words. However, as \textit{Presentation Speed} increases, less time is available to cognitively process words. Previous studies found significant differences in text comprehension, whereas text comprehension and \textit{Presentation Speed} are correlated negatively~\cite{doi:10.1177/154193120104500614}. Previous studies have confirmed that the preferred reading speed is a factor that depends on individual preferences and text type~\cite{kroll1980comprehension}.

\begin{figure*}
\centering
\subfloat[][]{
\includegraphics[width=0.33\columnwidth]{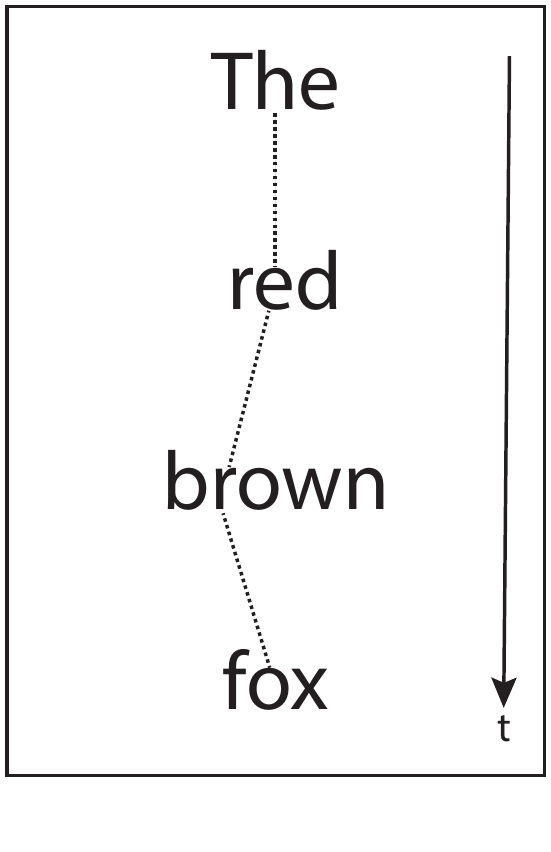}
\label{fig:centered}}
\subfloat[][]{
\includegraphics[width=0.33\columnwidth]{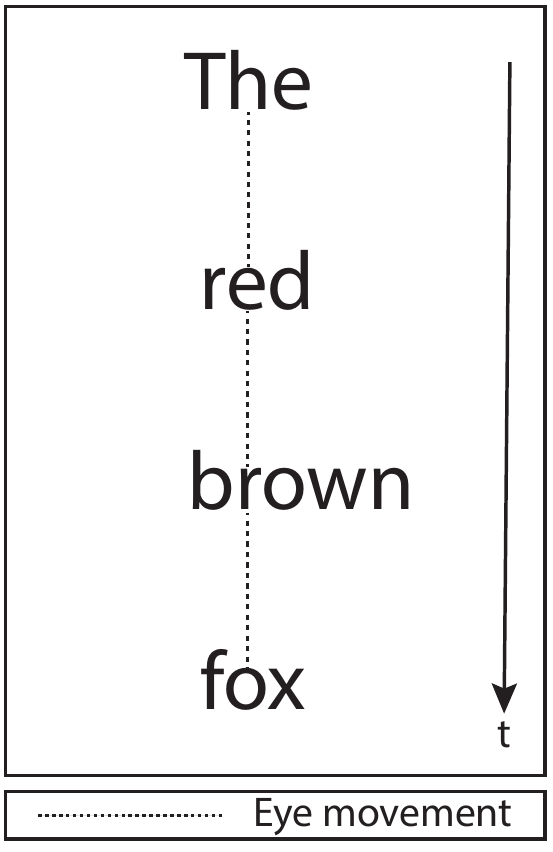}
\label{fig:optBW}}
\subfloat[][]{
\includegraphics[width=0.33\columnwidth]{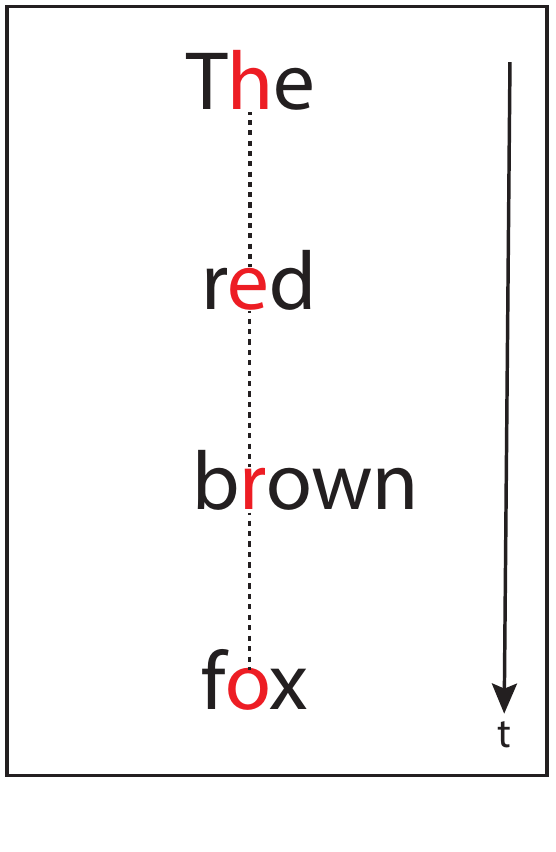}
\label{fig:opt}}
\caption[Different RSVP word positioning strategies]{Word position alignment for each text representation. \textbf{(a):} Centered word positioning. \textbf{(b):} ORP-centered word without color. \textbf{(c):} ORP-centered word with color.}
\label{fig:centeredvsspritz2}
\end{figure*}

\subsection{Cognitive Processing during Reading}
Reading elicits different strains of working memory regarding information retention. Daneman and Carpenter~\cite{DANEMAN1980450} revealed that reading performance and comprehension is individual among users. A reader with poor processing and storage functions may retain information for a shorter time compared to more experienced readers. However, at the cost of time, readers that take more time to process information can increase their overall text comprehension~\cite{doi:10.1080/01638537809544443}.

Various researchers presented approaches to quantify the exert of working memory for different RSVP design parameters that mainly include \textit{Text Alignment} and \textit{Presentation Speed}. The awareness about the workload placed on readers enables to compare RSVP design parameters in terms of time-savings and information retention. For example, Dingler et al.~\cite{Dingler:2015:ASS:2735711.2735796} presented the use of the NASA-TLX questionnaire~\cite{doi:10.1177/154193120605000909, HART1988139} to measure several facets of experienced workload when reading with different RSVP design parameters. Further subjective measures, such as the Continuous Subjective Workload Assessment Technique (C-SWAT)~\cite{luximon1998continuous}, have been employed to measure the placed cognitive workload for various RSVP parameters~\cite{doi:10.1177/1541931213601265}. However, the aforementioned measures are susceptible to subjective biases. When comparing RSVP design parameters in a study, participants might suggest a better reading performance due to previous knowledge about the presented text or lack of recall regarding their performance during the self-assessment itself.

To overcome the previous gaps from subjective assessments, physiological measurements have been proposed as an alternative to evaluating the suitability of RSVP design parameters regarding reading and information retention performance. Oliveira et al. present how EEG has been used to detect reading activity~\cite{Oliveira:2010:ERD:1774088.1774349, oliveira2009reading, Oliveira:2013:TMW:2459236.2459260}. They incorporate several frequency bands for each electrode to train a k-nearest neighbor classifier. However, they apply their approach for regular text reading only and do not evaluate reading activity during different RSVP design parameters. A similar approach was taken by Yuan et al.~\cite{Yuan:2014:TUM:2567574.2567624}, where reading comprehension was evaluated instead of reading activity during regular reading. Finally, Lees et al.~\cite{Lees_2018} provide a literature review about RSVP and EEG. In their review, they present past work that utilized an event-related potential-based brain-computer interface paradigm during the RSVP of images and words for analysis or input. The aforementioned research investigates frequency bands as an indicator of reading activity and comprehension. Specific EEG bandwidths, such as the frontal theta bandwidth (4 Hertz to 8 Hertz) or alpha bandwidth (8 Hertz to 12 Hertz), are correlated with executive cognitive functions and brain resting states that whether synchronize (\textit{i.e.}, theta power increases) or desynchronize (\textit{i.e.}, alpha power decreases) when cognitive workload is raised~\cite{klimesch2005functional}.

Cognitive functions correlate with changes in the power of theta and alpha frequency bands of EEG measurements. In particular, high alpha power is associated with the default mode of a resting brain, which is exemplified by synchronized neural activity in the alpha bandwidth~\cite{mantini2007electrophysiological}. Hence, cortical processing of perceived stimuli disrupts this default mode, resulting in alpha-desynchronization that manifests as reduced alpha power. For instance, inter-individual differences in alpha power have shown to reflect memory performance at the occipital lobe~\cite{KLIMESCH1999347}. In line with this, reading and text comprehension elicits lower alpha power as reading requires cognitive resources to memorize recently read propositions, text coherence, and information to put sentences into their context~\cite{doi:10.1080/01638539809545041, graesser1994constructing}. The theta frequency bandwidth, on the other hand, is related to executive functions and correlates negatively with the default mode of a resting state~\cite{scheeringa2008frontal}. An increase in theta power is indicative of focus or higher task engagement~\cite{berka2007eeg} across vigilance, learning, and memory tasks. In the context of the current work, high theta power is also prominent during the effective semantic processing of language~\cite{doi:10.1162/0898929053279469}. Thus, theta power is expected to rise with increasing language processing activity.

Some researchers also adopt the theta-alpha ratio~\cite{klimesch2005functional} as a hybrid metric for cortical activity that normalizes frontal theta power relative to alpha power. This entangles previous findings of frequency bands on brain resting states (\textit{i.e.}, alpha) and semantic processing (\textit{i.e.}, theta) into a metric that expresses cognitive processing or engagement during reading. Thus, a small theta-alpha quotient could correlate with low reading effort or no engagement in reading at all while a large theta-alpha quotient can indicate high reading engagement or increased cognitive demand.

\subsection{Summary}
Past research investigated how different settings of RSVP influence individual information processing, working memory, and engagement. However, previous work did not address how RSVP parameters can be implicitly evaluated without the need for individual user adjustments or reading interruptions, hence utilizing fixed RSVP reading parameters that do not take the mental demand into account. We close this gap by presenting a study that investigates the impact on EEG measures from three different \textit{Text Alignments} and \textit{Presentation Speeds}. We compare the alpha and theta bandwidth, subjective workload, and text comprehension of different RSVP parameter settings to regular reading and examine the placed cognitive workload. Finally, we evaluate alpha and theta bandwidths as a predictive metric for the current gain in speed relative to regular reading, subjectively experienced workload, and text comprehension.

\section{Study}
We designed and conducted a controlled user study to evaluate the influence of the two RSVP parameters \textit{Text Alignment} and \textit{Presentation Speed} on text comprehension, subjective workload, and measured cortical activity. The experiment consisted of a repeated-measures design comprising two factors with three levels each. They were \textit{Text Alignment} (\textit{centered}, \textit{ORP}, \textit{ORP} with colored letter) and \textit{Presentation Speed} in words per minute (WPM) (200 WPM, 350 WPM, and 500 WPM). Text comprehension, subjective workload, and cortical activity (\textit{i.e.}, power in the theta and alpha frequency band) served as the primary measures in our study. Comparisons are drawn to the baseline of normal non-RSVP reading from a PDF document and across the test conditions. We describe the independent variables in the following.

\subsection{Text Alignment}
RSVP can vary in terms of the alignment of presented words, relative to the display and each other. A \textit{centered} condition presented words within a bounding box that is \textit{centered} in the display (see Figure~\ref{fig:centered}). An \textit{ORP} condition centered each word around the \textit{ORP} (see Figure~\ref{fig:optBW}). An \textit{ORP} with colored letter condition colored the \textit{ORP} red in addition (see Figure~\ref{fig:opt}). This approach is similar to the algorithm employed by Spritz Inc.~\cite{BENEDETTO2015352}.

\begin{figure}
\centering
\subfloat[][]{
\includegraphics[width=0.33\columnwidth]{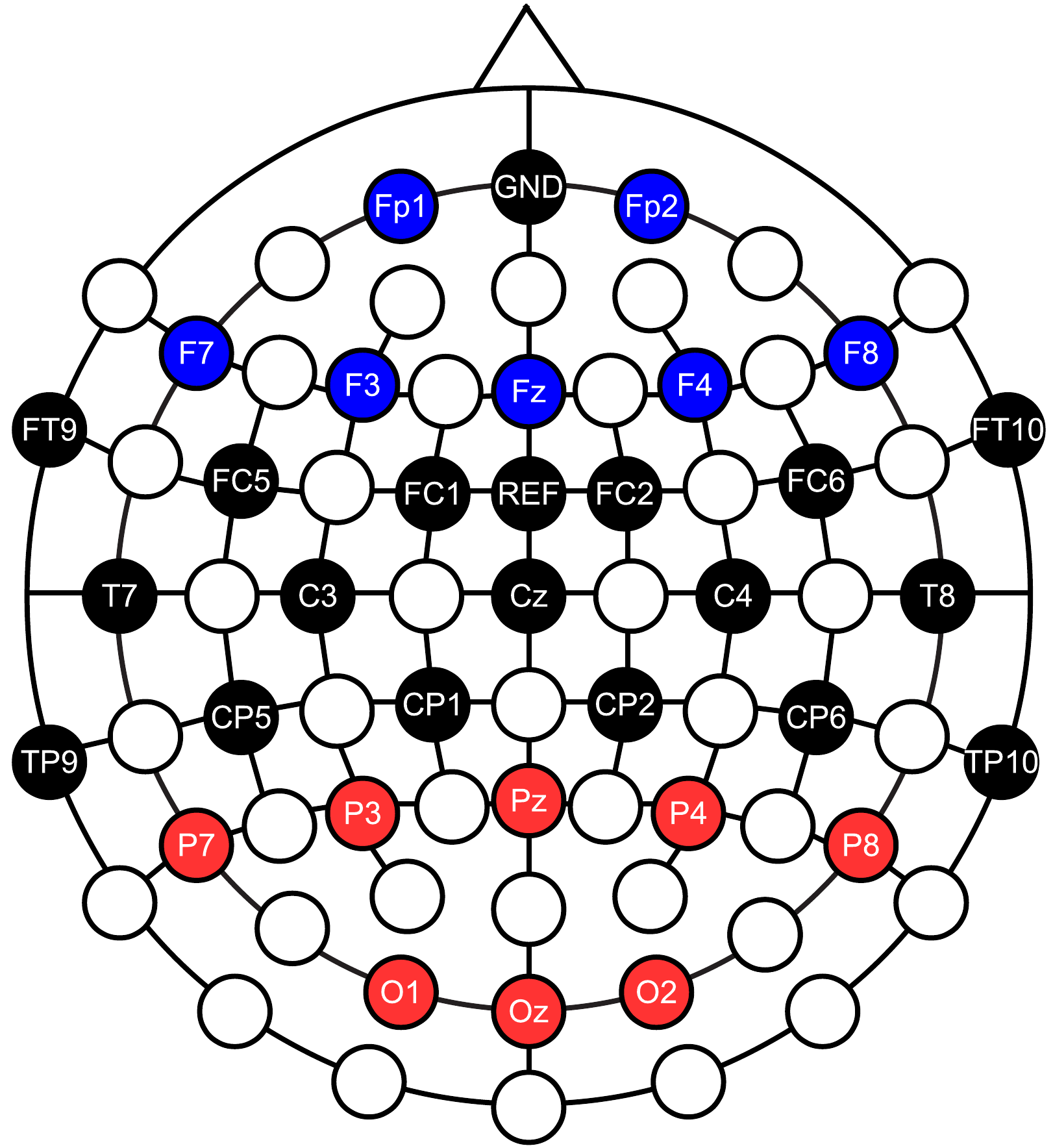}
\label{fig:electrode_layout}}
\subfloat[][]{
\includegraphics[width=0.47\columnwidth]{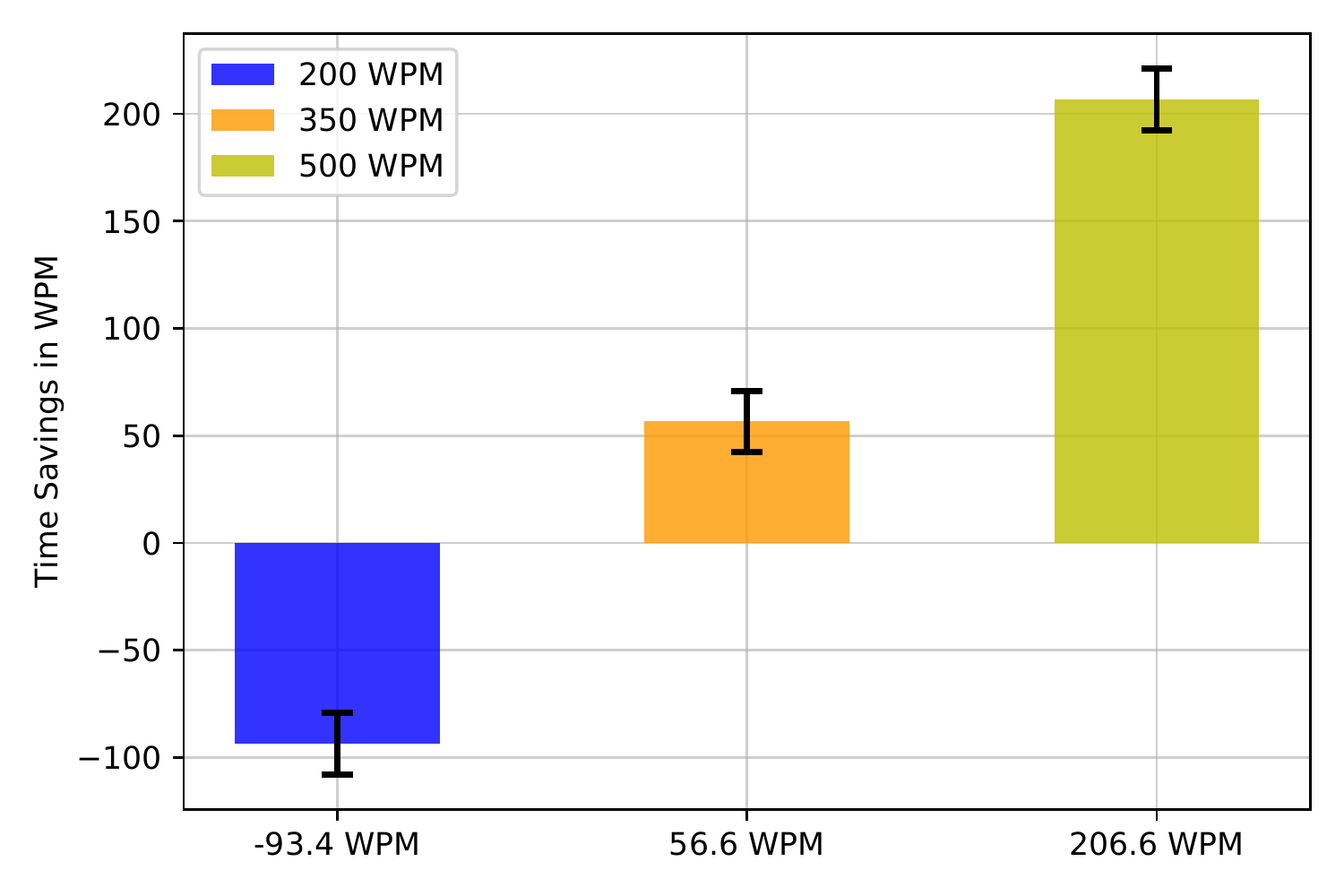}
\label{fig:mean_time_savings}}
\caption[Electrode layout and time savings of RSVP]{\textbf{(a):} 32 electrode layout used in the experiment. The ground electrode was placed on \textit{Fpz} and the reference electrode on \textit{FCz}. The red marked electrodes were used to analyze alpha frequencies while the blue electrodes were used to analyze theta frequencies. \textbf{(b):} Time difference of using RSVP compared to reading a PDF. Using 200 WPM slowed participants down, while 350 WPM and 500 WPM accelerated the reading times. Hence, 200 WPM did not provide effect time-savings compared to 350 WPM and 500 WPM. The error bars depict the standard error.}
\label{fig:electrode_layout_mean_time_savings}
\end{figure}

\subsection{Presentation Speed}
RSVP can vary in terms of how quickly they are presented measured in words per minute (WPM). Three \textit{Presentation Speeds} were chosen with equal intervals, namely 200, 350, and 500 WPM. Previous work suggests that readers can understand RSVP content well at 200 WPM~\cite{chen2007effects}. The conditions of 350 WPM and 500 WPM were chosen to investigate the influence of higher \textit{Presentation Speeds} on cortical activity, subjective workload, and cortical activity~\cite{cain2004children, doi:10.1177/0022219409338742}. Hence, we hypothesized workload to increase with higher \textit{Presentation Speeds}, albeit not necessarily at the cost of reading comprehension.

\subsection{Participants}
We recruited 18 participants (7 females, 11 males; age $M=27$, $SD = 3.7$). All participants had a normal or corrected-to-normal vision and none of them reported a history of neurological disorders. Two participants have used RSVP once before. Participants were recruited through university mailing-lists. They received 15 Euro or lecture study points for their participation. We removed one participant from our analysis due to technical difficulties during the experiment.

\subsection{Apparatus and Stimuli}
The study was conducted in a self-contained room with constant luminance throughout experimentation. The room was divided into two areas, a test area for the participant and a control area for the experimenter. The experiment was controlled remotely with a laptop while participants were presented with stimuli via an LCD monitor (Dell U2715H; 27 inches diagonal; $2560\,\times\,1440$ pixel resolution; 60 Hertz refresh rate).

The reading distance approximated 50 cm from the participants' head to the display. As suggested by previous research, words were displayed with a font size that did not strain the participants' eyes to avoid confounding measures during the experiment~\cite{doi:10.1177/154193120104500614}. These were consistently 20 points during the PDF trial and 55 points during the RSVP trials. The font size was adjusted to match a similar size relative to the screen resolution. EEG data was collected using a 32 active gel electrodes wireless system (BrainVision LiveAmp\footnote{\url{www.brainproducts.com/productdetails.php?id=63} - last access \urldate}, BrainProducts GmbH; bandpass: 0.1-1000 Hz, no notch filter). Electrodes were placed in accordance to the International 10-20 layout (ground electrode: \textit{Fpz}, reference electrode: \textit{FCz}; see Figure~\ref{fig:electrode_layout}). Before testing, the impedance of all electrodes was kept to less than 10 $k\si{\ohm}$. EEG data was collected at the occipital and frontal lobe. Changes in alpha power at the occipital lobe are associated with respective changes in brain resting states, where high alpha power is associated with a distinct resting state. Respectively, low alpha power is associated with low brain resting states~\cite{doi:10.1093/cercor/7.4.374}. Theta power is an indicator of task engagement~\cite{1999169} when measured at the frontal lobe. Thereby, high theta power corresponds to high task engagement whereas low theta power suggests the opposite. EEG data were annotated with unique triggers for the start and end of the experiment as well as for each word presentation according to the relevant test condition. We adopted a set of ten short excerpts from the book \textit{Speed Reading: A Course for Learners of English}~\cite{quinn1974speed, Rzayev:2018:RSG:3173574.3173619}. The presented excerpts had a mean average of $M=547.9$ words ($SD = 3.7$) and were accompanied by a unique set of ten comprehension questions.

\subsection{Procedure}
Before testing, participants were required to perform some tasks to establish baseline measures. First, we derived the individual alpha frequency (IAF) by recording one minute of resting-state EEG activity for when the participants' eyes were opened and closed~\cite{1999169}. Next, we requested participants to read a one-page PDF text excerpt that was presented all at once via a PDF viewer. With this, we established the baseline for each participants' reading time and individual cortical activity. Ten comprehension questions and a raw NASA-TLX questionnaire were conducted after each reading trial. Testing consisted of nine RSVP presentation conditions which were counter-balanced according to the balanced Latin square~\cite{cochran1950experimental}. After each condition, participants were presented with ten questions that tested for reading comprehension and a NASA-TLX questionnaire for evaluating subjective workload~\cite{HART1988139}. The full study took approximately 90 minutes to complete. Briefing and debriefing of the study purpose were performed before and after the study respectively. All participants provided signed informed consent.

\subsection{Data Preprocessing}
Four seconds of data were removed from the beginning and end of each recording to remove signals that are unrelated to cortical activity. The IAF bandwidth~\cite{1999169} was determined for each participant based on the separate spectral analyses of the EEG recording when the participants' opened their eyes and closed them; the mean power around the maximum difference peak ($\rpm$2 Hz) was submitted to a spatio-spectral decomposition filter~\cite{NIKULIN20111528, proakis2001digital}. The spatio-spectral decomposition derives the bandwidth power based on neighboring electrodes. A decomposition of recordings maximizes the signal power of the chosen frequency while simultaneously minimizing it at the neighboring electrodes, thus mitigating the appearance of noise for frequencies outside the alpha frequency bandwidth. We performed a short-time Fourier transform by dividing the signal into one-second slices with an overlap of half a second. After estimating the IAF from each of those electrodes, we averaged the individual frequency band per electrode. This procedure was applied to the electrodes \textit{Pz}, \textit{P3}, \textit{P7}, \textit{O1}, \textit{Oz}, \textit{O2}, \textit{P4}, and \textit{P8} located at the occipital lobe~\cite{doi:10.1093/cercor/7.4.374}. (see Figure~\ref{fig:electrode_layout}). A similar procedure was applied to extract theta power. The signal was filtered using a spatio-spectral decomposition (5 $\rpm$ 2 Hz). Again, a short-time Fourier transform was carried out by dividing the signal into one-second slices with an overlap of half a second and were mean averaged to obtain the final theta power. This procedure was applied to the electrodes \textit{Fp1}, \textit{Fp2}, \textit{F7}, \textit{F3}, \textit{Fz}, \textit{F4}, and \textit{F8} (see Figure~\ref{fig:electrode_layout}).

\begin{figure}
\centering
\subfloat[][]{
\includegraphics[width=0.47\columnwidth]{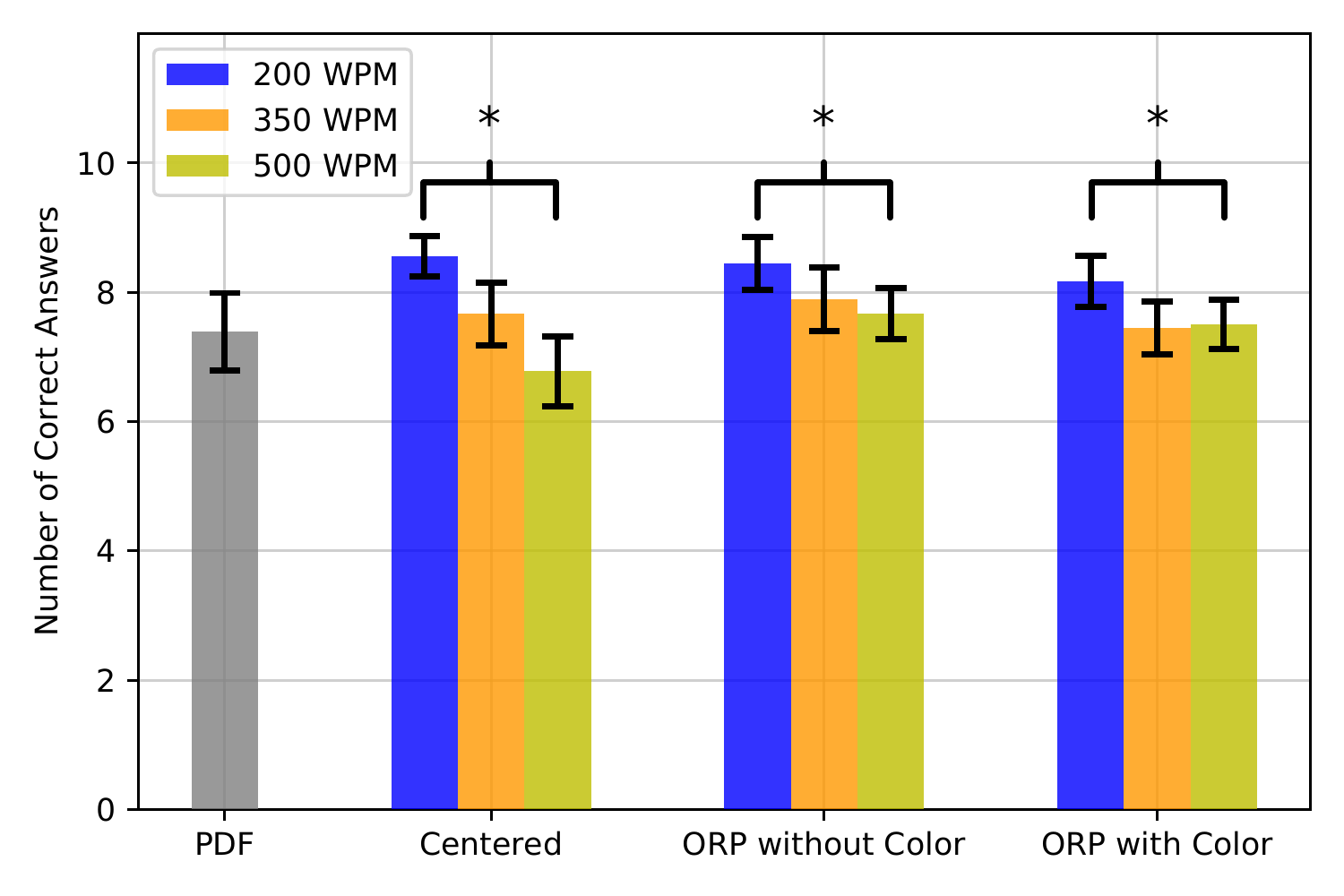}
\label{fig:mean_comprehension}}
\subfloat[][]{
\includegraphics[width=0.47\columnwidth]{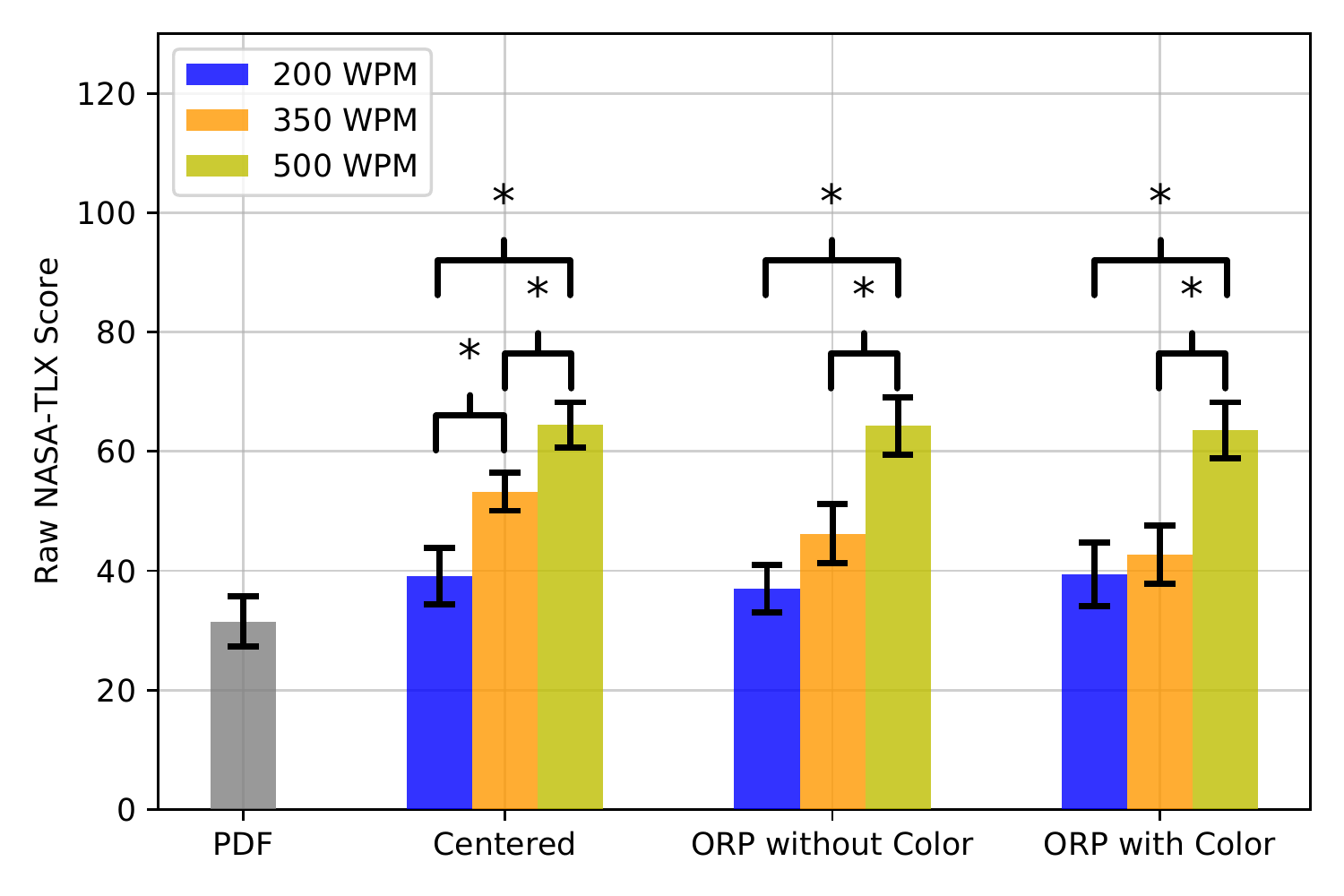}
\label{fig:mean_tlx}}
\caption[Mean comprehension and mean raw NASA-TLX scores]{\textbf{(a):} Mean comprehension scores for the showed questionnaires. Increasing the \textit{Presentation Speed} reduces the number of correct answers and may impact the overall comprehension. The error bars depict the standard error. Brackets indicate significant differences. \textbf{(b):} Mean raw NASA-TLX scores per condition. Less WPM induce less subjective compared to higher WPM. The error bars depict the standard error. Brackets indicate significant differences.}
\label{fig:mean_comprehension_mean_tlx}
\end{figure}

\begin{table}
\centering

\begin{tabular}{lllll}
\toprule
& Parameter & $F$ & $p$ & $\omega^2$ \\
\midrule
\multirow{3}{*}{Reading Comprehension} & \textit{Presentation Speed} & 3.68 & .04 & 0.08\\
& \textit{Text Alignment} & 0.85 & .44 & 0.00\\
& \makecell{\textit{Text Align.}\,$\times$\,\textit{Pr. Sp.}} & 0.60 & .67 & 0.00\\
\cline{2-5}

\multirow{3}{*}{Subjective Workload} & \textit{Presentation Speed} & 18.5 & $<$.001 & 0.27\\
& \textit{Text Alignment} & 1.39 & .26 & 0.00\\
& \makecell{\textit{Text Align.}\,$\times$\,\textit{Pr. Sp.}} & 1.02 & .40 & 0.00\\
\cline{2-5}

\multirow{3}{*}{IAF Power} & \textit{Presentation Speed} & 491.19 & $<$.001 & 0.61\\
& \textit{Text Alignment} & 2.37 & .13 & 0.00\\
& \makecell{\textit{Text Align.}\,$\times$\,\textit{Pr. Sp.}} & 2.48 & .11 & 0.00\\
\cline{2-5}

\multirow{3}{*}{Theta Power} & \textit{Presentation Speed} & 338.09 & $<$.001 & 0.58\\
& \textit{Text Alignment} & 1.00 & .38 & 0.00\\
& \makecell{\textit{Text Align.}\,$\times$\,\textit{Pr. Sp.}} & 0.42 & .79 & 0.00\\

\bottomrule
\end{tabular}

\caption[Summary of the confirmatory ANOVA results]{Summary of the confirmatory ANOVA results.}
\label{tab:summary_results}
\end{table}
\vspace{-1em}
\section{Results}
The RSVP test conditions provided three measurements for each participant that we submitted for further analyses with a 3\,$\times$\,3 repeated measures analysis of variance (ANOVA) for the factors of \textit{Text Alignment} (\textit{centered}; \textit{ORP}; \textit{ORP} with red color) and \textit{Presentation Speed} (200, 350, 500). They were reading comprehension, raw NASA-TLX workload score, and IAF power. They respectively represented performance, subjective workload, and cortical activity. A Shapiro-Wilk test did not reveal a deviation of normality from the measures ($p \geq .05$).

We adopted an alpha-level of 0.05 for statistical significance testing and report the $\omega^2$ for effect sizes of significant results. Greenhouse-Geisser corrections are reported when violations of sphericity are detected. Post hoc Bonferroni-corrected tests were performed between test conditions to investigate the significant main effect of \textit{Text Alignment}. Planned linear contrasts were performed to understand significant main effects or interactions that included \textit{Presentation Speed}. Post hoc pair-wise comparisons to baseline measures, derived from normal reading, was performed when applicable. A summary of the results can be found in Table~\ref{tab:summary_results}.

\subsection{Time-savings of RSVP}
To evaluate the time-savings of RSVP, we subtracted the normal non-RSVP reading times of our participants from 200, 350, and 500 WPM. The mean reading time was 293.4 WPM ($SD=59.3$). Hence, the mean time-savings were -93.4 WPM, 56.6 WPM, and 206.6 WPM, for the \textit{Presentation Speed} levels of 200 WPM, 350 WPM, and 500 WPM respectively. Figure~\ref{fig:mean_time_savings} illustrates these time-savings. One-sample t-tests show that participants were significantly slower at the \textit{Presentation speed} of 200 WPM ($t(16) = -6.69, p < .001, d = -1.58$) and significantly faster at 350 and 500 WPM ($t(16) = 4.05, p < .001, d = 0.95$; $t(16) = 14.8, p < .001, d = 3.49$), relative to their normal reading times.

\subsection{Reading Comprehension}
Reading comprehension was operationalized in terms of the number of correct answers out of the ten questions that were posed during each questionnaire. Figure~\ref{fig:mean_comprehension} illustrates the mean number of correct answers per participant and condition. We found a significant main effect for \textit{Presentation Speed} with a medium effect size, whereas no significant effect was found for \textit{Presentation Speed} and \textit{Text Alignment}\,$\times$\,\textit{Presentation Speed}.

\begin{figure}
\centering
\subfloat[][]{
\includegraphics[width=0.47\columnwidth]{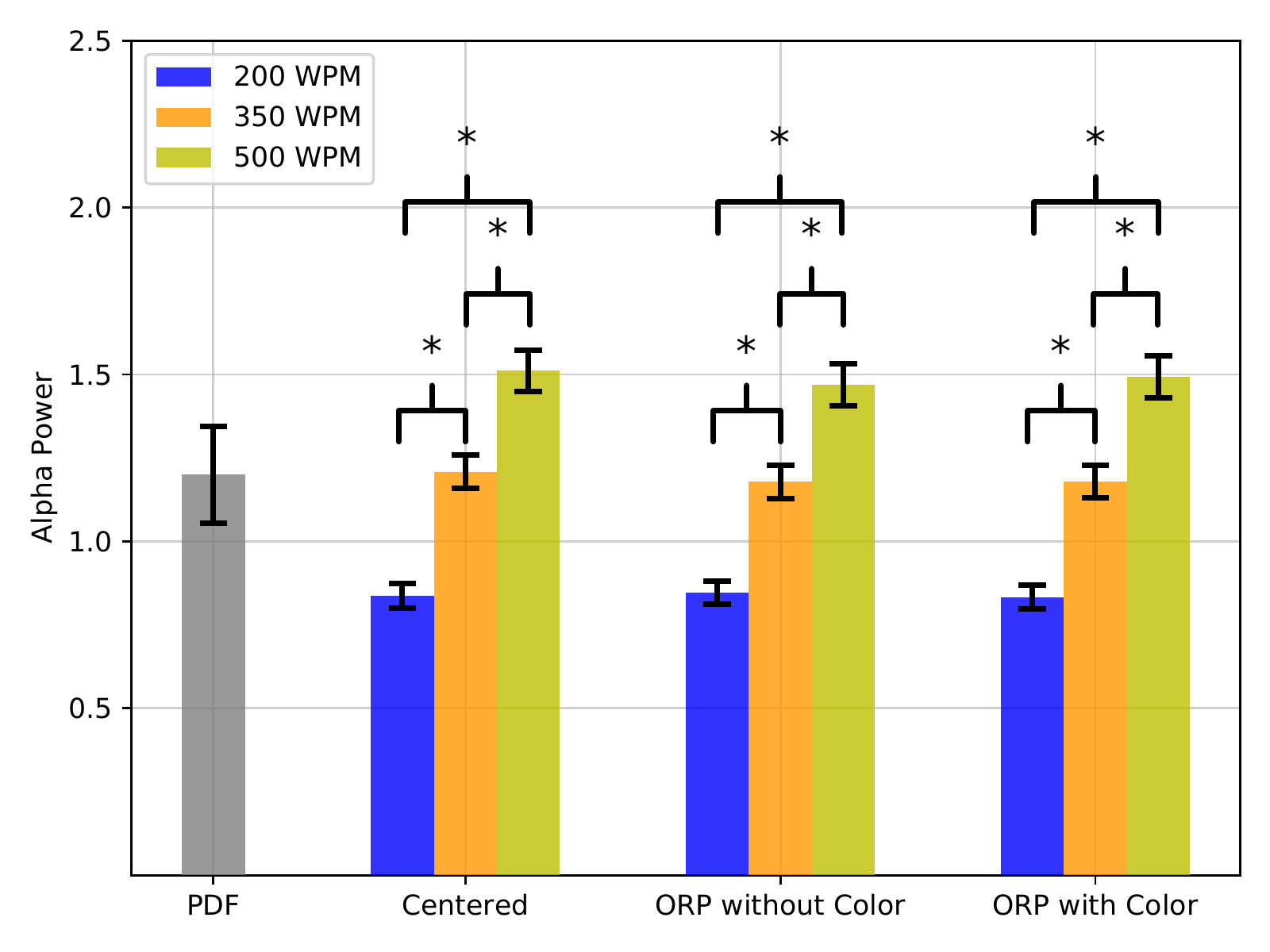}
\label{fig:mean_alpha}}
\subfloat[][]{
\includegraphics[width=0.47\columnwidth]{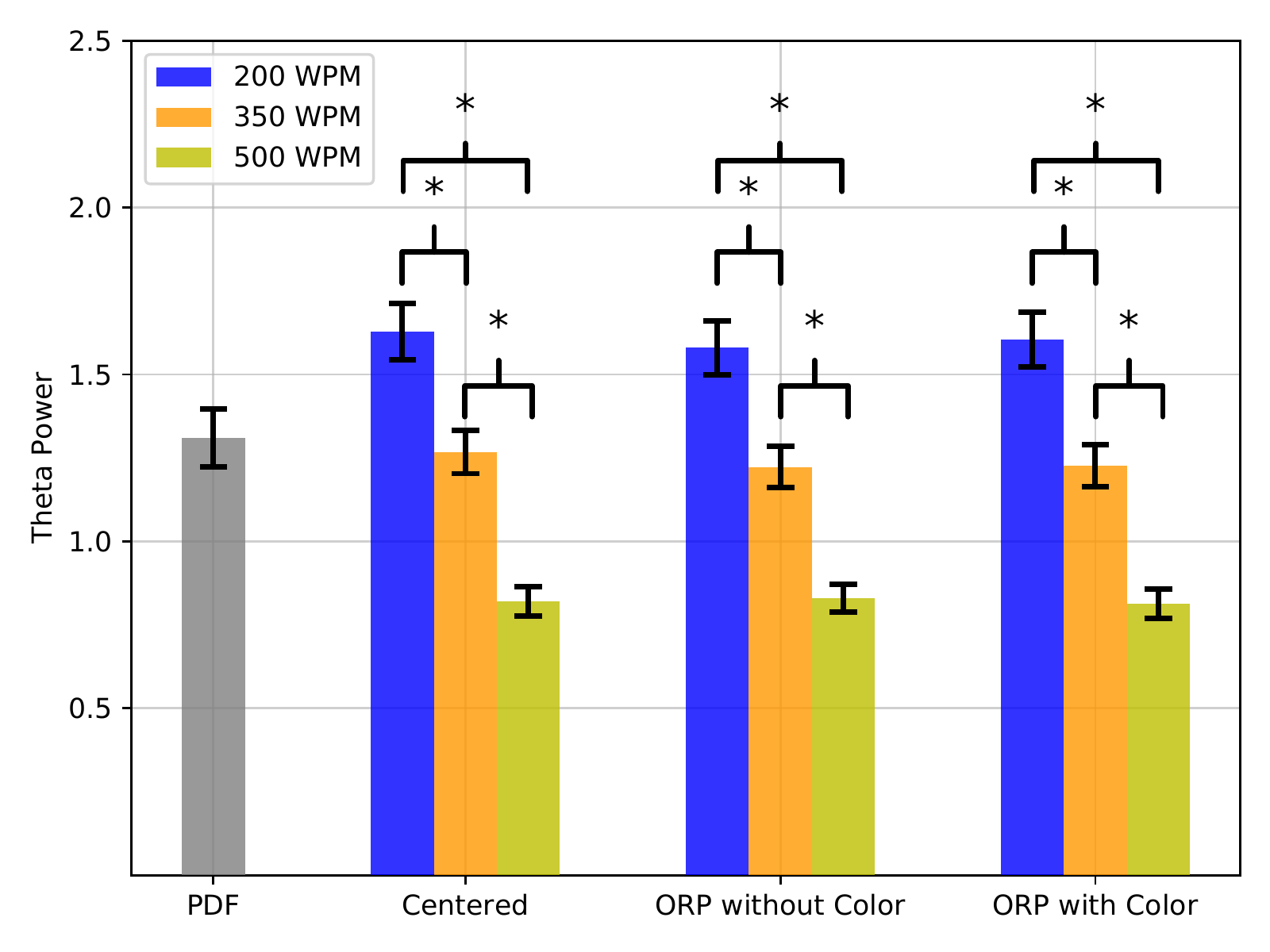}
\label{fig:mean_theta}}
\caption[Normalized IAF and theta power per condition]{\textbf{(a):} Normalized mean IAF band power per condition. Increasing the WPM yields an increase in alpha power band while slower speeds elicits lower alpha powers. The error bars depict the standard error. Brackets indicate significant differences. \textbf{(b):} Normalized mean theta power per condition. Lower \textit{Presentation Speeds} lead to higher theta power. The error bars depict the standard error. Brackets indicate significant differences.}
\label{fig:mean_comprehension_mean_tlx}
\end{figure}

\subsection{Subjective Workload}
The subjective workload was operationalized as raw NASA-TLX scores. Figure~\ref{fig:mean_tlx} shows the mean averaged raw NASA-TLX for each condition. There was a significant main effect of \textit{Presentation Speed} with a large effect size. No significant effect was found for \textit{Presentation Speed} and \textit{Text Alignment}\,$\times$\,\textit{Presentation Speed}.

\subsection{Cortical Activity for Cognitive Workload}
Cortical activity was operationalized in terms of the mean power within the IAF bandwidth (ca. $10\rpm2$ Hz) and theta bandwidth (ca. $5\rpm2$ Hz). Deviations from resting-state EEG activity result in lower alpha power and higher theta power~\cite{klimesch2005functional, Klimesch1993}. This is referred to as alpha-desynchronization and theta-synchronization. This is a reliable classification feature for working memory~\cite{baddeley1992working}, which describes the cognitive processing of information that is kept in short-term memory. The alpha bandwidth, characterized by a unimodal peak in spatio-spectral power, varies across individuals and age~\cite{1999169}. Thus, we determined the IAF bandwidth for each participant based on their peak frequency value across the eyes-opened and eyes-closed pre-test conditions. We defined $\rpm2$ Hz around the peak frequency as the individual alpha band. The mean alpha peak was $M = 9.5$ Hz ($SD= 1.95$), which provides a mean bandwidth of $7.5-11.5$ Hz. Since alpha and theta power are individual metrics which are different among participants, we normalized the bandwidths by their individual bandwidth power that is elicited during the PDF reading trial before aggregating the data.
\begin{table*}
\centering
\resizebox{\textwidth}{!}{%
\begin{tabular}{rrrrrrrrrrrr}
\toprule
&&& \multicolumn{1}{c}{IAF} && &&&&\multicolumn{1}{c}{Theta} \\
\cline{2-6}
\cline{8-12}
&\multicolumn{1}{c}{F} & \multicolumn{1}{c}{$p$} &\multicolumn{1}{c}{$R$} &\multicolumn{1}{c}{$R^2$} & \multicolumn{1}{c}{RMSE} && \multicolumn{1}{c}{F} & \multicolumn{1}{c}{$p$}&\multicolumn{1}{c}{$R$} & \multicolumn{1}{c}{$R^2$} & \multicolumn{1}{c}{RMSE}\\
\multicolumn{1}{l}{Reading Speed Gain}&3681.95& < .001& \textbf{.980}&\textbf{.961} &\textbf{0.11}&& 2266.56 & < .001 & \textbf{.984}&\textbf{.968} & \textbf{0.10}\\
\multicolumn{1}{l}{Subjective Workload}&25.24& < .001&.378&.143&20.88&&28.95& < .001&.401&.161&20.66\\
\multicolumn{1}{l}{Text Comprehension}&7.47&.007&.217&.047&1.83&&8.43&.004&.230&.053&1.82\\
\bottomrule
\end{tabular}}
\caption[Performance results of the predictive model for gains in reading speed, subjective workload, and text comprehension]{Results of the predictive models using a linear regression. All dependent variables derive a significant regression equation ($p$ < .05). The normalized IAF and theta powers are reliable predictors for the current gain in reading speed according to the large $R^2$ value and low Root Mean Square Errors (RMSE). However, subjective workload and text comprehension provide less accurate predictive models. Bold values represent the most efficient results.}
\label{tab:regression_results}
\end{table*}

\subsubsection{IAF Power}
There was a significant main effect for \textit{Presentation Speed} with a large effect size. This was a significant linear trend whereby IAF power increased with increasing \textit{Presentation Speed}. The main effect of \textit{Text Alignment} and \textit{Text Alignment}\,$\times$\,\textit{Presentation Speed} interaction were not statistically significant. Figure~\ref{fig:mean_alpha} shows the normalized mean IAF power for the employed \textit{Text Alignments} and \textit{Presentation Speeds}. Interestingly, the linear trend for cortical activity contradicts the percept of our participants for subjective workload. The current results suggest that there is less cognitive workload with increasing \textit{Presentation Speed} based on cortical activity. This raises the possibility that high \textit{Presentation Speed} might introduce subjective workload while restricting our participants' neuro-cognitive ability to process the presented words. To evaluate this possibility, we performed a post hoc tests to find a significant effect between 200 and 350 WPM ($t(16)=-21.77, p<.001, d=-5.28$), 200 and 500 WPM ($t(16)=-22.72, p<.001, d=-5.5$) as well as 350 and 500 WPM ($t(16)=-20.45, p<.001, d=-4.96$).

\subsubsection{Theta Power}
The analysis of theta power results in a significant main effect for \textit{Presentation Speed} with a large effect size. No significant main effect was found for \textit{Text Alignment} and \textit{Text Alignment}\,$\times$\,\textit{Presentation Speed}. Similar to the analysis in the IAF bandwidth, a significant linear trend presented here indicates the possibility to restrict the participants' possibility to process words. Figure~\ref{fig:mean_theta} depicts the normalized mean theta power per condition.

\subsubsection{Number of Eye Blinks}
We counted the number of eye blinks to investigate if the number of eye blinks significantly change with different \textit{Presentation Speeds}. An increased number of eye blinks is an indicator of visual fatigue~\cite{doi:10.1080/01449290110069400} that could arise from different presentation speeds. Since eye blinks introduce noise in EEG measures, we want to ensure that our EEG recordings result from cortical activity and not from noise. We use Python MNE to automatically count eye blinks using the electrodes \textit{Fp1} and \textit{Fp2}~\cite{10.1007/978-3-642-12197-5_38} to detect EOG artifacts using a threshold-based approach which was set to $200\,\mu V$. We found that 200 WPM elicited the highest number of eye blinks ($M=44.88$, $SD=33.12$) followed by 350 WPM ($M=23.55$, $SD=8.99$) and 500 WPM ($M=18.8$, $SD=7.6$). Normalizing the results to blinks per minute shows $16.38$ blinks per minute for 200 WPM, $15.04$ blinks per minute for 350 WPM, and $17.16$ for 500 WPM. Comparing each \textit{Presentation Speed} using pairwise t-tests did not yield a significant effect. Due to minimal differences in blinks per minute and non-significant differences for each \textit{Presentation Speed}, we do not assess distorted alpha or theta measures due to eye closure.

\subsection{Presentation Speed Compared to Regular Reading}
The main effect of \textit{Presentation Speed} was consistently significant across all three measures. Therefore, we performed pair-wise comparisons between normal reading and the three levels of \textit{Presentation Speed} on each measure. The median score of the three \textit{Text Alignment} levels were treated as the representative score of the corresponding level. Subjective workload was the only measure with significant differences, whereby the \textit{Presentation Speed} of 350 and 500 WPM resulted in significantly higher values of subjective workload compared to normal reading ($t(16) = 4.06, p < .001, d = 0.96; t(16) = 7.35, p < .001, d = 1.73$). There were no significant differences between the \textit{Presentation Speed} levels and normal reading for reading comprehension and cortical activity. To summarize, the time-savings of RSVP at 350 and 500 WPM was associated with the cost of increasing subjective workload, without significant improvements in reading comprehension or changes in cortical activity.

\section{Evaluating EEG as Predictive Metric}

The results show significant differences in alpha and theta power for reading with different \textit{Presentation Speeds}. We evaluate the efficiency of models that utilize EEG frequency bands to predict current gains in reading speed, subjective workload, and text comprehension using a linear regression analysis.

\subsection{Independent and Dependent Variables}
We use the IAF and theta power as independent variables for the regression analysis. Similar to the previous analysis and to mitigate person-dependent differences in the alpha and theta bandwidths, both bandwidths are normalized relative to the full-text baseline reading trial. For each RSVP condition concerning \textit{Presentation Speeds} and \textit{Text Alignments}, we calculated the mean bandwidth for the IAF and theta power for each RSVP reading trial. This resulted in nine data points for each bandwidth and each participant. Hence, a total of 153 data points were used to fit a function for the IAF and theta power. We use these data points to evaluate a predictive model using linear regression to evaluate the forecasting efficiency for the current gain in reading speed relative to the full-text reading trial, subjectively perceived workload, and text comprehension.

\subsection{Predictive Performance}
We describe the results of the regression analysis in the following. We report the significance of the regression equations and the accuracy of the fitted model. A summary of the results can be obtained from Table~\ref{tab:regression_results}.

A significant regression equation was found for gains in reading speed ($F(1, 151) = 3681.95, p < .001, R^2=.961$) with a linear trend for increasing \textit{Presentation Speeds} and IAF power. In contrary, theta power resulted in a significant regression equation ($F(1, 151)=2266.56, p <.001, R^2=.968$), where theta power shows a decreasing linear trend with higher \textit{Presentation Speeds}. This indicates a strong linear trend between the EEG frequency bands and gains in reading speed as suggested by the previous results. Similarly, a significant regression equation ($F(1, 151) = 25.24, p < .001, R^2=.143$) between increasing raw NASA-TLX scores and IAF power was confirmed. Furthermore, we find a significant regression equation ($F(1, 151) = 28.95, p < .001, R^2=.161$) between decreasing theta power and increasing raw NASA-TLX scores. However, the low $R$ and $R^2$ scores show that the variance of the raw NASA-TLX scores is large, thus indicating a low predictive performance. Finally, we apply the same regression analysis for the text comprehension scores. Again, this results in a significant equation ($F(1, 151) = 7.47, p = .007, R^2=.047$) with decreasing text comprehension for increasing \textit{Presentation Speeds}. Theta power also shows a significant equation ($F(1, 151) = 8.43, p < .001, R^2=.161$), where text comprehension decreases with decreasing theta power. Similar to the raw NASA-TLX scores, the $R$ and $R^2$ denote large variances among the data points and, therefore, a low predictive performance.

\begin{figure}
\centering
\subfloat[][]{
\includegraphics[width=0.47\columnwidth]{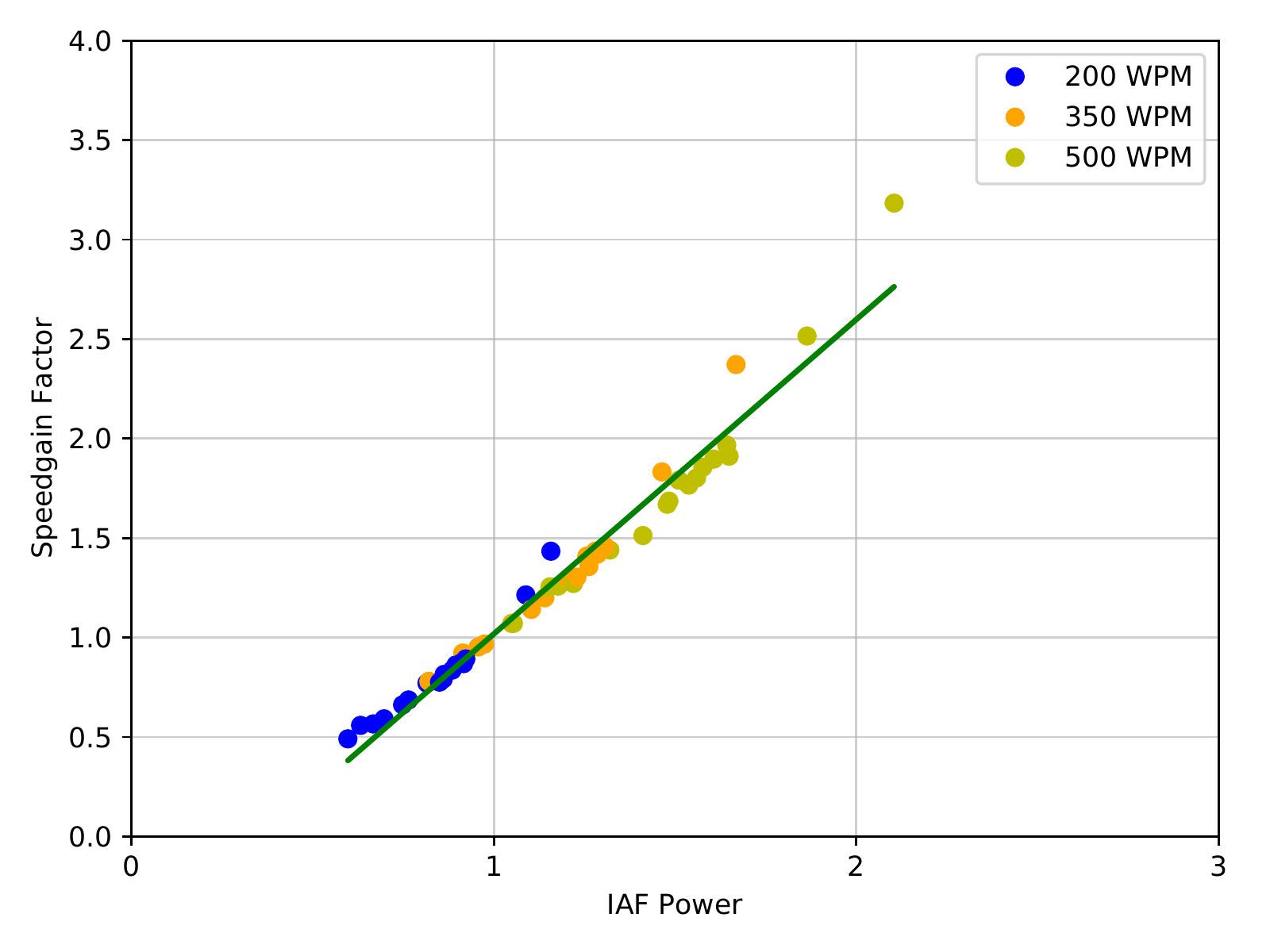}
\label{fig:regression_speed_gain_iaf}}
\subfloat[][]{
\includegraphics[width=0.47\columnwidth]{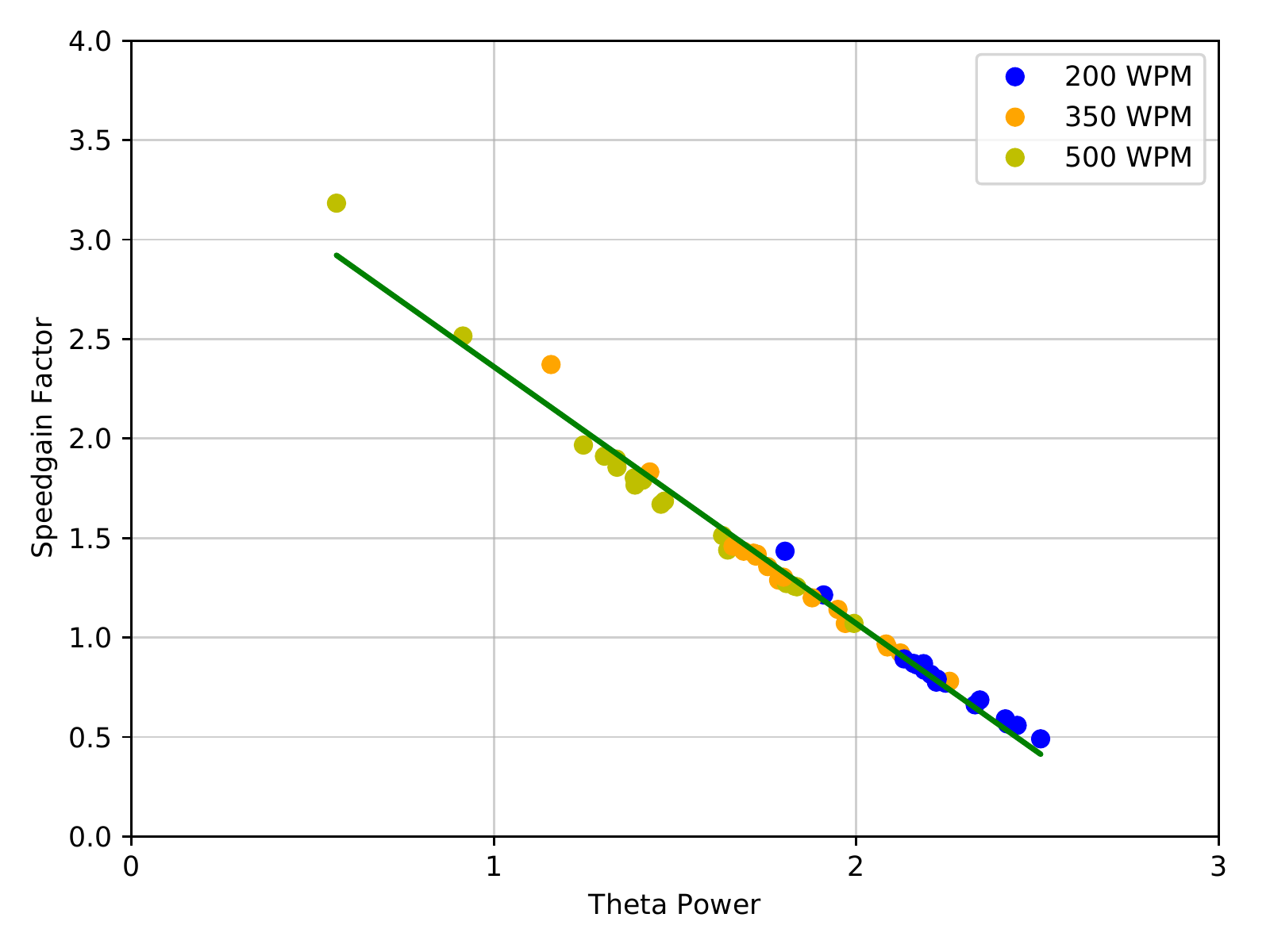}
\label{fig:regression_speed_gain_theta}}
\caption[Normalized linear regression for gains in reading speed and potential use cases that utilize dynamic RSVP parameter selections based on EEG]{\textbf{(a):} A linear regression of the normalized IAF power results in an efficient prediction for gains in reading speed. The green line denotes the regression function. The colors resemble the \textit{Presentation Speed}. \textbf{(b):} A linear regression of the normalized theta power results in an efficient prediction for gains in reading speed. The green line denotes the regression function. The colors resemble the \textit{Presentation Speed}.}
\label{fig:regression_speed_gain_iaf_theta}
\end{figure}

\subsection{Towards a General Model for Predicting Reading Speed}
The IAF and theta power show a strong linear trend regarding the current gain in reading speed. These results suggest a functional relationship between gains in reading speed and the employed EEG frequency bands that generalize to the individual user, hence obviating a dedicated calibration phase. First, we averaged the \textit{Presentation Speeds} for each \textit{Text Alignment}, resulting in 51 data points for all participants. We then performed a leave-one-participant-out validation between gains in reading speed and the measured frequency bandwidths. More specifically, we iteratively derived a linear regression model using all participants except one and used the remaining one for the validation. This resulted in a low averaged error rate for the IAF (RMSE = 0.12, $R^2$ = .94) and theta (RMSE = 0.11, $R^2$ = .97) power, indicating that the model fits interpersonal differences that may not require individual user calibration. Figure~\ref{fig:regression_speed_gain_iaf_theta} shows the regression lines for the IAF and theta power relative to the gains in reading speed for each participant and each \textit{Presentation Speed}.

\section{Discussion}
We conducted a study to investigate the influence of \textit{Text Alignments} and \textit{Presentation Speeds} on text comprehension, subjective workload, and cortical activity. Our results show that time-savings are achieved from the \textit{Presentation Speeds} of 350 WPM onwards which corresponds with increased alpha power and decreased theta power in EEG. Faster \textit{Presentation Speeds} significantly increases subjective workload and impairs text comprehension. Interestingly, \textit{Text Alignment} did not influence any of our measurements. Thus, it is clear that \textit{Presentation Speed} will continue to be a limiting factor. Until this is resolved, spatial factors such as the ideal presentation of text, which includes font readability could be relatively less important and should only be considered after optimizing \textit{Presentation Speeds}. It is worth noting that \textit{Text Alignment} could have minimal impact on wearable displays, which are small and unlikely to subtend to a larger visual angle than the fovea (\textit{i.e.}, $2^\circ$). Thus, we focus on the implications for \textit{Presentation Speeds} on future developments of RSVP readers.

\subsection{Cognitive Workload of RSVP}
Our results show that the physical manipulation of \textit{Presentation Speeds} does not simply bring gains in reading speed. It has a notable impact on our limited capacity for information processing at the cortical level. While 350 and 500 WPM induced significant gains in reading speeds, relative to regular reading, this was accompanied by a significant cost in subjective workload as well as comprehension scores. These performance findings agree with previous findings~\cite{doi:10.1177/154193120104500614}. We show that this is measurable in terms of EEG correlates with established cognitive processes, whereby higher \textit{Presentation Speeds} commensurate with reduced working memory and engagement as respectively measured by IAF and theta power~\cite{romei2008resting}. At 200 WPM, our participants were forced to read slower than their regular speed, which increased text comprehension scores that were commensurate with the EEG correlates of higher working memory load and engagement~\cite{romei2008resting}. Thus, we show that physical manipulation of the speed of RSVP readers exerts an influence on how we process information for comprehension that can be measured with EEG. Thus, we show that ongoing measurements of cortical activity can reliably indicate the rate at which we can process and comprehend presented stimuli. In many cases, it could be desirable to manipulate RSVP speeds to facilitate information processing, in a way that could be verified by EEG measurements, as long as fast reading is not a pressing requirement.

\subsection{Evaluating EEG as Measure for Predictive Models}
We use the mean power in the IAF and theta bands to predict gains in reading speed, subjective workload, and text comprehension scores. We found that only gains in reading speed could be reliably predicted with both IAF and theta power. This implies a linear relationship between \textit{Presentation Speeds} and the measured IAF and theta bandwidths. This forecast can support the dynamic selection of RSVP parameters according to the current context (\textit{i.e.}, reading in private or mobile spaces, reading a novel or a scientific article) to maximize time-savings, supporting the development of adaptive brain-computer interfaces. In contrast, subjective workload and text comprehension were not reliably associated with our EEG measurements. Non-linear models might be necessary to estimate subjective workload and text comprehension, given that NASA-TLX questionnaires are susceptible to differences in subjective perception. 

\subsection{Real-Time Assessment of RSVP Parameter Selections}
EEG affords high sampling rates and can be employed as a real-time indicator for the evaluation of RSVP reader designs~\cite{fairclough2009fundamentals}. Since RSVP can be ubiquitously employed on devices with limited screen space, the reading performance can be evaluated in short time frames using EEG. Our predictive model shows that workload-aware computer interfaces can implicitly sense the current gain in reading speed depending on the bandwidth powers relative to a reading baseline. The \textit{Presentation Speed} can be adjusted to a suitable reading level depending on the current level of cortical activity. Adaptive RSVP \textit{Presentation Speeds} can then be deployed on-the-go, where textual information is available at refresh rates suitable for mental processing. With the ubiquitous availability of RSVP-enabled devices, baselines of regular reading speeds can be collected in any context to evaluate the current RSVP reading speed using EEG. RSVP systems can then dynamically select \textit{Presentation Speeds} which implicitly suits the users' reading ability.

\subsection{Limitations}
The current study has several design limitations. We limited our manipulations to three \textit{Presentation Speeds}. Therefore, it is not clear how cortical activity is affected for slower \textit{Presentation Speeds} than outlined in the study. However, due to our findings, we expect a similar engagement of mental resources for slower \textit{Presentation Speeds}. Furthermore, the comprehension tests may be confounded by knowledgeable participants who were aware of the answers without reading the text. Although we made sure that the participants were not knowledgeable about the used text excerpts, we can not exclude the use of common knowledge during the tests. Finally, our sample consisted of people who exclusively did not use RSVP regularly. Therefore, our results only confirm the observed effects of non-trained RSVP users.

\subsection{Outlook}
Although our findings show the feasibility of relying on EEG measurements to predict gains in reading speed. Any gains due to \textit{Presentation Speed} depends on the baseline reading speed of the individual user. Being able to predict gains means that RSVP readers can adaptively determine the tradeoff of reading speed gains to ensure that \textit{Presentation Speed} is maintained at a level that allows for meaningful text comprehension. Aspects, such as mobile use, reading interruptions, outdoor scenarios, or different viewing postures have to be determined before RSVP-based brain-computer interfaces can be implemented in a way that allows for good signal acquisition. These are open research questions that were not addressed in our experimental design. However, our findings provide a first step towards the adaptive selection of \textit{Presentation Speed} for RSVP readers, which would need to be manually determined otherwise.

\section{Study Conclusion}
We present a user study that investigates the cognitive workload raised by the RSVP design parameters \textit{Text Alignment} and \textit{Presentation Speeds} using Electroencephalography (EEG). We find that a \textit{Presentation Speed} of 350 WPM increases the reading speed compared to regular reading while preserving a similar level of cortical activity and text comprehension. However, faster \textit{Presentation Speeds} increase the subjectively perceived workload with a decrease in the overall text comprehension and cortical activity. No effect was observed for different \textit{Text Alignments}, making \textit{Presentation Speed} the critical design parameter that needs to be optimized first before modifying \textit{Text Alignments}. Due to linear trends in the EEG measures and \textit{Presentation Speeds}, we perform a linear regression analysis to evaluate the robustness of predictive models for gains in reading speed, subjective workload, and text comprehension. While we find that the current individual gains in reading speed can be reliably forecasted using EEG, subjective workload and text comprehension are less suitable variables for reliable predictions. Our results show that future RSVP interface designer benefit from the presented approach to design workload-aware user interfaces that dynamically select RSVP parameters to suit the individuals' cognitive workload using EEG measures. We publish the data set to foster and encourage research in this area\footnote{\url{www.github.com/hcum/one-does-not-simply-rsvp} - last access \urldate}.

\section{Chapter Conclusion}
In this chapter, we investigated if \ac{EEG} provides suitable metrics to assess cognitive workload during information consumption. Thereby, we were able to derive a predictive model that can predict the current gain in reading speed using \ac{EEG} metrics. This enables an approximation of the reading speed towards the workload that is elicited during regular reading.

The human mind possesses a limited capacity for attention and working memory. These two factors can change during a reading session. During tedious passages of a manuscript, the attention level can go down, hence being a marker for annotating the displayed text as \textit{unread} and slowing down the reading speed. In contrast, exciting or easy parts of a manuscript can be sped up to increase the information processing capacities. The individual circadian cycle is a variable that contributes as well to this circumstance. Different reading performances can be expected when reading at different times. Thus, a method for setting adaptive reading speeds benefits the user. To answer \textbf{RQ7}, we conducted a study that utilized \ac{EEG} as a real-time measure to evaluate different \ac{RSVP} parameters. Going a step further towards workload-aware interfaces, we evaluate a predictive model that uses \ac{EEG} bandwidth powers to efficiently forecast the current gain in reading speed.

In conclusion, there is a tradeoff between reading speed and comprehension. In this chapter, we found that faster reading speeds provide time-savings although text comprehension is negatively affected. This is not detrimental per se since the human mind is efficient in filtering and discarding irrelevant information. Workload-awareness can be used to proactively regulate the users' current cognitive state and match the optimal \ac{RSVP} reading mode.

\part{Conclusion\label{part:guidelines_and_conclusion}}







\chapter{Conclusion and Future Work}
\label{ch:conclusion}

\epigraph{\textit{Here on this road of no release \\ Pale reign of misery \\ That even time cannot erase \\ On through my sorrows \\ Life will go on}}{André Olbrich, Hansi Kürsch}

In this thesis, we investigated how workload-aware systems can sense and utilize mental demand to provide cognitive assistance. Cognitive workload serves as an extrinsic modality which represents an additional dimension in the paradigm of context-aware computing. Such workload-aware interfaces are envisioned in the domain of assistive computing where cognitive workload is sensed and ``offloaded'' to a system. Through a series of lab studies and field experiments, we explored opportune moments for diurnal cognitive assistance, evaluated \ac{EEG} and eye tracking metrics as a measure for cognitive workload, and showcased several applications that benefit from real-time workload measurements. We provide a summary of the presented contributions and implications of this thesis as well as the limitations and future work in the following.

\section{Summary}
With the recent information growth and rising availability of ubiquitous computing systems, devices are competing for our attention and cognitive capabilities with regards to indirect as well as direct interaction. However, cognitive resources that map human cognition, such as working memory, are generally limited by the speed at which we can memorize information in the short-term memory, process or modify it, and provide the respective output (\textit{e.g.}, phonological loop)~\cite{baddeley1992working}. If too many information chunks have to be remembered, the cognitive system cannot keep up with additional incoming stimuli and information, and the likeliness of human errors and cognitive exhaustion rises. Hence, the treatment of the available cognitive resources is a crucial component to consider when designing workload-aware interfaces.

In Chapter~\ref{ch:smart_kitchen_requirements} we started to investigate \textit{how} cognitive assistance can be delivered regarding opportune moments that require cognitive assistance and pointed out potential strategies for cognitive assistance during a cooking session. Posing the question ``\textit{What are the design requirements for systems that provide opportune cognitive augmentation?}'' (\textbf{RQ1}) we conducted a qualitative inquiry in a living facility sheltering persons with cognitive impairments. This user group is used to report the need for cognitive assistance at every opportunity to avoid dangerous situations and injuries. These explicit statements allowed us to analyze a series of observations of a cooking process and conduct interviews with caretakers. This resulted in five design implications for future assistive computing systems that provide cognitive support. Thereby, persons with cognitive impairments revealed their need for cognitive assistance explicitly. After becoming aware of how users encounter situations that require cognitive assistance, we investigated the research question ``\textit{What are suitable in-situ feedback modalities for cognitive support?}'' (\textbf{RQ2}). We evaluated three in-situ feedback modalities regarding their acceptance as well as perception assuming that visual, auditory, and tactile sensation represent primary sources of information consumption. Using an assembly task designed to provoke errors, we studied the acceptance of visual, auditory, and tactile in-situ feedback. Visual in-situ feedback provided the highest user preference and performance in terms of task completion time and the overall number of errors. This was followed by auditory and tactile feedback, indicating a clear preference for visual feedback. This is potentially linked to the visual attention capabilities of humans which are responsible for recognizing up to 80$\,\%$ of the presented information. Hence, notifying users and augmenting interfaces visually anticipates the highest benefit for cognitive support.

Chapter~\ref{ch:smart_kitchen_requirements} and Chapter~\ref{ch:support_modalities} provide us with situations that elicit the need for cognitive augmentation and their potential feedback modalities. These can be used to augment or ``disaugment'' the user depending on the current level of perceived cognitive workload level. Our prior approach required users to actively state their need for cognitive assistance. However, it is still unclear \textit{when} cognitive assistance should be provided without stating an explicit need for it. Post hoc measures, such as questionnaire-based methods, provide measures for cognitive demand but lack real-time insights and are susceptible to subjective biases. To address this issue, we investigated objective \ac{EEG} and eye tracking metrics as a real-time indicator for cognitive demand. With the research question, ``\textit{Does electroencephalography provide measures of cognitive workload for user interface evaluation?}'' (\textbf{RQ3}) we investigated \ac{EEG} signals emitted by the human brain as a direct physiological measure for cognitive workload. During an assembly task with two different instruction modalities, where both varied in their difficulty, we found that \ac{EEG} is a suitable real-time measurement modality for working memory assessments. Hence, the cognitive workload generated by an interface can be synchronized with the user interaction to reveal user interface designers which aspects of their interface design induces workload. Finally, a system can use this measure to provide real-time interface adjustments. However, \ac{EEG} headsets are required to be body-worn. Thus, we further investigated eye gaze properties for cognitive workload assessment. Contrary to \ac{EEG}, eye gaze can be recorded contactless without the need for body-worn sensors. In this context, we leverage eye gaze to investigate the research question ``\textit{Do eye gaze metrics enable the classification of cognitive workload states?}'' (\textbf{RQ4}). First, we exploited changes in smooth pursuit eye movements during different levels of cognitive workload. We found that changes in eye movements can be efficiently measured between different task difficulties. Hence, a \ac{SVM}-based classifier revealed a classification accuracy of $99.5\%$ for person-independent classification between two workload states and a $84.5\%$ person-dependent classification accuracy between four workload states using smooth pursuit eye movements. This shows that eye movements are a suitable indicator for different workload levels. We showcase use cases that may exploit natural eye movements to estimate workload states. Since situations exist that do not allow the integration of moving objects to elicit smooth pursuit eye movements, we extended our research to pupil diameter measures. Since changes in pupil dilation inform about cognitive workload states, we leveraged a math multiplication task to induce workload. Using a \ac{SVM}-based classifier, we presented an adaptive system that alternates between simple and difficult task complexities based on the pupil diameter alone. We assess that an adaptive system enhances the overall task engagement while reducing the experienced frustration. The presented approach indicates the viability of using physiological sensing for implicit workload-aware measures.

The last part of the thesis looks at potential applications \textit{which} make use of the prior presented research. We learned that visual in-situ feedback during a cooking session may help persons with cognitive impairments for autonomous meal preparation. We designed and prepared an assistive system that utilizes visual and auditory feedback to support people with cognitive impairments through a cooking process, thus investigating the research question ``\textit{Does in-situ feedback provide cognitive support during a cooking task?}'' (\textbf{RQ5}). A two-week case study in a sheltered living facility was carried out where tenants with cognitive impairments cooked with caretakers and the assistive system. Although being slower using visual in-situ feedback, the tenants appreciated being able to independently prepare their meal through cognitive assistance. This underpins the physiological user interface evaluation pipeline presented in Chapter~\ref{ch:workloadbyeeg}. However, the interpretation of complex physiological data can become overwhelming for inexperienced users, posing the research question ``\textit{How can complex physiological signals be visualized to be utilized by non-expert users?}'' (\textbf{RQ6}). We developed and evaluated a visualization that displays the generation of brain activity inside the brain. Such visualizations are suitable to understand how stimuli affect the generation of cortical activity and, to this end, the impact of neurofeedback loops on neuronal processes. Finally, we investigated how cognitive workload is affected by different \ac{RSVP} reading parameters. Since the \ac{RSVP} parameter space is large in terms of \textit{Presentation Speed} and \textit{Text Alignments}, we pose the question \textit{``How can RSVP reading parameters be selected based on cognitive workload?''} (\textbf{RQ7}) using \ac{EEG}. The results of the study show that larger \textit{Presentation Speed} increases the subjectively perceived workload while decreasing cortical activity. However, different \textit{Text Alignments} did not show significant changes in cognitive workload, making \textit{Presentation Speed} the sole critical factor. A predictive model using a regression analysis shows that the results are applicable when predicting the current reading speed.

\section{Contribution and Results}
During this thesis, we followed an experimental approach to answer seven research questions that investigated how ubiquitous computing technologies can be used to sense cognitive workload to tailor task engagement while avoiding frustration and boredom. In the following we will briefly summarize the three key contributions:

\subsection{Identifying Opportune Moments for Cognitive Assistance}
Our studies in Chapter~\ref{ch:smart_kitchen_requirements} and~\ref{ch:support_modalities} showed \textit{how} and \textit{when} cognitive assistance is required during the accomplishment of daily routine tasks. Phases that require constant attention might be prone to user-related errors, increase the likeliness for injuries, and deteriorate the concentration level. Such situations render opportunities for visual, auditory, or tactile feedback to alert and notify users. After encountering a favorable moment for cognitive assistance, our results revealed that visual feedback was most preferred followed by auditory and tactile feedback (see Chapter~\ref{ch:support_modalities}). We evaluated visual in-situ feedback during a cooking scenario and found that mental workload is decreased on a cognitive and behavioral level (see Chapter~\ref{ch:smart_kitchen_deployment}).

\subsection{Methods for Quantifying Cognitive Workload}
We investigated eye tracking and \ac{EEG} as sensing modalities for cognitive workload detection. We investigated \ac{EEG} signals as an implicit marker for cognitive workload (see Chapter~\ref{ch:workloadbyeeg}). Thereby, \ac{EEG} showed a clear distinction of workload states between two assembly instruction designs. We describe a method that analyzes deviations in smooth pursuit eye movements to reliably classify different levels of cognitive workload on a person-dependent and person-independent basis between $88.1\%$ and $99.5\%$ (see Chapter~\ref{ch:smooth_pursuit}). Further investigations showed that the pupil diameter is a pivotal indicator for cognitive workload. Finally, we employed a math task with two task difficulties to show that classifications between $64\%$ and $99\%$ can be achieved (see Chapter~\ref{ch:pupil_dilation}). However, person-dependent differences and changing lighting conditions are biasing factors that are subject to future research when classifying based on pupil dilation. This resulted in physiological design pipelines to implicitly assess cognitive workload in real-time and rich data sets that pave the way for future research in this area.

\subsection{Tools for Workload-Aware User Interfaces}
After showcasing workload sensing methods, we presented use cases and their reference implementations for workload-aware systems. Specifically, we evaluated the validity of a physiological evaluation pipeline in Chapter~\ref{ch:workloadbyeeg}, self-awareness through visualizations in Chapter~\ref{ch:visualization_reflection}, and potential interfaces that adapt to cognitive workload levels in Chapter~\ref{ch:workload_information_consumption}. Hence, ubiquitous systems are evaluated by physiological measures with complementary physiological feedback. This tailors cognitive engagement to the individual user. With these results in mind, we presented a visualization to foster self-awareness regarding the perceived cognitive workload and present predictive models that can be used by future \ac{RSVP} reading interfaces for automatic parameter selection. Complemented by the data sets presented in the Chapters~\ref{ch:workloadbyeeg},~\ref{ch:smooth_pursuit},~\ref{ch:visualization_reflection}, and~\ref{ch:workload_information_consumption} as well as a reference implementation of an assistive system in Chapter~\ref{ch:smart_kitchen_deployment} we share tools for the development of further workload-aware systems.

\section{Limitations}
Throughout our studies, we have employed empirical approaches to answer the stated research questions. We initially conducted lab studies (\textit{e.g.}, assessing mental load through physiological sensing) before applying our methods within in-the-wild deployments. This was necessary to isolate confounding variables that might have influenced the validity of our results. After confirming the results, we transitioned our prototypes and methods into in-the-wild deployments. However, this transition resulted in the loss of control for variables that might lead to different results depending on the environment and individual user.

During our studies, we involved persons with cognitive impairments. These were included in the design process of workload-aware user interfaces since they explicitly require the need for cognitive assistance and supervision. The perception of task difficulty and support modality depends on the level of cognitive impairment. To investigate this, we included a large variety of cognitive impairments (\textit{i.e.}, ranging from mild cognitive impairments to autism and Trisomy 21). Here, we did not investigate how individual cognitive impairments perceive the design of the presented interfaces.

We employed physiological sensing to sense cognitive workload states. Such measures enable automatic activation of assistive functions. While these can provide cognitive augmentation to support efficient task accomplishment, unsuitable assistance can be detrimental for the user experience. We assess \ac{EEG} and eye tracking as an efficient marker for cognitive workload estimation in real-time. However, the analyzed features and space of physiological modalities presented in this thesis are far from exhaustive. The combination of multiple sensors for additional physiological responses may provide different results based on the used modalities. Another limitation is the susceptibility of noise during physiological sensing. Our studies were conducted in controlled lab environments to minimize noise. While this supports the validity of our results, we are not yet able to confirm similar results within in-the-wild settings. Thus, this is considered as part of future work.

Finally, individual differences in \textit{cognitive constitution} and \textit{physiological responses} vary per individual. While some approaches showed robust classification for cognitive workload (see Chapter~\ref{ch:smooth_pursuit}), they are limited to the classification of two workload states (\textit{i.e.}, binary classification). The fine-grained distinction of workload states often requires calibration to the individual user. While this might be acceptable in private spaces, interaction in public spaces (\textit{e.g.}, interaction in the context of smart cities) benefits from immediate and calibration-free interaction. However, we believe that with the further integration of sensing devices (\textit{e.g.}, smartphones) and ubiquitous activity tracking (\textit{e.g.}, fitness trackers) implicit calibration without the users' awareness will be possible.

\section{Future Work}
This thesis presented design implications, sensors, and applications for systems that utilize workload as input for user interaction. While we focus on the design and the measurement modalities to support workload-related tasks, our methods can be transferred to numerous applications. In the following, we present future research that can be continued by picking up the topics presented in this thesis.

The use of machine learning provides a huge potential for classifying cognitive states in real-time. The presented methods are far from exhaustive and advanced methods, such as deep learning, become common to automate the extraction of features from physiological data. \acp{FBN}~\cite{Park1238411} and \acp{CNN}~\cite{Steil:2019:PPH:3314111.3319913} are becoming popular methods for robust human state classification regardless of individual differences in cognition. However, the scarce availability of efficient methods for classification through deep learning and the need for huge amounts of data are a problem and provide possibilities for future research which includes the combination of sensors to enhance the overall classification performance. With the computing power available nowadays, these research topics can be picked up \textbf{immediately}.

The integration of physiological sensing into devices such as smartphones or fitness trackers has become a trend in the last few years since physiological sensors became smaller in size and are seamlessly integrated into wearables. This rise in devices that collect physiological data is credited to the increased acceptance of users to \textit{share} their physiological data for certain \textit{benefits}. For example, in the context of fitness trackers, activity data is shared for insights into health and activity states. After researching the area of \textit{learning} the physiological behavior of humans, wearable devices require a seamless integration into the user's internal or external environment. Research on the unobtrusive integration of physiological sensing into the users' environments represents a major research question that can be picked up in a \textbf{mid-term} time horizon, \textit{i.e.}, between three to five years.

Finally, ethics and user acceptance pose another important aspect of future workload-aware interfaces. With systems that can ``gaze'' into our cognitive states and mind, responsible handling of the underlying data and their interpretation must be established. Context-aware devices communicate sensitive data with vendors to enhance the user experience. Cognition represents such an additional sensitive variable where its availability to third parties should be carefully considered. This research area requires a reliable physiological workload model of the user with hardware that is already able to reliably sense physiological signals. Therefore, we expect this research to be picked up in the \textbf{long-term} horizon, \textit{i.e.}, ten years from now.

\section{Final Remarks}
With more and more tools becoming available to support our cognition, the question of what happens when removing the tool remains: Do humans depend on their tools for cognitive support? Can they still cope with daily chores when these tools are removed?

Technologies strive to make difficult tasks easier for humans, whether it be on a physical or mental level. Douglas Engelbart~\cite{engelbart1962augmenting, engelbart1995toward} elaborated in the \textit{60s} that technologies will play a pivotal role in augmenting the human intellect. Removing these tools will likely result in the use of support mechanisms that have been used before. This includes the integration of objects in the environment which is known as \textit{active externalism}~\cite{doi:10.1111/1467-8284.00096}, a concept that has accompanied humankind throughout history. Support for working memory was once achieved by, for example, using fingers to count or memorization of road forks for navigation. While most of these tasks were substituted by technologies, a symbiosis between the environment and technology still exists. We believe that technologies make humans more efficient by combining computing efficiency with human flexibility. While these advantages are likely to fade away when removing such tools, humans will compensate for this lack of skill by other means.

A final remark remains on the ethical aspects of workload-aware systems that rely on physiological data collection. What are the caveats to consider assistive technologies that \textit{gaze into the mind}?. Ubiquitous technologies, in general, collect huge amounts of data to provide benefits to the user. Including workload-awareness into a system extends the already large data collection corpus. This data analysis intends to provide a \textit{benefit} at a certain \textit{cost}. For example, the benefit might be an enhanced user experience at the cost of data sharing. In particular, physiological data that is transferred to third parties enables the derivation of patterns regarding, health, intellect, cognition, or stress levels. Transparent data collection and processing pipelines need to be disclosed by the respective stakeholders. The responsibility lies in the collecting institution to follow the guidelines of modern ethics~\cite{doi:10.1177/2053951714559253} that inform and protect the user.

\part{Listings and Bibliography\label{part:listings_and_bibliography}}

\cleardoublepage

\backmatter

\listoffigures	
\cleardoublepage
\listoftables
\cleardoublepage


\renewcommand{\bibname}{Bibliography}

\bibliographystyle{plain}
\bibliography{thesis}




\cleardoublepage
\phantomsection

\selectlanguage{english}
\markboth{Declaration}{Declaration}

\selectlanguage{ngerman}

~\vfill

\begin{Large}Eidesstattliche Versicherung\end{Large}

\begin{small}(Siehe Promotionsordnung vom 12.07.11, § 8, Abs. 2 Pkt. 5)\end{small}

Hiermit erkl\"{a}re ich an Eidesstatt, dass die Dissertation von mir selbstst\"{a}ndig und ohne unerlaubte Beihilfe angefertigt wurde.

\submissiondate

\vspace{2cm}

\hfill \myname
~\vfill

\selectlanguage{english} 

\end{document}